Editors: Mickael Beaufils, Kathi Schleidt, Hylke van der Schaaf, Dan Ponti, Neil Chadwick, Derrick Dasenbrock

# OGC Geotech Interoperability Experiment Engineering Report

**Warning**

This document is not an OGC Standard. This document is an OGC Public Engineering Report created as a deliverable in an OGC Interoperability Initiative and is not an official position of the OGC membership. It is distributed for review and comment. It is subject to change without notice and may not be referred to as an OGC Standard.

Further, any OGC Engineering Report should not be referenced as required or mandatory technology in procurements. However, the discussions in this document could very well lead to the definition of an OGC Standard.

| |
|---|
| Document type: OGC® Engineering Report |
| Document subtype: |
| Document stage: Approved |
| Document language: English |

## License Agreement

# Table of Contents

# Chapter 1. Abstract

This Engineering Report (ER) describes the outcomes of the Open Geospatial Consortium (OGC) Geotech Interoperability Experiment (IE). The objective of this IE was to develop a common conceptual model for describing geotechnical engineering data that bridges existing specifications for encoding those data and which could be integrated across OGC and buildingSMART International Standards,

This ER is directly imported from the project wiki found here: https://github.com/opengeospatial/Geotech/wiki.

# Chapter 2. Introduction

## 2.1. Motivation

BIM, GIS & Digital Twins introduce or emphasize several requirements regarding the data. This mostly concern the built environment but also orientate the way the geoscience data are expected to be delivered.

### 2.1.1. Semantic coherence for Geotechnics and the necessity of models federation

The main concern is about semantics. As also identified as the main criteria of interoperability in the FAIR principles, data shall be described in a non-ambiguous way. Those definitions shall be shared by the community independently from the standard that is used. For that purpose, a federation of model is needed. Then OGC and bSI will derive logical models for OGC based and bSI based (IFC) standards.

> ***This effort would fit into the ongoing collaboration between OGC initiatives and bSI projects on georeferencing, infrastructures alignments, procedural geometries, voxels, etc…***
>
> — Richard Petrie (bSI), Scott Simmons (OGC)

Although, this project is the opportunity to align with other standards. A non-exhaustive list of them include AGS/AGSi, DIGGS, Geo3DML, GeoValML, BoreholeML, ResqML, etc.

On the OGC side, this project is identified as the main contributor to leverage the OGC based standards. For the moment, the IFC schema extension for Geotechnics is mainly supported by the GeoSubgroup of the IFC Tunnel but it aims to serve all the IFC infrastructure projects.

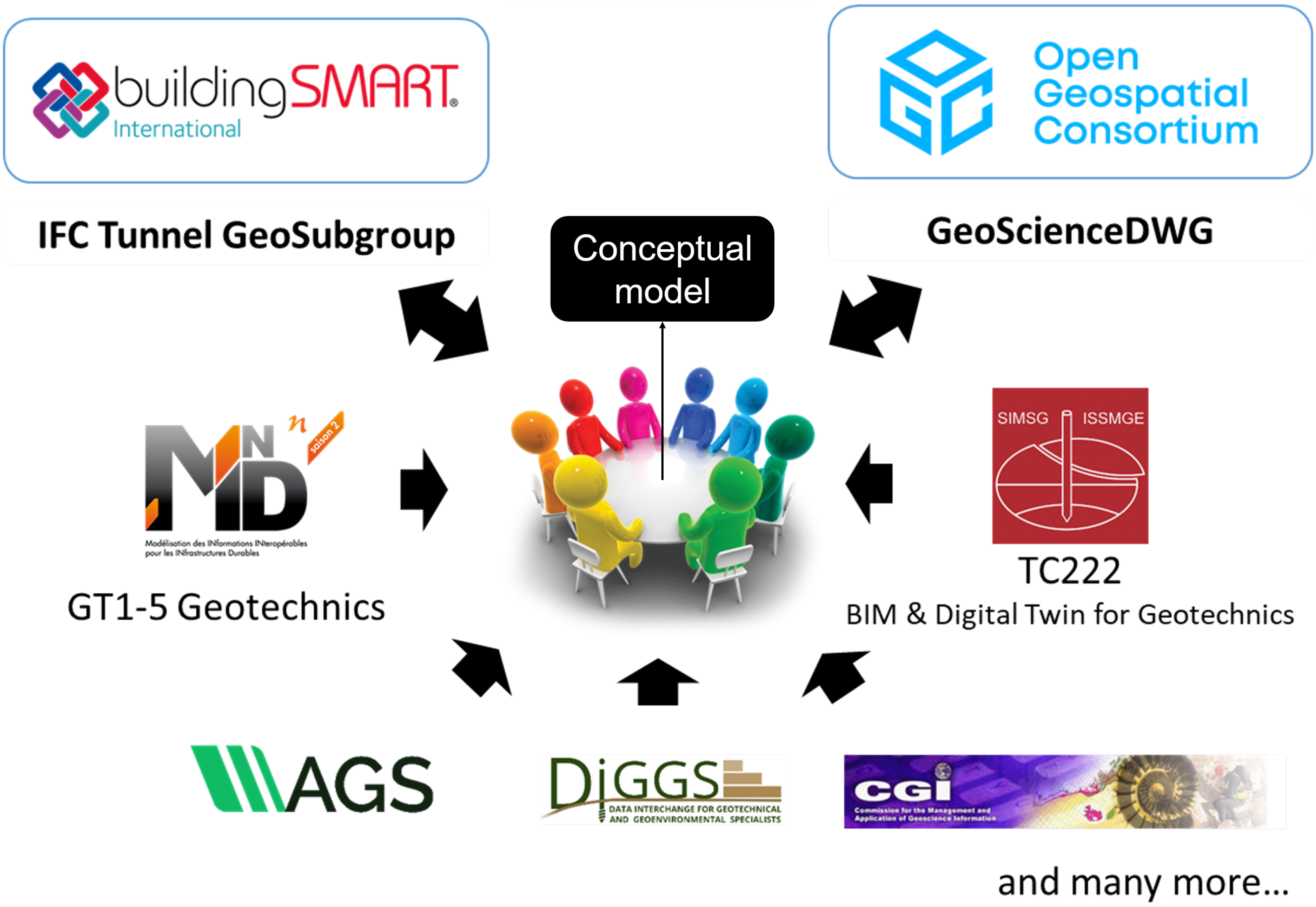

*Figure 1. Geotech Interoperability Experiment*

## 2.2. Participating standard communities

This GeotechIE project is operated thanks to the collaboration of people from multiple communities.

### 2.2.1. Open Geospatial Consortium

The Open Geospatial Consortium (OGC) is an international industry consortium participating in a consensus process to develop publicly available interface standards. The geoscience standards, especially GeoSciML have been designed jointly with CGI-IUGS, giving them the status of reference standards for geology. Efforts from OGC have also been made to extend those standards to address the topic of hydrogeology. The result was the standard GroundwaterML2 that is an extension of GeoSciML. The geoscience standards also rely on the ISO19156 standard that is a cross domain model for Observations, Measurements and Sampling.

Contributors from OGC include:

- Mickaël Beaufils, GeoScienceDWG Chairman, GeotechIE Leader, BRGM
- Scott Simmons, OGC Chief Standards Officer
- Kathi Schleidt, OMS SWG co-chair, DataCove
- Hylke van der Schaaf, SensorThingsAPI SWG co-chair, Fraunhofer IOSB

### 2.2.2. buildingSMART International

buildingSMART International (bSI) is an international organization committed to creating and developing open digital ways of working for the built asset industry. buildingSMART standards help asset owners and the entire supply chain work more efficiently and collaboratively through the entire project and asset lifecycle.The Industry Foundation Classes (IFC), ISO 16739, are the official standards for OpenBIM. Major evolutions have been made on them, especially with IFC v4.x, to extend their capacities: IFC 4.0 introduces the extension of IFC from building to infrastructure. IFC 4.3 is having another step forward in the direction of environmental modeling. It introduces the capacity of describing non-man-made objects and also georeferencing. Two major and necessary improvements to envisage the description of geotechnical objects or even earthworks. The IFC Common Schema is identified to offer a cross-infrastructure support to address those domains. Major evolutions in that direction are made through the IFC Tunnel project.

Contributors from bSI include:

- Jonas Weil, IFC Tunnel Geo-Subgroup Leader, IC Group
- Rie Wada, Oyo Company

### 2.2.3. AGS

AGS is the Association of Geotechnical and Geoenvironmental Specialists, a UK based trade association established to improve the profile and quality of geotechnical and geoenvironmental engineering. In 1991, the AGS set up a method for transferring data between industry organizations. This is known to many simply as ‘AGS format’ or ‘AGS data format’ and provides a standard way to transfer ground investigation, laboratory testing and monitoring data between the contributing parties of a project which involves geotechnical or geoenvironmental elements. The AGS format has been improved and enhanced over the years and is now widely used in the UK. It has also been adopted and adapted for use in Hong Kong, Singapore, Australia, New Zealand and parts of the Middle East. More recently, AGS has developed and released AGSi, a new data format for the exchange of ground model and interpreted data in the ground domain. Work is also being carried out on a new format for piling data.

Contributors from AGS include:

- Neil Chadwick, AGS Data Management Working Group
- Tony Daly, AGS Data Management Working Group

### 2.2.4. DIGGS

DIGGS (Data Interchange for Geotechnical and Geoenvironmental Specialists) is a special project of the Geo-Institute of the American Society of Civil Engineers, and the interchange standard has been adopted as the “Provisional Standard Practice for Digital Interchange of Geotechnical Data” by AASHTO, the association that represents highway and transportation departments across the USA and serves as a liaison between State departments of transportation and the Federal government. DIGGS builds off of the AGS data dictionary for exploratory data, and utilizes existing technologies and standards (eg. XML, GML, WITSML, OMS) in its schema design. Although initially focused on exploratory data collected in US practice, DIGGS supports 1D, 2D, and 3D sampling geometries, any language, and can reference any codes or language conventions, and thus is designed to be extensible and compatible with non-US practice. Beyond exploratory data obtained from boreholes, trenches and outcrops, DIGGS is being extended to exchange field survey geophysical data, construction data from grouting and pile load testing activities, performance monitoring instrumentation, and geo-environmental site monitoring.

Contributors from DIGGS include:

- Dan Ponti, DIGGS Steering Committee
- Derrick Dasenbrock

## 2.3. Supporting community

### 2.3.1. ISSMGE TC222

The International Society for Soil Mechanics and Geotechnical Engineering (ISSMGE) is the pre-eminent professional body representing the interests and activities of Engineers, Academics and Contractors all over the world that actively participate in geotechnical engineering. The aim of the International Society is the promotion of international co-operation amongst engineers and scientists for the advancement and dissemination of knowledge in the field of Geotechnics, and its engineering and environmental applications.

Contributors from ISSMGE TC222 include:

- Magnus Romoen, ISSMGE TC222 Chairman, NGI
- Mats Kalstrom, NGI

### 2.3.2. ITA WG22

The International Tunnelling and Underground Space Association (ITA) aims are to encourage the use of the subsurface for the benefit of public, environment and sustainable development and to promote advances in planning, design, construction, maintenance and safety of tunnels and underground space, by bringing together information thereon and by studying questions related thereto.

WG22 is the Working Group dedicated to information modeling in tunelling.

Contributors from ITA WG22 include:

- Florent Robert, ITA WG22 Animator, CETU

### 2.3.3. CGI-IUGS

The Commission for the Management and Application of Geoscience Information (CGI) is a working subcommittee of the International Union of Geological Sciences. Its mission is to enable the global exchange of knowledge about geoscience information and systems. CGI is the governing body responsible for the XML-based exchange languages Geoscience Markup Language (GeoSciML - in collaboration with the Open Geospatial Consortium) and EarthResource Markup Language (EarthResourceML). The CGI and its members also play a significant role in the OneGeology initiative.

Contributors from ITA WG22 include:

- Harvey Thorleifson, CGI-IUGS Chairman
- Edd Lewis, BGS

### 2.3.4. EGS UGEG

The Urban Geology Expert Group (UGEG) from EuroGeoSurvey (EGS) delivers high quality scientific information and expertise to the EU's urban decision-makers and European Institutions in the areas of sustainable urban development, urban resilience, future climate proofing of cities, SMART cities and safe construction.

Contributors from EGS UGEG include:

- Philip Wehrens, SwissTopo
- Teemu Lindqvist, GTK

### 2.3.5. MINnD

The MINnD National Project in France is since its creation in 2014 is the French initiative to push openBIM extensions for infrastructure description. Majors contributions from MINnD have been brought to the development of IFC Bridge, IFC Road, IFC Rail. IFC Tunnel is largely influenced by the work made by the MINnD Tunnel (WG1-4) and MINnD Geotechnics (WG1-5) projects, that are themselves part of the wider MINnD Underground Infrastructure. MINnD Geotechnics emphasized the importance of building on and connecting existing standards.

Contributors from MINnD include:

- Isabelle Halfon, BRGM
- Sylvie Bretelle, ANTEA
- Elodie Vautherin, Fondasol
- Pierre Garnier, WSP
- Alexis Serieys, SETEC Terrasol
- François Robida, MINnD

# Chapter 3. Geotechnical concepts

## 3.1. What is the conceptual model about?

A conceptual schema or conceptual data model is a map of concepts and their relationships. This describes the semantics of a domain and represents a series of assertions about its nature. Specifically, it describes the things of significance to a domain (entity classes), about which it is inclined to collect information, and its characteristics (attributes) and the associations between pairs of those things of significance (relationships). This model’s perspective is independent of any underlying data format or application.

The Conceptual Model proposes: * Definition and properties for relevant concepts identified by the geotechnical community, and * Mapping with existing standards

## 3.2. Which standards and versions were considered in this conceptual model?

### 3.2.1. ISO 19148

Linear Referencing Systems enable the specification of positions along linear objects. The approach is based upon the Generalized Model for Linear Referencing[12] first standardized within ISO 19133:2005. This document extends that which was included in ISO 19133, both in functionality and explanation.

ISO 19109 supports features representing discrete objects with attributes with values which apply to the entire feature. ISO 19123 allows the attribute value to vary, depending upon the location within a feature, but does not support the assignment of attribute values to a single point or length along a linear feature. Linearly located events provide the mechanism for specifying attribution of linear objects when the attribute value varies along the length of a linear feature. A Linear Referencing System is used to specify where along the linear object each attribute value applies. The same mechanism can be used to specify where along a linear object another object is located, such as guardrail or a traffic accident.

It is common practice to segment a linear object with linearly located events, based upon one or more of its attributes. The resultant linear segments are attributed with just the attributes used in the segmentation process, ensuring that the linear segments are homogeneous in value for these segmenting attributes.

### 3.2.2. OGC Observations, Measurements, and Samples / ISO 19156

This standard specifies an XML implementation for the OGC and ISO Observations and Measurements (O&M) conceptual model (OGC Observations and Measurements v2.0 also published as ISO 19156, as well as OGC Abstract Specification Topic 20: Observations, Measurements, and Samples), including a schema for Sampling Features. This encoding is an essential dependency for the OGC Sensor Observation Service (SOS) Interface Standard. More specifically, this standard defines XML schemas for observations, and for features involved in sampling when making observations. These provide document models for the exchange of information describing observation acts and their results, both within and between different scientific and technical communities.

More details here: https://www.ogc.org/standards/om

### 3.2.3. OGC GeoSciML 4.1

GeoSciML is a model of geological features commonly described and portrayed in geological maps, cross sections, geological reports, and databases. The model was developed by the IUGS CGI (Commission for the Management and Application of Geoscience Information) and version 4.1 is the first version officially submitted as an OGC standard. This specification describes a logical model and GML/XML encoding rules for the exchange of geological map data, geological time scales, boreholes, and metadata for laboratory analyses. It includes a Lite model, used for simple map-based applications; a basic model, aligned on INSPIRE, for basic data exchange; and an extended model to address more complex scenarios.

The specification also provides patterns, profiles (most notably of Observations and Measurements - ISO19156), and best practices to deal with common geoscience use cases.

More details here: http://geosciml.org/

### 3.2.4. OGC GroundWaterML 2.2

This standard describes a conceptual and logical model for the exchange of groundwater data, as well as a GML/XML encoding with examples.

More details here: https://www.ogc.org/standards/gwml2

### 3.2.5. OGC LandInfra

The scope of the Land and Infrastructure Conceptual Model is land and civil engineering infrastructure facilities. Anticipated subject areas include facilities, projects, alignment, road, railway, survey, land features, land division, and "wet" infrastructure (storm drainage, wastewater, and water distribution systems). The initial release of this standard is targeted to support all of these except wet infrastructure.

More details here: https://www.ogc.org/standard/infragml/

### 3.2.6. bSI IFC Tunnel proposal

IFC is a standardized, digital description of the built asset industry. IFC4.3 is an open, international standard (ISO 16739-1:2018) and promotes vendor-neutral, or agnostic, and usable capabilities across a wide range of hardware devices, software platforms, and interfaces for many different use cases. More about IFC, its uses and adoption may be found here.

While geotechnics was first addressed in IFC4.3, major changes are discussed in the context of the IFC Tunnel project. The mapping that is described in those pages is based on this discussed proposal and do not apply to IFC4.3.

### 3.2.7. AGS 4.1

AGS is a text file format used to transfer data reliably, between organizations in the site investigation industry, independent of software, hardware or operating system.

In 1991, the AGS set up a method for transferring data between industry organizations. This is known to many simply as 'AGS Format' or 'AGS Data Format' and provides a standard way to transfer ground investigation, laboratory testing and monitoring data between the contributing parties of a project which involves geotechnical or geoenvironmental elements.

More details here: https://www.ags.org.uk/data-format/ags4-data-format/ags-4-1/

### 3.2.8. DIGGS 2.5.a

DIGGS is an XML-based standard for the transfer of geotechnical data. This current version is limited to ground exploration data derived from boreholes, soundings, trenches and outcrops, including in-situ testing and monitoring, and laboratory test data. It is built from the AGS data dictionary, extending it for US practice, and leverages GML ad WITSML XML standards for describing spatial and geometric and unit of measurement symbology, respectively.

More details here: https://www.geoinstitute.org/special-projects/diggs/schema-tools

## 3.3. Origin of Books A, B, C concept

In order to separate the factual data from the investigation phases, the analysis/interpretation documents and the design documents, the AFTES (French Tunnelling and Underground Space Association), initially instituted in the recommendation GT32R1F1 of 2004 (then in the recommendation GT32R3F1 of 2016 which abrogates GT32R1F1), the structuring of the Tendering Documents, around three successive sub-folders or "books," using the results of the previous book to define the characteristics of the next book.

## 3.4. Definition

### 3.4.1. Book A: Factual data

Factual Data Collection: This is a non-binding compilation of project-specific data and results. It is further divided into three sub-books:

- A1: Geotechnical input data;
- A2:Input data for neighbouring buildings; and
- A3: Input data natural and human environment.

### 3.4.2. Book B: Models and their components

Geotechnical report: this contractual document gives the interpretation of the project manager on the whole of the data of the book A. It is also sub-divided into three sub-books:

- B1: Geotechnical synthesis memorandum;

- B2: Sensitivity memorandum for neighboring constructions; and
- B3: Comments on environmental constraints.

Each book B1, B2 and B3 ends with a register of uncertainties.

### 3.4.3. Book C: Design report

Design brief: this explanatory and non-contractual document produced by the project manager explains and justifies the design of the civil engineering works, the execution methods and the sizing of the works.

## 3.5. Book A Principles

The main idea of the Book A phase is to collect factual data and express their content. In Book A, one will find all the documents produced in the past on the studied area (geological maps, collection of geologists, etc.). Generally, it is necessary to complete these elements with specific investigations.

### 3.5.1. Acts

Geotechnical engineers and team can perform several acts in order to better understand the environment:

- Observation
- Observing Procedure
- Observable Properties
- Sampling and Preparation

### 3.5.2. Proxies

Observation, measurement, and sampling activities are applied on features of interest that act as proxies.

These include:

- Borehole; and
- Trial Pit

Some proxies can be collected on the field, to be studied elsewhere (e.g., in a laboratory), they are called:

- MaterialSample

Note: A borehole core is considered as a Material Sample.

### 3.5.3. FAQ

**What do observations and measurements include?**

All kind of observation and measurements that can be performed.

This include:

- Geological observations,
- Geotech in-situ tests,
- Geotech laboratory tests,
- Environmental analysis, and
- Geophysics.

**Why focusing on acts and not just providing results?**

The proposed conceptual model follows the principles of the OGC Standard for Observations, Measurements, and Samples.

In geotechnics and in many other domains results must be contextualized and can only be used in a specific context.

Acts of observation, measurement or sampling provide this context and the necessary metadata to enable the use and reuse of the data.

### 3.5.4. Borehole

A Borehole is the generalized term for any narrow shaft drilled in the ground, either vertically, horizontally, inclined, or deviated.

**Realizations**

| Data model | Concept name | Definition |
|---|---|---|
| OGC GeoSciML | Borehole | A Borehole is the generalized term for any narrow shaft drilled in the ground, either vertically, horizontally, or inclined. |
| IFC | IFCBorehole | A Borehole is the generalized term for any narrow shaft drilled in the ground, either vertically, horizontally, or inclined. |
| AGS | LOCA | AGS LOCA includes exploratory holes of any type, e.g., borehole, trial pits, CPT |

| Data model | Concept name | Definition |
|---|---|---|
| DIGGS | Borehole | A sampling feature feature that is a deep, narrow excavation made by drilling and/or extraction of earth material. Boreholes are constructed typically for the purpose of investigating subsurface geologic or geotechnical conditions, exploring for water or oil, for installation of wells or other downhole monitoring installations. |
| DIGGS | Sounding | A sampling feature that is created by a measurement or construction activity through insertion of a probe or tool into the ground. |

**Properties**

A proposed list of borehole properties is as follows.

| Name | Definition |
|---|---|
| id | Identifier |
| name | A human-readable display name for the borehole. |
| description | A human-readable description for the borehole. |
| metadata | A URI referring to a metadata record describing the provenance of data. |
| remarks | Any general remarks about the borehole. |
| boreholePurpose | The purpose for which the borehole was drilled. |
| boreholeUse | The current use of the borehole which could differ from the purpose for which the borehole was initially drilled. |
| status | The current status of the borehole.( eg. planned, completed, destroyed) |
| source | Details and citations to source materials for the borehole and, if available, providing URLs to reference material and publications describing the borehole. This could be a short text synopsis of key information that would also be in the metadata record referenced by metadata_uri. |

| Name | Definition |
|---|---|
| associatedFile | Identifies external files associated with this borehole. This allows for the feature to be further elaborated with information that cannot be represented by ASCII text, such as a photograph or other media, binary data, or a formatted report. |
| whenDestroyed | The time period during which the hole was destroyed or forever abandoned. This date must occur after all related observations, measurements or activities had been performed in the borehole, and should reflect the time after which no observations or activities are possible in the hole. |
| drillingMethod | The method(s) used to construct the hole. |
| drillingEquipment | The equipment used to construct the hole. |
| operator | The organization or agency responsible for commissioning of the borehole (as opposed to the agency which drilled the borehole). |
| driller | The organization responsible for drilling the borehole (as opposed to commissioning the borehole). |
| drillStartDate | The datetime of the start of drilling formatted according to ISO8601 (e.g., 2012-03-17T00:00:00). |
| drillEndDate | The date of the end of drilling formatted according to ISO8601 (e.g., 2012-03-28T00:00:00). |
| BoreholeLengthPlanned | The planned length of a borehole e.g., in a site investigation program. |
| BoreholeLengthDrilled | The total length of a borehole as drilled and logged. Length may have different sources (e.g., driller’s measurement, logger’s measurement, survey measurement). |
| startPoint | The position relative to the ground surface that marks the origin for measuring distance along a borehole’s trajectory. |
| collarGeometry | 2D or 3D point geometry that represents the location of the borehole at the ground surface in geographic space. |

| Name | Definition |
| --- | --- |
| srsName | The URI of a spatial reference system. If collarGeometry is a 2D point, this is the horizontal SRS. If collarGeometry is a 3D point, this must be a either a 3D SRS, a known compound SRS, or the EPSG code of the horizontal component of a compound SRS. |
| elevation_srsName | The URI of a spatial reference system of the elevation of the collarGeometry. (e.g., mean sea level). Mandatory if collarGeometry and srsName is the EPSG code of the horizontal component of a compound SRS. elevation_srsName shall be a one dimensional vertical SRS (i.e., EPSG code in the range 5600-5799). |
| positionalAccuracy | An estimate of the accuracy of the location of the collarGeometry. Ideally, this would be a quantitative estimate of accuracy (e.g., 20 metres). |
| locality | Non-coordinate location information for a borehole. |
| localCoordinates | A geometry object that holds the values of local coordinates for the borehole collar location. This object is used to carry information about the location of a feature in its original local reference system if not originally recorded in a well-known SRS. |
| azimuth | For a straight but inclined borehole, the azimuth of the borehole’s trajectory as a plane angle measurement, where 0 is true north, incrementing clockwise. |
| inclination | For a straight but inclined borehole, the inclination of the borehole’s trajectory as a plane angle measurement, where 0 is vertical and 90 is horizontal. |
| trajectoryType | Describes the general character of the borehole’s trajectory. (vertical, inclined, deviated) |

| Name | Definition |
|---|---|
| trajectoryGeometry | A geometry that represents and locates the borehole shape. It is represented as a lineString in 3D space and is of the same SRS as collarGeometry. A borehole can have multiple centerlines to represent sidetracks or to serve as a linear referencing element for observations within the borehole that may be located relative to different datums (eg. ground surface vs rig table). |
| linearReferencing | Defines the linear spatial reference system for the borehole. This LSRS can then be used to define the location of observations within the borehole as a 1D position along the borehole's trajectory, rather than in geographic coordinates. |
| boreholeType | Type of borehole. (core drilling, destructive drilling, trial pit, sounding) |

#### FAQ

##### What about Borehole core?

A borehole core is considered as a MaterialSample.

See MaterialSample for its description.

#### What about Borehole Logs and Observations?

All observations, measurements and test descriptions from a Borehole are considered as Observations.

See Observation for their description.

##### GitHub issue

https://github.com/opengeospatial/Geotech/issues/10

### 3.5.5. Trial Pit

A relatively shallow excavation into the earth's surface, dug either manually or by a mechanical excavator. Samples,observations and tests in the trial pit are referenced in a linear referencing system only (1D). This is a legacy sampling feature to support AGS trial pit constructs.

#### Realizations

| Data model | Concept name | Definition |
|---|---|---|
| OGC GeoSciML | TrialPit (candidate) | Same as DIGGS. |

| Data model | Concept name | Definition |
| --- | --- | --- |
| IFC | Borehole | |
| AGS | LOCA | AGS LOCA includes exploratory holes of any type, e.g., borehole, trial pit, CPT |
| DIGGS | TrialPit | A relatively shallow excavation into the earth's surface, dug either manually or by a mechanical excavator. Samples, observations and tests in the trial pit are referenced in a linear referencing system only (1D). This is a legacy sampling feature to support AGS trial pit constructs. The TrenchWall sampling feature should be used to represent more detail on walls of pits or trenches in 2D. |

**Properties**

A proposed list of trial pit properties follows.

| Name | Definition |
| --- | --- |
| id | Identifier |
| name | A human-readable display name for the trial pit. |
| description | A human-readable description for the trial pit. |
| metadata | A URI referring to a metadata record describing the provenance of data. |
| remarks | Any general remarks about the trial pit |
| pitPurpose | The purpose for which the trial pit was created. |
| pitUse | The current use of the trial pit which could differ from the purpose for which the trial pit was initially created. |
| status | The current status of the trial pit (e.g., planned, completed, destroyed. |
| source | Details and citations to source materials for the trial pit and, if available, providing URLs to reference material and publications describing the trial pit. This could be a short text synopsis of key information that would also be in the metadata record referenced by metadata_uri. |

| Name | Definition |
| --- | --- |
| associatedFile | Identifies external files associated with this trial pit. This allows for the feature to be further elaborated with information that cannot be represented by ASCII text, such as a photograph or other media, binary data, or a formatted report. |
| whenDestroyed | The time period during which the trial pit was destroyed or forever abandoned. This date must occur after all related observations, measurements or activities had been performed in the trial pit, and should reflect the time after which no observations or activities are possible.. |
| constructionMethod | The method(s) used for this trial pit (manual, mechanical) |
| constructionEquipment | The equipment used to construct the trial pit |
| operator | The organization or agency responsible for commissioning of the trial pit (as opposed to the agency which drilled the trial pit). |
| excavator | The organization responsible for digging the trial pit (as opposed to commissioning the trial pit). |
| pitStartDate | The datetime of the start of excavation formatted according to ISO8601 (e.g., 2012-03-17T00:00:00). |
| pitEndDate | The date of the end of excavation formatted according to ISO8601 (e.g., 2012-03-28T00:00:00). |
| pitDepthPlanned | The planned depth of a trial pit e.g., in a site investigation program. |
| pitDepthConstructed | The total depth of a trial pit as constructed and logged. |
| pitLength | The length of the long horizontal dimension of the pit. |
| pitWidth | The length of the short horizontal dimension of the pit. |
| startPoint | The position relative to the ground surface that marks the origin for measuring distance along a trial pit's trajectory |
| pitShoring | Description of shoring equipment and method. |
| pitBackfill | Information on construction of the trial pit backfill. |

| Name | Definition |
| --- | --- |
| locationGeometry | 2D or 3D point geometry that represents the location of the trial pit at the ground surface in geographic space. |
| srsName | The URI of a spatial reference system. If collarGeometry is a 2D point, this is the horizontal SRS. If collarGeometry is a 3D point, this must be a either a 3D SRS, a known compound SRS, or the EPSG code of the horizontal component of a compound SRS. |
| elevation_srsName | The URI of a spatial reference system of the elevation of the collarGeometry. (e.g., mean sea level). Mandatory if collarGeometry and srsName is the EPSG code of the horizontal component of a compound SRS. elevation_srsName shall be a one dimensional vertical SRS (i.e., EPSG code in the range 5600-5799). |
| positionalAccuracy | An estimate of the accuracy of the location of the collarGeometry. Ideally, this would be a quantitative estimate of accuracy (e.g., 20 metres). |
| locality | Non-coordinate location information for a trial pit. |
| localCoordinates | A geometry object that holds the values of local coordinates for the trial pit location. This object is used to carry information about the location of a feature in its original local reference system if not originally recorded in a well-known SRS |
| trajectoryGeometry | A geometry lineString in 3D space that represents the top and base of the trial pit. This object is of the same SRS as locationGeometry. |
| linearReferencing | Defines the linear spatial reference system for the trial pit. This LSRS can then be used to define the locations of observations within the trial pit as a 1D position along the trial pit's trajectory, rather than in geographic coordinates. |

### FAQ

#### How are observed properties that vary laterally in a trial pit handled in 1D?

Observations of the same property where results vary laterally in the pit at the same depth are assigned different stratum codes. A stratum code is a letter or number code that links the stratum shown on a face sketch of the trial pit to the observation. In OMS, this would be accomplished by having different features of interest identified for the same 1D location.

See Github issue: https://github.com/opengeospatial/Geotech/issues/18

### 3.5.6. Observation

An observation is an act associated with a discrete time instant or period through which a number, term or other symbol is assigned to a characteristic. This act involves application of a specified procedure, such as a sensor, instrument, algorithm or process chain. The procedure may be applied in-situ, remotely, or ex-situ with respect to the sampling location. The result of an observation is an estimate of the value of a property of some feature; an observation is a property-value-provider for a feature-of-interest. Use of a common model allows observation data using different procedures to be combined unambiguously.

In conventional measurement theory the term “measurement” is used. However, a distinction between measurement and category-observation has been adopted in more recent work so the term “observation” is used here for the general concept. “Measurement” may be reserved for cases where the result is a numerical quantity.

The observation itself is also a feature, since it has properties and identity. Observation details are important for data discovery and for data quality estimation. The observation could be considered to carry “property-level” instance metadata, which complements the dataset-level and feature-level metadata commonly provided via catalogue services (e.g., ISO 19115 or other community agreed one).

**Realizations**

| Data model | Concept name | Definition |
| --- | --- | --- |
| OGC OMS | Observation | An observation is an act associated with a discrete time instant or period through which a number, term or other symbol is assigned to a characteristic. This act involves application of a specified procedure, such as a sensor, instrument, algorithm or process chain. The procedure may be applied in-situ, remotely, or ex-situ with respect to the sampling location. The result of an observation is an estimate of the value of a property of some feature; an observation is a property-value-provider for a feature-of-interest. Use of a common model allows observation data using different procedures to be combined unambiguously. |
| IFC | GeoscienceObservation | Detailed collected information, including measured parameters, descriptions etc. related to geoscientific observations that can be assigned to physical or spatial elements using IfcRelAssignsToProduct. |
| AGS | | No specific concept as such as all data is treated similarly in AGS data, whether it be location data, hole/test/other metadata or results<br>(i.e. observations in this context) |
| DIGGS | Observation | An interval or region defined at a sampling feature that contains qualitative observations or interpretations. |

| Data model | Concept name | Definition |
| --- | --- | --- |
| DIGGS | Measurement | An act or event whose results are quantitative estimates of the values of properties of the target of an investigation. DIGGS currently has three specialized measurement objects, 1) a Test, which is a measurement made over a spatial domain, such as laboratory tests on samples collected in the field, or in-situ tests where measurements are made directly on site, 2) monitoring activities, which are measurements made over a temporal domain, such as water level measurements or inclinometer readings, and 3) a Material Test, which is for measurements made on material samples that are manufactured such that the result pertains only to the sample and not to any associated location. |

### FAQ

#### Observations and / or measurements?

Several distinctions can be made between Observations and Measurements. A common one being that Observations provide qualitative results whereas Measurements provide quantitative results.

The proposed conceptual model does not make this distinction. Yet soft typing can be applied if this distinction is considered necessary.

The DIGGS standard proposes an explicit distinction between Observations and Measurements.

#### How to deal with multiple Observations / measurements?

It is very common for a procedures or test (especially for measurements) to provide a result composed of multiple values. This includes time series, z-series.

The proposed conceptual model offers a fine granularity which aims at being able to get any value separately. Each single observation or measurement (e.g., at a specific time of a timeseries and/or at a specific location of a z-series) is then contextualized and associated to its procedure and other metadata.

#### I don't want to retrieve single observations / measurements but have them grouped

Grouping of observations and measurements having some similarities is a basic function that Observation APIs shall be able to manage.

Common grouping criteria include: ObservingProcedure, ObservedProperty, Time, Location and/or FeatureOfInterest.

OGC APIs like SensorThingAPI offer such capacity.

#### GitHub issue

https://github.com/opengeospatial/Geotech/issues/19

## 3.5.7. Observing Procedure

Description of steps performed in order to determine the value of an observableProperty by an Observer.

### A typology of ObservingProcedure

- In-situ and remote observing procedure: the observation is made on the field, either directly or remotely.
- Ex-situ observing procedure: a sample is collected on the field and studied in a different place, generally a laboratory.

### Realizations

| Source | Type | Link | Definition |
|---|---|---|---|
| BRGM | Registry | https://data.geoscience.fr/ncl/Proc | This register lists the different methods and processes that can be used to acquire knowledge. |
| DIGGS | XML Schema | https://www.diggsml.org/schemas/2.5.a/Geotechnical.xsd | This schema document contains the currently adopted test procedure methods (objects), their properties and property definitions. In instance documents, these objects are contained within a parent Test or MaterialTest measurement object |

A non-exhaustive list of ObservingProcedure for geotechnics includes the following.

- Particle size distribution'@en|'Analyse granulométrique'@fr
- Cone penetration test'@en|'Essai de pénétration statique au cône'@fr
- Mechanical cone penetration test'@en|'Essai de pénétration statique pointe mécanique'@fr
- Electrical cone penetration test'@en|'Essai de pénétration au cône électrique'@fr
- piezocone penetration test'@en|'Essai de pénétration au piézocone'@fr
- Dynamic penetration test'@en|'Essai de pénétration dynamique'@fr
- Standard Penetration test'@en|'Essai de pénétration dynamique standard (SPT)'@fr
- Dynamic probing'@en|'Essai de pénétromètre dynamique'@fr
- Ménard pressuremeter test'@en|'Essai au pressiomètre Ménard'@fr
- Self-boring pressuremeter test'@en|'Essai pressiométrique autoforé'@fr
- Full displacement pressuremeter test'@en|'Essai au pressiomètre refoulant'@fr
- Dilatometer test'@en|'Essai au dilatomètre'@fr
- Flexible dilatometer test'@en|'Essai au dilatomètre flexible'@fr
- Borehole jack test'@en|'Essai au dilatomètre rigide diamétral'@fr
- Field vane test'@en|'Scissomètre de chantier'@fr
- Flat jack test'@en|'Mesure des contraintes au vérin plat'@fr
- Hydraulic Testing of pre-existing fractures (HTPF)'@en|'Mesure de contrainte par fracturation hydraulique (Essai HTPF)'@fr
- Water permeability test in a borehole using open systems'@en|'Essai de perméabilité à l'eau dans un forage ouvert (essai Lefranc)'@fr
- Water pressure test in rock (Lugeon packer test)'@en|'Essai de pression d'eau dans les roches (essai Lugeon)'@fr
- Pumping test'@en|'Essai de pompage'@fr
- Infilirometer test'@en|'Essai d'infiltration'@fr
- Water permeability test in a borehole using closed systems'@en|'Essai de perméabilité à l'eau dans un forage en tube fermé'@fr
- Water content measure'@en|'Mesure de la teneur en eau pondérale'@fr
- Bulk density'@en|'Determination de la masse volumique du sol'@fr
- Particle density'@en|'Détermination de la masse volumique de particules solides'@fr
- Incremental loading oedometer test'@en|'Essai de compressibilité à l'oedométre par paliers'@fr
- Unconfined compression test'@en|'Essai de compression uniaxiale'@fr
- Unconsolidated undrained triaxial test'@en|'Essai triaxial non consolidé non drainé'@fr
- Consolidated undrained triaxial compression test on water saturated soils'@en|'Essai triaxial consolidé non drainé'@fr
- Consolidated drained triaxial compression test on water saturated soils'@en|'Essai triaxial consolidé drainé'@fr

- Direct shear test'@en|'Essai de cisaillement direct'@fr
- Atterberg limits : liquid ans plastic limits'@en|'Détermination des Limites d'Atterberg'@fr
- Shrinkage test'@en|'Essai de dessiccation'@fr
- Measuring of the methylene blue adsorption capacity of a rocky soil — Determination of the methylene blue of a soil by means of the stain test'@en|'Mesure de la capacité d'adsorption de bleu de méthylène d'un sol ou d'un matériau rocheux Détermination de la valeur de bleu de méthylène d'un sol ou d'un matériau rocheux par l'essai à la tache'@fr
- Determination of the organic matter content - Ignition method'@en|'Détermination de la teneur pondérale en matières organiques d'un matériau méthode par calcination'@fr
- Determination of the organic matter content - Soil chemical test'@en|'Détermination de la teneur pondérale en matières organiques d'un sol - méthode chimique'@fr
- Determination of the decomposition state (humification) of organic soils - Von Post test'@en|'Détermination de l'état de décomposition (humification) des sols organiques - Essai Von Post'@fr
- Determination of the carbonate content - Calcimeter method'@en|'Détermination de la teneur en carbonate - Méthode du calcimètre'@fr
- Determination of minimal and maximal density of cohesionless soils'@en|'Détermination des masses volumiques minimale et maximale des sols non cohérents'@fr
- Laboratory vane test'@en|'Essai scissométrique en laboratoire'@fr
- Oedometer swelling test - Swelling determination by testing on several specimens'@en|'Essai de gonflement à l'oedomètre - Détermination des déformations par chargement de plusieurs éprouvette.'@fr
- Huder-Amberg swelling test'@en|'Essai de gonflement Huder-Amberg'@fr
- Soil quality - Determination of water-soluble and acid-soluble sulfate'@en|'Qualité du sol - Dosage du sulfate soluble dans l'eau et dans l'acide'@fr
- Determination of the compaction reference values of a soil type — Standard Proctor Test — Modified Proctor Test'@en|'Détermination des références de compactage d'un matériau - Essai Proctor Normal - Essai Proctor modifié'@fr
- Indice CBR après immersion. Indice CBR immédiat. Indice Portant Immédiat - Mesure sur échantillon compacté dans le moule CBR.'@fr
- Rock - Determination of water content of rock - Oven-drying method'@en|'Roches - Détermination de la teneur en eau pondérale - Méthode par étuvage'@fr
- Rock - Tests for physical properties of rock - Determination of density - Cutting curb - Water immersion methods'@en|'Roches - Essais pour déterminer les propriétés physiques des roches - Détermination de la masse volumique - Méthodes géométriques et par immersion dans l'eau'@fr
- Rock - Tests for physical properties of rock - Determination of porosity'@en|'Roches - Essais pour déterminer les propriétés physiques des roches - Détermination de la porosité'@fr
- Rock - Determination of the ultrasonic waves velocity in laboratory - Transmission method'@en|'``Vitesse de propagation des ondes ultrasonores

- méthode par transparence'@fr
- Rock - Determination of the uniaxial compression strength'@en|'`Roches
    - Détermination de la résistance à la compression uniaxiale'@fr
- Rock - Determination of the tensile strength - Indirect method - Brazil test'@en|'`Roches - Détermination de la résistance à la traction
    - Méthode indirecte - Essai brésilien'@fr
- Rock - Determination of the triaxial compressive strength'@en|'`Roches
    - Détermination de la résistance à la compression triaxiale'@fr
- Rock - Direct shear testing along a rock joint - Normal surface joint, constant load testing'@en|'Roches - Cisaillement direct selon une discontinuité de roche - Essai sous un effort constant, normal à la surface de discontinuité'@fr
- Rock - Determination of the Young modulus and the Poisson ratio'@en|'Roches - Détermination du module de Young et du coefficient de Poisson'@fr
- Rock - Point load strength test - Franklin test'@en|'Roches - Résistance sous charge ponctuelle - Essai Franklin'@fr
- Rock - Determination of the drill penetration index'@en|'Roches - Détermination de l'indice de résistance à la pénétration par un foret'@fr
- Rock - Determination of the rock abrasiveness - Part 1 : Scratching-test with a pointed tool'@en|'Roches - Détermination du pouvoir abrasif d'une roche - Partie 1 : essai de rayure avec une pointe'@fr
- Rock - Determination of the rock abrasiveness - Part 2 : Test with a rotating tool'@en|'Roches - Détermination du pouvoir abrasif d'une roche - Partie 2 : essai avec un outil en rotation'@fr
- Tests for mechanical and physical properties of aggregates — Part 2: Methods for the determination of resistance to fragmentation'@en|'Essais pour déterminer les caractéristiques mécaniques et physiques des granulats — Partie 2 : Méthodes pour la détermination de la résistance à la fragmentation'@fr
- Tests for mechanical and physical properties of aggregates — Part 1: Determination of the resistance to wear (micro-Deval)'@en|'Essais pour déterminer les caractéristiques mécaniques et physiques des granulats - Partie 1 : détermination de la résistance à l'usure (micro-Deval)'@fr
- Expert estimates or Engineering judgment'@en|'Dire d'expert ou jugement de l'ingénieur'@fr
- Correlation'@en|'Corrélation'@fr
- Back analysis'@en|'Calcul de calage, méthode inverse'@fr

**FAQ**

**GitHub issue**

https://github.com/opengeospatial/Geotech/issues/34

## 3.5.8. Observable Properties

A quality (property, characteristic) of the feature-of-interest that can be observed.

**Realizations**

| Source | Type | Link | Definition |
| --- | --- | --- | --- |
| BRGM | Registry | https://data.geoscience.fr/ncl/ObsProp | This register contains the observable properties. |
| DIGGS | XML file | https://www.diggsml.org/dictionaries/DIGGSTestPropertyDefinitions.xml | This is an XML dictionary file (extension of GML Dictionary) containing the controlled property code, name, definition and associated test procedures for currently adopted observed properties for DOGGS Test and Monitor measurement objects |

**Result types**

A non-exhaustive list of Observable Properties for geotechnics includes the following.

| Name | Definition EN | Definition FR |
| --- | --- | --- |
| natural water content | Ratio of the mass of free water to the mass of dry soil | Rapport de la masse d'eau sur la masse sèche |
| dry density | Ratio of the mass of solid to the unit total volume | Rapport de la masse solide sur le volume total |
| dry unit weight | Product of the dry density times the gravity acceleration g | Produit de la masse volumique sèche par l'accélération de la pesanteur g |
| bulk density | Ratio of the total mass of the soil to the unit total volume | Rapport de la masse totale sur le volume total |
| bulk unit weight | Product of the bulk density times the gravity acceleration g | produit de la masse volumique humide par l'accélération de la pesanteur g |
| particle density or density of solid particles | Ratio of the mass of solid to the volume of solid | Rapport de la masse solide sur le volume de grains solides |
| specific gravity | Product of the particle density times the gravity acceleration g | Produit de la masse volumique des grains par l'accélération de la pesanteur g |
| total porosity | Ratio of the volume of void to the total volume | Rapport entre le volume de vide et le volume total |

| Name | Definition EN | Definition FR |
|---|---|---|
| void index | Ratio of the volume of voids to the volme of solid | Rapport entre le volume de vide et le volume des grains solides |
| saturation ratio or degree of saturation | Ratio of the volume of water to the volume of voids | Rapport entre le volume de l'eau et le volume des vides |
| lithology classification | The value that describes the lithology as a controlled term (classification code or controlled name), rather than a simply descriptive one. | Valeur qui décrit la lithologie comme un terme contrôlé (code de classification ou nom contrôlé), plutôt que simplement descriptif. |
| lithology description | Descriptive information about the soil or rock lithology and should be used in conjunction with lithology classification. A lithology description typically includes a description of color, grain size disctribution, constituents and field properties (eg. well-graded sand (SW), fine-grained, dark grey, massive, dry, hard). | Informations descriptives sur la lithologie du sol ou des roches et doivent être utilisées conjointement avec la classification lithologique. Une description lithologique comprend généralement une description de la couleur, de la distribution granulométrique, des constituants et des propriétés du terrain (par exemple, sable bien classé (SW), à grains fins, gris foncé, massif, sec, dur). |
| lithology symbol | A string or numeric value that is used to define a graphic pattern that may be used to symbolize the unit on a borehole log or map. | Une chaîne ou une valeur numérique utilisée pour définir un motif graphique pouvant être utilisé pour symboliser l'unité sur un journal ou une carte de forage. |
| unit name | The name of the geological or geotechnical unit | |
| ***Atterberg limits*** | | |
| liquid limit | Water content at which the soil changes from the liquid state to the plastic state | Teneur en eau pour laquelle le sol passe de l'état liquide à l'état plastique |
| plastic limit | Water content at which the soil changes from the plastic state to a semi-solid state | Teneur en eau pour laquelle le sol passe de l'état plastique à l'état semi-solide |
| plasticity index | Difference between the liquid limit and the plastic limit | Différence entre la limite de liquidité et la limite de plasticité |

| Name | Definition EN | Definition FR |
| --- | --- | --- |
| shrinkage limit | Water content at which the soil changes from a solid state without shrinkage and a solid state with shrinkage | Teneur en eau pour laquelle le sol passe d'un état solide sans retrait et un état solide avec retrait |
| ***Particle size parameters*** | | |
| methylene blue test value | weight of methylene blue adsorbed (fixed) to 100g of the fraction 0/50mm of the soil particles | Masse de bleu de méthylène adsorbée pour 100g de la fraction 0/50mm du sol étudié |
| maximal particle diameter | maximum diameter of particles | diamètre maximum des grains |
| passing | Mass percentage of grain particles passing through a given sieve size opening, to the total dry mass of the tested sample. | Pourcentage massique de particules d'un sol traversant un tamis d'ouverture donnée, rapporté à la masse sèche totale de l'échantillon. |
| retained | Mass percentage of grain particles retained on a given sieve size opening, to the total dry mass of the tested sample. | Pourcentage massique de particules de grains retenues sur une ouverture de tamis de taille donnée, par rapport à la masse sèche totale de l'échantillon testé. |
| particle diameter d60 | Particle size such that 60% of the particles by weight are smaller than size | Diamètre de grains tel que 60% des grains en poids sont plus petits que ce diamètre |
| median particle diameter d50 | Particle size such that 50% of the particles by weight are smaller than size | Diamètre de grains tel que 50% des grains en poids sont plus petits que ce diamètre |
| particle diameter d30 | Particle size such that 30% of the particles by weight are smaller than size | Diamètre de grains tel que 30% des grains en poids sont plus petits que ce diamètre |
| particle diameter d10 | Particle size such that 10% of the particles by weight are smaller than size | Diamètre de grains tel que 10% des grains en poids sont plus petits que ce diamètre |
| boulder content | Mass percentage of the sample with particles larger than 300 mm or 256 mm depending on classification system used. | Pourcentage massique de l'échantillon avec des particules supérieures à 300 mm ou 256 mm selon le système de classification utilisé. |

| Name | Definition EN | Definition FR |
|---|---|---|
| cobble content | Mass percentage of the sample with particle sizes between 75 and 300 mm or between 64 and 256 mm depending on classification system used. | Pourcentage massique de l'échantillon avec des granulométries comprises entre 75 et 300 mm ou entre 64 et 256 mm selon le système de classification utilisé. |
| pebble content | Mass percentage of the sample with particle sizes between 4 and 64 mm. | Pourcentage massique de l'échantillon avec des tailles de particules comprises entre 4 et 64 mm. |
| gravel content | Mass percentage of the sample with particle sizes between 4.75 and 75 mm, or greater than 2 mm, depending on the classification system used. | Pourcentage massique de l'échantillon dont la granulométrie est comprise entre 4,75 et 75 mm, ou supérieure à 2 mm, selon le système de classification utilisé. |
| granule content | Mass percentage of the sample with particle sizes between 2 and 4 mm | Pourcentage massique de l'échantillon dont la granulométrie est comprise entre 2 et 4 mm. |
| sand content | Mass percentage of the sample with particle sizes between 0.075 and 4.75 mm, or between than 0.625 and 2 mm, depending on the classification system used. | Pourcentage massique de l'échantillon avec des tailles de particules comprises entre 0,075 et 4,75 mm, ou entre 0,625 et 2 mm, selon le système de classification utilisé. |
| un | Mass percentage of the sample with particle sizes between 2 and 4.75 mm. | Pourcentage massique de l'échantillon avec des tailles de particules comprises entre 2 et 4,75 mm. |
| medium sand content | Mass percentage of the sample with particle sizes between 0.425 mm and 2 mm. | Pourcentage massique de l'échantillon avec des tailles de particules comprises entre 0,425 mm et 2 mm. |
| fine sand content | Mass percentage of the sample with particle sizes between 0.075 and 0.425 mm | Pourcentage massique de l'échantillon avec des tailles de particules comprises entre 0,075 et 0,425 mm |
| silt content | | |
| clay content | Mass percentage of the sample smaller than 0.002 mm | Pourcentage massique de l'échantillon inférieur à 0,002 mm |

| Name | Definition EN | Definition FR |
|---|---|---|
| fines content | Mass percentage of the sample smaller than 0.075 mm or 0,0625 mm depending on the classification system used | Poucentage massique de particule passant par le tamis de 0,063 mm |
| uniformity coefficient (or hazen coefficient) | Ratio of d60/d10. | Rapport des diamètres d60/d10 |
| ***Geochemical parameters*** | | |
| carbonate content | Percentage of equivalent calcium carbonate obtained from the amount of CO2, as percentage of dry weight. | Pourcentage massique de la fraction carbonatée contenue dans un sol rapportée au poids du sol sec |
| sulfate content | Amount of sulfate (expressed as SO42-) in milligrams per kilogram of dry soil. | Concentration en sulfate (exprimée sous la forme SO42-), en milligrames par kilogramme de sol séché à l’air. |
| organic matter content | Ratio of the mass of organic matter in a soil sample by the total solid particle mass. | Quotient de la masse de matière organique contenues dans un échantillon de sol par la masse des particules solides |
| gtr class | Soil class as per French GTR guidelines or NF P11-300 standard | classe de sol au sens du GTR (Guide des terrassements routiers) et de la norme NF P 11-300 |
| ***Mechanical parameters of soils*** | | |
| uniaxial compressive strength | The maximum axial compressive stress that a right-cylindrical sample of material can withstand under unconfined conditions—the confining stress is zero. Also known as unconfirmed compressive strength. | Contrainte de compression axiale maximale à laquelle un échantillon de matériau cylindrique droit peut résister dans des conditions non confinées : la contrainte de confinement est nulle. Également connue sous le nom de résistance à la compression non confirmée. |

| Name | Definition EN | Definition FR |
|---|---|---|
| shear strength | Maximum value of the shear stress that a soil can support before it fails along a shear plane, when it is submitted to a deviator of stresses. The shear strength is usually expressed by the Coulomb criterion $\tau = c + \sigma.tg\phi$, with c, the cohesion, and $\phi$, the friction angle. | Valeur de la contrainte de cisaillement maximale que le sol peut supporter, avant la rupture par glissement le long d'un plan, lorsqu'il est soumis à un déviateur des contraintes. Selon le critère de Coulomb, la résistance au cisaillement d'un sol est $\tau = c + \sigma.tg\phi$, avec c, la cohésion, et $\phi$, l'angle de frottement. |
| undrained shear strength | Value of the shear stress at failure of a cohesive soil, under undrained loading conditions. | Valeur de la contrainte ou taux de cisaillement à la rupture dans un sol cohérent, en conditions de chargement non drainées |
| undrained friction angle | Value of the friction angle of a soil under undrained loading condtions (null for a saturated cohesive soil) | Valeur de l'angle de frottement d'un sol, en conditions de chargement non drainées (égal à 0 pour un sol cohérent saturé) |
| drained cohesion | Value of the shear strength of a soil, for a null normal stress, under drained loading conditions | Valeur de la résistance au cisaillement d'un sol pour une contrainte normale nulle, en conditions de chargement drainées |
| drained friction angle | Value of the friction angle of a soil under drained loading conditions | Valeur de l'angle de frottement d'un sol, en conditions de chargement drainées |
| residual cohesion | Value of cohesion for high displacements on a failure plane. | Valeur de la cohésion après de grands déplacements sur une surface de rupture. |
| residual friction angle | Value of friction angle for high displacements on a failure plane. | Valeur de l'angle de frottement après de grands déplacement sur une surface de rupture |
| soil compressibility parameters | | |

| Name | Definition EN | Definition FR |
|---|---|---|
| compression index cc | Gradient of the linear portion of the oedometer test curve (void ratio versus the logarithm of vertical effective stress), beyond the pressure of pre-consolidation . | Pente de la partie linéaire de la courbe oedométrique (indice de vides en fonction du logarithme de la contrainte verticale effective), lors du chargement, pour une contrainte verticale appliquée supérieure à la pression de consolidation. |
| unloading / recompression index cs (or cg) | Gradient of the linear portion of the oedometer test curve (void ratio versus the logarithm of vertical effective stress ), before the pressure of pre-consolidation. | Pente de la partie linéaire de la courbe oedométrique (indice de vides en fonction du logarithme de la contrainte verticale effective), lors du chargement pour une contrainte verticale appliquée inférieure à la pression de consolidation ou lors du déchargement. |
| consolidation pressure (or yield stress) | Highest value of vertical effective stress applied on a soil in its geological history. In an oedometer test, vertical effective pressure at the intersection of the reloading and first loading linear portions. | Contrainte effective verticale la plus élevée à laquelle un sol a été soumis au cours de son histoire géologique. Dans un essai oedométrique, contrainte effective verticale à l'intersection de la droite de recompression réelle et de la droite de compression normale. |
| coefficient of vertical consolidation | Parameter which relates the degree of consolidation to time from the start of consolidation. | Paramètre qui relie le degré de consolidation au temps écoulé depuis le début de la consolidation. |
| creep coefficient | The ratio of the change in height to the inital height over one log cycle of time during the secondary compression phase. | Rapport entre la variation de hauteur et la hauteur initiale de l'éprouvette au cours de la compression secondaire. |
| swelling index cg | Gradient of the linear portion of the void ratio versus the logarithm of vertical effective stress curve, during unloading. | Pente de la courbe oedométrique (indice de vides en fonction du logarithme de la contrainte verticale effective), lors du déchargement. |

| Name | Definition EN | Definition FR |
|---|---|---|
| swelling pressure | Vertical pressure necessary to maintain a constant volume (i.e to prevent absorption of water) when a soil is saturated. | Pression nécessaire pour maintenir un volume constant (c'est-à-dire pour empêcher l'absorption d'eau) dans un sol saturé. |
| pressuremeter test parameters | | |
| menard limit pressure | Pressure that characterizes the failure of the soil during the Ménard pressuremeter test. It corresponds conventionally to the pressure that leads to double the volume of the tested cavity. | Pression caractérisant la rupture du sol lors de l'essai pressiométrique Ménard. Par convention, elle correspond à la pression qui entraine le doublement du volume de la cavité initiale. |
| net limit pressure | Ménard limit pressure to which the total horizontal stress at tested depth is substracted. | Pression limite de l'essai pressiométrique à laquelle on déduit la contrainte totale horizontale à la profondeur de l'essai. |
| menard creep pressure | Pressure that characterizes the limit between the pseudo-elastic eand the plastic behaviour of the soil, during the Ménard pressuremeter test. | Pression caractérisant la limite entre le comportement élastique et le comportement plastique du sol, lors de l'essai pressiométrique. |
| net creep pressure | Ménard creep pressure to which the total horizontal stress at tested depth is substracted. | Pression de fluage à laquelle on déduit la contrainte horizontale totale à la profondeur de l'essai |
| menard pressuremeter modulus | Deformation modulus during the pseudo-elastic phase of the Ménard pressuremeter test. | Module de déformation du sol mesuré lors de l'essai pressiométrique pendant la phase pseudo-élastique de l'essai. |
| ***Static penetration test parameters*** | | |
| cone resistance | Ratio of the axial load Qc divided by the total area of the cone basis Ac. | Rapport de la force axiale mesurée Qc divisée par l'aire totale de la base du cône Ac. |
| sleeve friction | The sleeve friction is determined by the force required to push the sleeve through the soil | Frottement latéral mesuré sur le manchon |

| Name | Definition EN | Definition FR |
|---|---|---|
| friction ratio | Ratio of the lateral side friction divided by the cone resistance, both measured at the same depth. | Rapport entre le frottement latéral mesuré sur le manchon et la résistance de pointe, les deux étant réalisés à la même profondeur. |
| piezocone pore pressure | Fluid pressure measured in the filter element placed an the cone basis, during the cone penetration. | Pression mesurée dans l'élément filtrant au cours des essais de pénétration au cône. |
| ***Dynamic penetration test parameters*** | | |
| N value | Value of a Standard Penetration Test (SPT) defined as the number of blows exerted by the hammer to achieve a penetration of the last 30 cm, on a total penetration of 45 cm. | Nombre de coups de battage d'un carottier SPT pour s'enfoncer des 30 derniers cm, lors d'une passe de 45 cm. |
| N60 | Calculated value that corrects the value of the Standard Penetration Blow Count (N-value), to account for energy of the hammer (specifically normalized to 60% energy). | Valeur calculée qui corrige la valeur du nombre de coups de pénétration standard (valeur N), pour tenir compte de l'énergie du marteau (spécifiquement normalisée à 60 % d'énergie). |
| N1,60 | Calculated value that corrects the value of the Standard Penetration Blow Count (N), to account for energy of the hammer (specifically normalized to 60% energy) and specific overburden conditions (specifically to an overburden stress of 1 ton per square foot). | Valeur calculée qui corrige la valeur du nombre de coups de pénétration standard (N), pour tenir compte de l'énergie du marteau (spécifiquement normalisée à 60 % d'énergie) et des conditions de surcharger spécifiques (en particulier pour une contrainte de surcharger de 1 tonne par pied carré). |
| first increment spt number of blows n1 | Number of blows exerted by the hammer to achieve a penetration of a first increment of 15 cm. | Nombre de coups de battage d'un carottier SPT pour s'enfoncer d'un premier incrément de 15 cm |
| second increment spt number of blows n2 | Number of blows exerted by the hammer to achieve a penetration of a second increment of 15 cm. | Nombre de coups de battage d'un carottier SPT pour s'enfoncer d'un deuxième incrément de 15 cm |

| Name | Definition EN | Definition FR |
|---|---|---|
| third increment spt number of blows | Number of blows exerted by the hammer to achieve a penetration of a third increment of 15 cm. | Nombre de coups de battage d’un carottier SPT pour s’enfoncer d’un troisième incrément de 15 cm |
| driveset blowcount | Number of blows exerted by the hammer in this driveset to achieve a given penetration into the soil | |
| driveset penetration | Penetraton distance into the soil that is achieved given a specific driveset blowcount | |
| driveset index | A number that defines the specific increment number for a driveset, which consist of a blowcount paired with the penetration achieved. The first driveset has an index of 1, the second, 2 and so on. | |
| ***Rock mechanics parameters*** | | |
| unconfined compressive strength | Ratio between the force applied during fracture of the cylindrical test piece and the area of the cross section determined before the test | Rapport entre la force appliquée lors de la rupture de l’éprouvette cylindrique et l’aire de la section transversale déterminée avant essai |
| young modulus | The Young modulus is the ratio between the normal stress and the axial strain in a linear elastic and isotropic medium. | Le module d’Young est le rapport entre la contrainte normale et la déformation axiale, dans un matériau élastique linéaire isotrope. |
| poisson coefficient | The Poisson’s ratio is a measure of the deformation of a material in direction perpendicular to the specific direction of loading. | Le coefficient de Poisson est le rapport entre la déformation latérale et la déformation dans l’axe du chargement |
| tensile strength (brazilian test) | Ratio between the force applied during fracture of the cylindrical test piece and the area of the cross section determined before the test. It is an indirect measurement of tensile strength. | Rapport entre l’effort de compression appliqué le long d’un diamètre d’une éprouvette cylindrique, et la surface du plan de rupture. |

| Name | Definition EN | Definition FR |
| --- | --- | --- |
| tensile strength (franklin point load test) | Ratio between the punctual force applied on both sides of a cylindrical test piece and the area of the cross section determined before the test. It is an indirect measurement of tensile strength. | Rapport entre l’effort de compression ponctuel de part et d’autre du diamètre d’une éprouvette cylindrique, et la surface du plan de rupture. Il s’agit d’une mesure indirecte de résistance à la traction. |
| hoek & brown coefficient mi | Constant introduced by Hoek & Brown ranging from 4 (for some clayey rocks) to more than 30 (for igneous and some metamorphic rocks), used in the Hoek & Brown parabolic failure criterion. | Constante introduite par Hoek 1 Brown, variant entre 4 (pour les roches argileuses) et plus de 30 (pour les roches ignées et métamorphiques), utilisée dans le critère de rupture de Hoek & Brown. |
| hardness (cerchar) | Index characterising the resistance to penetration of a rock by a drill bit under standard test conditions. | Indice traduisant la résistance à la pénétration d’un foret dans une roche sous des conditions de chargement normalisées |
| abrasivity index (cerchar) | Index characterinsing the ability of a rock to cause wear of cutting tool. | Indice caractérisant la capacité d’une roche à causer l’usure des outils de coupe du rocher. |
| los angeles index | Index characterising the fragmentability of a rock material under mechanical actions. Percentage of the initial sample reduced to a size less than 1.6mm during its rotation in a cylinder loaded with steel balls | Indice traduisant la fragmentabilité d’une roche sous l’effet d’une usure mécanique. Il correspond au rapport de la masse initiale d’un échantillon de roche sur la masse des fragments de moins de 1,6 mm de diamètre, produit par la rotation de l’échantillon dans un cylindre rempli de billes d’acier. |
| micro-deval index | Index characterising the fragmentability of a rock material under mechanical actions. Percentage of the initial sample reduced to a size less than 1.6mm during its rotation in a cylinder loaded with an abrasive charge | Indice traduisant la fragmentatibilité d’une roche sous l’effet d’une usure mécanique. Il correspond au rapport de la masse initiale d’un échantillon de roche sur la masse des fragments de moins de 1,6 mm de diamètre, produit par la rotation de l’échantillon dans un cylindre avec une charge abrasive. |

| Name | Definition EN | Definition FR |
| --- | --- | --- |
| core recovery | Ratio in percent of the length of core recovered to the total length of the core drilled on a given run. | Rapport en pourcent de la longueur de carotte récupérée à la sortie du carottier sur la longueur de la passe de carottage. |
| rqd rock quality designation | The cumulative length of cores with a length greater than 10 cm, divided by the total length of the core pass with a length greater than or equal to 1m, in percent. | Longueur cumulée des tronçons de carottes de longueur supérieure à 10cm, divisée par la longueur totale de la passe carottée de longueur supérieure ou égale à 1m, exprimée en pourcent. |
| weathering grade | Level of weathering of a rock mass determined by visual observation, on a scale between I (sound rock) to VI (completely weathered rock - residual soil). | Niveau d’altération d’un massif rocheux déterminé par observation visuelle, sur une échelle de I (roche saine) à VI (roche totalement décomposée - sol résiduel) |
| gsi : geological strength index | Empirical index introduced by E.Hoek, derived from the RMR and Barton-Q indices, and that characterises the quality of the rock mass. | Indice empirique introduit par E.Hoek déterminé à partir des indices RMR et Q-Barton, permettant de caractériser la qualité du massif rocheux. |
| rmr : rock mass rating | Index developped by Bieniawski, to provide an quantitative estimate of the quality of a rock mass. RMR is equal to the sum of the six following ratings : 1. Strength of intact rock rating 2. RQD rating 3. Spacing of discontinuities rating 4. Conditions of dicontinuities rating 5 . Groundwater condition rating 6. Adjustment for orientation of discontinuities Bieniawski (1989) | Indice proposé par Bieniawski, permettant de fournir une évaluation quantitative de la qualité d’un massif rocheux. Le RMR est la somme des six notes suivantes : 1. Note “résistance de la matrice rocheuse” 2. Note RQD 3. Note “espacement des discontinuités” 4. Note “Conditions des dicontinuités” 5. Note “Conditions hydrauliques” 6. Ajustement selon l’orientation des discontinuités. Bieniawski (1989) |

| Name | Definition EN | Definition FR |
|---|---|---|
| q-barton index | Index introduced by Barton, tha tprovides an quantitative estimate of the quality of rock mass. It is calculated from 6 parameters : . Le RQD . Jn : Joint set number . Jr Joint roughness number . Ja : Joint alteration number . Jw : Joint water reduction factor . SRF : Stress reduction factor | Indice introduit par Barton qui fournit une estimation quantitative de la qualité du massif rocheux à partir de 6 paramètres : . Le RQD . Jn : Joint set number . Jr Joint roughness number . Ja : Joint alteration number . Jw : Joint water reduction factor . SRF : Stress reduction factor |
| discontinuity | In rock mechanics, any mechanical crack or fissure in a rock mass having a null or a much lower strength than the rock matrix. | En mécanique des roches, toute fissure ou joint au sein d’un massif rocheux ayant une résistance nulle ou faible comparativement à la résistance de la matrice rocheuse. |
| type of discontinuity set | Type of the discontinuity set, for example : bedding plane, schistosity, etc. | Type de discontinuité. Par exemple : joint de stratification, diaclase, schistosité, etc. |
| strike | The geographic direction of the discontinuity plane in space with respect to the North. | Direction géographique d’une ligne créée par l’intersection du plan de la discontinuité avec le plan horizontal. |
| dip angle | The angle that a rock unit, fault or other structure makes with a horizontal plane. The angle is measured in a plane perpendicular to the strike of the rock structure. | Angle d’une surface (formation géologique, faille, ou autre structure) avec le plan horizontal. Sa mesure est celle du plongement de la ligne de plus grande pente de cette surface. |
| spacing | The perpendicular distance between adjacent dicontinuities in a set. | Distance entre deux discontinuités les plus proches d’une même famille mesurée perpendiculairement à celles-ci. |
| interval between discontinuity (id index) | The mean of the intact rock legnth between successive discontinuities along a survey line whose length and orientation must be recorded. | Moyenne des intervalles découpés par les discontinuités successives le long d’une ligne de mesure, dont il convient de préciser la longueur et l’orientation. |

| Name | Definition EN | Definition FR |
|---|---|---|
| extension | The total area of the discontinuity in all directions. | L'extension ou la persistence des discontnuités correspond à la surface totale de la discontinuité dans l'espace. |
| aperture | Size of the gap between the joint walls of the discontinuity, measured perpendicularly to the joint plane | Distance entre épontes comptée perpendiculairement au plan de discontinuité. |
| roughness | Qualifies the surface irregularities on the joint walls of a discontinuity, from very rough to slickensided. | La rugosité qualifie les irrégularités de surface des épontes d'une discontinuité. Elle varie de : surface très rugueuse à lustrée. |
| joint roughness coefficient (jrc) | Dimensionless coefficient relating to joint roughness and size, ranked in ascending order from 0 for a flat smooth discontinuity to 20 for a wavy rough discontinuity, according to Barton's standard profiles (1977). | Coefficient sans dimension relié à la rugosité et à la taille des épontes ; il varie entre 0 (pour une discontinuité planne et lisse) à 20 (pour une discontinuité ondulée et rugueuse), selon la classification établie par Barton (1977). |
| infill | The nature of the material filling the discontinuity or coating the walls. | Nature du matériau de remplissage ou de l'enduit des épontes. |
| shear strength | Value of the shear stress at failure. For a discontinuity, the shear strength is usually expressed by the Coulomb criterion $\tau = c + \sigma.tg\phi$, with c, the cohesion, and $\phi$, the friction angle. | Valeur de la contrainte de cisaillement à la rupture. Pour une discontinuité, la résistance au cisaillement est exprimée par le critère de Coulomb : $\tau = c + \sigma.tg\phi$, avec c, la cohésion, et $\phi$, l'angle de frottement. |
| cohesion | Shear stress at failure, for a normal stress equal to zero. For a discontinuity, appraent cohesion which does not express an intrisic property of the joint wall material but the influence of irregularites in the walls on shear behaviour. | Contrainte de cisaillement à la rupture pour une contrainte normale nulle. Pour une discontinuité, il s'agit d'une cohésion apparente et non une propriété intrinsèque du materiau de remplissage du joint, cette cohésion caractérisant l'influence des irrégularités de surfaces des épontes. |

| Name | Definition EN | Definition FR |
|---|---|---|
| friction angle | Friction angle of the discontinuity that depends on the rock nature, the roughness of joint walls and the weathering grade on the walls. | Angle de frottement du joint dépendant de la nature pétrographique, de la rugosité de la surface et du dégré d'altération des épontes. |
| normal stiffness | In an uniaxial compression tests on joints oriented perpendicular to the direction of load application, slope of the curve of normal stress versus normal displacement, . | Lors d'un essai de compression simple, dirigé perpendiculairement au joint de discontinuité, pente de la courbe de la contrainte normale en fonction du déplacement normal. |
| tangential stiffness | In a shear test on joints, slope of the curve of tangential stress versus tangential displacement. | Lors d'un essai de cisaillement direct le long d'une discontinuité, pente de la courbe contrainte tangentielle en fonction du déplacement tangentiel. |
| compression wave velocity | Ratio of the propagation time of a compression wave in an elastic body medium between two points divided by the distance between these points | Rapport de la durée de propagation d'une onde de compression entre deux points dans un milieu élastique, divisé par la distance entre ces deux points |
| shear wave velocity | Ratio of the propagation time of a shear wave in an elastic body medium between two points divided by the distance between these points | Rapport de la durée de propagation d'une onde de cisaillement entre deux poinst dans un milieu élastique, divisé par la distance entre ces deux points |
| ***Geo-hydraulics parameters*** | | |
| darcy's permeability | Parameter that characterizes the ability of a porous (continous) medium to be crossed by a fluid, to a flow from the same direction | Paramètre caractérisant l'aptitude d'un milieu poreux (continu) à se laisser traverser par un fluide, vis-à-vis d'un flux de même direction. |

| Name | Definition EN | Definition FR |
|---|---|---|
| intrinsic permeability | Parameter characterizing a porous medium isotropic, which measures its permeability to a homogeneous fluid arbitrary, regardless of the characteristics of the fluid. Volume of kinematic viscosity unit fluid flowing through in a unit of time, under the effect of a unit of potential gradient, one unit surface orthogonal to the direction of flow. It can be expressed in darcy s | Paramètre caractérisant un milieu poreux isotrope, qui mesure sa perméabilité vis-à-vis d'un fluide homogène quelconque, indépendamment des caractéristiques du fluide. Volume de fluide d'unité de viscosité cinématique qui traverse en une unité de temps, sous l'effet d'une unité de gradient de potentiel, une unité de surface orthogonale à la direction du flux. ll est exprimable en darcy s. |
| packer test - lugeon unit | 1 Lugeon unit is equivalent to 1 liter of water flow per meter of tested zone, per minute, under a pressure of 1 MPa. | 1 unité Lugeon équivaut à 1 litre d'eau prélevé par mètre de longueur d'essai, par minute, sous une pression de 1 MPa. |
| transmissivity | Parameter governing the flow of water flowing per unit width of the saturated zone of a continuous aquifer (measured in a direction orthogonal to that of flow), and per unit of hydraulic gradient | Paramètre régissant le débit d'eau qui s'écoule par unité de largeur de la zone saturée d'un aquifère continu (mesurée selon une direction orthogonale à celle de l'écoulement), et par unité de gradient hydraulique. |
| storage coefficient | Ratio of the volume of water released or stored per unit area of an aquifer, to the corresponding charge in hydraulic head, without reference to time | Rapport du volume d'eau libérée ou emmagasinée par unité de surface d'un aquifère, à la variation de charge hydraulique correspondante, sans référence au temps (ou en un temps illimité). |
| specific storage coefficient | Volume of water released or stored per unit area of an aquifer, to the corresponding charge in hydraulic head, without reference to time | Volume d'eau libérée ou emmagasinée par unité de volume du milieu aquifère, par unité de variation de charge hydraulique correspondante, sans référence au temps. |

| Name | Definition EN | Definition FR |
| --- | --- | --- |
| piezometric level | Upper level of the static liquid column that balances the hydrostatic pressure at the point to which it relates. It is materialized by the free level of the water in a vertical tube open at th epoint considered (piezometer). | Niveau supérieur de la colonne liquide statique qui équilibre la pression hydrostatique au point auquel elle se rapporte. Il est matérialisé par le niveau libre de l’eau dans un tube vertical ouvert au point considéré (piézomètre). |

## 3.5.9. MaterialSample

A MaterialSample is a physical, tangible Sample.

**Realizations**

| Data model | Concept name | Definition |
| --- | --- | --- |
| OGC OMS | MaterialSample | A MaterialSample is a physical, tangible Sample. |
| IFC | MaterialSample | Same as OGC OMS |
| AGS | SAMP | SAMP includes information for both the sample and the sampling activity |
| DIGGS | Sample | A material sample, either solid, fluid, or gas that is obtained as a result of a sampling activity, for the purpose of observation and/or testing. This is a concrete representation of AbstractSample, which serves as the head of a substitution group for this and other sample specialties. |
| DIGGS | TrialGroutBatch | A specialization of Sample that also contains properties describing the characteristics of a batch of grout developed for testing to determine a grout design mix. |

| Data model | Concept name | Definition |
|---|---|---|
| DIGGS | GroutBatch | A specialization of Sample that contains properties describing the characteristics of a batch of grout used for production purposes. This object must reference a DesignGroutMix object defined within the GroutingProgram feature. |
| DIGGS | AbstractSpecimen | AbstractSpecimen serves as the head of a substitution group for specimen specializations based on state or type: eg. soil, liquid, gas, rock. A specimen derives from one or more samples, is created as part of a test procedure, and exists only in the context of the test procedure that creates it. |
| DIGGS | GroutSpecimen | A specialization of a specimen that describes the physical conditions of a grout or concrete sample subjected to testing. |
| DIGGS | FluidSpecimen | A specialization of a specimen that describes the physical conditions of a fluid sample or samples subjected to testing. |
| DIGGS | RockSpecimen | A specialization of a specimen that describes the physical conditions of a rock sample or samples subjected to testing. |
| DIGGS | SoilSpecimen | A specialization of a specimen that describes the physical conditions of a soil sample or samples subjected to testing. |

### Properties

### FAQ

#### What about Specimen?

In geotechnics, a distinction may appear between the MaterialSample collected on the field and another MaterialSample made from it by sub-sampling. The first one is often called Sample, while the second(s) is/are often called Specimen(s).

Here in this conceptual model, the term MaterialSample is generic and can be used for both. Attributes can be used to explain the lineage between two (or more) MaterialSample.

Standards like DIGGS yet propose an explicit terminology.

#### What about Sampling?

The sampling activity can be described with [Sampling]

#### GitHub issue

https://github.com/opengeospatial/Geotech/issues/11

### 3.5.10. Sampling and Preparation

Sampling is an act applying a SamplingProcedure to create or transform one or more MaterialSamples.

Sampling Procedure is a description of steps performed by a Sampler in order to extract a Sample from its sampledFeature in the frame of a Sampling.

Preparation Procedure is the description of preparation steps performed on a Sample.

**Realizations**

| Data model | Concept name | Definition |
|---|---|---|
| OGC OMS | Sampling | Sampling is an act applying a SamplingProcedure to create or transform one or more Sample(s). |
| OGC OMS | SamplingProcedure | Sampling Procedure is a description of steps performed by a Sampler in order to extract a Sample from its sampledFeature in the frame of a Sampling. |
| OGC OMS | PreparationProcedure | Sampling is an act applying a SamplingProcedure to create or transform one or more Sample(s). |
| AGS | SAMP | SAMP includes information for both the sample and the sampling activity |

| Data model | Concept name | Definition |
| --- | --- | --- |
| DIGGS | SamplingActivity | The action taken to obtain or produce material samples even if the activity fails to produce a sample (eg. a core run that produces no recovery). The activity type must be specified, and indicates if the produced sample(s) are a result of collection at a sampling feature, sub-sampling, aggregation of two or more samples, or a sample created as a test or blank sample. All samples must refer to a SamplingActivity feature. SamplingActivity is restricted to the collection or creation of a new material sample, or a subsampling or aggregation of existing collected or created samples. The activity of transforming a material sample into a specimen that is being tested is described in the preparationProcedure property of a specimen object that derives from an existing material sample. |
| DIGGS | AbstractSpecimen | AbstractSpecimen serves as the head of a substitution group for specimen specializations based on state or type: eg. soil, liquid, gas, rock. A specimen derives from one or more samples, is created as part of a test procedure, and exists only in the context of the test procedure that creates it. |

**Properties**

## 3.6. Book B Principles

In their tentative to appreciate the environment of a projected infrastructure, geotechnical engineers pay attention to several aspects:

- Geology,
- HydroGeology,
- Geotechnics,
- Surrounding constructions,
- Geohazards, and
- Environmental impact (polluted site).

Models are built in order to reflect an understanding of this environment.

Each Book B ends with a register of uncertainties: what is known, what is unknown.

### 3.6.1. Concepts list

The Book B is organized in Geomodels which themselves are components of different Features based on their thematic content.

### 3.6.2. Primary models and their components

The following components are included in geology, hydrogeology, and geotechnics models.

- Geomodel
- GeologicUnit
- Fault
- Contact
- Fold
- HydroGeoUnit
- FluidBody
- FluidBodySurface
- GeotechUnit
- DiscreteDiscontinuity
- Void
- HazardArea
- SurroundingConstruction

### 3.6.3. GeoModel

A Geomodel is a simplified view of a model. A model is defined as the result of geoscience data processing or interpretation. Proposing either a spatial distribution of geoscientific objects with properties of interest (feature) or attempt of retranscribing natural/man-made behavior through mathematical functions or algorithms (process).

#### Realization

| Data model | Concept name | Definition |
|---|---|---|
| OGC EPOS | ModelView | Simplified view of a Model. A model is defined as the result of geoscience data processing or interpretation. Proposing either a spatial distribution of geoscientific objects with properties of interest (feature) or attempt of retranscribing natural/man-made behavior through mathematical functions or algorithms (process). |
| IFC | | |
| AGSi | agsiModel | A digital geometric (1D, 2D or 3D) representation of the ground. There are potentially many different types of model covering different categories (conceptual, observational, analytical) and domains (geological, geotechnical, hydrogeological, geoenvironmental, etc.). |

#### Github issue

https://github.com/opengeospatial/Geotech/issues/26

### 3.6.4. GeologicUnit

Conceptually, a GeologicUnit may represent a body of material in the Earth whose complete and precise extent is inferred to exist (e.g., North American Data Model GeologicUnit, Stratigraphic unit in the sense of NACSN, or International Stratigraphic Code ), or a classifier used to characterize parts of the Earth (e.g., lithologic map unit like ‘granitic rock’ or ‘alluvial deposit’, surficial units like ‘till’ or ‘old alluvium’). It includes both formal units (i.e., formally adopted and named in an official lexicon) and informal units (i.e., named but not promoted to a lexicon) and unnamed units (i.e., recognizable, described and delineable in the field but not otherwise formalized). In simpler terms, a geologic unit is a package of earth material (generally rock).

#### Realizations

| Data model | Concept name | Definition |
|---|---|---|
| OGC GeoSciML | GeologicUnit | Conceptually, a GeologicUnit may represent a body of material in the Earth whose complete and precise extent is inferred to exist (e.g., North American Data Model GeologicUnit, Stratigraphic unit in the sense of NACSN, or International Stratigraphic Code ), or a classifier used to characterize parts of the Earth (e.g., lithologic map unit like ‘granitic rock’ or ‘alluvial deposit’, surficial units like ‘till’ or ‘old alluvium’). It includes both formal units (i.e., formally adopted and named in an official lexicon) and informal units (i.e., named but not promoted to a lexicon) and unnamed units (i.e., recognizable, described and delineable in the field but not otherwise formalized). In simpler terms, a geologic unit is a package of earth material (generally rock). |
| IFC | GeologicUnit | Same as OGC GeoSciML |
| DIGGS | StratigraphyObservation | Descriptions of ordered bodies of rock or soil, such as formations, biostratigraphic units or aquifers. |
| AGSi | agsiModelElement | In AGSi a model is collection of elements (agsiModelElement object) and this may include geological units, identified as such (or as a specialization) using the elementType attribute. |

**Specializations (Types of GeologicUnit)**

| Source | Type | Link | Definition |
|---|---|---|---|
| BRGM | Registry | https://data.geoscience.fr/ncl/GeolUnitType | This register lists all the geologic units. |
| CGI | Registry | https://cgi.vocabs.ga.gov.au/object?uri=http%3A//resource.geosciml.org/classifier/cgi/geologicunittype | This register lists all the geologic units. |

- allostratigraphic unit
- alteration unit
- artificial ground
- biostratigraphic unit
- chronostratigraphic unit
- deformation unit
- excavation unit
- geomorphologic unit
- geophysical unit
- lithodemic unit
- lithogenetic unit
- lithologic unit
- lithostratigraphic unit
- lithotectonic unit
- magnetostratigraphic unit
- mass movement unit
- pedoderm
- pedostratigraphic unit
- polarity chronostratigraphic unit

**Properties**

| PropertyName | Definition |
|---|---|
| Identifier | The identifier should have the same value as the corresponding GeoSciML MappedFeature identifier, if available. |

| Pro pert yNa me | Definition |
|---|---|
| Name | name is a display name for the GeologicUnit. |
| Description | description is a description of the GeologicUnit, typically taken from an entry on a geological map legend. |
| | |
| geologicUnitType | geologicUnitType contains the type of GeologicUnit (as defined in GeoSciML). To report an identifier from a controlled vocabulary, geologicUnitType_uri shall be used. |
| rank | rank contains the rank of GeologicUnit (as defined by ISC. e.g., group, formation, member). |
| representativeLithology | description of the GeologicUnit’s lithology, possibly formatted with formal syntax (see 8.9.2.3). The description can be language-dependent. To report an identifier from a controlled vocabulary, representativeLithology_uri shall be used. |
| representativeAge | description of the age of the GeologicUnit (where age is a sequence of events and may include process and environment information). To report an identifier from a controlled vocabulary, representativeAge_uri, representativeOlderAge_uri, representativeYoungerAge_uri shall be used. |
| numericOlderAge | a numerical representation of the unit’s older age in million years (Ma). |
| numericYoungerAge | a numerical representation (Primitive::Number) of the unit’s younger age in million years (Ma). |
| observationMethod | a metadata snippet indicating how the spatial extent of the feature was determined. ObservationMethod is a convenience property that provides a simple approach to observation metadata when data are reported using a feature view (as opposed to observation view). |
| | |

| PropertyName | Definition |
|---|---|
| source | If present, the property source:Primitive::CharacterString is human readable text describing feature-specific details and citations to source materials, and if available provides URLs to reference material and publications describing the geologic feature. This could be a short text synopsis of key information that would also be in the metadata record referenced by metadata_uri. |
| metadata_uri | If present, the property metadata_uri:Primitive::CharacterString contains a URI referring to a metadata record describing the provenance of data. |
| | |
| shape | The property shape:GEO::GM_Object contains a geometry defining the extent of the feature of interest. |
| positionalAccuracy | If present, the property positionalAccuracy:Primitive::CharacterString is a quantitative value (a numerical value and a unit of length) defining the radius of an uncertainty buffer around a MappedFeature (e.g., a positionalAccuracy of 100 m for a line feature defines a buffer polygon of total width 200 m centred on the line). |
| | |
| GeolFormation | ISO14689:2017 |
| FaciesLithostratigraphy | The characteristics of a rock or a sediment unit that reflect its environment of deposition and allow it to be distinguished from rock or sediment deposited in an adjacent environment. |
| FaciesPetrophysics | A rock mass that can be recognized by its composition, structures or fossil content and mapped on the basis of those characteristics. |

## FAQ

### What about Hydrogeologic Units?

They are a type of GeologicUnit and are described in HydrogeoUnit

**GitHub issue**

https://github.com/opengeospatial/Geotech/issues/7

### 3.6.5. Fault

A fault includes all brittle to ductile style structures along which displacement has occurred, from a simple, single ‘planar’ brittle or ductile surface to a fault system comprised of tens of strands of both brittle and ductile nature. This structure may have some significant thickness (a deformation zone) and have an associated body of deformed rock that may be considered a deformation unit (which geologicUnitType is ‘DeformationUnit’) which can be associated to the ShearDisplacementStructure using GeologicFeatureRelation from the GeoSciML Extension package

**Realizations**

| Data model | Concept name | Definition |
| --- | --- | --- |
| OGC GeoSciML | ShearDisplacementStructure | A shear displacement structure includes all brittle to ductile style structures along which displacement has occurred, from a simple, single ‘planar’ brittle or ductile surface to a fault system comprised of tens of strands of both brittle and ductile nature. This structure may have some significant thickness (a deformation zone) and have an associated body of deformed rock that may be considered a deformation unit (which geologicUnitType is ‘DeformationUnit’) which can be associated to the ShearDisplacementStructure using GeologicFeatureRelation from the GeoSciML Extension package. |
| IFC | Fault | Same as OGC |
| AGSi | agsiModelElement | In AGSi a model is collection of elements (agsiModelElement object) and this may include elements representing geological structure such as faults, identified as such using the elementType attribute. |

**Properties**

| PropertyName | Definition |
|---|---|
| identifier | Globally unique identifier:Primitive::CharacterString shall uniquely identifies a tuple within the dataset and be formatted as an absolute URI conformant to RFC 3986. |
| name | name contains a display name for the ShearDisplacementStructure. |
| description | description contains a human readable text description of the ShearDisplacementStructure, typically taken from an entry on a geological map legend. |
| | |
| faultType | faultType contains a human readable description of the type of ShearDisplacementStructure (as defined in GeoSciML). To report an identifier from a controlled vocabulary, faultType_uri shall be used. |
| movementType | movementType contains a human readable summary of the type of movement (e.g., dip-slip, strike-slip) on the ShearDisplacementStructure. To report an identifier from a controlled vocabulary, movementType_uri shall be used. |
| deformationStyle | deformationStyle contain a human readable description of the style of deformation (e.g., brittle, ductile etc.) for the ShearDisplacementStructure. To report an identifier from a controlled vocabulary, deformationStyle_uri shall be used. |
| displacement | displacement contains a text summarizing the displacement across the ShearDisplacementStructure. |
| geologicHistory | geologicHistory contains a text, possibly formatted with formal syntax, describing the age of the ShearDisplacementStructure (where age is a sequence of events and may include process and environment information). |

| PropertyName | Definition |
|---|---|
| observationMethod | observationMethod contains a metadata snippet indicating how the spatial extent of the feature was determined. ObservationMethod is a convenience property that provides a quick and dirty approach to observation metadata when data are reported using a feature view (as opposed to observation view). |
| | |
| numericOlderAge | numericOlderAge reports the older age of the fault/shear structure, represented million years (Ma). |
| numericYoungerAge | numericYoungerAge reports the younger age of the fault/shear structure, represented million years (Ma). |
| | |
| source | source contains a text describing feature-specific details and citations to source materials, and if available providing URLs to reference material and publications describing the geologic feature. This could be a short text synopsis of key information that would also be in the metadata record referenced by metadata_uri. |
| metadata_uri | If present, the property metadata_uri:Primitive::CharacterString contains a URI referring to a metadata record describing the provenance of data. |
| | |
| shape | The property shape:GM_Object contains a geometry defining the extent of the feature of interest. |
| positionalAccuracy | If present, the property positionAccuracy:Primitive::CharacterString contains quantitative representation defining the radius of an uncertainty buffer around a MappedFeature (e.g., a positionalAccuracy of 100 m for a line feature defines a buffer polygon of total width 200 m centred on the line). |
| | |

| PropertyName | Definition |
|---|---|
| FaultActivity | Qualitative rating of the fault activity. A fault that has slipped in historic time and which is likely to slip again in the future. |
| FaultAperture | GT1R1A1 |
| FaultLength | Fault length (L) is the longest horizontal or subhorizontal dimension along the fault plane |
| FaultDepth | Fault depth is the depth where no further relative displacement is recorded |
| FaultInfilling | Substance interspersed between the wall surfaces of a discontinuity (JGS) generally, the material occupying the space between joint surfaces, faults, and other rock discontinuities. The filling material may be clay, gouge, various natural cementing agents, or alteration products of the adjacent rock.(ASTM, ISRM) |
| DipDirection | The geographic direction of a line created by the intersection of a plane and the horizontal plane. If non specific convention is used, this angular value is in the range 0 to 18 degrees. In this case, there is an ambiguity on the dip orientation. For example, a plane with an orientation of 90 degrees from the north could either have a dip direction to the North or to the South. Thus, to avoid this ambiguity, a strike value is generally completed with an indication of the dip orientation (Quadrant). |
| FaultSlipDirection | Observed relative displacement direction of the hanging wall (top) to the footwall, e.g., up. Down, left, right or combinations (like top down - left) |
| FaultPitchAngle | angle between the strike of the slip surface and the slip vector (striation) |

| PropertyName | Definition |
|---|---|
| FaultThrow | the vertical displacement between the Hanging wall and Footwall |
| FaultHeave | the horizontal displacement between the Hanging wall and Footwall. |
| FaultOffset | The horizontal displacement between points on either side of a fault, which can range from millimeters to kilometers. |
| FaultDipSeparation | offset of beds parallel to fault |
| FaultStrikeSeparation | offset of beds parallel to dip |

**GitHub issue**

https://github.com/opengeospatial/Geotech/issues/20

### 3.6.6. Contact

A contact is a general concept representing any kind of surface separating two geologic units, including primary boundaries such as depositional contacts, all kinds of unconformities, intrusive contacts, and gradational contacts, as well as faults that separate geologic units.

**Realizations**

| Data model | Concept name | Definition |
|---|---|---|
| OGC GeoSciML | Contact | A contact is a general concept representing any kind of surface separating two geologic units, including primary boundaries such as depositional contacts, all kinds of unconformities, intrusive contacts, and gradational contacts, as well as faults that separate geologic units. |
| IFC | Contact | Same as OGC GeoSciML |

**Properties**

| Property Name | Definition |
|---|---|
| identifier | Globally unique identifier:Primitive::CharacterString shall uniquely identifies a tuple within the dataset and be formatted as an absolute URI conformant to RFC 3986. |
| name | name reports the display name for the Contact. |
| description | description reports the description of the Contact, typically taken from an entry on a geological map legend. |
| | |
| contactType | contactType reports the type of Contact (as defined in GeoSciML) as a human readable label. To report an identifier from a controlled vocabulary, contactType_uri shall be used. |

| Property Name | Definition |
| --- | --- |
| observationMethod | observationMethod reports a metadata snippet indicating how the spatial extent of the feature was determined. ObservationMethod is a convenience property that provides a quick and simple approach to observation metadata when data are reported using a feature view (as opposed to observation view). |
| positionalAccuracy | positionalAccuracy reports quantitative values defining the radius of an uncertainty buffer around a MappedFeature (e.g., a positionalAccuracy of 100 m for a line feature defines a buffer polygon of total width 200 m centred on the line). |
| source | source contains a text describing feature-specific details and citations to source materials, and if available providing URLs to reference material and publications describing the contact feature. This could be a short text synopsis of key information that would also be in the metadata record referenced by metadata_uri. |
| metadata_uri | metadata_uri reports a URI referring to a metadata record describing the provenance of data. |
| genericSymbolizer | genericSymbolizer contains an identifier for a symbol from standard (locally or community defined) symbolization scheme for portrayal. |
| | |

| Property Name | Definition |
|---|---|
| Shape | The property shape:GM_Object contains a geometry defining the extent of the contact feature. |
| | |
| DipDirection | The geographic direction of a line created by the intersection of a plane and the horizontal plane. If non specific convention is used, this angular value is in the range 0 to 18 degrees. In this case, there is an ambiguity on the dip orientation. For example, a plane with an orientation of 90 degrees from the north could either have a dip direction to the North or to the South. Thus, to avoid this ambiguity, a strike value is generally completed with an indication of the dip orientation (Quadrant). |

**GitHub issue**

https://github.com/opengeospatial/Geotech/issues/8

### 3.6.7. Fold

A fold is formed by one or more systematically curved layers, surfaces, or lines in a rock body. A fold denotes a structure formed by the deformation of a geologic structure, such as a contact which the original undeformed geometry is presumed, to form a structure that may be described by the translation of an abstract line (the fold axis) parallel to itself along some curvilinear path (the fold profile). Folds have a hinge zone (zone of maximum curvature along the surface) and limbs (parts of the deformed surface not in the hinge zone). Folds are described by an axial surface, hinge line, profile geometry, the solid angle between the limbs, and the relationships between adjacent folded surfaces if the folded structure is a Layering fabric.

**Realizations**

| Data model | Concept name | Definition |
|---|---|---|
| OGC GeoSciML | Fold | A fold is formed by one or more systematically curved layers, surfaces, or lines in a rock body. A fold denotes a structure formed by the deformation of a geologic structure, such as a contact which the original undeformed geometry is presumed, to form a structure that may be described by the translation of an abstract line (the fold axis) parallel to itself along some curvilinear path (the fold profile). Folds have a hinge zone (zone of maximum curvature along the surface) and limbs (parts of the deformed surface not in the hinge zone). Folds are described by an axial surface, hinge line, profile geometry, the solid angle between the limbs, and the relationships between adjacent folded surfaces if the folded structure is a Layering fabric. |
| IFC | Fold | Same as OGC GeoSciML |

**Properties**

| PropertyName | Definition |
|---|---|
| profileType | FoldProfileType contains a term from a controlled vocabulary specifying the concave/convex geometry of fold relative to earth surface, and relationship to younging direction in folded strata if known. (e.g., antiform, synform, neutral, anticline, syncline, monocline, ptygmatic). |
| amplitude | The amplitude property reports the length from line segment connecting inflection points on adjacent fold limbs to the intervening fold hinge. |

| PropertyName | Definition |
| --- | --- |
| axialSurfaceOrientation | The property axialSurfaceOrientation is used to characterize the geometry of a fold. The axial surface of a particular fold may be located based on observations of the folded geologic structure, but in general it has no direct physical manifestations. As a geologic surface, it has geometric properties, including orientation, which may be specified by observations at one or more locations, or generalized using terminology (upright, inclined, reclined, recumbent, overturned). Dip and Dip Direction are one approach to specifying the value. |
| geneticModel | The property geneticModel contains a term from a controlled vocabulary describing the specification of genetic model for fold, e.g., flexural slip, parallel. |
| hingeLineCurvature | The hingeLineCurvature property contains a term from a controlled vocabulary that describes the variation in orientation of fold hinge along trend of fold, distinguishing sheath from cylindrical folds (e.g., sheath, dome, basin, cylindrical.). |
| hingeLineOrientation | The property hingeLineOrientation reports the specification of the hinge line orientation for fold. GSML_LinearOrientation allows for a term value specification or a numeric specification of either or both the trend and plunge of hinge line. Hinge plunge term examples: sub-vertical, steeply plunging, sub-horizontal, reclined and vertical for special cases in which hinge plunge is close to axial surface dip. 0..* cardinality allows for both a numeric specification and a term specification. |
| hingeShape | The property hingeShape reports a term from a controlled vocabulary describing the hinge shape, e.g., Rounded vs. angular hinge zones. This property has to do with the proportion of the wavelength that is considered part of hinge. |
| interLimbAngle | The property interLimbAngle contains a term from a controlled vocabulary describing the interlimb angle using a tightness term (e.g., gentle (120-180°), open (70-120°), close (30-70°), tight (10-30°), isoclinal (0-10°)). |
| limbShape | The limbShape property contains a term from a controlled vocabulary describing the shape of the limb (e.g., straight vs curved limbs, kink, chevron, sinusoidal, box). |
| span | The span property reports a value describing the linear distance between inflection points in a single fold. |

| PropertyName | Definition |
| --- | --- |
| symmetry | The symmetry property contains a term from a controlled vocabulary describing the concordance or discordance of bisecting surface and axial surface, or the ratio of length of limbs. The folded surface may have asymmetry defined by limb length ratio if inflection points are defined. The definition based on bisecting surface/axial surface angle depends on having multiple surfaces defined such that the axial surface may be identified (symmetric, asymmetric). |
| DipDirection | The geographic direction of a line created by the intersection of a plane and the horizontal plane. If non specific convention is used, this angular value is in the range 0 to 18 degrees. In this case, there is an ambiguity on the dip orientation. For example, a plane with an orientation of 90 degrees from the north could either have a dip direction to the North or to the South. Thus, to avoid this ambiguity, a strike value is generally completed with an indication of the dip orientation (Quadrant). |

**FAQ**

**GitHub issue**

https://github.com/opengeospatial/Geotech/issues/9

### 3.6.8. HydrogeoUnit

> These are distinct volumes of earth material that serve as containers for subsurface fluids. The boundaries of a unit are typically discriminated from those of another unit using properties related to the potential or actual ability to contain or move water. The properties can be geological or hydraulic, and typically include influences from the surrounding hydrological environment. More specifically, the conceptual model delineates two types of hydrogeological units, with slightly different orientations: aquifer-related units have boundaries delimited by the hydrogeological properties of the rock body, while groundwater basins have boundaries delimited by distinct flow regimes. Aquifer-related units are subdivided into aquifer systems, which are collections of aquifers, confining beds, and other aquifer systems. Confining beds are units that impede water flow to surrounding units, and supersede notions such as aquitards, aquicludes, and aquifuges, which are not included herein, as it is difficult to differentiate these in practice.

### Realizations

| Data model | Concept name | Definition |
|---|---|---|
| OGC GWML2 | HydrogeoUnit | Any soil or rock unit or zone that by virtue of its hydraulic properties has a distinct influence on the storage or movement of groundwater. |
| IFC | HydrogeoUnit | Same as OGC GWML2 |
| AGSi | agsiModelElement | In AGSi a model is collection of elements (agsiModelElement object) and this may include hydrogeological units, identified as such using the elementType attribute. |

### Properties

#### Inherited properties from the GeologicUnit concept

See GeologicUnit

#### Specific properties

| Property Name | Definition |
|---|---|
| Identifier | Globally unique identifier shall uniquely identifies a tuple within the dataset and be formatted as an absolute URI conformant to RFC 3986. |
| Name | name contains a display name for the HydroGeoUnit. |
| | |
| gwUnitDescription | Description of the unit. |
| gwUnitMetadata | Metadata for the unit. |
| gwUnitName | Name of the unit (common local name or formal name). |
| | |
| gwUnitThickness | Typical thickness of the unit. |
| | |
| gwUnitMedia | Type of material or, by proximity, type of voids (e.g., granular, fracture, karstic, or mixed). |
| gwUnitRecharge | Volumetric flow rate of water that enters an hydrogeologic unit, at potentially multiple locations. |

| Property Name | Definition |
|---|---|
| gwUnitDischarge | Volumetric flow rate of water that goes out of an hydrogeologic unit, at potentially multiple locations. |
| gwUnitWaterBudget | Sum of water input and output of a hydrogeologic unit, at a particular point in time, with a description of inflows and outflows. |
| gwUnitVulnerability | The susceptibility of the aquifer to specific threats such as various physical events (earthquakes), human processes (depletion), etc. |
| | |
| gwUnitShape | The geometry of the unit. |
| | |
| LinkToAnObservationAPI | |
| | |
| PermeabilityHoz | value of horizontal permeability |
| Permeabilityvert | value of vertical permeability |
| Transmissivity | Parameter governing the flow of water flowing per unit width of the saturated zone of a continuous aquifer (measured in a direction orthogonal to that of flow), and per unit of hydraulic gradient |
| Storage coefficient | Ratio of the volume of water released or stored per unit area of an aquifer, to the corresponding charge in hydraulic head, without reference to time |
| Specific storage | Volume of water released or stored per unit area of an aquifer, to the corresponding charge in hydraulic head, without reference to time |
| EffectivePorosity | Ratio of the volume of gravitational water that a medium porous may contain in a state of saturation then release under the effect of complete drainage (laboratory drainage on a sample), to its total volume |
| | |
| IntrinsicPermeabilityDirection | direction of intrinsic permeability |
| IntrinsicPermeabilityValue | value of intrinsic permeability |
| HydraulicFlowVelocity | Fictitious macroscopic speed of a water flow in uniform movement through a saturated aquifer medium (speed vector of Darcy's law) deduced from the flow rate referred to the total section of the aquifer crossed by the flow |
| InitialWaterSaturation | In situ state, initial water saturation level (before construction) |

**FAQ**

**GitHub issue**

https://github.com/opengeospatial/Geotech/issues/13

### 3.6.9. FluidBody

These are distinct bodies of fluid (liquid or gas) that fill the voids in hydrogeological units. Fluid bodies are made of biologic (e.g., organisms), chemical (e.g., solutes), or material constituents (e.g., sediment). While it is expected that the major constituent of a fluid body will be water, the conceptual model allows for other types of major constituents such as petroleum. Minor constituents are not necessarily fluids, but can be gases, liquids, or solids (including organisms), and are included in the fluid body in various forms of mixture, such as solution, suspension, emulsion, and precipitates. Fluid bodies can also have other fluid bodies as parts, such as plumes or gas bubbles. Surfaces can be identified on a fluid body, such as a water table, piezometric or potentiometric surface, and some such surfaces can contain divides, which are lines projected to the fluid surface denoting divergence in the direction of flow systems within the fluid.

**Realizations**

| Data model | Concept name | Definition |
|---|---|---|
| OGC GWML2 | FluidBody | These are distinct bodies of fluid (liquid or gas) that fill the voids in hydrogeological units. Fluid bodies are made of biologic (e.g., organisms), chemical (e.g., solutes), or material constituents (e.g., sediment). While it is expected that the major constituent of a fluid body will be water, the conceptual model allows for other types of major constituents such as petroleum. Minor constituents are not necessarily fluids, but can be gases, liquids, or solids (including organisms), and are included in the fluid body in various forms of mixture, such as solution, suspension, emulsion, and precipitates. Fluid bodies can also have other fluid bodies as parts, such as plumes or gas bubbles. Surfaces can be identified on a fluid body, such as a water table, piezometric or potentiometric surface, and some such surfaces can contain divides, which are lines projected to the fluid surface denoting divergence in the direction of flow systems within the fluid. |
| IFC] | FluidBody | Same as OGC GWML2 |
| AGSi | agsiModelElement | In AGSi a model is collection of elements (agsiModelElement object) and this may include fluid bodies, identified as such using the elementType attribute. |

**FAQ**

#### GitHub issue

https://github.com/opengeospatial/Geotech/issues/15

### 3.6.10. FluidBodySurface

A surface on a fluid body within a local or regional area, e.g., piezometric, potentiometric, water table, salt wedge, etc.

#### Realizations

| Data model | Concept name | Definition |
|---|---|---|
| OGC GWML2 | FluidBodySurface | A surface on a fluid body within a local or regional area, e.g. piezometric, potentiometric, water table, salt wedge, etc. |
| IFC | PiezometricWaterLevel | Same as OGC GWML2 |
| AGSi | agsiModelElement | In AGSi a model is collection of elements (agsiModelElement object) and this may include a fluid body surface, e.g. piezometric surface, identified as such using the elementType attribute. |

#### FAQ

#### GitHub issue

https://github.com/opengeospatial/Geotech/issues/16

### 3.6.11. GeotechnicalUnit

A surface or a volume in which the mechanical behavior and other design-relevant characteristics are characterized using the same geotechnical parameters values. Several alternative classifications can be required in a project for different design tasks.

#### Realizations

| Data model | Concept name | Definition |
|---|---|---|
| IFC | GeotechUnit | A surface or a volume in which the mechanical behavior and other design-relevant characteristics are characterized using the same geotechnical parameters values. Several alternative classifications (=GeotechModels) can be required in a project for different design tasks. |
| OGC GeoSciML | GeotechUnit (as a GeologicUnit with Type = GeotechnicalUnit) | Same as IFC 4.4+ |
| DIGGS | GeoUnitObservation | Descriptions of a soil or rock where the unit is primarily defined by the value(s) or a value range of one or more physical or engineering properties. The criteria for distinguishing such units may vary depending on the engineering application. |
| AGSi | agsiModelElement | In AGSi a model is collection of elements (agsiModelElement object) and this may include geotechnical units, identified as such using the elementType attribute. |

### Properties

The GeotechUnit concept distinguish SoilLikeMaterial and IntactRockMaterial. Each having its own characteristics. The tables below list relevant properties for those types of GeotechUnit. Some being interpreted, calculated or derived from measurements.

#### Inherited properties from the GeologicUnit concept

See GeologicUnit

#### Relevant properties for soil like material

#### Relevant properties for intact rock

#### Discontinuities in rock

| Property name | Definition |
|---|---|
| RQD | Quotient of the cumulative length of cores with a length greater than 10 cm, by the total length of the core pass with a length greater than or equal to 1m |
| DiscontinuityType | pattern of bedding, folds, faults and discontinuities in rock masses, which subdivide the mass into individual domains or rock blocks (ISO14689) |
| DiscontinuityStrikeDirection | The geographic direction of a line created by the intersection of a plane and the horizontal plane. If non specific convention is used, this angular value is in the range 0 to 18 degrees. In this case, there is an ambiguity on the dip orientation. For example, a plane with an orientation of 90 degrees from the north could either have a dip direction to the North or to the South. Thus, to avoid this ambiguity, a strike value is generally completed with an indication of the dip orientation (Quadrant). |
| DiscontinuityDipAngle | The dip is the steepest angle of descent of a géological plane to a horizontal plane. It's value its in the range 0 to 90 degrees. |
| DiscontinuitySpacing | The term "spacing" refers to the mean or modal spacing of a set of discontinuities and is the perpendicular distance between adjacent discontinuities. The number of discontinuity sets, the differences in spacing and the angles between the sets shall be reported as these determine the block shape. The discontinuity spacing should be measured in millimeters and can be classified using the terms in Table 8. (ISO14689) |
| DiscontinuityPersistence | The linear extent of discontinuities from their inception to their termination in solid rock mass or against other discontinuities or outside the exposure shall be reported. The size of the exposure shall also be recorded. If possible and appropriate, measurements should be made in two or preferably three orthogonal directions. (ISO14689) |
| DiscontinuityAperture | The perpendicular distance between the two surfaces of a discontinuity is referred to as the aperture. (ISO14689) |

| Property name | Definition |
|---|---|
| DiscontinuityInfilling | The infilling material between discontinuity surfaces shall be identified and described (e.g., soil,minerals such as calcite, quartz, epidote, chlorite, anhydrite, clay gouge, rock gouge or breccia). (ISO14689) |
| DiscontinuityRoughness | The surface condition and the shape of discontinuities shall be described on the basis of three scales of observation, respectively, and using the terms given in Table 9 and illustrated in Figure 2: a) small scale (several millimeters) — smooth or rough; b) medium scale (several centimeters) — planar, stepped or undulating; c) large scale (several metres) — straight, curved or wavy. (ISO14689) |
| DiscontinuityWaterPresence | Descriptive of estimated waterinflow in excavation, to be used to define Jw(Barton), or RMR groudnwater inflow rating. |
| Cohesion | Cohesive shear strength of a rock or soil that is independent of interparticle friction. |
| FrictionAngle | Derived from the Mohr-Coulomb failure criterion and used to describe the friction shear resistance of ground materials, together with the normal effective stress. |

**Properties for rock mass**

**Extra properties**

| Property name | Definition |
|---|---|
| PwaveVelocity | Value of S wave (Shear wave (S)) measured through the unit. |
| SwaveVelocity | Value of P wave (pressure or primary elastic body wave (P)) measured through the unit. |
| Resistivity | Electrical resistivity of a rock or soil (Ohm-m). |
| GroundwaterTemperature | Temperature measured or assumed in the groundwater |
| WetBulkDensity | Ratio of the total mass to the unit total volume (material at its natural moisture content) |
| DryBulkDensity | Ratio of the mass of solid to the unit total volume (dry material) |

**FAQ**

**GitHub issue**

https://github.com/opengeospatial/Geotech/issues/22

## 3.6.12. DiscreteDiscontinuity

Any interruption of the continuity in the rock material with its attendant mechanical, hydraulic and thermal properties.

**Realizations**

| Data model | Concept name | Definition |
|---|---|---|
| IFC | DiscreteDiscontinuity | Any interruption of the continuity in the rock material with its attendant mechanical, hydraulic and thermal properties. |
| OGC GeosciML | Joint | Fracture across which there is no displacement at the scale of interest. |
| DIGGS | DiscontinuityObservation | Descriptions of faults, fractures and joints and their spacing, with attendant mechanical and hydraulic properties. |

**Properties**

| Property name | Definition |
|---|---|
| DiscontinuityType | pattern of bedding, folds, faults and discontinuities in rock masses, which subdivide the mass into individual domains or rock blocks (ISO14689) |
| DipDirection | The azimuth of the dip (dip direction) shall be measured in degrees in the range 0° to 360° counted clockwise from true north and expressed as a three-digit number, e.g., 240 or 015. (ISO14689) |
| DipAngle | The maximum declination (dip) of the mean plane of the discontinuity from the horizontal shall be measured with the clinometer in the range 0° to 90° and should be expressed in degrees as a two-digit number, e.g., 50. (ISO14689) |

| Property name | Definition |
| --- | --- |
| DiscontinuitySpacing | The term “spacing” refers to the mean or modal spacing of a set of discontinuities and is the perpendicular distance between adjacent discontinuities. The number of discontinuity sets, the differences in spacing and the angles between the sets shall be reported as these determine the block shape. The discontinuity spacing should be measured in millimeters and can be classified using the terms in Table 8. (ISO14689) |
| DiscontinuityPersistence | The linear extent of discontinuities from their inception to their termination in solid rock mass or against other discontinuities or outside the exposure shall be reported. The size of the exposure shall also be recorded. If possible and appropriate, measurements should be made in two or preferably three orthogonal directions. (ISO14689) |
| DiscontinuityRoughness | The surface condition and the shape of discontinuities shall be described on the basis of three scales of observation, respectively, and using the terms given in Table 9 and illustrated in Figure 2:<br>a) small scale (several millimeters) — smooth or rough;<br>b) medium scale (several centimeters) — planar, stepped or undulating;<br>c) large scale (several metres) — straight, curved or wavy. (ISO14689) |
| DiscontinuityAperture | The perpendicular distance between the two surfaces of a discontinuity is referred to as the aperture. (ISO14689) |
| DiscontinuityInfilling | The infilling material between discontinuity surfaces shall be identified and described (e.g., soil,minerals such as calcite, quartz, epidote, chlorite, anhydrite, clay gouge, rock gouge or breccia). (ISO14689) |
| DiscontinuityWaterSeepage | Free moisture or water flow visible at individual spots or from discontinuities |
| Cohesion | Cohesive shear strength of a rock or soil that is independent of interparticle friction. |

| Property name | Definition |
| --- | --- |
| FrictionAngle | Derived from the Mohr-Coulomb failure criterion and used to describe the friction shear resistance of ground materials, together with the normal effective stress. |

#### FAQ

##### How to deal with DiscontinuitySets in a GeotechnicalUnit?

A description of the sets of discontinuities can be described as properties of a GeotechnicalUnit.

##### GitHub issue

https://github.com/opengeospatial/Geotech/issues/23

### 3.6.13. Void

Voids are the spaces inside a unit (e.g., aquifer) or its material (e.g., the sandstone material of an aquifer), and might contain fluid bodies. Voids are differentiated from porosity, in that porosity is a ratio of void volume to total volume of unit plus voids, while voids are the spaces themselves. It is important to conceptually differentiate voids from units and their containers, in order to represent, for example, the volume of fractures, caves, or pores in a particular unit or its portion.

#### Realizations

| Data model | Concept name | Definition |
| --- | --- | --- |
| OGC GWML2 | HydroGeoVoid | Voids are the spaces inside a unit (e.g., aquifer) or its material (e.g., the sandstone material of an aquifer), and might contain fluid bodies. Voids are differentiated from porosity, in that porosity is a ratio of void volume to total volume of unit plus voids, while voids are the spaces themselves. It is important to conceptually differentiate voids from units and their containers, in order to represent, for example, the volume of fractures, caves, or pores in a particular unit or its portion. |

| Data model | Concept name | Definition |
|---|---|---|
| IFC 4.4+ | Void | Same as OGC GWML2 |

**Specializations (Types of Voids)**

| Source | Type | Link | Definition |
|---|---|---|---|
| BRGM | Registry | https://data.geoscience.fr/ncl/CaviTy | This register lists all the types of cavities. By cavity is meant an empty space in a rocky environment, of millimeter size (microcavities) to multidecametric (cave), filled with a gas or liquid phase. |

**FAQ**

**GitHub issue**

https://github.com/opengeospatial/Geotech/issues/24

## 3.6.14. HazardArea

Discrete spatial objects representing a natural hazard.

**Realizations**

| Data model | Concept name | Definition |
|---|---|---|
| OGC / INSPIRE Theme III: Natural Risk Zones | Hazard Area | Discrete spatial objects representing a natural hazard. |
| IFC | HazardArea | Same as OGC INSPIRE Natural Risk Zone |

**Properties**

| PropertyName | Definition |
|---|---|
| Hazard Type | A generic classification and a specific classification of the type of hazard. |

| PropertyName | Definition |
|---|---|
| LikelihoodOfOccurence | Likelihood is a general concept relating to the chance of an event occurring. |
| LevelOfIntensity | An expression of the magnitude or the intensity of a phenomenon. |
| validityPeriod | Future finite time frame where the hazard applies |
| DeterminationMethod | Different ways to delineate the perimeter of a hazard |

#### Existing codelist for HazardType

| Source | Type | Link | Definition |
|---|---|---|---|
| INSPIRE | Codelist | https://inspire.ec.europa.eu/codelist/NaturalHazardCategoryValue | A generic classification of types of natural hazards. |

#### FAQ

##### GitHub issue

https://github.com/opengeospatial/Geotech/issues/43

### 3.6.15. Surrounding Construction

Any other construction that may be impacted by the construction of the building or infrastructure.

#### Realizations

| Data model | Concept name | Definition |
|---|---|---|
| OGC LandInfra | Facility | Facilities include buildings and civil engineering works and their associated siteworks. Civil engineering works, or infrastructure facilities, are construction works comprising a structure, such as a dam, bridge, road, railway, runway, utilities, pipeline, or sewerage system, or are the result of operations such as dredging, earthwork, and geotechnical processes. A facility has a life cycle, including planning, design, construction, maintenance, operation, and removal phases The design and construction phases are typically performed as part of a project. There may be multiple such projects during the life cycle of the facility to enable phased construction and incremental improvement. |
| IFC | IFC_Bridge | |
| IFC | IFC_Tunnel | |
| IFC | IFC_Road | |
| IFC | IFC_Building | |

And all other relevant facilities and construction the construction would like to pay attention too.

**Properties**

| PropertyName | Definition |
|---|---|
| assessmentOfVulnerability | Assessment of the vulnerability of the exposed element. |

**Existing codelist for HazardType**

| Source | Type | Link | Definition |
| --- | --- | --- | --- |
| INSPIRE | Codelist | https://inspire.ec.europa.eu/codelist/NaturalHazardCategoryValue | A generic classification of types of natural hazards. |

**FAQ**

**GitHub issue**

https://github.com/opengeospatial/Geotech/issues/43

## 3.7. Book C Principles

Book C starts with a detailed description of the feedback (in civil engineering terms) from prior structures built near or adjacent to similar terrain.

Book C then goes on to summarize the decisive elements – among all those set out in Books B – that have dictated the foundational design choices.

Book C also details the recommended execution methods for each part of the structure 'Book C' should also include details of the overall strategy suggested by the project manager for all matters relating to excavated materials A specific chapter should be devoted to neighboring structures Lastly, Book C should present a broad overview of the monitoring methods.

Book C end with the Risk register resulting from the intersection between the uncertainties listed in Book B and the envisaged construction methods.

### 3.7.1. Application

The book C is (linear) infrastructure focused. It provides a summary of the expected conditions and possible risks that may occur during the project construction.

The information that are provided are deeply linked to the targeted infrastructure and its design. Changes on those aspects may considerably alter the validity of the results that are provided.

Remembering AFTES guidelines are tunnel focused, slight adaptation were adopted to make the proposal also applicable to other kind of infrastructure.

### 3.7.2. Concepts list

The core of Book C is the GeotechSynthesisModel

That is is mainly composed of:

- Alignment
- GeotechTypicalSection
- RiskZone

### 3.7.3. GeotechSynthesisModel

Link between the design and modeled geology and geotechnical conditions: summarized interpretation with regard to building, construction method,… in relation to a section of the alignment or building structure. Typical definition of “baseline conditions” as usually included in a geotech.

#### Realization

| Data model | Concept name | Definition |
|---|---|---|
| OGC EPOS | ModelView | Same as IFC |
| IFC | GeotechSynthesisModel | Link between the design and modeled geology and geotechnical conditions: summarized interpretation with regard to building, construction method, in relation to a section of the alignment or building structure. Typical definition of "baseline conditions" as usually included in a geotech. longitudinal section |
| AGSi | agsiModel | A digital geometric (1D, 2D or 3D) representation of the ground. There are potentially many different types of model covering different categories (conceptual, observational, analytical) and domains (geological, geotechnical, hydrogeological, geoenvironmental, etc.). |

#### FAQ

##### Github issue

https://github.com/opengeospatial/Geotech/issues/26

### 3.7.4. Alignment

PositioningElement which provides a Linear Referencing System for locating PhysicalElements. An Alignment shall be continuous, non-branching, and non-overlapping. If it is a Project Alignment, it is for a single alternative, as specified by its owning ProjectPart.

Note: this concept has been commonly defined between OGC LandInfra and bSI.

#### Realizations

| Data model | Concept name | Definition |
| --- | --- | --- |
| OGC LandInfra | Alignment | PositioningElement which provides a Linear Referencing System for locating PhysicalElements. An Alignment shall be continuous, non-branching, and non-overlapping. If it is a Project Alignment, it is for a single alternative, as specified by its owning ProjectPart. |
| IFC 4.4+ | Alignment | Same as OGC LandInfra |

#### FAQ

##### GitHub issue

https://github.com/opengeospatial/Geotech/issues/44

### 3.7.5. GeotechTypicalSection

Interval along the alignment/building structure with similar ground conditions, as part of the GeotechSynthesis model that represents the connection between the ground model and the building. Includes key-properties like expected distribution of ground types (reference to GeotechUnits) and baseline-definition of expected ground conditions and potential hazards, and may also include key-information on design like excavation measures, distribution of support types etc.

#### Realizations

| Data model | Concept name | Definition |
|---|---|---|
| IFC | GeotechTypicalSection | Interval along the alignment/building structure with similar ground conditions, as part of the GeotechSynthesis model that represents the connection between the ground model and the building. Includes key-properties like expected distribution of ground types (reference to GeotechUnits) and baseline-definition of expected ground conditions and potential hazards, and may also include key-information on design like excavation measures, distribution of support types etc. |
| OGC LandInfra | FacilityPart | Land and Infrastructure facility, such as a road or bridge. A facility has a life cycle, including planning, design, construction, maintenance, operation, and removal phases The design and construction phases are typically performed as part of a project. There may be multiple such projects during the life cycle of the facility to enable phased construction and incremental improvement. |
| ISO 19148 | FeatureEvent | The FeatureEvent interface is used to linearly reference a feature along a LinearElement and possibly further qualified as applying at an instant in, or during a period of, time. |

Note: the GeotechTypicalSection can be seen on different aspect and then have multiple realizations: - a part of a tunnel (a part of a facility) - an event along an alignment.

**Properties**

| PropertyName | Definition |
|---|---|
| ExpGeotechUnit | Representative GeotecUnit expected in the subject typical section |
| ExpDistGeotechUnits | Expected distribution of project related “geotechnical units”(= ground types) in subject tunnel section |
| DiscontinuitySettings | Expected Sets of discontinuities, with typical orientation and properties like spacing, persistence etc. |
| Overburden | Range of overburden (distane from tunnel to ground surface) in the subject section |
| GroundwaterDesignPressure | Groundwater pressure to be considered in the design |
| GroundwaterInflowSum | Groundwater inflow measured or expected during a certain time span |
| GroundwaterAgressiveness | Rating of the aggressiveness e.g., towards concrete |
| GroundwaterTemperature | Groundwater temperature observed at specific location or expected in the subject section |
| Karst | Description of expected or observed karst features |
| Boulders | Description of expected or observed boulders features |
| GasType | Type of natural gas occurrence in subject section |
| GasConcentration | Expected concentration of gas to be encountered during excavation |
| DumpCategory | Project-related category for re-use and storage of the excavated material |
| Contamination | Expected anthropogenic contaminations |
| InSituStress | Anisotropic stress regime, with direction and magnitude of principal stresses |
| Swelling | Expected swelling behavior |
| GroundTemperature | Ground temperature expected in the subject section |
| CloggingPotential | Description or rating of the potential for clogging equipment in sticky ground |
| AsbestosPotential | Description or rating of the potential presence of asbestos in the ground |
| HeavyMetalsPotential | Description or rating of the potential presence of heavy metals in the ground |
| RiskIdentification | Name and Description of the risk |
| RiskNature | Description of the nature of the risk |
| | |
| RiskSource | Description of the source of the risk related to a specific uncertainty |

| PropertyName | Definition |
| --- | --- |
| Event | Occurrence or change of a particular set of circumstances |
| Likelihood | chance of something happening |
| Consequences | outcome of an event affecting the objectives |
| LevelOfRisk | Magnitude of a risk or combination of risks, expressed in terms of the combination of consequences and their likelihood |
| PreventiveTreatment | Description of the preventive treatment |
| LevelOfResidualRisk | Description of the preventive treatment |
| CurativeTreatment | Description of the curative treatment |

#### FAQ

##### GitHub issue

https://github.com/opengeospatial/Geotech/issues/44

### 3.7.6. RiskZone

Discrete spatial objects representing the spatial extent of a combination of the consequences of an event (hazard) and the associated probability/likelihood of its occurrence.

#### Realizations

| Data model | Concept name | Definition |
| --- | --- | --- |
| OGC / INSPIRE Theme III: Natural Risk Zones | RiskZone | Discrete spatial objects representing the spatial extent of a combination of the consequences of an event (hazard) and the associated probability/likelihood of its occurrence. |
| IFC | GeotechTypicalSection | |

Note : In IFC, Risks are infrastructure (tunnel, bridge) focused and their expression is limited to the GeotechTypicalSection

#### Properties

| Property Name | Definition |
|---|---|
| LevelOrIntensity | The level of risk is an assessment of the combination of the consequences of an event (hazard) and the associated probability/likelihood of the occurrence of the event. |

### FAQ

#### GitHub issue

https://github.com/opengeospatial/Geotech/issues/43

# Chapter 4. OGC and ISO models

## 4.1. General mapping

These figures propose an overview of how the geotech concepts are proposed to be realized with existing ISO and OGC standards (with a blue border) but also IFC (with a red border).

Details about the mapping are then provided concept per concept.

### 4.1.1. Model Typology

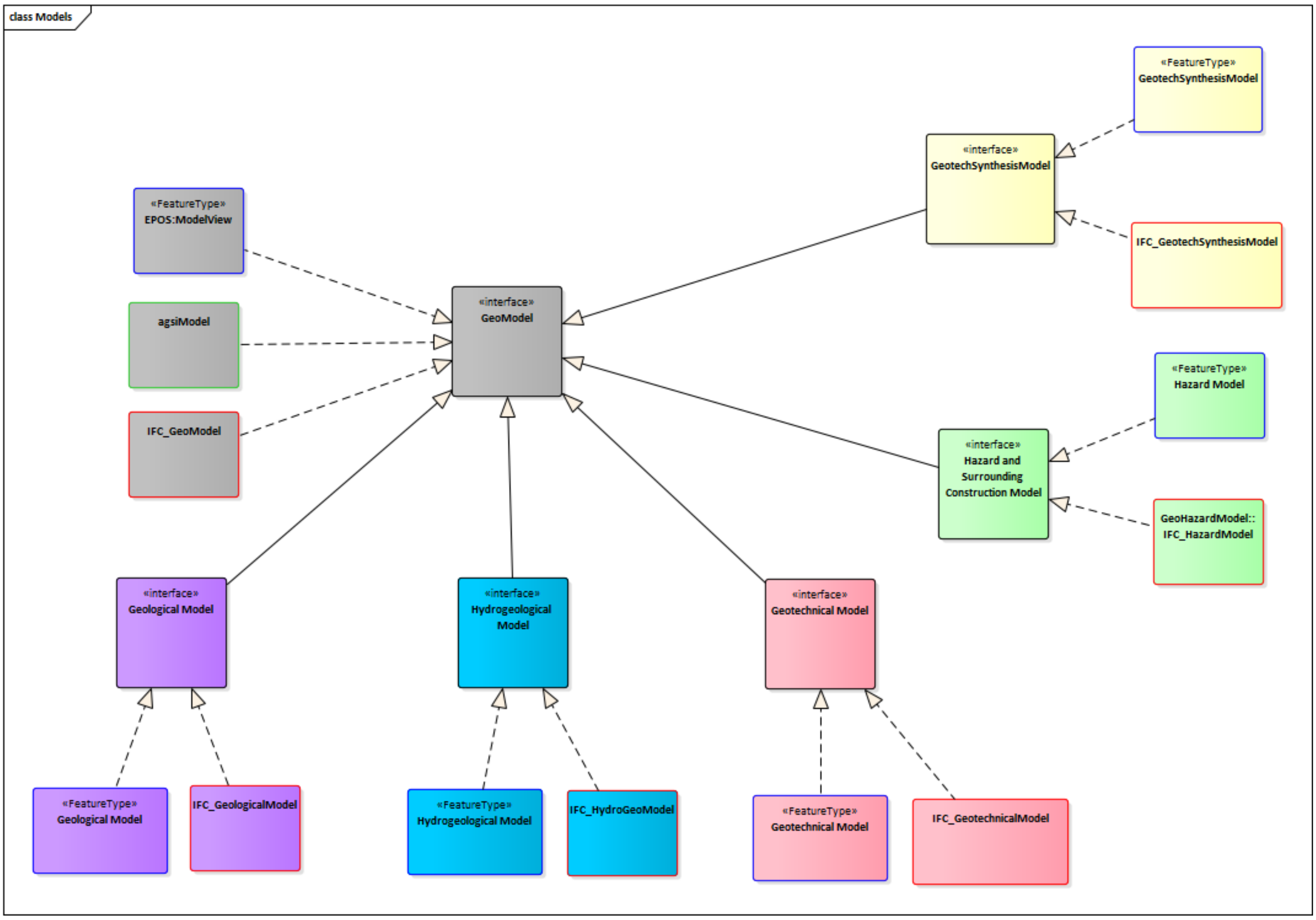

*Figure 2. Models*

### 4.1.2. For Geological Modeling realization

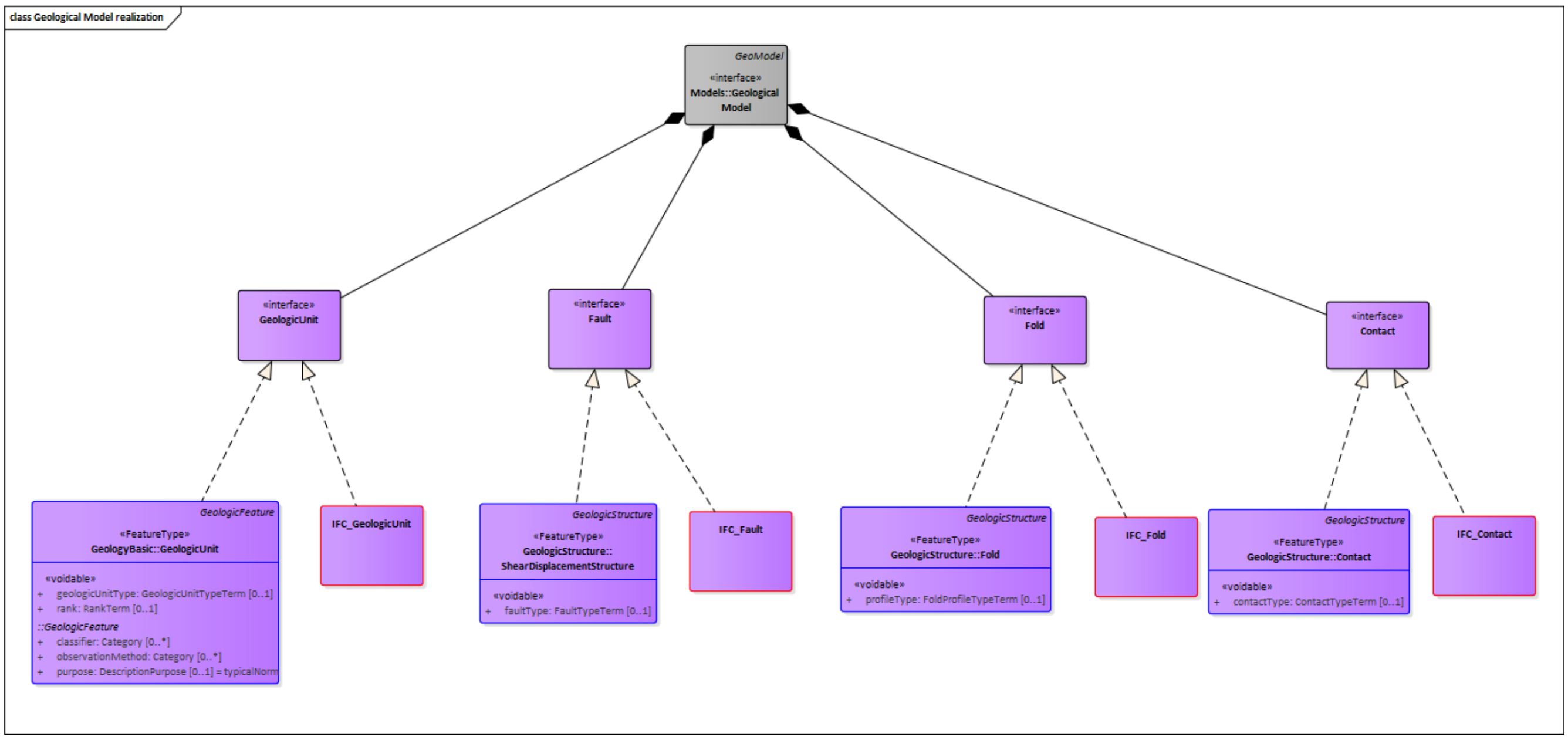

*Figure 3. Geological Model realization*

## 4.1.3. For Hydrogeological Modeling

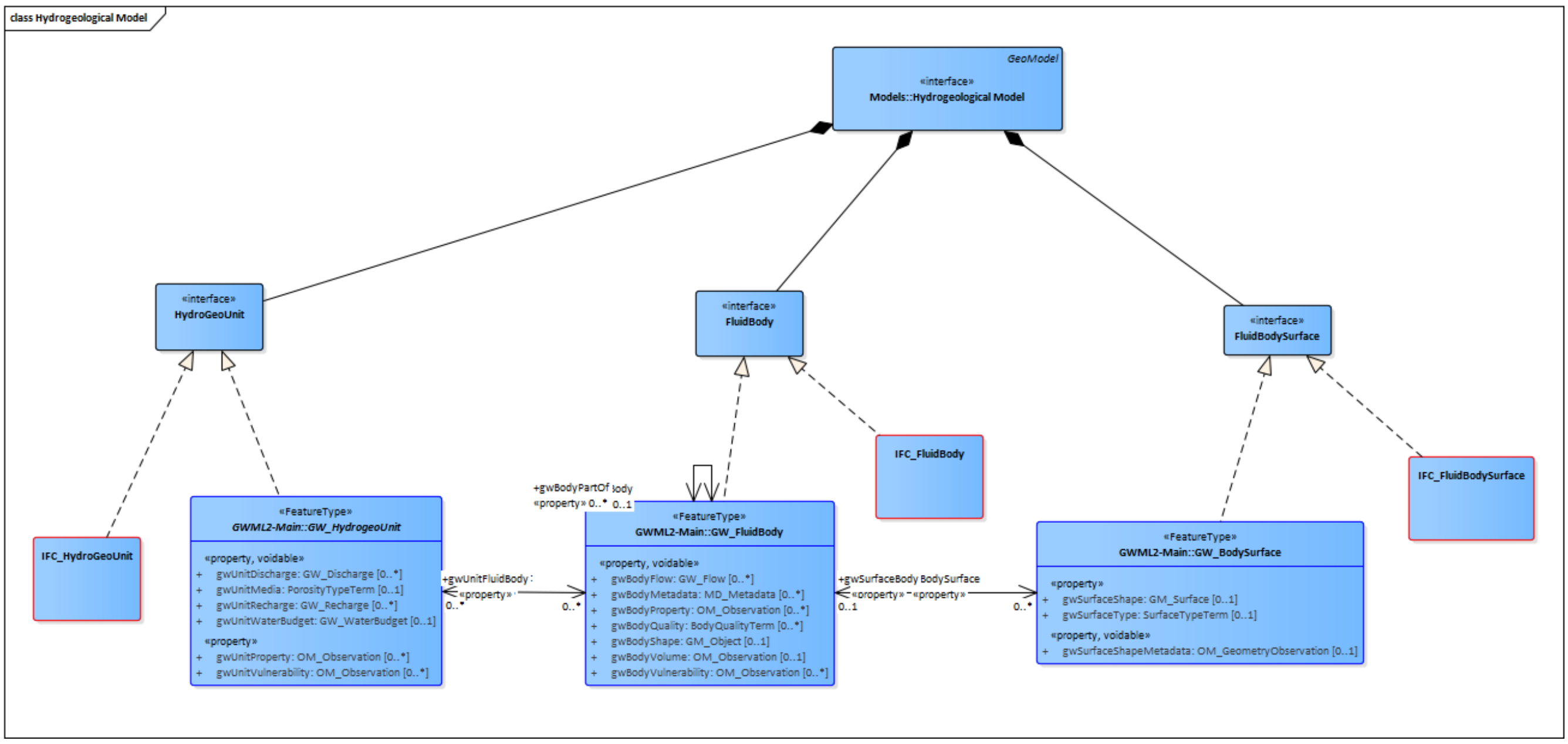

*Figure 4. Hydrogeological Model*

## 4.1.4. For Geotechnical Modeling

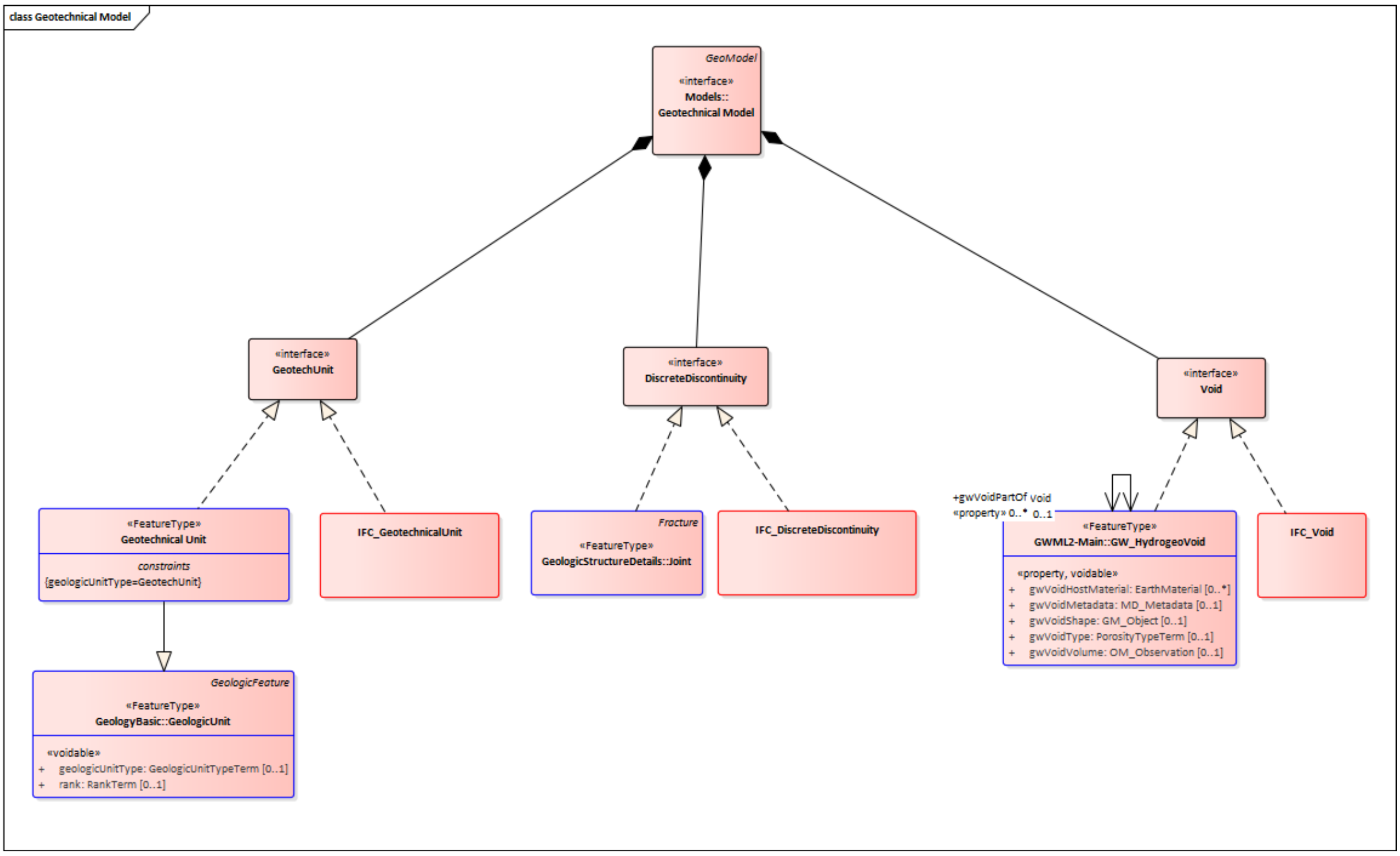

*Figure 5. Geotechnical Model*

## 4.1.5. For Hazard and Surrounding Construction Modeling

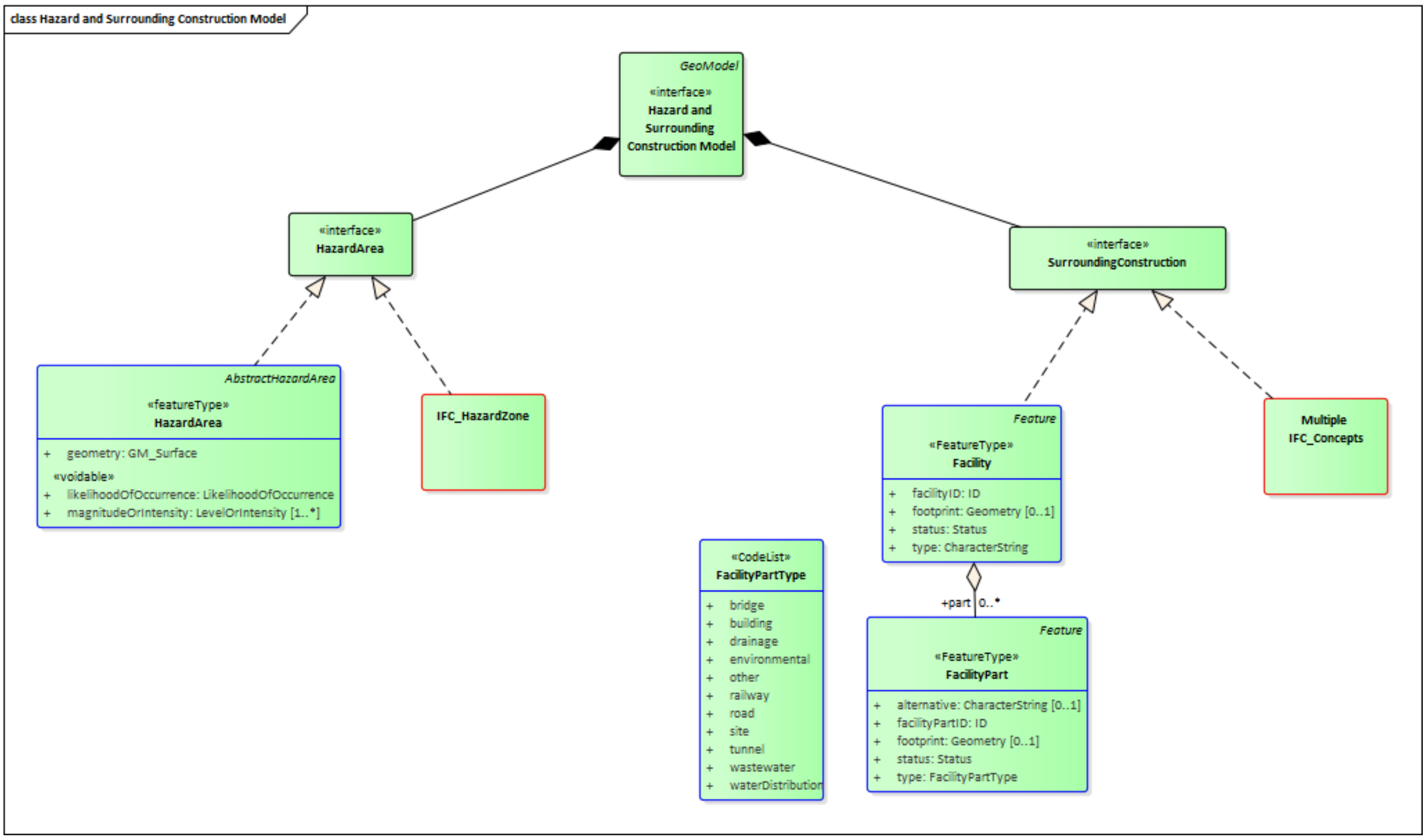

*Figure 6. Hazard and Surrounding Construction Model*

## 4.1.6. For Book C

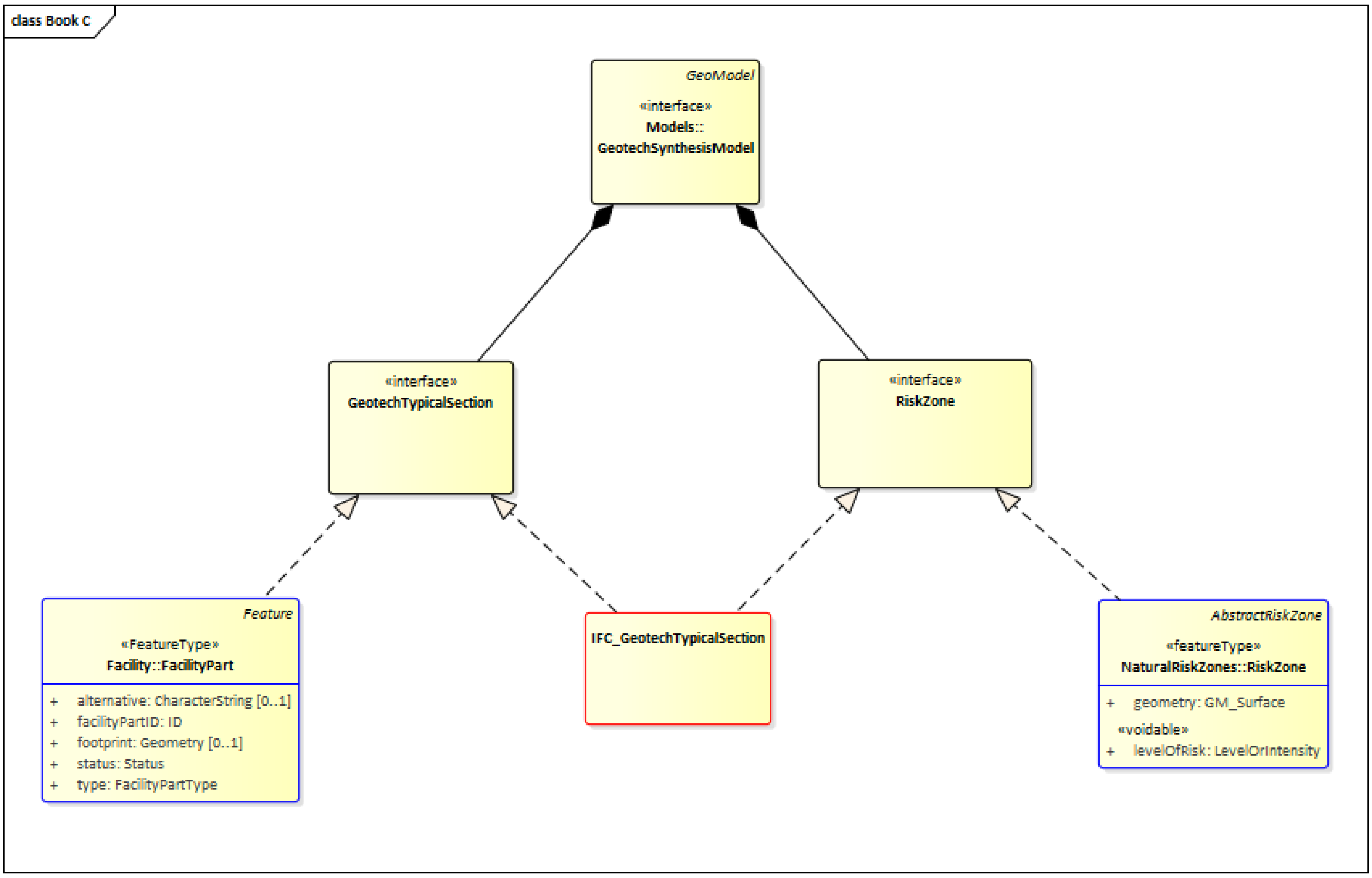

*Figure 7. Book C*

## 4.1.7. Mapping table summary

| AFTES Book | Geotech concept | OGC + EPOS + INSPIRE | With OGC SensorThingsAPI |
|---|---|---|---|
| A | Borehole | gsml:Borehole | Either BhCollarThing or BhTrajectoryThing |
| A | BoreholeCollar | gwml2:BoreCollar | BhCollarThing |
| A | BoreholeTrajectory | | BhTrajectoryThing |
| A | The whole borehole | | BhFeatureOfInterest (with BhFeatureType = Hole & Entirety) |
| A | A part of the borehole | | BhFeatureOfInterest (with BhFeatureType = Hole & Segment) |
| A | A location in the borehole | | BhFeatureOfInterest (with BhFeatureType = Hole & Point) |
| A | MaterialSample | oms://MaterialSample/[oms:MaterialSample] | BhFeatureOfInterest |
| A | Specimen | oms://MaterialSample/[oms:MaterialSample] | BhFeatureOfInterest (with BhFeatureType = Specimen) |
| A | BoreholeCore | oms://MaterialSample/[oms:MaterialSample] | BhFeatureOfInterest (with BhFeatureType = Core & Entirety) |

| AFTES Book | Geotech concept | OGC + EPOS + INSPIRE | With OGC SensorThingsAPI |
|---|---|---|---|
| A | A part of the borehole core | oms://MaterialSample/[oms:MaterialSample] | BhFeatureOfInterest (with BhFeatureType = Core & Segment or Point) |
| A | A location of the borehole core | oms://MaterialSample/[oms:MaterialSample] | BhFeatureOfInterest (with BhFeatureType = Core & Point) |
| A | EnvironmentalMonitoringFacility | ef:environmentalMonitoringFacility | Sensor |
| A | Observation | oms:observation | |
| A | Geotech test | oms:observationCollection | Sensor |
| A | Series of observations | oms:observationCollection | DataStream |
| A | Individual observation | oms:observation | Observation |
| A | Sampling | oms://Sampling/[oms:Sampling] | BhSampling |
| B | Geomodel | epos:ModelView | |
| B | GeologicUnit | gsml:GeologicUnit (with GeologicUnitType = LithoStratigraphicUnit)] | BhFeatureOfInterest |
| B | Contact | gsml:Contact | BhFeatureOfInterest |
| B | Fold | gsml:Fold | BhFeatureOfInterest |
| B | Hydrogeounit | gwml2:HydroGeoUnit | BhFeatureOfInterest |
| B | FluidBody | gwml2:FluidBody | BhFeatureOfInterest |
| B | FluidBodySurface | gwml2:FluidBodySurface | BhFeatureOfInterest |
| B | Fault | gsml:ShearDisplacementStructure | BhFeatureOfInterest |
| B | GeotechnicalUnit | gsml:GeologicUnit (with GeologicUnitType = GeotechUnit)] | BhFeatureOfInterest |
| B | DiscreteDiscontinuity | gsml:Joint | BhFeatureOfInterest |
| B | Void | gwml2:HydroGeoVoid | BhFeatureOfInterest |
| B | HazardArea | nz:HazardArea | BhFeatureOfInterest |
| C | Alignment | landinfra:Alignment | BhFeatureOfInterest |
| C | GeotechSynthesisModel | epos:ModelView | BhFeatureOfInterest |

| AFTES Book | Geotech concept | OGC + EPOS + INSPIRE | With OGC SensorThingsAPI |
|---|---|---|---|
| C | GeotechTypical Section | landinfra:FacilityPart | BhFeatureOfInterest |
| C | RiskZone | nz:RiskZone | BhFeatureOfInterest |

## 4.2. General considerations

### 4.2.1. Borehole Interoperability Experiment

Following the birth of the OGC GeoScience DWG in September 2017, an Interoperability Experiment about Borehole data was launched.

Starting in March 2018, during the Orleans OGC TC, the GeoScience DWG work on how connect several existing borehole definitions.

More details about this initiative can be found on the Borehole IE Github.

#### Engineering report

This document describes a conceptual model, logical model, and GML/XML encoding schema for the exchange of borehole related data and especially all the elements that are positioned along a borehole trajectory. In addition, this document provides GML/XML encoding instances documents for guidance

Link to the OGC engineering report: https://hal.science/hal-03943388/

#### Sampling Boreholes

Before one can provide data for various tests to be performed on boreholes, one must determine what spatial features must be defined to serve as features-of-interest for these tests. We sStart by differentiating between the Hole (the absence) and the Core in their entirety. Both the Hole and the Core can then be subdivided, either as Points or as Segments. Dedicated featureTypes have been created for all six of these variants, as illustrated in the diagram below.

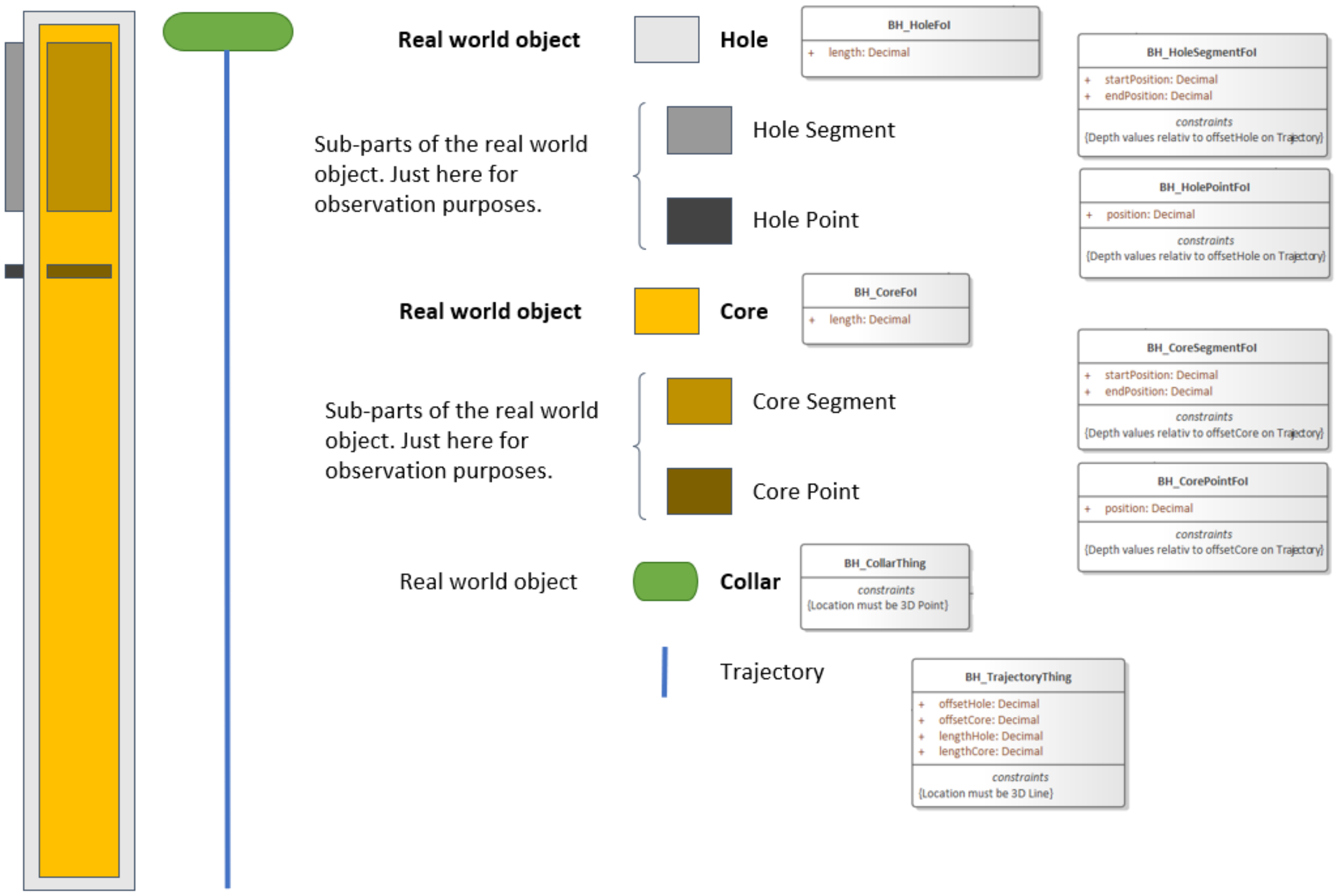

*Figure 8. SF_BoreholeClasses*

As linear referencing is a key function for the provision of borehole observation, it was then proposed to connect ISO19148 (linear referencing) and ISO 19156 (observations, measurements and samples).

### 4.2.2. Geometries in OGC

This page is about how geometries are currently handled by OGC standards. OGC Standards consider two- and three-dimensional (2D and 3D) geometries. All OGC feature (vector) geometries ultimately are based on the OGC Abstract Specification Topic 1 - Spatial Schema (published as ISO 19107:2019).

#### Simple Features (2D)

OGC's oldest and most-implemented Standard, Simple Features, describes how to define and encode 2D feature geometries. Simple Features representation is used in all other OGC and similar Standards for feature geometry, including the list below. Simple Features is also the native geometry for geospatial-enabled commercial and open source database systems.

#### GML (2D and 3D)

As written here, the Geography Markup Language (GML) from OGC is

> an XML grammar for expressing geographical features. GML serves as a modeling language for geographic systems as well as an open interchange

> format for geographic transactions on the Internet. GML is also an ISO standard (ISO 19136:2007).

#### GeoJSON (2D)

Contrary to some belief, GeoJSON is not authored by OGC but by the IETF. One current limitation of GeoJSON is the fact that it is only handling one coordinate reference system (CRS): EPSG:4326. Yet the big interest of the internet for JSON based format makes it worth considering.

#### OGC Features and Geometries JSON (2D and 3D)

The work-in-progress OGC Features and Geometries JSON is designed to be the successor of GeoJSON. It is designed such that it can be used in a way that is fully backwards-compatible with GeoJSON. It does this by leaving all fields and structures defined by GeoJSON in place, and only defining new functionality in new fields that do not conflict with existing properties.

OGC Features and Geometries JSON will:

- include the ability to use Coordinate Reference Systems (CRSs) other than WGS84
- follow the OGC Axis Order Policy,
- allow the use of non-Euclidean metrics, in particular ellipsoidal metrics,
- support solids and multi-solids as geometry types, and
- provide guidance on how to represent feature properties, e.g., including temporal properties.

The development of FG-JSON happens on: https://github.com/opengeospatial/ogc-feat-geo-json

#### Geo3DML (3D)

In Geo3DML: A standard-based exchange format for 3D geological models, the China Geological Survey studied the topic of 3D space geomodels. They proposed extensions of OGC standards are proposed, especially for geometries.

#### Voxels (3D)

Voxels (or volumetric cells) are a 3D gridded representation of a volume, with attributed values. Voxels can be expressed directly as modeled cubes in Standards such as CityGML or as datacubes represented by coverages or in formats such as HDF5 or Zarr, amongst others.

#### The concept of gsml:MappedFeature in GeoSciML

The fact that models of the earth and their features can be represented in different way is something well understood and managed in GeoSciML thanks to the concept of MappedFeature that is defined this way:

> A MappedFeature is part of a geological interpretation. It provides a link between a notional feature (description package) and one spatial representation of it, or part of it (exposures, surface traces and intercepts,

> etc.). The mapped features are the elements that compose a map, a cross-section, a borehole log, or any other representation. The mappingFrame identifies the domain being mapped by the geometries. For typical geological maps, the mapping frame is the surface of the earth (the 2.5D interface between the surface of the bedrock and whatever sits on it; atmosphere or overburden material for bedrock maps). It can also be abstract frames, such as the arbitrary plane that forms a mine level or a cross-section, the 3D volume enclosing an ore body or the line that approximate the path of a borehole.

### 4.2.3. The distinction between feature properties and observations.

**Definitions**

A video introduction by Kathi Schleidt of the observation concept can be found here: https://www.youtube.com/watch?v=bYDSgs2fKLk

It notably proposes an explanation of what is an observation vs a property of a feature. The picture below makes it clear.

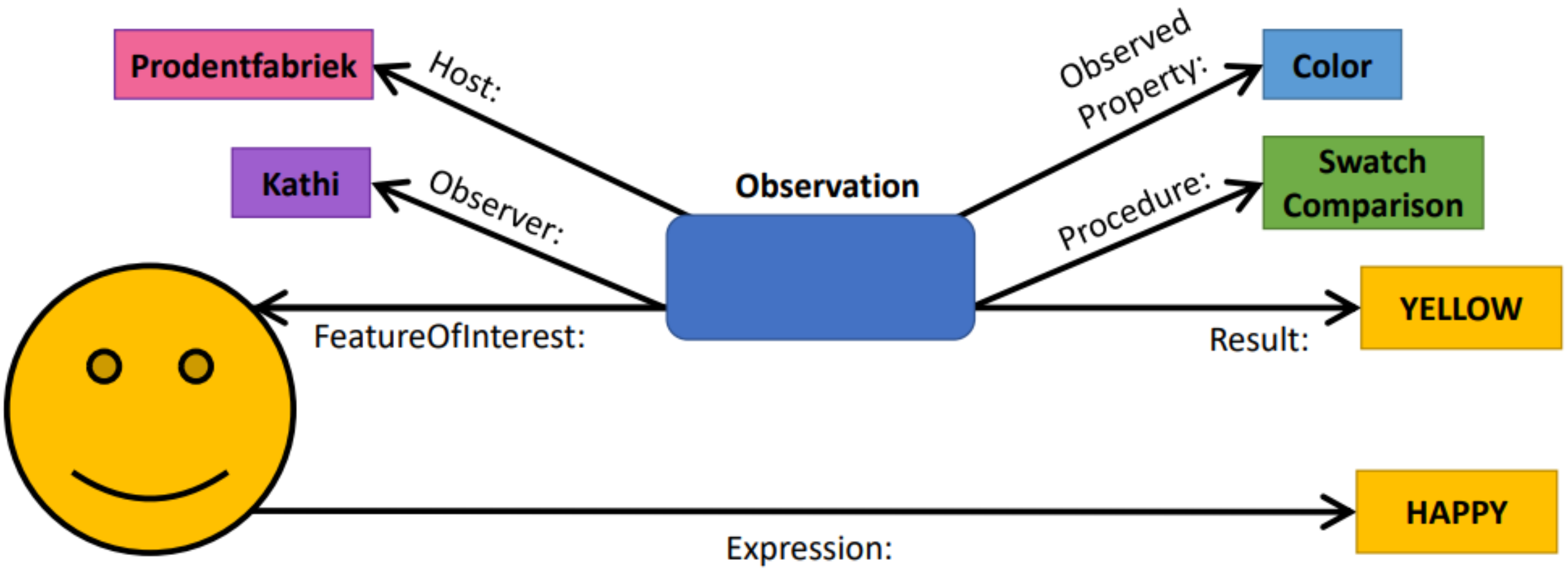

*Figure 9. Feature property and Observation*

**Feature property**: one object called the feature (here the smily face) is identified and is associated to a property (here Expression = Happy)

**Observation**: a dedicated object called the observation is set. It is more exactly an act which itself have some properties that enable to define the context and also the result of the observation act (here that the Smily Face looks to be of a yellow color according to Kathy’s experience).

**Feature properties vs observations**

Feature properties are a direct way to associate an information to an object. The association is “absolute” in the sense that it is always true, no matter when, who/what is involved in describing the object. For example, the name of a person, its date of birth.

On the contrary, the observation concept offer the capacity to contextualize an information that is provided. For example, the time, the location, the method, the observer: that is to say conditions in which the information was obtained and may have an important impact on the validity of the result.

For those reasons one can see that:

- feature properties perfectly fit domain where the knowledge of the objects are well mastered. This is typically the case in **industry** that propose to build from hand or machine made well known components.
- observation is very relevant in domain where the knowledge is incomplete, or when only a small part (either in terms of space or time) of an object is studied. This is typically the case in **science** where each information must be kept in its context before (possibly) deducting a more general law or rule.

#### When to use one or the other in geotechnics

Information obtained through a process of observation, measurement, interpretation or even calculus fit the concept of observation and shall be exposed that way.

#### How to associate observations to features?

A basic need for semanticized objects is to be able to associate them with information.

#### Approach 1: (Direct) Feature properties, as in GeoSciML Lite

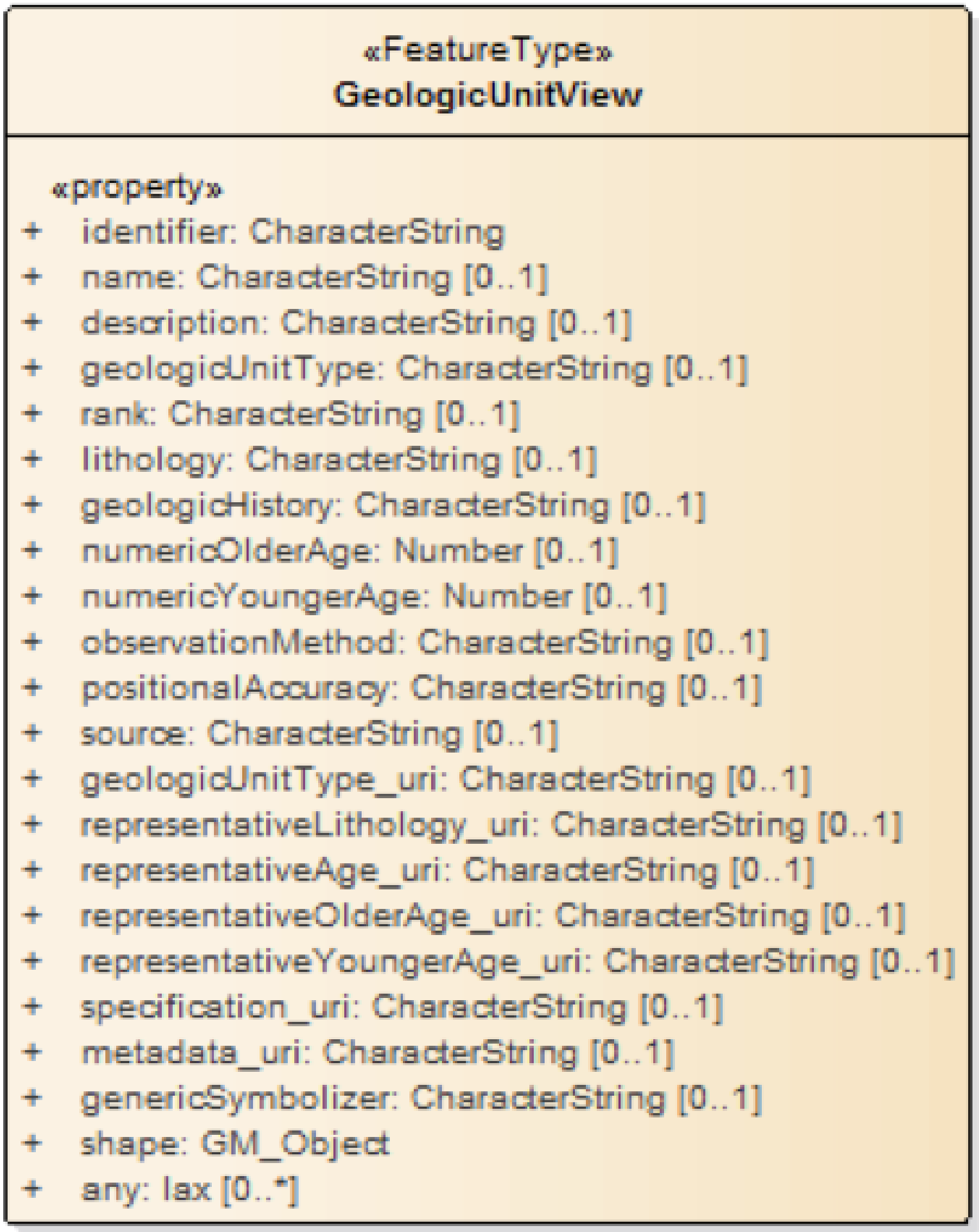

*Figure 10. In this simplified version of GeoSciML, a direct association between the "ObservedProperty" and the "Result" is provided.*

**Approach 2: Association of one ObservedProperty to a Measurement, as in GroundWaterML2**

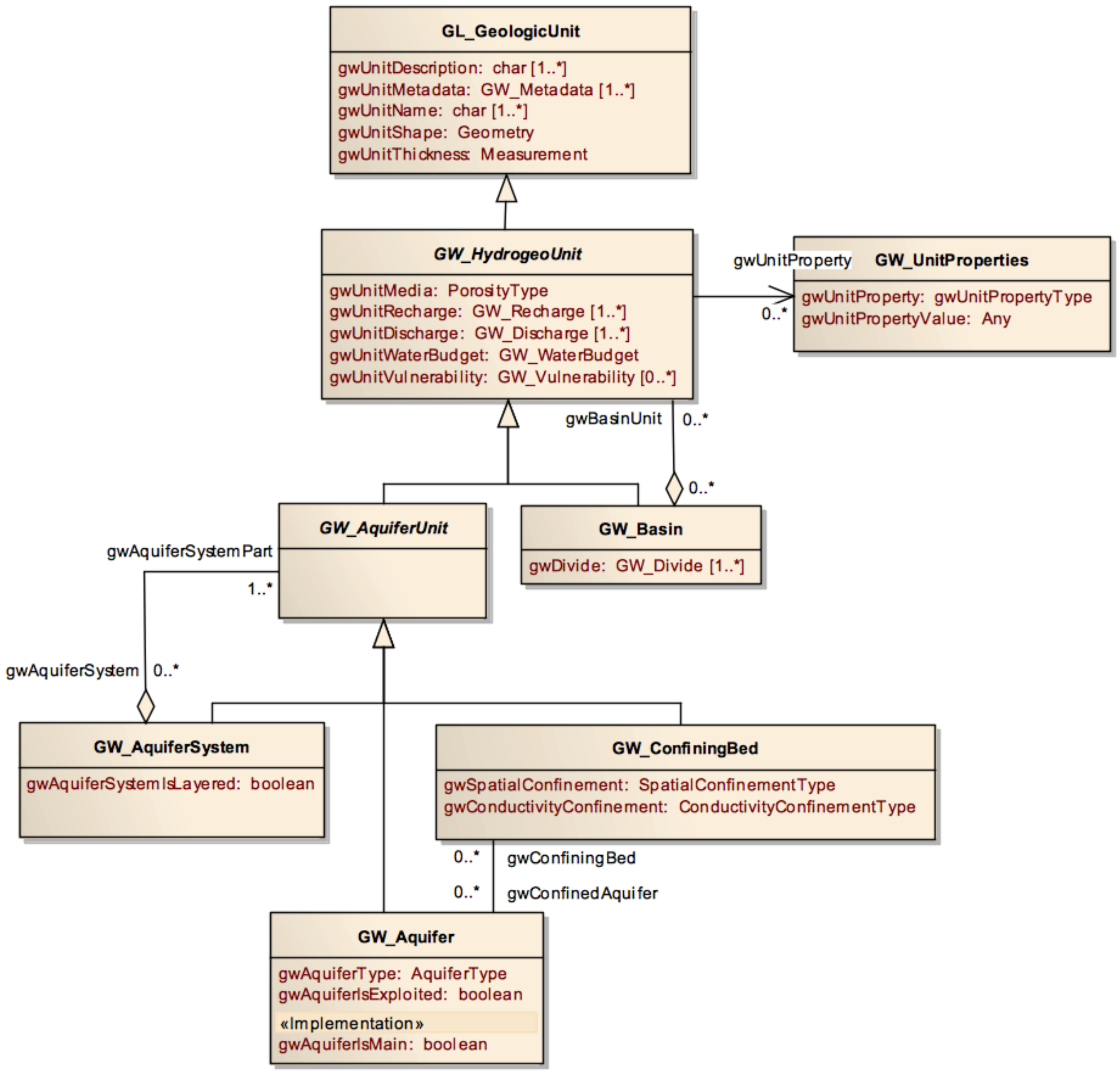

*Figure 11. Here the gwUnitThickness of the GL_GeologicUnit is intended to be provided as a Measurement, following ISO19156.*

To realize this association, the “ObservedProperty” shall be associated to the identifier of a Measurement Object. A clean way to do that is to have the measurement exposed through an OGC API and use the link to it.

The picture below illustrates measurement with a link to a WFS service.

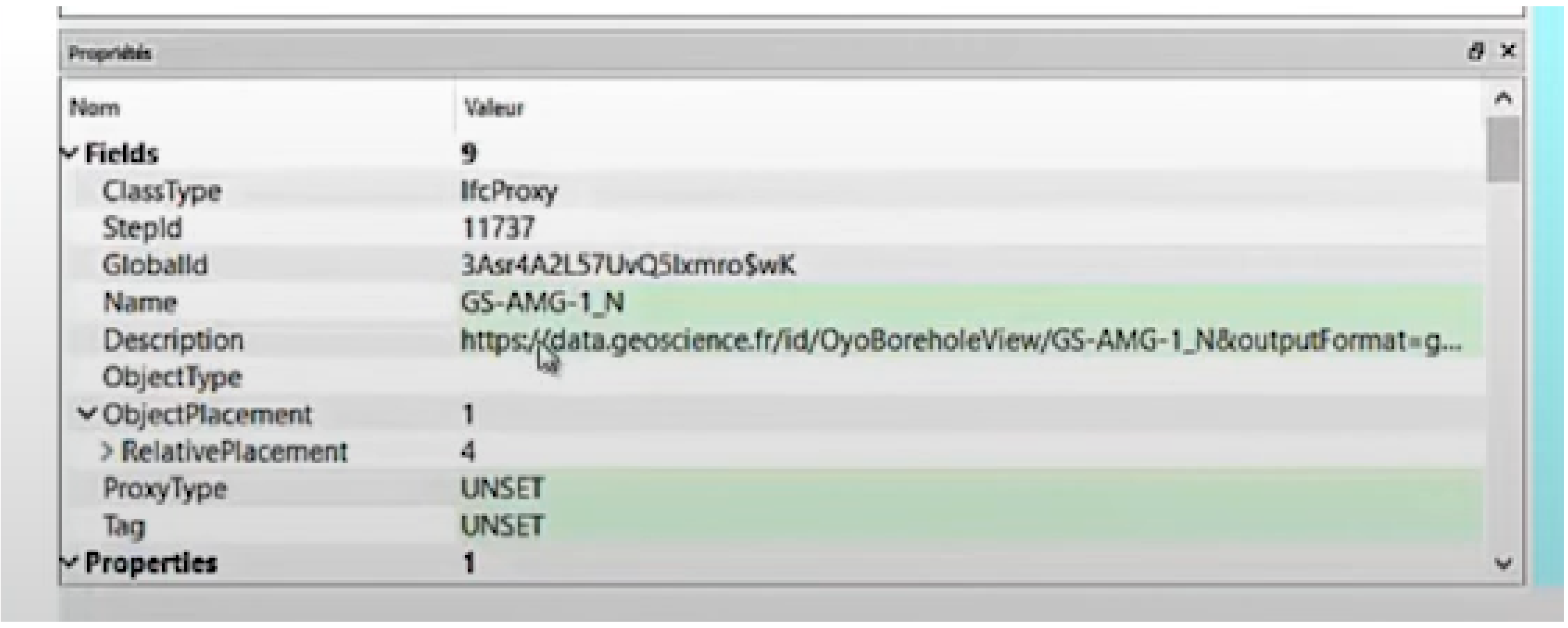

*Figure 12. Extracted from https://www.youtube.com/watch?v=L1L2u3u35cY*

**Approach 3 : Association to an OGC API**

A drawback of the second approach is that one needs to have all the ObservedProperties described in the feature and to associate the results one by one.

A solution to avoid this constraint is to directly associate the feature to an “Observation Oriented API” that will list the available observations associated to the feature.

The following request to an [Introduction-to-SensorThingsAPI-data-model[OGC SensorThingsAPI]] enable to list the “DataStreams”, that is to say collection of observations associated to the Object of interest (here a borehole or BhTrajectoryThing):

> https://ogc-demo.k8s.ilt-dmz.iosb.fraunhofer.de/FROST-GeoTech/v1.1/BhTrajectoryThings(11)/Datastreams

This is the preferred approach for this IE.

# 4.3. ISO 19148 and ISO 19156

## 4.3.1. Introduction

For the data modelling on the GeoTech IE, in addition to the basic OGC geospatial standards, the following two standards where of core interest, strongly guiding the final data model.

**ISO 19148**

Linear Referencing Systems enable the specification of positions along linear objects. The approach is based upon the Generalized Model for Linear Referencing[12] first standardized within ISO 19133:2005. This document extends that which was included in ISO 19133, both in functionality and explanation.

ISO 19109 supports features representing discrete objects with attributes with values which apply to the entire feature. ISO 19123 allows the attribute value to vary, depending upon the location within a feature, but does not support the assignment of attribute values to a single point or length along a linear feature. Linearly located events provide the mechanism for specifying attribution of linear objects when the attribute value varies along the length of a linear feature. A Linear

Referencing System is used to specify where along the linear object each attribute value applies. The same mechanism can be used to specify where along a linear object another object is located, such as guardrail or a traffic accident.

It is common practice to segment a linear object with linearly located events, based upon one or more of its attributes. The resultant linear segments are attributed with just the attributes used in the segmentation process, ensuring that the linear segments are homogeneous in value for these segmenting attributes.

**ISO 19156**

The Observations, Measurements, and Samples standard (OMS), jointly prepared and published by the Open Geospatial Consortium and ISO/TC 211 as OGC Abstract Specification Topic 20 (OGC 20-082r4) and ISO 19156:2023, defines a conceptual schema for observations, for features involved in the observation process, and for features involved in sampling when making observations. Models support the exchange of information describing observation acts and their results, both within and between different scientific and technical communities.

Observations commonly involve sampling of an ultimate feature-of-interest. OMS defines a common set of sample types according to their spatial, material (for ex situ observations), or statistical nature. The Standard's schema includes relationships between sample features (sub-sampling, derived samples). It also adds concepts that were deemed missing in the previous version and provides additional clarification to the provided concepts and their relationships while keeping the core data model mostly intact. In addition, a new fine-grained requirements class structure has been created, enabling implementations to unambiguously declare the parts of the standard they conform to.

The abstract data models described in OMS provide common concepts and logical structures for exchanging observational data and metadata between various information systems as well as for harmonized handling of such information from various heterogeneous sources. Technical implementation standards and profiles, on the other hand, provide concrete solutions tailored for storing, exchanging and processing OMS information in particular technical environments and use cases.

The work on revising existing OMS-related OGC Implementation Standards to fully comply with the OGC 20-082r4 requirements is in progress in OGC, including Sensor Things API and Timeseries Markup Language. Work is also underway for harmonization between the already closely related OMS and the W3C SSN/SOSA Standards. New OGC Implementation Standards are being pursued for OMS JSON encoding and OMS related dataset metadata led by the OGC OMS Standards Working Group.

The OMS family of Standards contribute to the FAIR (Findable, Accessible, Interoperable, Reusable) principles for various kinds of measurable information, including environmental monitoring as well as remote and in-situ sensing by providing a common conceptual framework for discovering, collecting, and analyzing related information from various sources and data providers.

**Geotech concepts which might use ISO 19148 and ISO 19156**

Those data models and the extensions designed for them are proposed in order to realize all the concepts listed in [Book-A-organization-and-components], including:

- Observation
- [Sampling-and-Preparatio]
- Borehole
- MaterialSample

### 4.3.2. Motivation

As described in the Brief introduction to ISO 19148 and ISO 19156,

> ISO 19148 was designed in order to express locations along a linear object,

and

> ISO 19156 was designed in order to describe observations, measurements and samples.

The motivation to connect ISO 19148 and ISO 19156 is to have the capacity to describe observations, measurements and samples that are linear referenced. In geotechnics, one major use case for that is borehole data.

#### Methodology

In order to explain the approach toward integrating the concepts from Linear Referencing (ISO 19148) with parallel concepts from the Sampling part of Observations, Measurements and Samples (ISO 19156), start with an example provided with the Linear Referencing standard. When specifying a feature located along a linear feature, 19148 utilizes the FeatureEvent Class. This Event references the following.

- the linear feature under consideration, association role: linearElement
- information where along this linear feature (distance relative to the origin of the linear feature) the linearly referenced feature is located (this can be a point or interval; dataTypes AtLocation and FromToLocation provided for this information). Association role: location
- the linearly referenced feature, located along the linear feature. Association role: locatedFeature.

In the example provided with 19148, a wildlife fence is located at a specific position along a road (from meter 0 to meter 5), leading to the following UML:

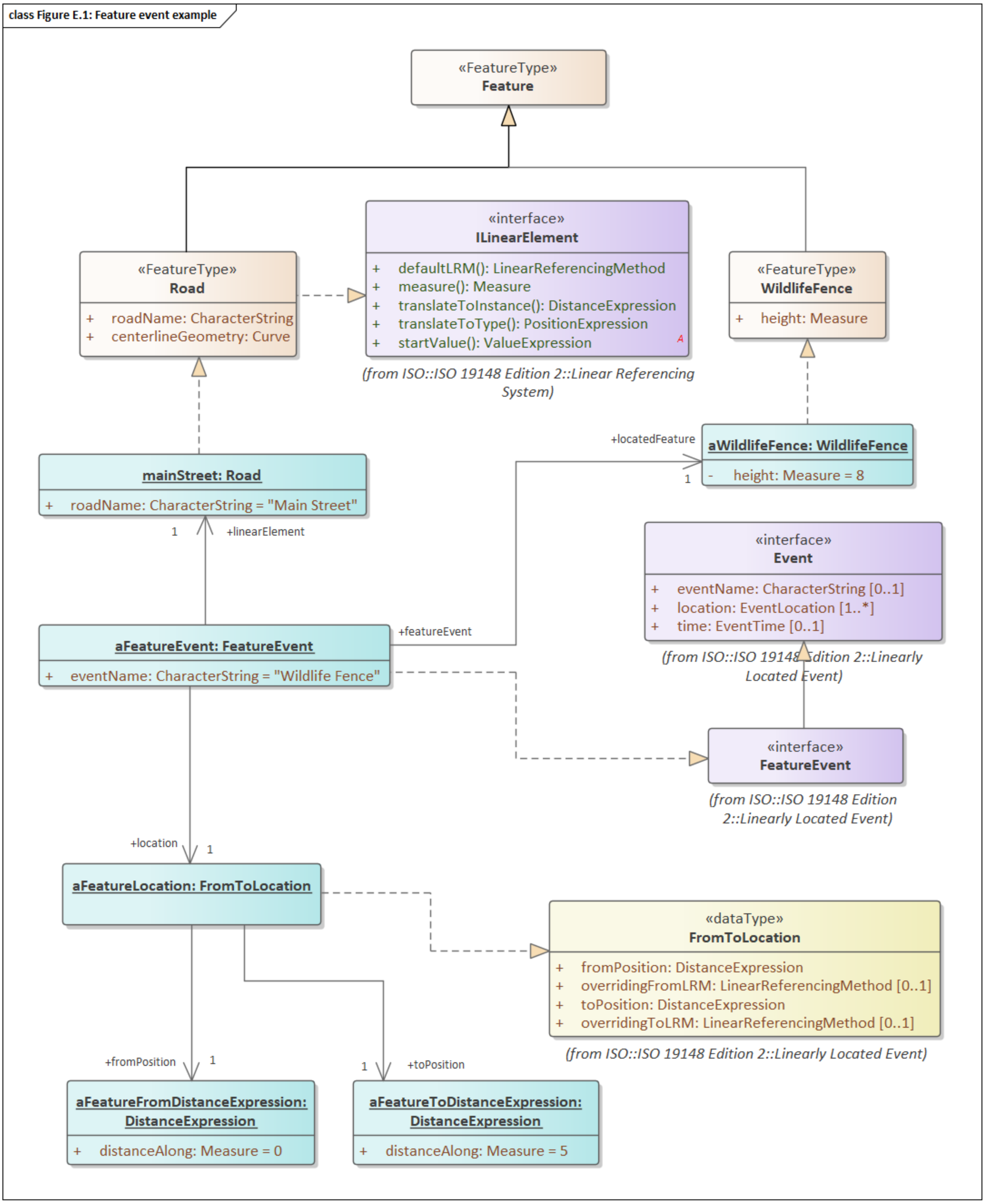

*Figure 13. Feature event example*

In a first step, the same structure was used, but the classes from the example were replaced by examples relating to boreholes and sampling a segment of this borehole. In addition, where relevant, Interfaces from the OMS model were added, in order to show the alignment between the interfaces from Linear Referencing with those from OMS. Class replacements were as follows:

- FT Road: BH_Trajectory
- FT WildlifeFence: BH_Segment

This brings us to the following UML:

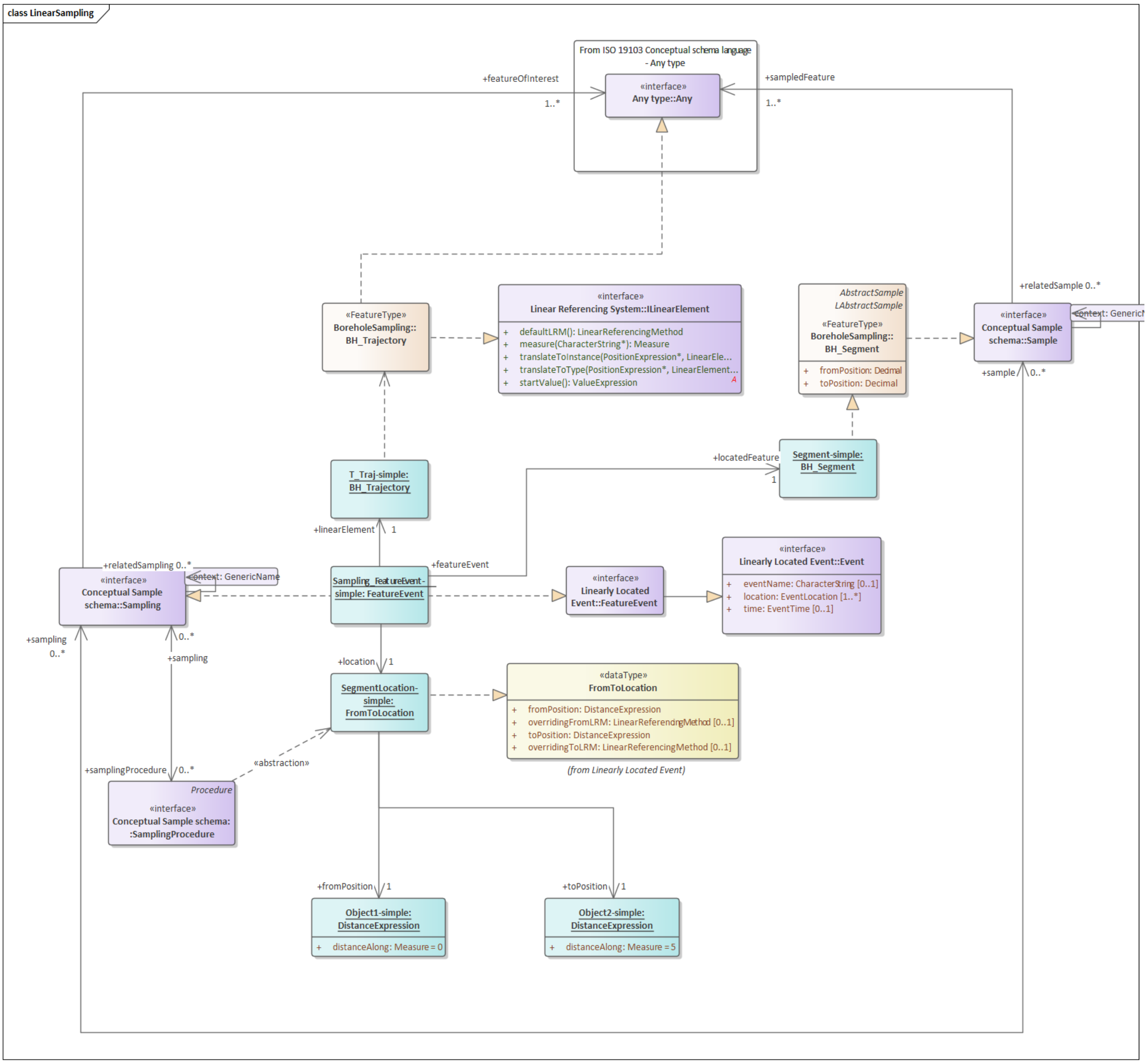

*Figure 14. Linear Sampling*

In this diagram, the parallel between the linear referencing FeatureEvent and the OMS Sampling become apparent, as well as the fact that the linear referencing locatedFeature corresponds with the OMS Sample. In addition, the location information is related to the OMS SamplingProcedure.

However, when dealing with boreholes, it is common to provide information on the borehole collar, to indicate the start of the borehole (as well as to provide additional information on the borehole). The linear referencing model foresees the Referent Interface to provide the origin of the linear feature. Integrating a BH_Collar featureType into the model above leads us to the following final level of the conceptual linear borehole model:

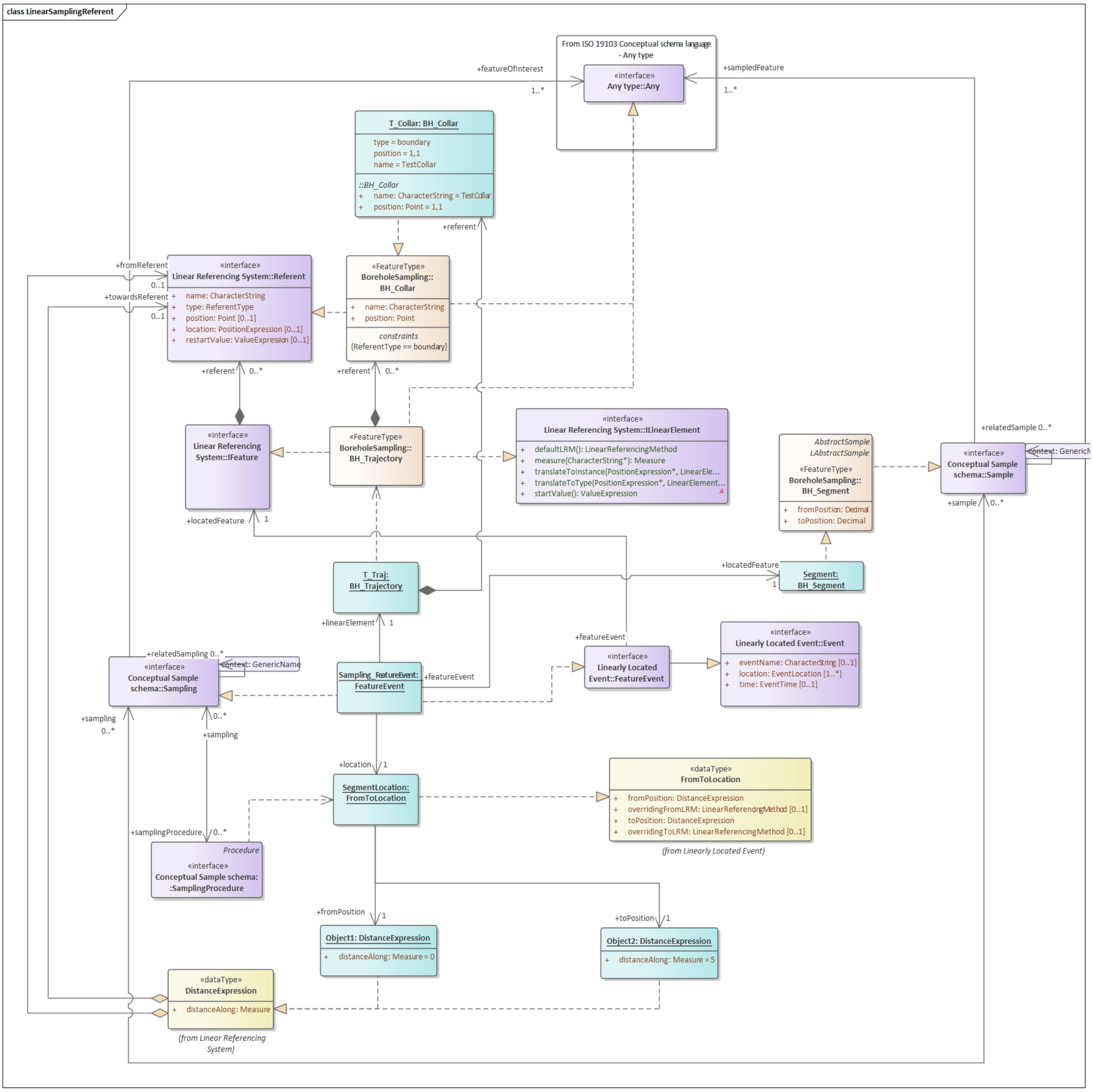

*Figure 15. Linear Sampling Reference*

## 4.3.3. Conceptual Borehole Model

The core conceptual model for boreholes consists of the following featureTypes.

- BH_Collar: Information on the location of the borehole collar, together with relevant attributes.
- BH_Trajectory: The trajectory of the borehole. Encoding of the Geometry TBD, as unclear if we use a 3D Linestring or local polar coordinates.
- BH_SamplingFeatureEvent: An act applying a SamplingProcedure to create or transform one or more Sample(s). Information on how and where the Sample of the borehole under investigation (Point or Segment) has been sampled from this borehole.
- BH FoI (Sample) Types: An object that is representative of a concept, real-world object or phenomenon.
- BH_Hole: A featureType representing the entire Hole of the borehole.

- BH_HolePoint: A featureType representing a point along a borehole Hole, as indicated by the attribute atPosition.
- BH_HoleSegment: A featureType representing a segment of a borehole Hole, as indicated by the attributes fromPosition and toPosition.
- BH_Core: A featureType representing the entire Core of the borehole.
- BH_CorePoint: A featureType representing a point along a borehole Core, as indicated by the attribute atPosition.
- BH_CoreSegment: A featureType representing a segment of a borehole Core, as indicated by the attributes fromPosition and toPosition.

*Note 1: Position values in the BH_FoIs are relative to the BH_Trajectory*

*Note 2: Point and Segment FoIs are sampled from the BH_Trajectory, not the entire Hole or Core. This modeling decision was reached due to technical constraints posed by STA. In order to maintain the link between Point and Segment FoIs and their entire Hole or Core feature, the sampledFeature association will be used.*

In addition, the diagram below still shows the dataTypes providing the EventLocation, namely AtLocation and FromToLocation. However, the relevant attributes have already been shifted to the BH_Point and BH_Segment featureTypes respectively, valid due to 1:1 associations between these types.

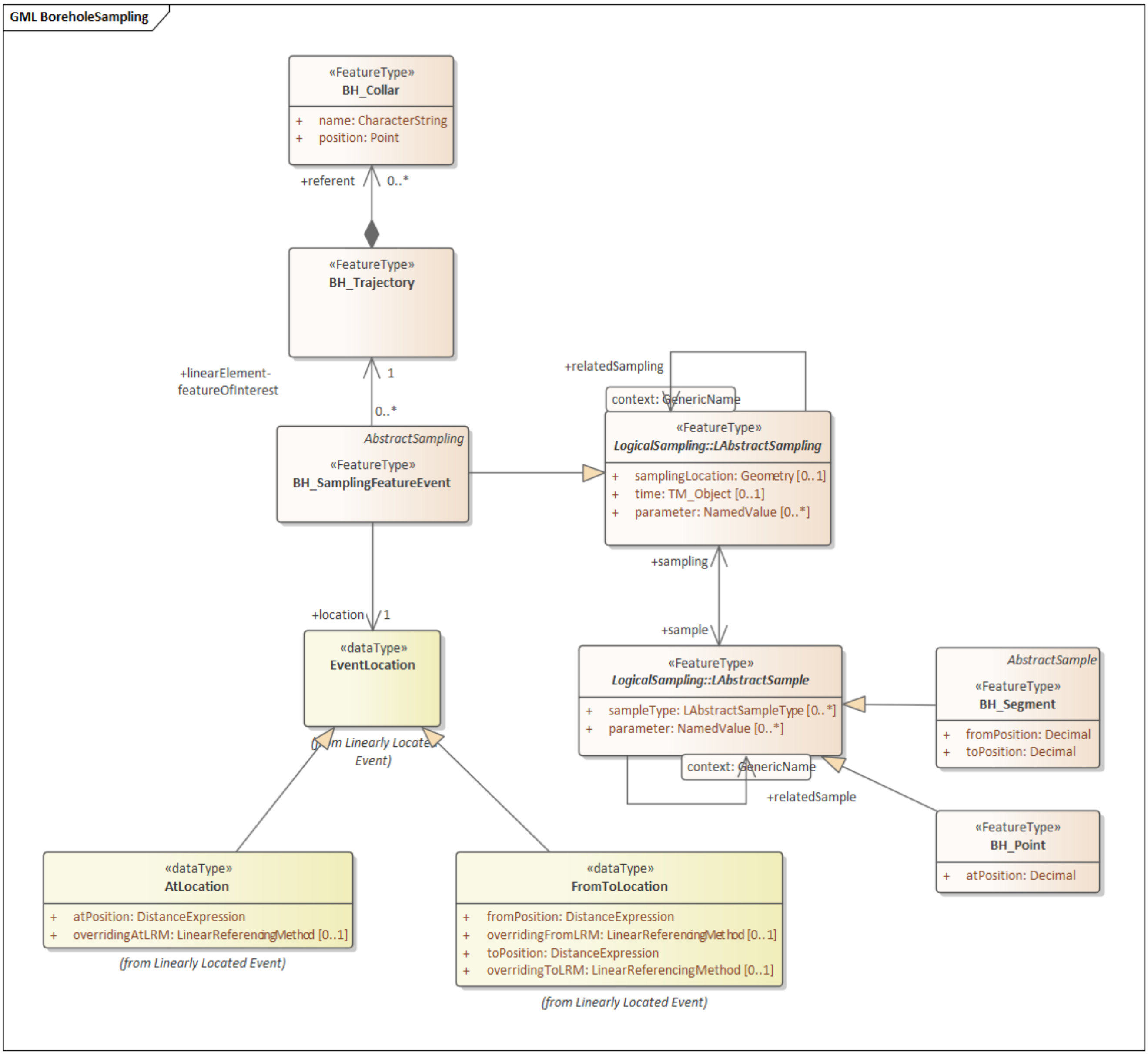

*Figure 16. Borehole Sampling*

### Conceptual Borehole Model with Interfaces

In addition, we provide the same image including the relevant interfaces from the Linear Referencing and OMS Standards.

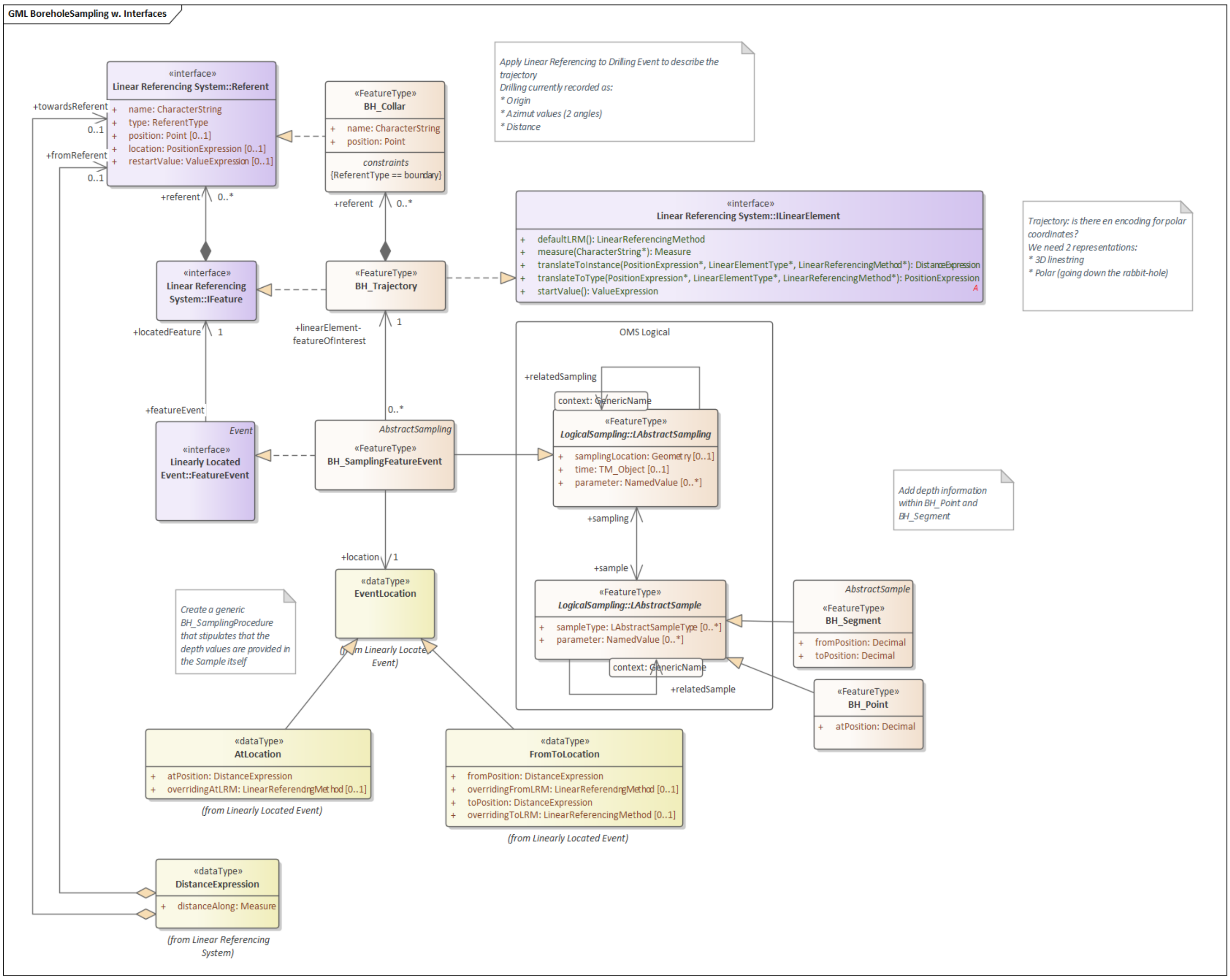

*Figure 17. Borehole Sampling Interfaces*

### Conceptual Borehole Model with all Sampling Classes

In order to provide detailed information on how the Samples have been created, as well as any additional processing steps performed on these samples, the additional featureTypes from the OMS Sampling model have also been added. The following classes have been added to the diagram below.

- BH_Sampler: a device or entity (including humans) that is used by, or implements, a SamplingProcedure to create or transform one or more Sample(s).
- BH_SamplingProcedure: description of steps used to create or transform a Sample.
- BH_PreparationStep: an individual step pertaining to a PreparationProcedure.
- BH_PreparationProcedure: description of preparation steps performed on a Sample that is being observed.

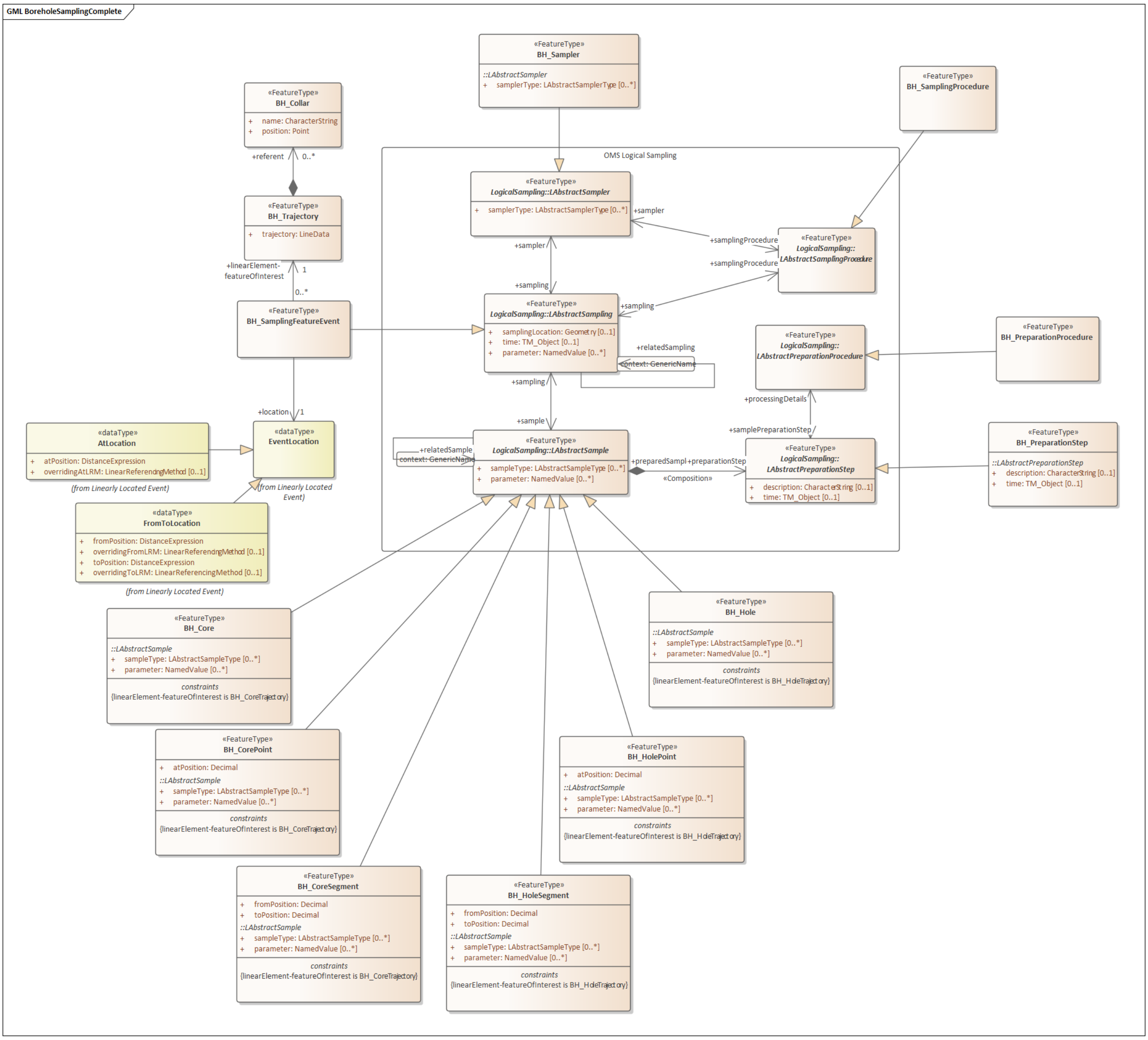

*Figure 18. Borehole Sampling Complete*

## 4.4. SensorThings API data model

### 4.4.1. Introduction to OGC SensorThings API

The OGC SensorThings API (STA) was originally designed to be an API for exchanging observational data in the IoT domain, but it quickly shown itself to be a full successor to the OGC Sensor Observation Service. The SensorThings API in short:

- A standard for exchanging sensor data and metadata
- Historic data & current data
- RESTful + JSON Encoded, with a powerful querying mechanism (based on OASIS OData)
- Supporting ISO MQTT messaging
- Easy to use & understandable: Discoverable with just a web browser

Here is a diagram showing the different classes from OGC SensorThings API Part 1: Sensing Version

1.1

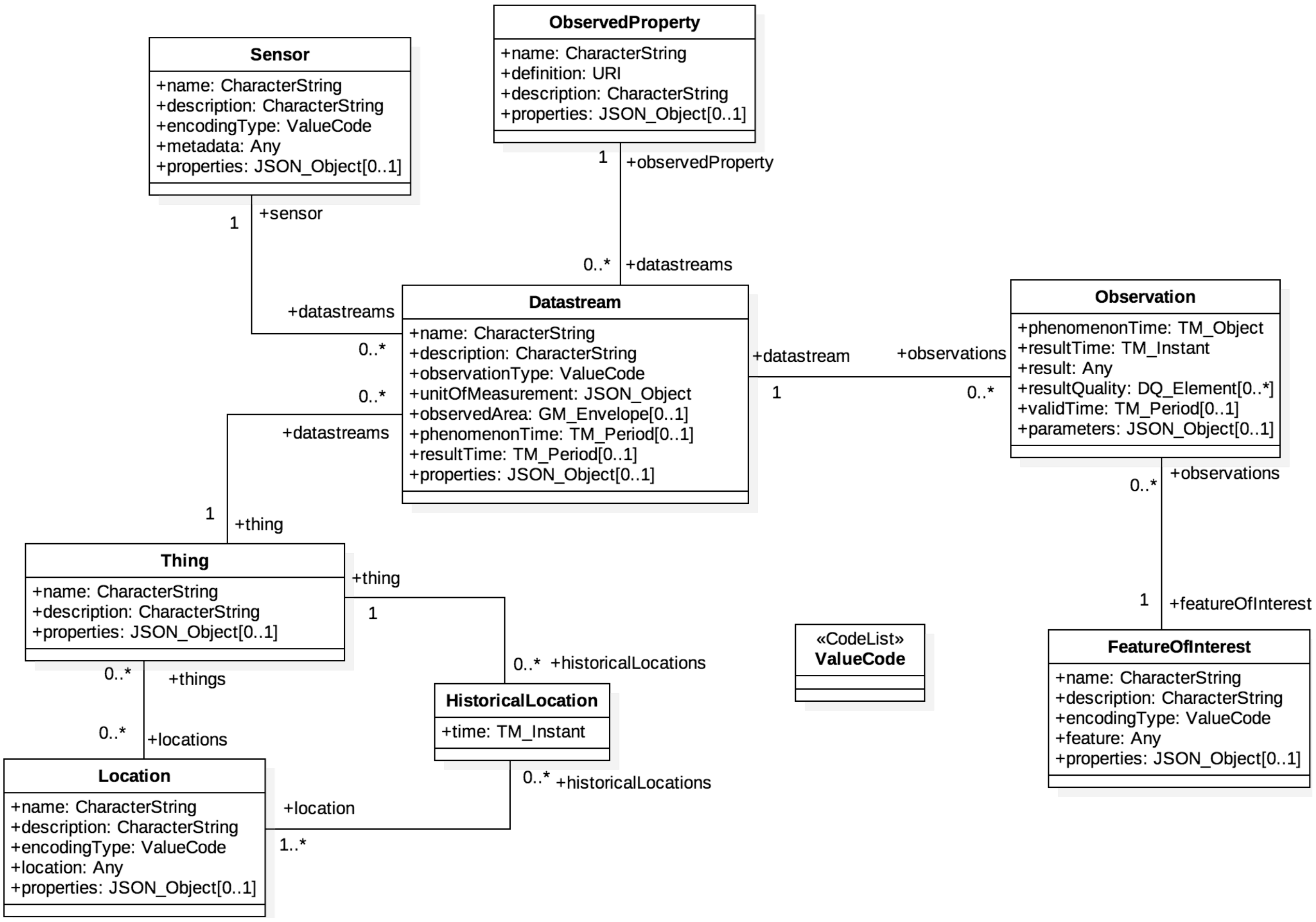

*Figure 19. SensorThings API Part 1*

For those who are familiar with ISO 19156 (aka OGC OMS), one will find those data models have a lot in common.

The main class to consider is the Observation that provides the result that was obtained at a specific date, location, following a particular procedure. The ObservedProperty, Sensor, Thing and FeatureOfInterest enable the contextualization of those parameters.

The main difference with OMS relies in the fact that STA introduces the Datastream concept as a way to group observations that "come together".

This is very common with repeated measurements like time series, or measurements made along a trajectory. The Datastream enables grouping those observations and avoids repeating parameters that remain unchanged like the Sensor or Observed Property.

This concept is explained further in this video: https://www.youtube.com/watch?v=bYDSgs2fKLk

Further documentation of the OGC SensorThings API:

- The specification of version 1.1: https://docs.ogc.org/is/18-088/18-088.html
- The GitHub page: https://github.com/opengeospatial/sensorthings
- Tutorials and examples made for INSPIRE: https://datacoveeu.github.io/API4INSPIRE/sensorthingsapi/1_Home.html

GeotechIE proposes to reuse and extend this data model to accommodate the provision of borehole data. See STA Borehole Model

## 4.4.2. STA Borehole Model

Based on the Conceptual Borehole Model, a physical STA-based model has been derived. In the diagram below, you see both the core STA data model as well as the various extensions proposed for the GeoTech IE.

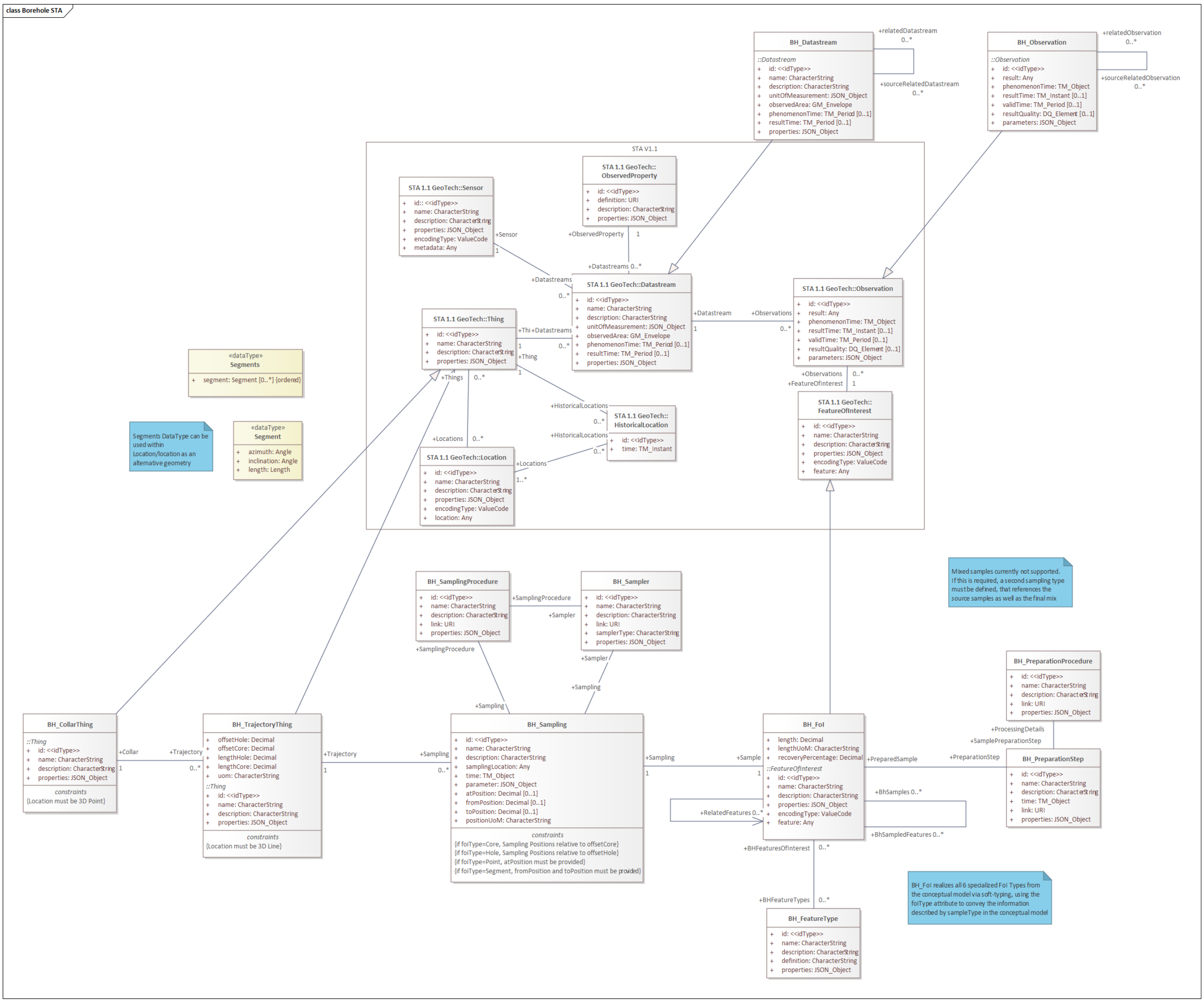

*Figure 20. Borehole STA*

In order to understand how this physical model has been created from the Conceptual Borehole Model, the following diagram sets the STA classes into their conceptual context, indicating which STA classes realize which classes from the Conceptual Borehole Model.

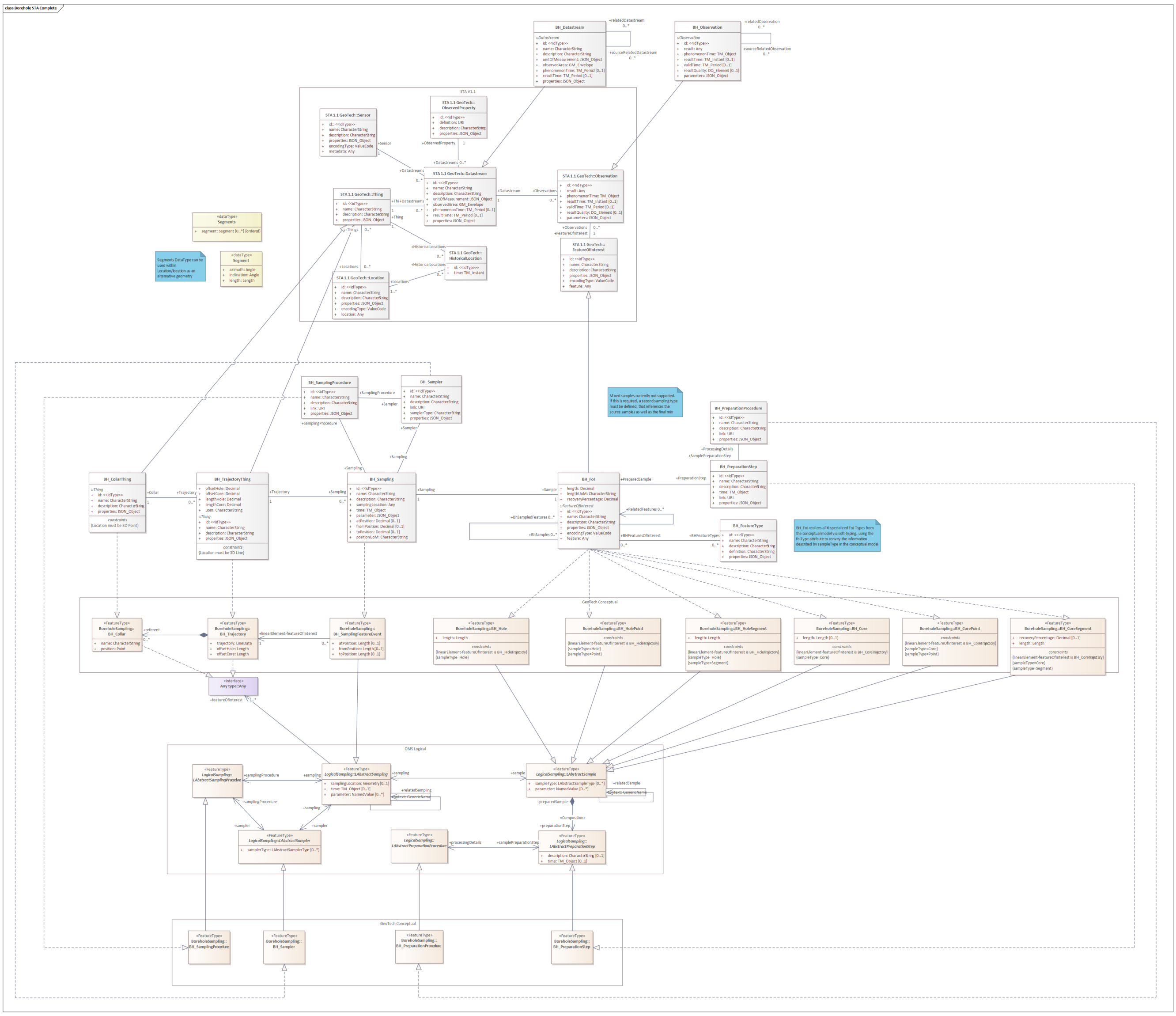

*Figure 21. Borehole STA Complete*

## 4.5. GeoSciML

### 4.5.1. Introduction to GeoSciML

From the CGI-IUGS website:

> **GeoSciML** is an XML–based data transfer standard for the exchange of digital geoscientific information. It accommodates the representation and description of features typically found on geological maps, as well as being extensible to other geoscience data such as drilling, sampling, and analytical data. **GeoSciML** provides a standard data structure for a suite of common geologic features (eg, geologic units, structures, earth materials) and artifacts of geological investigations (e.g., boreholes, specimens, measurements). Supporting objects such as the geologic timescale and vocabularies are also provided as linked resources, so that they can be used

> as classifiers for the primary objects in the GeoSciML standard.

The latest version of the specifications of that joint CGI-IUGS and OGC standard can be found here: https://docs.ogc.org/is/16-008/16-008.html

### 4.5.2. Existing concepts proposed for to reuse and extension for geotechnics

- gsml:GeologicUnit
- gsml:ShearDisplacementStructure
- gsml:Fold
- gsml:Contact
- gsml:Joint

### 4.5.3. Concepts proposed for addition to geotechnics?

- gsml:GeotechUnit

### 4.5.4. GeologicUnit according to GeoSciML

From GeoSciML v4.1:

> Conceptually, a GeologicUnit may represent a body of material in the Earth whose complete and precise extent is inferred to exist (e.g., North American Data Model GeologicUnit, Stratigraphic unit in the sense of NACSN, or International Stratigraphic Code ), or a classifier used to characterize parts of the Earth (e.g. lithologic map unit like 'granitic rock' or 'alluvial deposit', surficial units like 'till' or 'old alluvium'). It includes both formal units (i.e. formally adopted and named in an official lexicon) and informal units (i.e. named but not promoted to a lexicon) and unnamed units (i.e., recognizable, described and delineable in the field but not otherwise formalised). In simpler terms, a geologic unit is a package of earth material (generally rock).

#### Geotech concept that can use gsml:GeologicUnit

gsml:GeologicUnit is identified to realize the geotech concept called GeologicUnit.

More specifically, gsml:GeologicUnit definition is large and enable to address different kind of units.

In geotechnics, when speaking about a geologic unit, the most common understanding is "Lithostratigraphic Unit".

### 4.5.5. ShearDisplacementStructure according to GeoSciML

From GeoSciML v4.1:

> A **shear displacement structure** includes all brittle to ductile style structures along which displacement has occurred, from a simple, single 'planar' brittle or ductile surface to a fault system comprised of tens of strands of both brittle and ductile nature. This structure may have some significant thickness (a deformation zone) and have an associated body of deformed rock that may be considered a deformation unit (which geologicUnitType is 'DeformationUnit') which can be associated to the **ShearDisplacementStructure** using GeologicFeatureRelation from the GeoSciML Extension package (8.5.1.2).

#### Geotech concept that can use gsml:ShearDisplacementStructure

gsml:ShearDisplacementStructure is identified to realize the geotech concept called Fault.

### 4.5.6. Fold according to GeoSciML

From GeoSciML v4.1:

> A fold is formed by one or more systematically curved layers, surfaces, or lines in a rock body. A fold denotes a structure formed by the deformation of a geologic structure, such as a contact which the original undeformed geometry is presumed, to form a structure that may be described by the translation of an abstract line (the fold axis) parallel to itself along some curvilinear path (the fold profile). Folds have a hinge zone (zone of maximum curvature along the surface) and limbs (parts of the deformed surface not in the hinge zone). Folds are described by an axial surface, hinge line, profile geometry, the solid angle between the limbs, and the relationships between adjacent folded surfaces if the folded structure is a Layering fabric.

#### Geotech concept that can use gsml:Fold

gsml:Fold is identified to realize the geotech concept called Fold.

### 4.5.7. Contact according to GeoSciML

From GeoSciML v4.1:

> A **contact** is a general concept representing any kind of surface separating

> two geologic units, including primary boundaries such as depositional contacts, all kinds of unconformities, intrusive contacts, and gradational contacts, as well as faults that separate geologic units.

**Geotech concept that can use gsml:Contact**

gsml:Contact is identified to realize the geotech concept called Contact.

### 4.5.8. Adding GeotechUnit

The proposal is to have gsml:GeotechUnit defined as a specialization of gsml:GeologicUnit.

In GeoSciML, the concept of Geologic Unit is defined as "a package of earth material" (see: GeologicUnit). To offer some variations of the Geologic Unit concept, CGI introduced Geologic Unit types.

The proposal is then to add a Geologic Unit type entitled GeotechUnit which would follow this definition:

> A geologic unit defined by its geotechnical properties

**Geotech concept that can use gsml:GeotechUnit**

gsml:GeotechUnit is identified to realize the geotech concept called GeotechUnit.

**How representations / geometries should be handled for gsml:GeotechUnit?**

As gsml:GeotechUnit is proposed as a specialization of gsml:GeologicUnit which it itself a specialization of a GeologicFeature it inherits its capacity to be associated to (multiple) gsml:MappedFeature which are intended to each one to be possible representation of the concept.

**Geotechnical properties**

A GeotechUnit being a GeologicUnit, it inherits its properties.

### 4.5.9. Joint

The gsml:Joint concept is not mentioned in the GeosciML v4.1 specifications but is clearly identified in the UML model as part of the Geologic Structure Details.

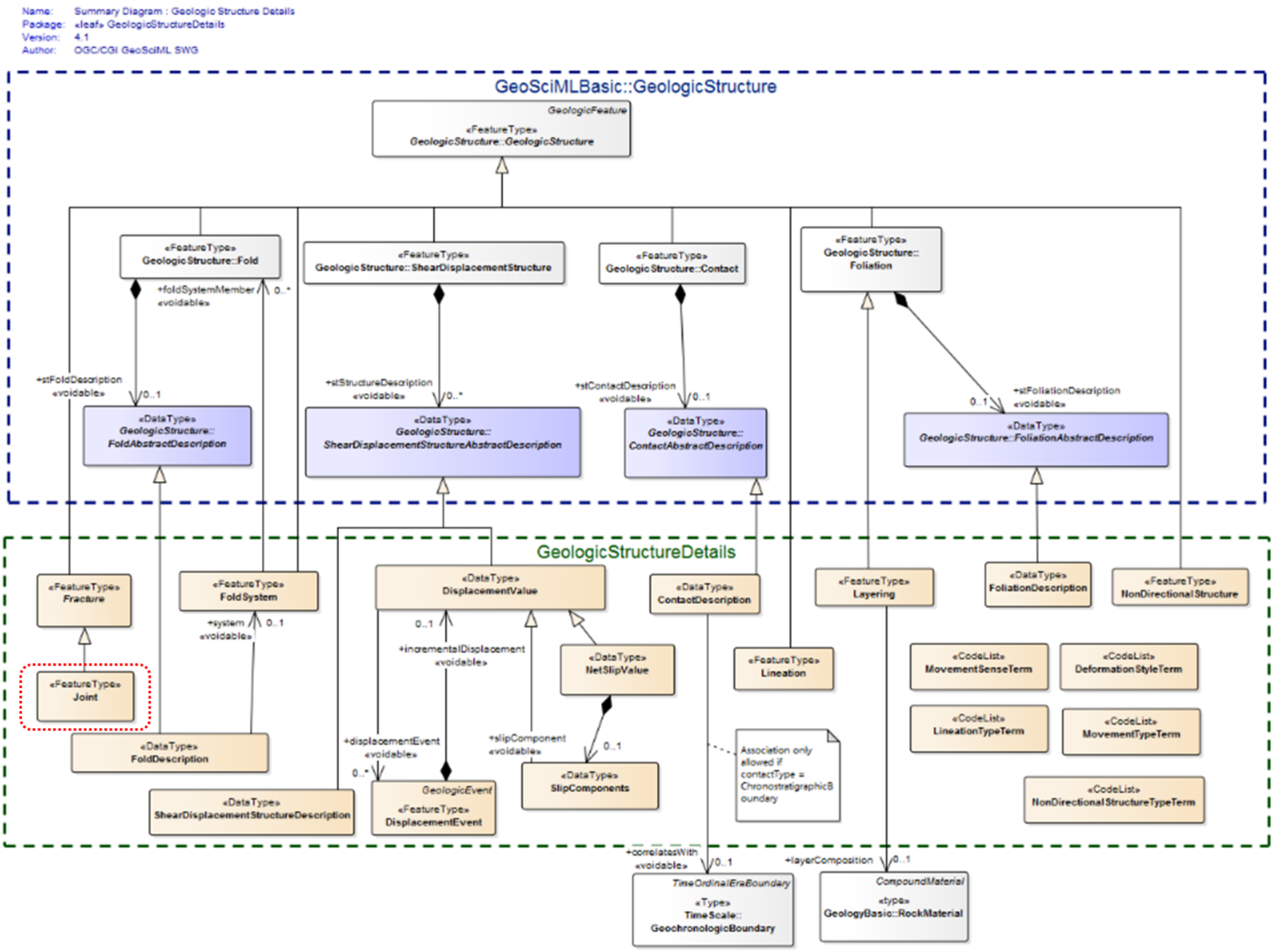

*Figure 22. Geologic Structure (including Joint)*

A gsml:Joint is defined as a:

> Fracture across which there is no displacement at the scale of interest.

gsml:Joint is as a specialization of a gsml:Fracture defined which is defined as:

> Fractures are cracks in the earth surface. If there is no displacement it is a joint. If there is displacement and you are in the brittle zone it is a fault. In the ductile zone, a fracture with displacement with fracture is called a shear.

#### Geotech concept that can use gsml:Joint

gsml:Joint is identified to realize the geotech concept called DiscreteDiscontinuity.

#### How representations / geometries should be handled for gsml:Joint?

As gsml:DiscreteDiscontinuity is proposed as a specialization of gsml:GeologicStructure which it itself a specialization of a GeologicFeature it inherits its capacity to be **associated to (multiple) gsml:MappedFeature** where each one are intended to be possible representation of the concept

#### Geotechnical properties

A GeotechUnit being a GeologicUnit, it inherits of its properties.

## 4.6. GroundWaterML2

### 4.6.1. Introduction to GroundWaterML2

From the dedicated page on the OGC Website:

```
This standard describes a conceptual and logical model for the exchange of groundwater data, as well as a GML/XML encoding with examples.
```

The latest version of the specifications of that OGC standard can be found here: https://docs.ogc.org/is/19-013/19-013.html

#### Existing concepts proposed for reuse and extension for geotechnics

- HydroGeoUnit
- FluidBody
- FluidBodysurface
- HydroGeoVoid

### 4.6.2. HydroGeoUnit according to GroundWaterML2

From GroundwaterML2:

> Any soil or rock unit or zone that by virtue of its hydraulic properties has a distinct influence on the storage or movement of groundwater (after ANS, 1980).

#### Geotech concept that can use gwml2:HydroGeoUnit

gwml2:HydroGeotUnit is identified to realize the geotech concept called HydrogeoUnit.

### 4.6.3. FluidBody according to GroundWaterML2

From GroundwaterML2:

> A distinct body of some fluid (liquid, gas) that fills the voids of a container such as an aquifer, system of aquifers, water well, etc. In hydrogeology this body is usually constituted by groundwater, but the model allows for other types of fillers e.g., petroleum.

#### Geotech concept that can use gwml2:FluidBody

gwml2:FluidBody is identified to realize the geotech concept called FluidBody.

### 4.6.4. FluidBodySurface according to GroundWaterML2

From GroundwaterML2:

> A surface on a fluid body within a local or regional area, e.g., piezometric, potentiometric, water table, salt wedge, etc.

#### Geotech concept that can use gwml2:FluidBodySurface

gwml2:FluidBodySurface is identified to realize the geotech concept called FluidBodySurface.

#### Proposed adjustments for geotech

Existing attributes:

| Attribute | Type and Multiplicity | Definition |
| --- | --- | --- |
| gwSurfaceShape | Surface | Geometry / position of the surface. |
| gwSurfaceType | SurfaceType | Type of fluid body surface, e.g., piezometric, potentiometric, water table, salt wedge, etc. |
| gwSurfaceMetadata | ObservationMetadata | Date, time, method, etc., of the observation or calculation of the surface. |

Existing relations:

| Relation | Source | Target | Description |
| --- | --- | --- | --- |
| Association | Entity: GW_Divide | Entity: GW_FluidBodySurface | Relates a fluid body surface to a line on e.g., a water table or piezometric surface, on either side of which the groundwater flow diverges. |
| | | | |
| | Role: gwSurfaceDivide | Role: gwDivideSurface | |
| Association | Entity: GW_FluidBodySurface | Entity: GW_FluidBody | Relates a fluid body to a surface hosted by the body, e.g., the top of the water table. |
| | | | |
| | Role: gwBodySurface | Role: gwSurfaceBody | |

### 4.6.5. HydroGeoVoid according to GroundWaterML2

From GroundwaterML2:

> Voids represent the spaces inside (hosted by) a unit or its material. E.g., the pores in an aquifer, or in the sandstone of an aquifer. Voids can contain fluid bodies. Voids are differentiated from ‘porosity’ in that porosity is the proportion of void volume to total volume, while voids are the spaces themselves. Voids are required in GWML2, for example, to capture the volume of fractures in an aquifer.

#### Geotech concept that can use gwml2:HydroGeoVoid

gwml2:HydroGeoVoid is identified to realize the geotech concept called Void.

The name HydroGeoVoid makes it very hydrogeology oriented, yet the definition is generic enough for the geotech purpose.

For man made voids (e.g., tunnels, gallery) the concept of Facility from LandInfra may also be used.

Existing attributes:

| Attribute | Type and Multiplicity | Definition |
|---|---|---|
| gwVoidDescription | char | General description of the void |
| gwVoidHostMaterial | EarthMaterial [0..*] | The material that hosts the void, if specified. Note voids can be hosted by a unit (an aquifer) or its material (e.g., sandstone). |
| gwVoidMetadata | GW_Metadata | Metadata for the void. |
| gwVoidShape | Geometry | Shape and position of the void. |
| gwVoidType | PorosityType | Type of void e.g., fractured, intergranular, etc. |
| gwVoidVolume | Measurement | Volume of the void. |

## 4.7. EPOS WP15

### 4.7.1. Introduction to EPOS

EPOS, the **European Plate Observing System**, is a multidisciplinary, distributed research infrastructure that facilitates the integrated use of data, data products, and facilities from the solid

Earth science community in Europe.

For a better overview of EPOS, please have a look to this paper: https://www.nature.com/articles/s41597-023-02697-9

**EPOS TCS GIM**

Within the EPOS initiative, the Thematic Core Service (TCS) **Geological Information and Modeling** develops and consolidates the information and data infrastructures produced by the international community, primarily the European Geological Surveys. Shared reference datasets on geological maps, boreholes (vertical shafts constructed for water, oil, or gas extraction), 3D geological models, and mineral resources are accessible to the community through the EPOS multidisciplinary platform provided from the European Geological Data Infrastructure.

**Existing concepts proposed for reuse and extension for geotechnics**

- eposl:ModelView

### 4.7.2. ModelView according to the EPOS datamodel

The ModelView approach is a creation by EPOS Thematic Core Service on Geological Information and Modelling aiming at providing descriptive information on 3D/4D models. It builds on GeoSciML Lite philosophy which is best thought of as a view of GeoSciML and O&M data that denormalizes the data and concatenates complex property values into single, human-readable, labels and returns single, representative, values from controlled vocabularies for properties multi-valued properties that can be used when generating thematic maps, or portrayals, of the data. The rationale targeted for ModelView is to have a complete corresponding UML model describing 3D/4D Models based on Observations and Measurements and the ModelView being a simplified View to support discovery of 3D/4D Models. ModelView triggered to move the original GeoSciML Lite philosophy from SF-0 to SF-1 to allow some element to appear multiple time.

**Definition**

From the EPOS dedicated schema:

> Simplified view of a Model. A model is defined as the result of geoscience data processing or interpretation. Proposing either a spatial distribution of geoscientific objects with properties of interest (feature) or attempt of retranscribing natural/man-made behavior through mathematical functions or algorithms (process).

**Geotech concept that can use eposl:ModelView?**

eposl:ModelView is intended as a realization of the Geomodel concept and especially one of its specialization the GeotechSynthesisModel

# 4.8. LandInfra and InfraGML

## 4.8.1. LandInfra and InfraGML

From the OGC website:

> The scope of the Land and Infrastructure (**LandInfra**) Conceptual Model is land and civil engineering infrastructure facilities. Anticipated subject areas include facilities, projects, alignment, road, railway, survey, land features, land division, and “wet” infrastructure (storm drainage, wastewater, and water distribution systems). The initial release of this standard is targeted to support all of these except wet infrastructure.

> The OGC **InfraGML** Encoding Standard presents the implementation-dependent, GML encoding of concepts supporting land and civil engineering infrastructure facilities specified in the OGC Land and Infrastructure Conceptual Model Standard (**LandInfra**), OGC 15-111r1. Conceptual model subject areas include land features, facilities, projects, alignment, road, railway, survey (including equipment, observations, and survey results), land division, and condominiums. **InfraGML** is published as a multi-part standard.
>
> - Part 0 addresses the Core Requirements Class from LandInfra.
> - Part 1 addresses the LandFeature Requirements Class from LandInfra.
> - Part 2 addresses the Facility and Project Requirements Classes from LandInfra.
> - Part 3 addresses the Alignment Requirements Class from LandInfra.
> - Part 4 addresses the Road and RoadCrossSection Requirements Classes from LandInfra.
> - Part 5 addresses the Railway Requirements Class from LandInfra.
> - Part 6 addresses the Survey, Equipment, Observations and Survey Results Requirements Classes from LandInfra.
> - Part 7 addresses the Land Division and Condominium Requirements Classes from LandInfra.

**Existing concepts proposed for reuse and extension for geotechnics**

- InfraGML:Alignment

- InfraGML:Facility and FacilityPart

### 4.8.2. Reusing LandInfra Alignment

The Alignment concept was developed in common between OGC and bSI in the LandInfra Domain Working Group.

It is part of the OGC InfraGML 1.0: Part 0 – LandInfra Core – Encoding Standard

Also a Logical model called InfraGML was developped on top of this Conceptual Model. This work is known as OGC InfraGML 1.0: Part 3 - Alignments - Encoding Standard.

An Alignment is defined as a:

> PositioningElement which provides a Linear Referencing System for locating PhysicalElements. An Alignment shall be continuous, non-branching, and non-overlapping. If it is a Project Alignment, it is for a single alternative, as specified by its owning ProjectPart.

#### Geotech concept that can use LandInfra:Alignment

LandInfra:Alignment is identified to realize the geotech concept called Alignment.

#### Geotechnical properties

No additionnal geotechnical properties identified.

### 4.8.3. Extending LandInfra Facility and FacilityPart

In InfraGML, a Facility is defined as:

> Buildings and civil engineering works and their associated siteworks. Civil engineering works, or infrastructure facilities, are construction works comprising a structure, such as a dam, bridge, road, railway, runway, utilities, pipeline, or sewerage system, or are the result of operations such as dredging, earthwork, and geotechnical processes.

In InfraGML, a FacilityPart is defined as:

> Land and Infrastructure facility, such as a road or bridge. A facility has a life cycle, including planning, design, construction, maintenance, operation, and removal phases The design and construction phases are typically performed as part of a project. There may be multiple such projects during the life cycle of the facility to enable phased construction and incremental improvement.

#### Geotech concepts that can use InfraGML:Facility and FacilityPart?

InfraGML:Facility is identified to realize the geotech concept called Surrounding Construction

InfraGML:FacilityPart is identified to realize the geotech concept called GeotechTypicalSection

#### Additionnal properties and constraints

See Surrounding Construction and GeotechTypicalSection for properties to be added.

## 4.9. INSPIRE Theme III: Natural Risk Zone

### 4.9.1. INSPIRE

The INSPIRE Directive aims to create a European Union Spatial Data Infrastructure (SDI) for the purposes of EU environmental policies and policies or activities which may have an impact on the environment. This European Spatial Data Infrastructure will enable the sharing of environmental spatial information among public sector organisations, facilitate public access to spatial information across Europe and assist in policy-making across boundaries.

INSPIRE is based on the infrastructures for spatial information established and operated by the Member States of the European Union. The Directive addresses 34 spatial data themes needed for environmental applications.

#### INSPIRE Theme III: Natural Risk Zone

The approach taken to model Natural Risk Zones is generic in its treatment of each of hazard, exposure, vulnerability and risk, but five use cases have been created to demonstrate the fit of the model with specific examples for different types of hazard: - Floods (calculation of flood impact, reporting and flood hazard/risk mapping) - Risk Management Scenario (an example from a national perspective) - Landslides (hazard mapping, vulnerability assessment and risk assessment) - Forest fires (danger, vulnerability and risk mapping) - Earthquake insurance

#### Existing concepts proposed for reuse and extend for geotechnics

- NZ:HazardArea
- NZ:RiskZone

### 4.9.2. Hazard Area according to INSPIRE:NZ

From the INSPIRE Data Specification on Natural Risk Zones – Technical Guidelines

> Discrete spatial objects representing a natural hazard.

#### Geotech concept that can use this concept

HazardArea is identified to realize the geotech concept called HazardArea.

#### Proposed adjustments for geotech

See HazardArea for properties to be added.

### 4.9.3. What is an RiskZone according to INSPIRE:NZ?

From the INSPIRE Data Specification on Natural Risk Zones – Technical Guidelines

> Discrete spatial objects representing the spatial extent of a combination of the consequences of an event (hazard) and the associated probability/likelihood of its occurrence.

#### Geotech concept that can use this concept

RiskZone is identified to realize the geotech concept called RiskZone.

#### Proposed adjustments for geotech

It is recommended to rely on ISO 19156 for the provision of the “LevelOrIntensity” estimation with the RiskZone being the FeatureOfInterest.

# Chapter 5. Implementation guide, resources, and examples

## 5.1. About implementation

Implementation is the state where concepts and ideas take a more concrete shape.

In data modeling, this means tools and technical solutions are provided in order to realize the conceptual model. Therefore implementation then participate in the validation of the conceptual model and of course enable to truly provide data (here geotech) in an interoperable way.

One must distinguish the capacity of exposing:

- vocabularies, codelists, or
- observations, measurements, and features description.

The GeotechIE experimented with some OGC APIs to provide geotech data: OGC APIs

## 5.2. OGC APIs

OGC API standards define modular API building blocks to spatially enable Web APIs in a consistent way. The OGC API family of standards is organized by resource type.

See https://ogcapi.ogc.org/ for more details.

### 5.2.1. OGC APIs considered for geotechnics

The GeotechIE proposes reliance upon:

- OGC SensorThings API for the provision of Observations, Measurements and Sampling description; and
- OGC API Features for the provision of observed features like Borehole, TrialPit, MaterialSample, but also interpreted featured like a GeologicUnit, Fault, HydrogeoUnit, etc.

#### OGC SensorThings API

The OGC SensorThings API provides an open, geospatial-enabled and unified way to interconnect the Internet of Things (IoT) devices, data, and applications over the Web. At a high level the OGC SensorThings API provides two main functionalities and each function is handled by a part. The two parts are the Sensing part and the Tasking part. The Sensing part provides a standard way to manage and retrieve observations and metadata from heterogeneous IoT sensor systems. The Tasking part is planned as a future work activity and will be defined in a separate document as the Part II of the SensorThings API.

See https://ogcapi.ogc.org/sensorthings/ for more details.

**OGC API - Features**

OGC API - Features specifies the fundamental API building blocks for interacting with features. The spatial data community uses the term ‘feature’ for things in the real world that are of interest.

See https://ogcapi.ogc.org/features/ for more details.

### 5.2.2. OGC APIs & INSPIRE

Both OGC SensorThingsAPI and OGC API - Features have been assessed as suitable to implement the INSPIRE Directive.

See:

- Extending INSPIRE to the Internet of Things through SensorThings API
- INSPIRE SensorThings API Good Practice
- additional materials are available at https://datacoveeu.github.io/API4INSPIRE

## 5.3. Introduction to FROST

The FRaunhofer Opensource SensorThings-Server (FROST) is an implementation of the OGC SensorThings API. It can be deployed as a Docker image, or directly on an existing Tomcat or similar server. Since version 2 it has a flexible data model implementation that allows the data model to be extended or even completely replaced with little effort. This makes FROST an ideal platform for data model experiments.

For the GeoTech IE, a plugin has been made for FROST that extends the STA core data model with extra entities. See FROST plugin for Geotech.

More details can be found at the FROST GitHub page or the FROST Documentation site.

## 5.4. FROST Plugin for geotech

### 5.4.1. A short intro to the OGC SensorThingsAPI data model

For those who are not (yet) familiar with the data model from the OGC SensorThings API, please visit: Introduction to SensorThingsAPI data model.

### 5.4.2. Extending STA for Geotech

Thanks to the DataStream concept, that is part of the core STA data model, STA is very suited for use cases where repeated observations and measurements are made. The support for time series is native and is the essence of STA. When it comes to address measurements along a trajectory, STA functions similarly to a time series, except the variable is space and not time.

In practice, measurements along a trajectory within a borehole are a kind of time series because the results are obtained one after the other as the hole is advanced. The difference is that the time parameter is not important whereas the location parameter is the important domain for the measurement.

## Extended data model

To improve the support for boreholes, the core SensorThings data model has been extended. The conceptual background of this extension is explained in STA Borehole Model. This data model extension has been implemented as a FROST-Server plugin, together with a security extension that allows fine-grained access control to borehole data:

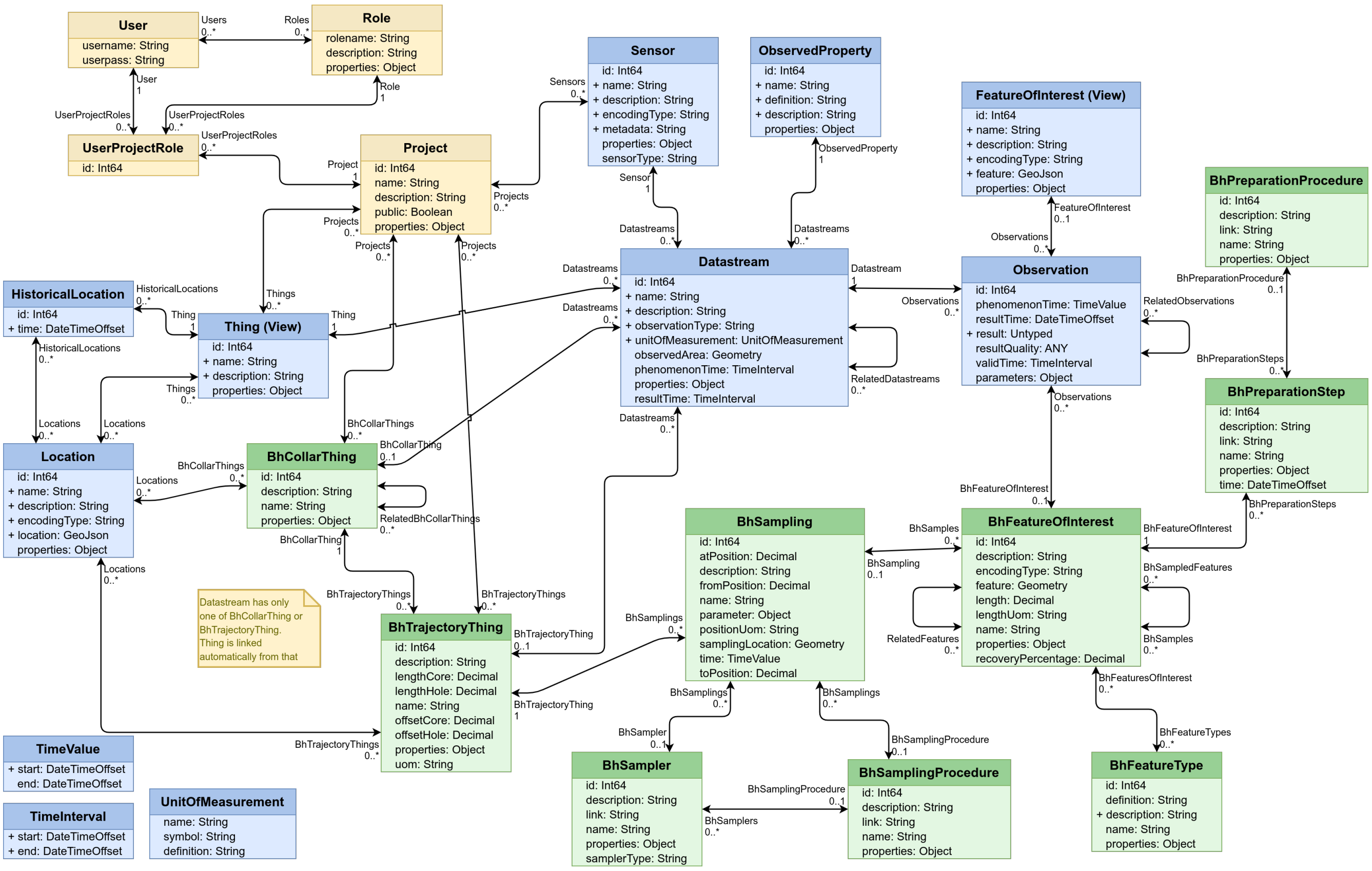

*Figure 23. Core STA data model in blue, with the GeoTech data model extension in green and the security extension in yellow*

## Map geotech data to the STA data model

### Test (geotechnical in-situ)

Act (Sensor) carried out as part of a (Project)

Associated with a borehole (BhCollarThing) that follows a trajectory (BhTrajectoryThing)

### Observation / observation group

Obtained during a test (Sensor)

One or more parameter value(s) (ObservedProperty) measured at a given location (FeatureOfInterest)

These given locations are defined by sampling (BhSampling) along the trajectory (BhTrajectoryThing)

#### Geotech data considered

This implementation is for delivering observations & measurements (aka Book A data. See: [Book-A-organization-and-components] for more details).

The study mainly focused on in-situ tests, including:

- Cone Penetration Test (CPT),
- Standard Penetration Test (SPT),
- Menard Pressuremeter Test.

It also addresses laboratory tests, including discussions about:

- Atterberg limits test.

#### Download FROST GeoTech plugin

The FROST GeoTech plugin, with installation help, can be found at the FROST GeoTech Plugin GitHub page. The plugin can be used as a docker image. An example DockerCompse file is provided with the plugin. To try it out without installing anything, a demo service is available at: https://ogc-demo.k8s.ilt-dmz.iosb.fraunhofer.de/FROST-GeoTech/v1.1/ (use username/password read/read for read-only access)

#### Entity creation order

Since some entities in the data model require other entities, like the Datastream requiring a Thing, there is a certain order in which these entities should be created. The general order is:

1. FeatureType (re-used often)
2. ObservedProperty (re-used often)
3. BhCollarThing + Location
4. BhTrajectoryThing + Location
5. BhSampler + BhSamplingProcedure (if present)
6. BhSampling
7. BhFeatureOfInterest
8. BhPreparationProcedure (if present)
9. BhPreparationStep (if present)
10. Sensor
11. Datastream
12. Observation

There are some things to take into account when populating the data model.

- The FeatureTypes and ObservedProperties are re-used often, and should only be created once per server.

- Things and FeaturesOfInterest are not directly created in the GeoTech data model extension. These entities are automatically populated from the BhCollarThing, BhTrajectoryThing and BhFeatureOfInterest.
- A Datastream is linked to either a BhCollarThing or a BhTrajectoryThing, but not to both.
- When creating an Observation, the id of the BhFeatureOfInterest should be given as FeatureOfInterest, otherwise FROST will try to generate a FeatureOfInterest, and fail. For example: `{ "phenomenonTime": "2017-12-31T23:00:00.000Z/2018-01-01T00:00:00.000Z", "result": 12.6, "Datastream": { "@iot.id": 42 }, "FeatureOfInterest": { "@iot.id": 99 } }`

A demonstration Batch Request is available that creates examples of all entity types, with their relations, in the correct order.

**Security Extension**

The security extension that is part of the FROST GeoTech Plugin adds powerful project-management related access control features. The extension adds four extra EntityTypes to the data model:

- User
- Role
- Project
- UserProjectRole

The `User` entities represent users of the system. When combined with the BasicAuth authentication plugin (the Default in the docker image) users must be explicitly created with a username and a password. When combined with the KeyCloak plugin, User entities are automatically created, and don't have a password field. The password field can not be read by anyone, not even global admins.

The `Project` entities represent a unit of management that ties Users in one or more Roles to Entities. Projects are directly linked to Sensors and Things. All other entities in the data model are indirectly linked to Projects through a Thing. For instance, Observations belong to the Project of the Thing of their Datastream (Project → Thing → Datastream → Observation) and BhFeaturesOfInterest are linked to a Project through the Sampling (Project → BhTrajectoryThing → BhSampling → BhFeatureOfInterest). ObservedProperties are shared among all Projects and can only be created or modified by a global server admin. Finally, Projects can be either public or private through the property `public`. If the `public` property of a Project is true, then all users of the system can read entities associated to the Project. If the `public` property is false, then only Users that have the `read` right on that Project, or have global `read` rights, can read Observations associated with that Project.

The `Role` entities represent rights that a User can have. Roles can be directly linked to a User, or indirectly through a UserProjectRole Entity for a given Project. Roles that are directly connected to a User apply to the entire service. Roles that are linked to a user for a Project only apply to Entites of that Project. For instance, a User that has the global Role `update` can modify all Entities in the service. A user that has the `update` role for a given Project can only modify those Entities that are related to that specific Project.

A UserProjectRole entity gives a User a certain Role within a certain Project.

A normal user can only read the `User` Entity that belongs to their user account. Users that have a project role of `admin` can read all User entities, since a project admin must be able to add Users to their Project. The password of a user (when BasicAuth is used) can only be changed by the User themselves, or by a global admin.

The rights for the different user types:

| | **Admin (admin, c,r,u,d)** | **Geotech Expert (r,c,u,d)** | **Public (for open projects)** | **Public (for private projects)** | **Project manager (a,c,r,u,d)** | **Project contributor (r,c,u,d)** | **Project member (read)** |
|---|---|---|---|---|---|---|---|
| Project | CRUD | R | R | | R | R | R |
| User | CRUD | R (self) | | | R (all) | R (self) | R (self) |
| Role | CRUD | R (self) | | | R (all) | R (self) | R (self) |
| UserProjectRole | CRUD | R (self) | | | CRUD (project) | R (self) | R (self) |
| ObsProp | CRUD | CRUD | R | R | R | R | R |
| Sensor | CRUD | CRUD | R | | CRUD | R | R |
| Thing & Location | CRUD | CRUD | R | | CRUD | R | R |
| FOI, Sampling, Preparation | CRUD | CRUD | R | | CRUD | CRUD | R |
| DataStream | CRUD | CRUD | R | | CRUD | CRUD | R |
| Observation | CRUD | CRUD | R | | CRUD | CRUD | R |

- The admin user type has global Admin, Create, Read, Update and Delete rights.
- The Geotech Expert user type has global Create, Read, Update and Delete rights, and can thus create Observed Properties and access all data in the server, but not manage users.
- The Project user types do not have any global roles, only project-related roles.

# 5.5. Exposing geotech investigation data with OGC SensorThings API

## 5.5.1. Approach for borehole logs

A borehole log is a record detailing the in-situ conditions and aspects of geotechnical exploration activities associated with the drilling and sampling of a borehole. It is typically a graphical representation of the data collected from various locations within the hole, along with properties associated with the borehole, its construction, and associated metadata.

**Data presented on geotechnical borehole logs**

The specific kinds of data presented on geotechnical borehole logs varies greatly depending on the application, but most typically include descriptions of the earth materials encountered in the borehole at locations along the borehole's trajectory (lithologic descriptions, perhaps including geological units encountered), location and identification of any material samples collected, summary results of any tests performed in the borehole or on the samples, and related information about the borehole as a whole (eg. name, location), construction/destruction information that may vary along its trajectory, installations, and other metadata.

The following example shows a portion of a representative geotechnical borehole log produced for a landslide investigation by the US State of Ohio Department of Transportation (ODOT), and illustrates the types of information that may be included in such a log:

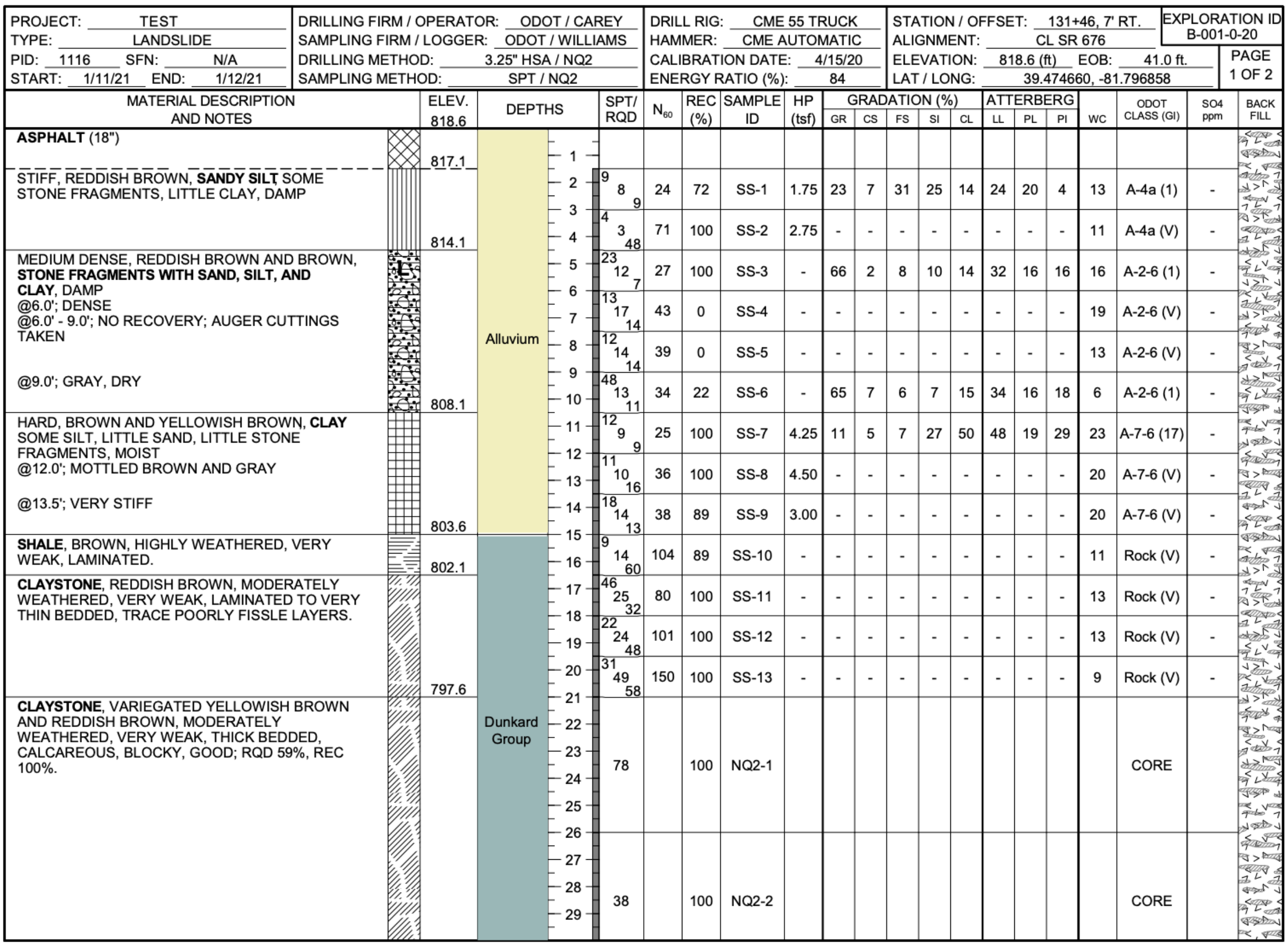

| | | | | |
|---|---|---|---|---|
| PROJECT: TEST | DRILLING FIRM / OPERATOR: ODOT / CAREY | DRILL RIG: CME 55 TRUCK | STATION / OFFSET: 131+46, 7' RT. | EXPLORATION ID B-001-0-20 |
| TYPE: LANDSLIDE | SAMPLING FIRM / LOGGER: ODOT / WILLIAMS | HAMMER: CME AUTOMATIC | ALIGNMENT: CL SR 676 | |
| PID: 1116 SFN: N/A | DRILLING METHOD: 3.25" HSA / NQ2 | CALIBRATION DATE: 4/15/20 | ELEVATION: 818.6 (ft) EOB: 41.0 ft. | PAGE 1 OF 2 |
| START: 1/11/21 END: 1/12/21 | SAMPLING METHOD: SPT / NQ2 | ENERGY RATIO (%): 84 | LAT / LONG: 39.474660, -81.796858 | |

| MATERIAL DESCRIPTION AND NOTES | ELEV. 818.6 | DEPTHS |
|---|---|---|
| **ASPHALT** (18") | 817.1 | Alluvium |
| STIFF, REDDISH BROWN, **SANDY SILT**, SOME STONE FRAGMENTS, LITTLE CLAY, DAMP | 814.1 | Alluvium |
| MEDIUM DENSE, REDDISH BROWN AND BROWN, **STONE FRAGMENTS WITH SAND, SILT, AND CLAY**, DAMP<br>@6.0'; DENSE<br>@6.0' - 9.0'; NO RECOVERY; AUGER CUTTINGS TAKEN<br>@9.0'; GRAY, DRY | 808.1 | Alluvium |
| HARD, BROWN AND YELLOWISH BROWN, **CLAY** SOME SILT, LITTLE SAND, LITTLE STONE FRAGMENTS, MOIST<br>@12.0'; MOTTLED BROWN AND GRAY<br>@13.5'; VERY STIFF | 803.6 | Alluvium |
| **SHALE**, BROWN, HIGHLY WEATHERED, VERY WEAK, LAMINATED. | 802.1 | Dunkard Group |
| **CLAYSTONE**, REDDISH BROWN, MODERATELY WEATHERED, VERY WEAK, LAMINATED TO VERY THIN BEDDED, TRACE POORLY FISSLE LAYERS. | 797.6 | Dunkard Group |
| **CLAYSTONE**, VARIEGATED YELLOWISH BROWN AND REDDISH BROWN, MODERATELY WEATHERED, VERY WEAK, THICK BEDDED, CALCAREOUS, BLOCKY, GOOD; RQD 59%, REC 100%. | | Dunkard Group |

| SPT/ RQD | $N_{60}$ | REC (%) | SAMPLE ID | HP (tsf) | GRADATION (%) GR | CS | FS | SI | CL | ATTERBERG LL | PL | PI | WC | ODOT CLASS (GI) | SO4 ppm | BACK FILL |
|---|---|---|---|---|---|---|---|---|---|---|---|---|---|---|---|---|
| 9 8 9 | 24 | 72 | SS-1 | 1.75 | 23 | 7 | 31 | 25 | 14 | 24 | 20 | 4 | 13 | A-4a (1) | - | |
| 4 3 48 | 71 | 100 | SS-2 | 2.75 | - | - | - | - | - | - | - | - | 11 | A-4a (V) | - | |
| 23 12 7 | 27 | 100 | SS-3 | - | 66 | 2 | 8 | 10 | 14 | 32 | 16 | 16 | 16 | A-2-6 (1) | - | |
| 13 17 14 | 43 | 0 | SS-4 | - | - | - | - | - | - | - | - | - | 19 | A-2-6 (V) | - | |
| 12 14 14 | 39 | 0 | SS-5 | - | - | - | - | - | - | - | - | - | 13 | A-2-6 (V) | - | |
| 48 13 11 | 34 | 22 | SS-6 | - | 65 | 7 | 6 | 7 | 15 | 34 | 16 | 18 | 6 | A-2-6 (1) | - | |
| 12 9 9 | 25 | 100 | SS-7 | 4.25 | 11 | 5 | 7 | 27 | 50 | 48 | 19 | 29 | 23 | A-7-6 (17) | - | |
| 11 10 16 | 36 | 100 | SS-8 | 4.50 | - | - | - | - | - | - | - | - | 20 | A-7-6 (V) | - | |
| 18 14 13 | 38 | 89 | SS-9 | 3.00 | - | - | - | - | - | - | - | - | 20 | A-7-6 (V) | - | |
| 9 14 60 | 104 | 89 | SS-10 | - | - | - | - | - | - | - | - | - | 11 | Rock (V) | - | |
| 46 25 32 | 80 | 100 | SS-11 | - | - | - | - | - | - | - | - | - | 13 | Rock (V) | - | |
| 22 24 48 | 101 | 100 | SS-12 | - | - | - | - | - | - | - | - | - | 13 | Rock (V) | - | |
| 31 49 58 | 150 | 100 | SS-13 | - | - | - | - | - | - | - | - | - | 9 | Rock (V) | - | |
| 78 | | 100 | NQ2-1 | | | | | | | | | | | CORE | | |
| 38 | | 100 | NQ2-2 | | | | | | | | | | | CORE | | |

*Figure 24. ODOT Borehole Log. Log courtesy of DataForensics, LLC and Ohio Department of Transportation. Project and location data have been anonymized, and some data altered or manufactured.*

From the log, it can be seen that the borehole was advanced by a 3.25" diameter hollow stem auger with continuous Standard Penetration Test sampling to a depth of 21 feet, at which point the advancement method was switched to rotary drilling with continuous rock coring (NQ2). Material descriptions and RQD results were made by visual inspection of samples recovered from the hole while it was being advanced, and other information resulted from laboratory testing of the recovered samples. The log contains the following information:

Properties and metadata that apply to entire borehole

1. Name, number and type of the project for which the borehole was drilled (eg. Project, PID, Type)
2. Beginning and end date for the drilling (Start, End)

3. Location data (Station/Offset, Alignment, Elevation, Lat/Long)
4. Borehole ID (Exploration ID)
5. Borehole length (EOB)

Properties and metadata that apply to points or segments of the borehole and describe construction/destruction activities or information that does not directly apply to the earth materials encountered

1. Depth-specific remarks (notes) taken during drilling that note drilling or sampling conditions
2. Drilling methods (Drilling Method)
3. Drill rig operators (Drilling Firm, Operator)
4. Drilling equipment (Drill Rig)
5. A graphic column illustrating the depth-interval defined materials used to backfill the hole (in this case a bentonite/cement slurry)

Observations of earth materials that occur while advancing the hole (in-situ observations)

1. Depth-interval defined material descriptions, including lithology and physical properties
2. Depth-specific remarks that note changes in material properties within a lithologic interval
3. A graphic column illustrating the depth-interval defined primary lithology (eg, Sandy Silt, Shale)
4. A graphic column illustrating depth-interval defined geologic units encountered (eg. Dunkard Group)
5. Drive set blow counts and N60 result values from SPT tests (SPT/RQD column above 21 ft depth, N60 column).

Sampling and drilling activities that occur while advancing the hole

1. Drilling and Sampling methods (Drilling Method, Sampling Method)
2. Samplers/Observers (Sampling Firm, Logger)
3. Observing equipment and associated metadata (Hammer, Calibration Date, Energy Ratio)
4. Sample identifiers with their locations down hole, and recovery percentages (Sample ID, Rec columns)

Observations from laboratory tests on samples that occur ex-situ

1. Results from the following tests on the samples: a) compressive strength from hand penetrometer (HP column), b) particle size gradation (Gradation columns), c) Atterberg Limits (Atterberg column, LL=liquid limit, PL=plastic limit, PI=plasticity index), d) natural water content
2. Soil classification of samples above 21 ft depth (ODOT Class (GI) column). This is a modified AASHTO classification assigned based on results of the particle size and Atterberg limits tests
3. RQD result from samples (SPT/RQD column below 21 ft)

#### Application to the STA model

A borehole log is typically a collection of several Observations and Sensors, along with data that are not directly observation-related, but all associated with one Thing (the borehole). To expose the data necessary to construct a borehole log, multiple queries to the FROST server related to the BhCollarThing of interest must be made to extract the information needed to populate the various components of the log. For the log components in the above example, below are recommended mappings to STA objects. With the exception of SPT and Atterberg limits discussed above, the other log information have not been considered in detail in this IE. It is also worth noting that the current Geotech data model and STA do not have objects specifically designed to handle activity information and installations that vary with position in the hole (eg. drill advancement data, backfill materials, installed casings, wells or instrument locations), or hole-point and hole-segment specific annotations that are not properties of earth materials. This information can be included within FROST as Observations or as properties of existing STA objects, but further evaluation of the data model and its STA implementation to accommodate these types of information is needed.

#### Project

Although developed as a security extension to represent a unit of management that ties Users in one or more Roles to Entities, the Project object in STA can be reasonably used to represent the geotechnical concept of a project, which is a business activity that encompasses a collection of activities, material samples, and observations that occur within boreholes and other sampling features. Example for the log above (item 1):

```
{
"@iot.id": 5,
"description": "Route 676 Improvements",
"name": "TEST",
"properties": {
    "pid": 1116,
    "type":"LANDSLIDE"
    }
}
```

**NOTE:** Keys within the properties object should come from a controlled list of property terms for the Project. Different organizations and existing transfer standards (such as AGS and DIGGS) have their own keys for Project properties; this IE has not attempted to harmonize them to develop a standard codelist for Project:properties keys. For full interoperability, this harmonization is important. A workaround at present could be to define the context for the properties keys using json-ld.

#### Sensor

A Sensor instance should be created for each observation procedure in the log:

- **Material Descriptions** (items 11, 12, 13 above)

Example:

```
    {
        "@iot.id": 15,
        "description": "Ohio Department of Transportation Soil and Rock Descriptions",
        "encodingType": "application/pdf",
        "metadata"
:"https://www.transportation.ohio.gov/working/engineering/geotechnical/manuals/geotech
nical-explorations/app/app-a",
        "name": "Visual Soil and Rock Classification",
        "sensorType": "Lithology Description",
        "Projects": [
            {"@iot.id": 5}
        ]
    }
```

- **Geologic Units** (item 14) Example:

```
    {
        "@iot.id": 16,
        "description": "Ohio Department of Transportation Soil and Bedrock
Classification",
        "encodingType": "application/html",
        "metadata": "https://mrdata.usgs.gov/geology/state/fips-unit.php?state=OH",
        "name": "Geologic units in Ohio (state in United States)",
        "sensorType": "Lithostratigraphy",
        "Projects": [
            {"@iot.id": 5}
        ]
    }
```

- **SPT Tests** (item 15 above) See this discussion for example.
- **Hand Penetrometer Tests** (item 20a above) Example

```
    {
        "@iot.id": 17,
        "description": "Method is used to evaluate consistency and approximate
unconfined compressive strength of soils by means of using a pocket penetrometer.",
        "encodingType": "application/html",
        "metadata": "https://www.astm.org/workitem-wk27337",
        "name": "Pocket Penetrometer",
        "sensorType": "Hand Penetrometer test",
        "Projects": [
            {"@iot.id": 5}
        ]
    }
```

- **Particle Size Tests** (item 20b above) Example:

```
    {
        "@iot.id": 18,
        "description": "Method used to separate particles into size ranges and to
determine quantitatively the mass of particles in each range. These data are combined
to determine the particle-size distribution (gradation).",
        "encodingType": "application/html",
        "metadata": "https://www.astm.org/d6913-04r09e01.html",
        "name": "Particle-Size Distribution (Gradation) of Soils Using Sieve
Analysis",
        "sensorType": "Particle size distribution",
        "Projects": [
            {"@iot.id": 5}
        ]
    }
```

- **Atterberg Limits Tests** (item 20c above) See this discussion for example.
- **Water Content Tests** (item 20d above) Example:

```
    {
        "@iot.id": 19,
        "description": "Method used to determine the water (moisture) content by mass
of soil, rock, and similar materials where the reduction in mass by drying is due to
loss of water, method A.",
        "encodingType": "application/html",
        "metadata": "https://www.astm.org/d2216-19.html",
        "name": "Natural water content, Method A",
        "sensorType": "Water content measure",
        "Projects": [
            {"@iot.id": 5}
        ]
    }
```

- **ODOT Classification** (item 21 above) Example:

```
    {
        "@iot.id": 20,
        "description": "Ohio Department of Transportation Soil Classification. ODOT
utilizes a modified AASHTO classification system based on gradation and plastic index
(PI). Percentages are based on: dry weight not volume.",
        "encodingType": "application/pdf",
        "metadata"
:"https://www.transportation.ohio.gov/working/engineering/geotechnical/manuals/geotech
nical-explorations/app/app-a",
        "name": "ODOT Soil Classification",
        "sensorType": "Soil Classification",
        "Projects": [
            {"@iot.id": 5}
```

```
        ]
    }
```

- **RQD Tests** (item 22 above) Example:

```
    {
        "@iot.id": 21,
        "description": "Method used to determine the rock quality designation (RQD) as a standard parameter in drill core logging of a core sample",
        "encodingType": "application/html",
        "metadata": "https://www.astm.org/d6032_d6032m-17.html",
        "name": "Rock Quality Designation (RQD) of Rock Core",
        "sensorType": "Rock Quality Designation",
        "Projects": [
            {"@iot.id": 5}
        ]
    }
```

**NOTES:**

- For Sensors, the required metadata key should be a pointer to a resource for the specific observing procedure used.
- The optional sensorType key should contain a value from a controlled list of terms for ObservingProcedure.
- The "Projects" property links the Sensor to the Project by its id. # BhCollarThing The BhCollarThing object is used to represent the borehole in its entirety. All information associated with the borehole, with the exception of observations, sampling and samples, its trajectory and locations are contained within this object. From the above log, these properties are in items: 2 and 4. Example:

```
    {
        "@iot.id": 10,
        "description": "Borehole B-001-0-20",
        "name": "B-001-0-20",
        "properties": {
            "drilltartDate": "2021-01-11T00:00:00",
            "drillEndDate": "2021-01-12T00:00:00"
         }
        "Projects": [
            {"@iot.id": 5}
        ]
    }
```

**NOTES:**

- The properties keys should be values from a controlled list of terms for Borehole properties.

- Projects links to the associated Project object (see above).

### BhTrajectoryThing

BhTrajectoryThing links to BhCollarThing and exposes the borehole lengths (item 5 above) and any offsets for downhole measurements. Example:

```
{
    "@iot.id": 11,
    "description": "Trajectory of Borehole B-001-0-20",
    "name": "B-001-0-20 Trajectory",
    "lengthCore": 41.0,
    "lengthHole": 41.0,
    "offsetCore": 0,
    "offsetHole": 0,
    "uom":"ftUS",
    "BhCollarThing": [
        {"@iot.id": 10}
    ],
   "Projects": [
        {"@iot.id": 5}
    ]
}
```

**NOTES:**

- Projects links to the associated Project object (see above).

### Location

The Location object provides the geographic locations of the BhCollarThing and its associated BhTrajectoryThing (item 3 above). Non-coordinate location information is provided in the Location instance associated with BhCollarThing. Examples:

#### Location for BhCollarThing

```
{
    "@iot.id": 10,
    "name": "B-001-0-20 Collar Location",
    "description": "Location of Borehole B-001-0-20 collar",
    "encodingType": "application/geo+json",
    "location": {
        "type": "Point",
        "coordinates": [
            -81.796858,
            39.47466,
            249.50928
        ]
    },
```

```
        "properties":{
            "station": "131+48",
            "offset": "7' RT",
            "alignment": "CL SR 676"
        },
        "BhCollarThings": [
            {"@iot.id": 10}
        ]
    }
```

**NOTES:**

- BhCollarThings links to the associated BhCollarThing object (see above).
- Keys within the properties object should come from a controlled list of property terms for the Location. Different organizations and existing transfer standards (such as AGS and DIGGS) have their own keys for Location properties; this IE has not attempted to harmonize them and develop a standard codelist for Location:properties keys. For full interoperability, this harmonization is important. A workaround at present could be to define the context for the properties keys using json-ld.
- Current implementation of STA only supports geoprocessing for geometries encoded as application/geo+json which requires coordinates to be in WGS84. Many geotech organizations maintain location data in projected CRS's (e.g., UTM, State Plane) and local vertical datums, and many at times use local (engineering) coordinate systems. Until STA is capable of working with additional geometry encodings, coordinate conversions may be necessary and in some cases may be impossible.

**Location for BhTrajectoryThing**

```
    {
        "@iot.id": 11,
        "name": "B-001-0-20 Trajectory Location",
        "description": "Location of Borehole B-001-0-20 trajectory",
        "encodingType": "application/geo+json",
        "location": {
            "type": "LineString",
            "coordinates": [
                [
                    -81.796858,
                    39.47466,
                    249.50928
                ],
                [
                    -81.795909,
                    39.475026,
                    237.01248
                ]
          ]
        },
```

```
        "BhTrajectoryThings": [
            {"@iot.id": 11}
        ]
    }
```

**NOTES:**

- BhTrajectoryThings links to the associated BhTrajectoryThing object (see above).

**BhSampler**

BhSampler implements the concept of a Sampler, which is a device or entity (including humans) that is used by, or implements, a SamplingProcedure to create or transform one or more samples (FeaturesOfInterest). In the above log example, borehole drilling and sampling of the Borehole-Hole only (e.g., SPT testing) is accomplished by the Drilling Firm/Operator (item 7 above) using a drill rig (item 9 above), whereas sampling of the borehole material itself (Core) is accomplished by the Sampling Firm/Logger (item 17). Examples:

```
    {
        "@iot.id": 15,
        "description": "Drilling Personnel",
        "name": "ODOT/CAREY",
        "samplerType":"Drilling Firm/Operator"
        "properties":{
            "equipment": [
                {
                    "name": "CME 55 TRUCK",
                    "class": "Drill Rig"
                }
            ]
        }

    }
```

and

```
    {
        "@iot.id": 16,
        "description": "Sampling Personnel",
        "name": "ODOT/WILLIAMS",
        "samplerType":"Sampling Firm/Logger"
   }
```

**NOTES:**

- STA allows only one Sampler per Sampling (sampling activity) and at present does not allow Samplers to relate to each other, which necessitates combining the drilling/sampling personnel and related equipment into a single Sampler instance. More flexibility in this regard should be considered in a future IE.

- The samplerType values and property keys should come from controlled lists of terms. Different organizations and existing transfer standards (such as AGS and DIGGS) have their own keys and allowable values for samplerType and Sampler properties; this IE has not attempted to harmonize them or develop standard codelists for these properties.

**BhSamplingProcedure**

BhSamplingProcedure is the process used perform a sampling activity. Both drilling and the collection of samples constitute sampling activities as both expose FeaturesOfInterest for observation. The example log defines four such procedures (item 15 above). Examples:

- Hollow stem augering (uppermost 21 ft of the hole). Cuttings samples produced by the drilling may be collected for testing or observed from the auger flights.

```
{
    "@iot.id": 13,
    "description": "Drilling with 3.25\" hollow stem auger",
    "name": "3.25\" HSA",
    "properties":{
        "type": "Drilling Method"
    }
}
```

- Rotary drilling using a NQ2 diamond core system (below 21 ft). Cuttings produced by the drilling.

```
{
    "@iot.id": 14,
    "description": "Drilling with NQ2 diamond core system",
    "name": "NQ2",
    "properties":{
        "type": "Drilling Method"
    }
}
```

- Soil sampling collected as a result of performing an SPT test (uppermost 21 ft of the hole), producing disturbed core samples.

```
{
    "@iot.id": 15,
    "description": "Sampling resulting from SPT testing",
    "name": "SPT",
    "properties":{
        "type": "Sampling Method"
    }
```

```
    }
```

- Rock core sampling using a double tube NQ core barrel below 21 ft. This procedure produces solid rock core

```
    {
        "@iot.id": 16,
        "description": "Core sampling using a double tube NQ core barrel",
        "name": "NQ2",
        "properties":{
            "type": "Sampling Method"
        }
    }
```

**NOTES:**

- The properties:type is defined to distinguish sampling procedures that specifically are designed to collect material samples (Sampling Method) from those that are primarily designed to advance the hole (Drilling Method).

**BhFeatureType**

BhFeatureType is used to classify different kinds of BhFeaturesOfInterest. A BhFeatureOfInterest may be associated with any number of BhFeatureTypes. From our previous discussion on Borehole Sampling, five dedicated generic feature types are defined for features of interest within a borehole (a BhFeatureOfInterest):

- Hole
- Core
- Point
- Segment
- Entirety

These feature types are used in combination to provide the appropriate context for the feature. Features that derive from the Hole (thus simply locations where observations are made in-situ), are assigned a feature type of Hole, plus a feature type of either Point, Segment, or Entirety depending on the feature’s linear extent. Material samples recovered from the hole (features of interest typically subjected to ex-situ observation), are assigned a type of Core, plus a Point, Segment, or Entirety feature type as with Hole features. Instances for these dedicated types are:

**Hole**

```
    {
        "@iot.id": 1,
        "definition": "https://ogc.org/Hole",
        "description": "A Sample from a Borehole-Hole",
```

```
        "name": "Hole"
    }
```

**Core**

```
    {
        "@iot.id": 2,
        "definition": "https://ogc.org/Core",
        "description": "A Sample from a Borehole-Core",
        "name": "Core"
    }
```

**Point**

```
    {
        "@iot.id": 3,
        "definition": "https://ogc.org/Point",
        "description": "A Sample obtained at a Point",
        "name": "Point"
    }
```

**Segment**

```
    {
        "@iot.id": 4,
        "definition": "https://ogc.org/Segment",
        "description": "A Sample obtained at a Segment",
        "name": "Segment"
    }
```

**Entirety**

```
    {
        "@iot.id": 5,
        "definition": "https://ogc.org/Entirety",
        "description": "A Sample taken from the Entirety",
        "name": "Entirety"
    }
```

For BhFeaturesOfInterest that are material samples, it is often important to distinguish by type the primary samples from those that are derived from collected samples either by subsampling (Subsample) or aggregation (Amalgamate). In addition, for the example borehole log, samples subject to laboratory testing are extracted from source samples and prepared in various ways prior to testing. These samples are commonly called specimens and should be identified as such by a BhFeatureType:

**Specimen**

```
{
    "@iot.id": 6,
    "definition": "https://ogc.org/Specimen",
    "description": "A Sample prepared from an existing Sample for testing purposes",
    "name": "Specimen",
}
```

The processing used to produce a specimen is described in BhPreparationProcedure and BhPreparationStep (see Atterberg limits)

**NOTES:** BhFeatureType names should come from a controlled list of terms. Different organizations and existing transfer standards (such as AGS and DIGGS) have their own keys and allowable values for sample types (eg. disturbed, remolded, soil, rock, etc.). Beyond the generic feature types discussed above, this IE has not attempted to harmonize other sample types or develop standard codelists for BhFeatureType names.

### BhSampling and BhFeatureOfInterest

BhSampling is used to expose the location of a sampling activity within the borehole and therefore must be associated with a borehole trajectory (BhTrajectoryThing). BhSampling is also optionally associated with both a BhSampler and a BhSamplingProcedure and produces one or more BhFeaturesOfInterest. BhFeatureOfInterest is associated with an Observation and can either be a material sample (BhFeatureType of Core, Core Point or Core Segment) or in the case of an in-situ observation, an Observation location (BhFeatureType of Hole, Hole Point or Hole Segment). Many BhSampling and BhFeatureOfInterest instances are required to represent the data in the example borehole log. Below are example instances of BhSampling and BhFeatureOfInterest for selected elements of the example log.

**MATERIAL DESCRIPTIONS FROM 4.5 TO 10.5 FT DEPTH**

**BhSampling** *(Sampler=“ODOT/CAREY”, Sampling Procedure = “3.25" HSA”)*

```
{
    "@iot.id": 299,
    "description": "Material description layer",
    "name": "Auger interval 3",
    "fromPosition": 4.5,
    "toPosition": 10.5,
    "positionUom": "ftUS",
    "BhTrajectoryThing": {"@iot.id": 11},
    "BhSampler": {"@iot.id": 15},
    "BhSamplingProcedure": {"@iot.id": 13}
}
```

**BhFeatureOfInterest** *(Feature Type=“CORE, SEGMENT”)*

```
{
    "@iot.id": 300,
    "description": "Cuttings observed from 4.5-10.5 ft",
    "name": "Cuttings 4-10.5 ft",
    "BhSampling": {"@iot.id": 299},
    "BhFeatureTypes": [{"@iot.id": 2},{"@iot.id": 4}]
}
```

**PHYSICAL PROPERTY CHANGE AT 6 FT DEPTH**

**BhSampling** *(Sampler="ODOT/CAREY", Sampling Procedure = "3.25" HSA")*

```
{
    "@iot.id": 300,
    "description": "Change in density remark",
    "name": "Physical Property Remark 1",
    "atPosition": 6,
    "positionUom": "ftUS",
    "BhTrajectoryThing": {"@iot.id": 11},
    "BhSampler": {"@iot.id": 15},
    "BhSamplingProcedure": {"@iot.id": 13}
}
```

**BhFeatureOfInterest** *(Feature Types="HOLE, POINT")*

```
{
    "@iot.id": 301,
    "description": "Observation location at 6 ft",
    "name": "Location at 6 ft",
    "BhSampling": {"@iot.id": 300},
    "BhFeatureTypes": [{"@iot.id": 1},{"@iot.id": 3}]
}
```

**SPT TEST FROM 1.5 TO 3 FT DEPTH** )*(Multiple features of interest are associated to one sampling location)*

**BhSampling** *(Sampler="ODOT/WILLIAMS", Sampling Procedure = "SPT")*

```
{
    "@iot.id": 301,
    "description": "SPT Test at 1.5 ft",
    "name": "SPT 1.5",
    "fromPosition": 1.5,
    "toPosition": 3.0,
    "positionUom": "ftUS",
    "BhTrajectoryThing": {"@iot.id": 11},
    "BhSampler": {"@iot.id": 16},
```

```
        "BhSamplingProcedure": {"@iot.id": 13}
    }
```

**BhFeatureOfInterest #1** *(Feature of interest for SPT observation itself, Feature Types="HOLE ,SEGMENT")*

```
    {
        "@iot.id": 302,
        "description": "Observation location 1.5 to 3 ft",
        "name": "Location from 1.5-3.0 ft",
        "BhSampling": {"@iot.id": 301},
        "BhFeatureTypes": [{"@iot.id": 1},{"@iot.id": 4}]
    }
```

**BhFeatureOfInterest #2** *(Sample collected from the SPT test, Feature Types="CORE, SEGMENT")*

```
    {
        "@iot.id": 303,
        "description": "SPT Sample 1.5 to 3 ft",
        "name": "SS-1",
        "length": 1.08,
        "lengthUom": "ftUS",
        "recoveryPercentage": 72,
        "BhSampling": {"@iot.id": 301},
        "BhFeatureTypes": [{"@iot.id": 2},{"@iot.id": 4}]
    }
```

**BhFeatureOfInterest #3** *(Specimen created for particle size testing, Feature Type="SPECIMEN", Source sample (BhSampledFeature) = "SS-1")*

```
    {
        "@iot.id": 304,
        "description": "Particle size specimen 1.5 to 3 ft",
        "name": "Acme 123",
        "BhSampling": {"@iot.id": 301},
        "BhFeatureTypes": [{"@iot.id": 2},{"@iot.id": 4},{"@iot.id": 6}],
        "BhSampledFeatures": [{"@iot.id": 303}]
    }
```

**NOTES:**

- In the example log, a Specimen BhFeatureOfInterest instance is required each time a source sample (eg. SS-1) is prepared for testing (eg. particle size, Atterberg limits, etc.).
- A Specimen is assumed to be subsampled from its source sample(s).
- A Specimens that constitute the entire spatial extent of its source sample may link to the same BhSampling entity as its source sample. However, if a specimen is extracted from only a portion of a source core sample, an additional BhSampling instance would be required to identify the specific location of the Specimen within the borehole trajectory.

- A Specimen BhFeatureOfInterest instance need not be created if the entire source sample is consumed in testing.

**ObservedProperty**

ObservedProperty identifies the properties of the BhFeatureOfInterest that is observed by the Sensor (observing procedure). One ObservedProperty instance should exist for each Observation result type that is obtained. For the example log above the observed properties are:

| Sensor/Observing procedure | Observed Properties (see Observable Properties) |
|---|---|
| Visual Soil and Rock Classification | lithology description<br>lithology classification<br>lithology symbol |
| Geologic units in Ohio (state in United States) | unit name |
| SPT test | see Approach for SPT |
| Pocket penetrometer | uniaxial compressive strength |
| Particle size distribution | gravel content<br>coarse sand content<br>fine sand content<br>silt content<br>clay content |
| Atterberg limits | see Approach for Atterberg limits |
| Natural water content, Method A | natural water content |
| ODOT Soil Classification | lithology classification |
| Rock Quality Designation | rqd rock quality designation |

Below are example instances for the observed properties associated with soil and rock classification and pocket penetrometer:

```
    {
        "@iot.id": 29,
        "name": "lithology classification",
        "definition":
  "https://github.com/opengeospatial/Geotech/wiki/ObservableProperties",
        "description": "The value that describes the lithology as a controlled term"
    }
```

```
    {
        "@iot.id": 30,
        "name": "lithology description",
```

```
        "definition":
"https://github.com/opengeospatial/Geotech/wiki/ObservableProperties",
        "description": "Descriptive information about the soil or rock lithology"
    }
```

```
    {
        "@iot.id": 31,
        "name": "lithology symbol",
        "definition":
"https://github.com/opengeospatial/Geotech/wiki/ObservableProperties",
        "description": "A string or numeric value that is used to define a graphic
pattern"
    }
```

```
    {
        "@iot.id": 32,
        "name": "uniaxial compressive stress",
        "definition":
"https://github.com/opengeospatial/Geotech/wiki/ObservableProperties",
        "description": "The maximum axial compressive stress that a right-cylindrical
sample of material can withstand under unconfined conditions"
    }
```

**NOTE:** Ideally, the definition property should link to a registry or online dictionary that provides a definition and full context for the observed property.

**Datastream**

The Datastream object serves to link individual observation results to their associated observed properties and observing procedures (Sensors). It also is used to provide additional context to the observation result (such as unit of measure) and to the observed property. One Datastream instance is required for each unique combination of BhTrajectoryThing, Sensor, and ObservedProperty. The following are a few example Datastream instances for the borehole log above:

- **Datastream for ObservedProperty = “lithology description” and Sensor = “Visual Soil and Rock Classification”**

```
    {
        "@iot.id": 29,
        "description": "Datastream for B-001-0-20 lithology description",
        "name": "B-001-0-20 Lithology description",
        "observationType": "http://www.opengis.net/def/observationType/OGC-
OM/2.0/OM_Measurement",
        "unitOfMeasurement": {},
        "Sensor": {"@iot.id": 15},
        "ObservedProperty": {"@iot.id": 30},
        "Thing": {"@iot.id": 11}
```

```
    }
```

- **Datastream for ObservedProperty = “uniaxial compressive stress” and Sensor = “Pocket Penetrometer”**

```
    {
        "@iot.id": 30,
        "description": "Datastream for B-001-0-20 Pocket penetrometer",
        "name": "B-001-0-20 Pocket penetrometer/uniaxaal compressive stress",
        "observationType": "http://www.opengis.net/def/observationType/OGC-
OM/2.0/OM_Measurement",
        "unitOfMeasurement": {
           "name": "tonf[US]/ft2",
           "symbol": "tonf[US]/ft2",
           "definition": "US tons force per square foot"
        },
        "Sensor": {"@iot.id": 17},
        "ObservedProperty": {"@iot.id": 32},
        "Thing": {"@iot.id": 11}
    }
```

- **Datastream for ObservedProperty = “lithology classification” and Sensor = “ODOT Soil Classification”**

```
    {
        "@iot.id": 31,
        "description": "Datastream for B-001-0-20 ODOT Soil Classification",
        "name": "B-001-0-20 ODOT Soil Classification",
        "observationType": "http://www.opengis.net/def/observationType/OGC-
OM/2.0/OM_Measurement",
        "unitOfMeasurement": {},
        "Sensor": {"@iot.id": 20},
        "ObservedProperty": {"@iot.id": 29},
        "Thing": {"@iot.id": 11}
    }
```

- **Datastream for ObservedProperty = “lithology classification” and Sensor = “Visual Soil and Rock Classification”**

```
    {
        "@iot.id": 32,
        "description": "Datastream for B-001-0-20 Soil Classification for Graphic
Log",
        "name": "B-001-0-20 Soil Classification for Graphic Log",
        "observationType": "http://www.opengis.net/def/observationType/OGC-
OM/2.0/OM_Measurement",
        "unitOfMeasurement": {},
```

```
        "Sensor": {"@iot.id": 15},
        "ObservedProperty": {"@iot.id": 29},
        "Thing": {"@iot.id": 11}
    }
```

**Observation**

The Observation object holds the result of an ObservedProperty for a BhFeatureOfInterest. Each Observation instance links directly to a BhFeatureOfInterest instance and indirectly to the result's ObservedProperty through the Observation's associated Datastream.

On the example borehole log, every numeric, text or graphic entry below the log's header (with the exception of depth and elevation information) is the result of an individual Observation instance. Below are a few examples: * **Observation for material description at depth interval between 4.5 and 10.5 feet**

```
    {
        "@iot.id": 881,
        "phenomenonTime": "2021-01-11T00:00:00-05",
        "result": "MEDIUM DENSE, REDDISH BROWN AND BROWN, STONE FRAGMENTS WITH SAND,
SILT, AND CLAY, DAMP",
        "FeatureOfInterest": {"@iot.id": 300},
        "Datastream": {"@iot.id": 29}
     }
```

- **Observation for graphic log symbolization at depth interval between 4.5 and 10.5 feet**

```
    {
        "@iot.id": 882,
        "phenomenonTime": "2021-01-11T00:00:00-05",
        "result": "STONE FRAGMENTS WITH SAND, SILT, AND CLAY",
        "FeatureOfInterest": {"@iot.id": 300},
        "Datastream": {"@iot.id": 32}
     }
```

- **Observation for material description (physical property change) at 6 ft depth**

```
    {
        "@iot.id": 883,
        "phenomenonTime": "2021-01-11T00:00:00-05",
        "result": "@6.0'; DENSE",
        "FeatureOfInterest": {"@iot.id": 301},
        "Datastream": {"@iot.id": 29}
     }
```

- **Observation of uniaxial compressive strength, from hand penetrometer test on sample**

**from 1.5 to 3 ft depth**

```
{
    "@iot.id": 884,
    "phenomenonTime": "2021-01-11T00:00:00-05",
    "result": 1.75,
    "FeatureOfInterest": {"@iot.id": 303},
    "Datastream": {"@iot.id": 30}
 }
```

- **Observation of ODOT Soil Classification, from sample collected from 1.5 to 3 ft depth**

```
{
    "@iot.id": 885,
    "phenomenonTime": "2021-01-11T00:00:00-05",
    "result": "A-4a (1)",
    "FeatureOfInterest": {"@iot.id": 303},
    "Datastream": {"@iot.id": 31}
 }
```

**NOTES:**

- The Observation property “phenomenonTime” is required and will default to the server time if the phenomenonTime is not provided when an Observation instance is created. In geotechnics, the time of an observation is often not reported for laboratory tests and many in-situ tests. If the observation time is unknown to the data provider, phenomenonTime must be estimated to circumvent the default behavior. In either case, an estimated or default phenomenonTime for an observation should be identified as such in an additional Observation property.
- Some kinds of tests, such as SPT and Atterberg limits, produce complex sets of Observation results that require that links be created to group related Observations and Datastreams together in order to distinguish interim or procedural observations from the final reported test results. Examples of these are shown in the detailed discussions for SPT and Atterberg limits tests.

## 5.5.2. Approach for CPT

Cone Penetration Testing (CPT) is a technique used to evaluate subsurface stratigraphy and material properties. An instrumented probe is advanced into the soil by direct-push techniques from a reaction system, which is often a heavy vehicle but may be an independent frame with soil reaction anchors. Measurements from probe sensors are recorded on the probe or transmitted in real-time to an up-hole data collection and processing system at frequent intervals. Unlike many other in-situ geotechnical investigation techniques, the system does not use rotary drilling. CPT is frequently used both independently and as a companion to borehole drilling. Cone penetration testing is used worldwide but is generally less common than rotary drilling and related testing (such as SPT) due to the ubiquity of drill rigs and the ability of drills to penetrate nearly all geomaterials for site characterization. CPT rigs may be mounted on skids, wheeled or tracked vehicles, or reaction frames which can be moved into place manually or using light vehicles and small equipment.

Vehicles and frames may use earth anchors to provide reaction for the probe push system. CPT systems can also be deployed in marine environments using specialty systems.
Most modern CPT probes are configured with sensors to measure the force acting on the cone tip, force acting on a nearby friction sleeve, pore water pressure, and probe vertical tilt (used primarily to evaluate the probe position and inclination for terminating a push if verticality is not well maintained).

The technique has several advantages.

- Probe sensors are generally robust and easy to check for proper calibration; measurements are often highly repeatable and less operator dependent than other techniques.
- Data is collected electronically at frequent time/depth intervals creating a near-continuous record.
- The technique is relatively fast with a standard advance rate of 20mm/sec.
- Probes are available in several shapes and sizes for targeted applications.
- The technique is well suited for both land and marine applications.
- Many systems are automated to varying degrees.
- Probes may be configured with a variety of additional sensors:
  - Multi axial or multiple location geophones (seismic wave speed; pressure and shear wave measurement)
  - Video cameras
  - Electrical conductivity or resistivity
  - Fuel fluorescence
  - Gamma radiation detectors
  - Magnetometers
  - Multiple friction sleeves
- Relative safety.
  - Rigs are often enclosed with protection from weather, traffic, dust, and other conditions.
  - There are no rotating parts in the operational area greatly minimizing operator hazards
- Reaction force may come from dead-weight or a variety of anchoring systems
- Rigs exist in a wide range of sizes and configurations
- The operation results in relatively little disturbed ground and spoil and is well suited to environmentally sensitive sites.

CPT systems record near-continuous readings of measured sensor output. These readings are then processed using calibration conversions to engineering units, usually in real time during the advance of the probe for operator observation and reaction. Some probes have an onboard memory system and the probes are pushed and retrieved without real- time information [limiting issues related to wires and wireless commination]. Proprietary output files usually contain a variety of metadata, calibration information, and probe measurements. Additional supporting files may be created if standard testing is paused to conduct specialty time-domain tests at specified depths, such as pore water pressure dissipation tests or shear wave velocity tests. Recorded measurements are usually post-processed to develop a variety of engineering parameters based on calculations, theoretical, or empirical relationships.

Basic measurement output from modern electronic CPT systems includes:

- Time
- Depth [measured by an up-hole sensor]
- Tip stress
- Sleeve friction
- Pore-water pressure*
- X-Y Inclination [measured by a pair of accelerometers]

Other measured real-time values could include the measured down pressure of the drive system from pressure transducers included in the hydraulic system. A common calculated real-time value is the Friction Ratio. The pore pressure is typically measured immediately behind the tip element at the cone shoulder (the u2 position). Some specialty cones have the pore pressure element at the cone nose tip (the u1 position) or behind the friction sleeve (the u3 position), though these are far less common in non-research [production] applications. The location of the pressure sensor element is usually included in the test metadata. The CPT test measurements are usually post-processed and information about the site location (coordinates and elevation) is often added into the file along with other site information, such as the sounding number, time, weather, rig idendification, etc. Complete depth-based post-processed output frequently includes corrected measurements and calculated parameter values:

- Corrected Tip stress
- Friction Ratio
- Resistivity/Conductivity
- Seismic arrival times (P, S1, S2 waves at intervals)
- Correlated parameters (by one or more methods)
  - Soil Behavior Type (SBT)
  - Soil Material Index (Ic)
  - Effective Friction Ratio (phi prime)
  - Undrained Shear Strength (su)
  - Unit weight
  - Preconsolidation Stress (sigma prime sub p)
  - Yield Stress Ratio (YSR)
  - Lateral Stress Coefficient (Ko)
  - Constrained Modulus (D')
  - Drained Young's Modulus (E')
  - Bulk Modulus (K')

- Subgrade Reaction Modulus (ks)
- Resilient Modulus (ks)
- Small Strain Shear Modulus (Gmax)
- Coefficient of Permeability (k)
- Coefficient of Consolidation (Cv)

Some specialty tests are performed at intervals during a regular CPT advance. Pore water pressure dissipation tests may include the time and a series of pore pressure measurements taken at a specific depth. These time-domain tests are typically conducted at selected depths (rather than at regular intervals) and may take a few minutes to several hours to complete. The test data is typically recoded in a separate CPT data file, at each depth where a test is conducted, with a different data structure, associated with the main data file by test ID.

Similarly, seismic tests may include the time and geophone response at a specific depth. Other specialty measurements (or data streams) may be recorded with the basic measurements with additional sensors on the probe, such as electrical conductivity or magnetometers. Similar to pore water pressure dissipation, the test data is typically recoded in a separate CPT data file at each depth where a test is conducted, with a different data structure, associated with the main data file by test ID.

A video feed from a camera mounted within the probe may be recorded as a separate data stream. Video cones are outfitted with depth encoders to include the depth on the video recording (which may be recorded separately as an analog or digital format companion file). Typically CPT logs contain the basic test data and dissipation and seismic data is processed separately and the calculated "results" (such as permeability (k) or pressure or shear wave velocity) are indicated on the log.

CPT operators often also have the capability to advance a soil sampler using the same reaction, advance system, and rods in an adjacent sounding. Samples are obtained and companion information from samples, such as photographs, moisture assessments, stratigraphic notes, soil identification and classification may accompany the digital record.

In addition to the advantages described for data acquisition in the field, the relative robustness of the data is useful for correlations and assessments of site variability. CPT data is often used to characterize site stratigraphy [with a high degree of confidence], model liquefaction potential, or to model the design performance of both deep and shallow foundation systems. Often, CPT measurements are correlated to traditional Mohr-Coulomb (phi-c) engineering design parameters, but some modern design techniques use direct-design from original measured CPT engineering values (without the correlation step). There are dozens of direct CPT design methods, making the exchange and interoperability of CPT data especially valuable for use when importing values into design software.

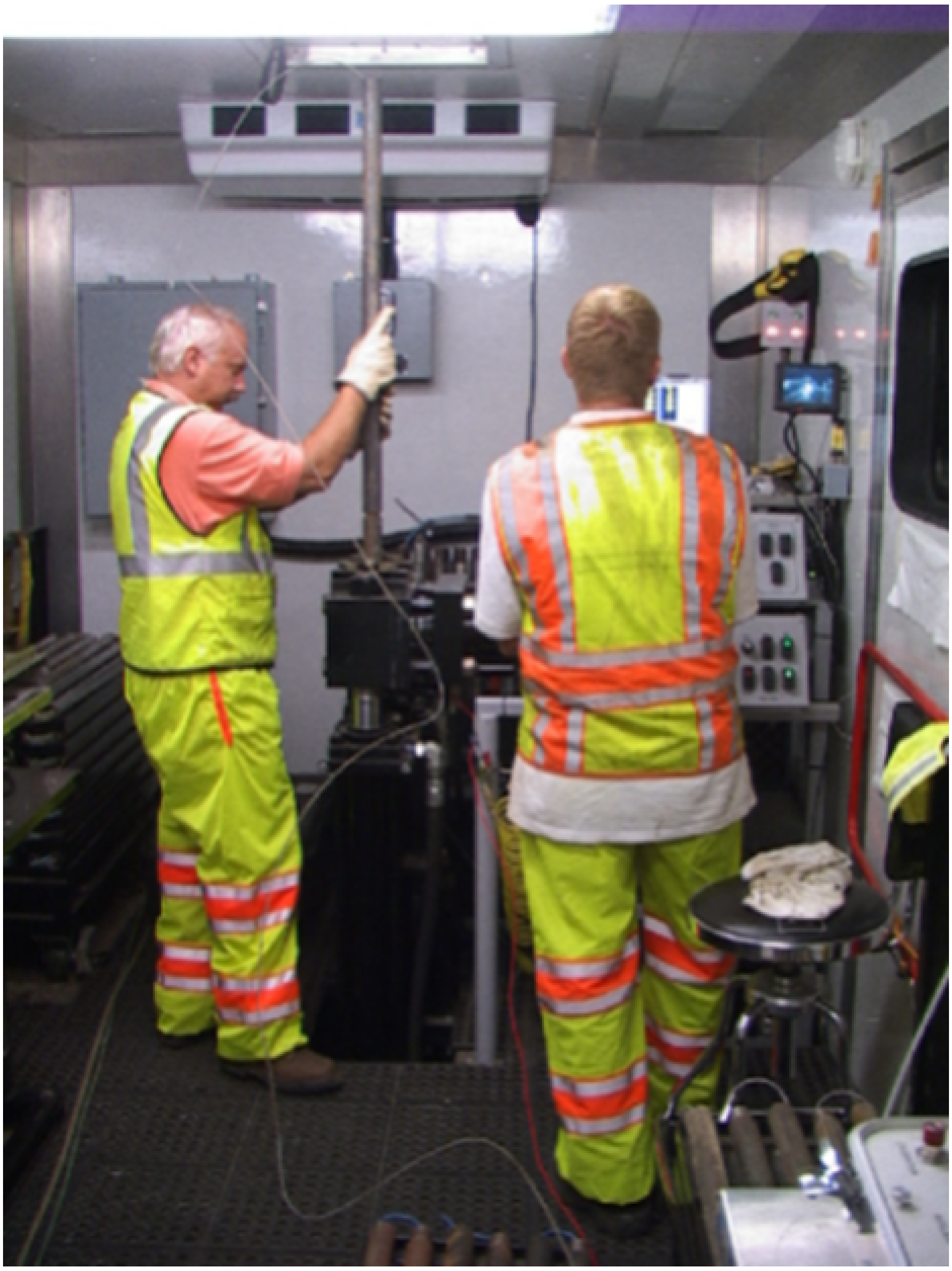

*Figure 25. CPT rods being added as a probe is advanced at a project site. Photo courtesy of Minnesota Department of Transportation.*

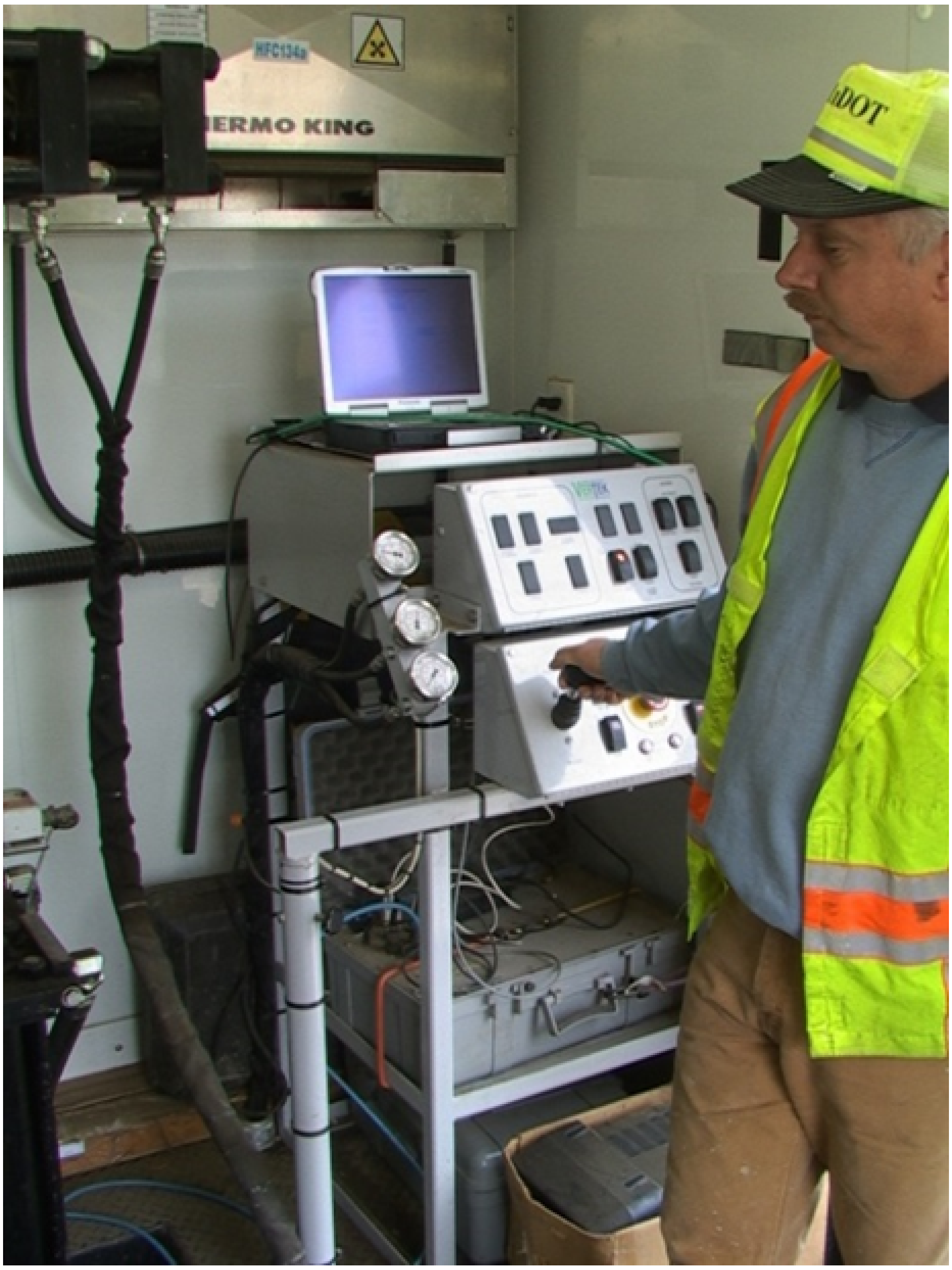

*Figure 26. Operator controlling the push rate as a CPT probe is advanced and data is automatically recorded by a data acquisition system. Photo courtesy of Minnesota Department of Transportation.*

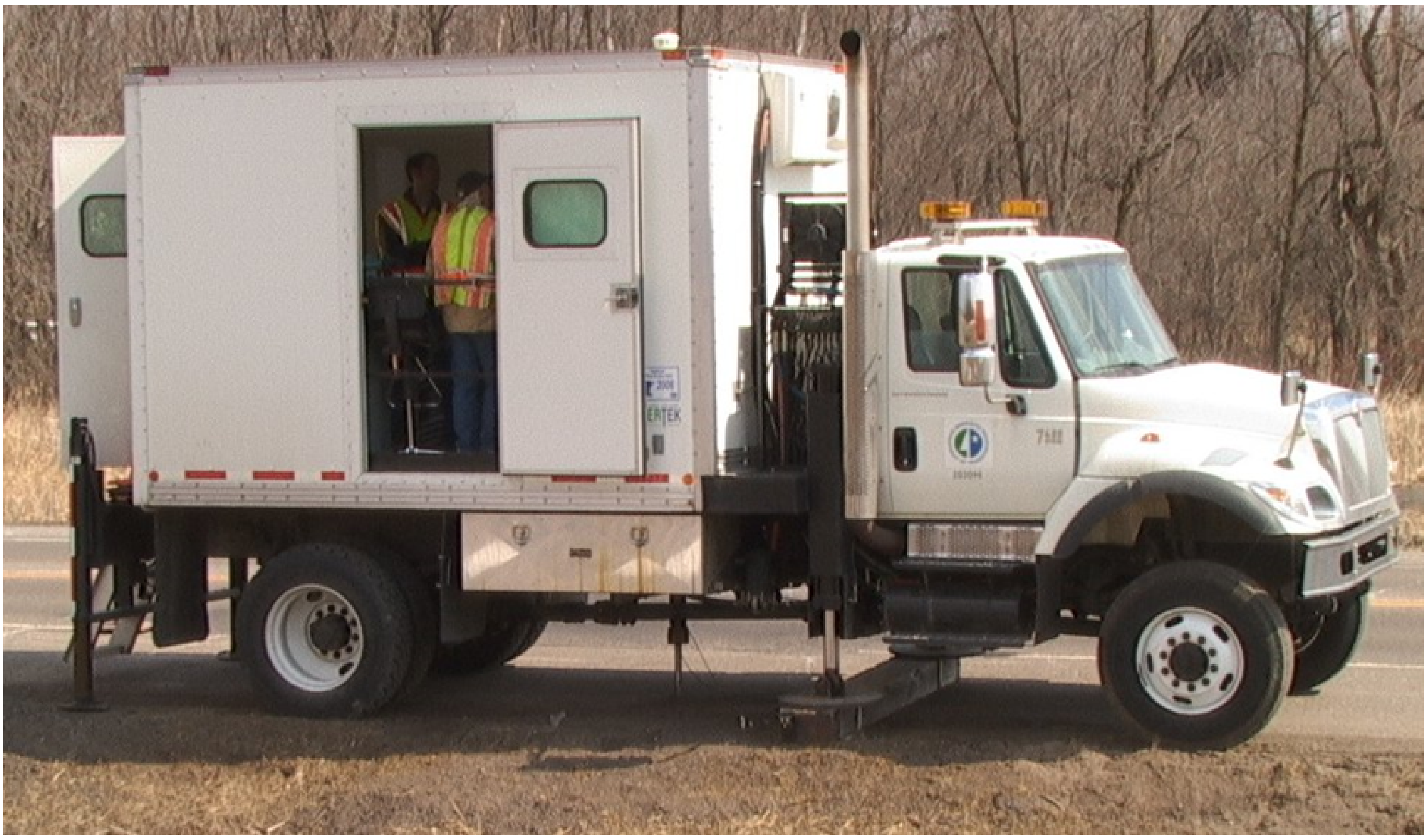

*Figure 27. A medium-sized truck-mounted CPT direct-push rig characterizing soils near an existing culvert. The truck is leveled on a set of jacks and the cone is pushed from a location near the center of gravity of the carrier. Photo courtesy of Minnesota Department of Transportation.*

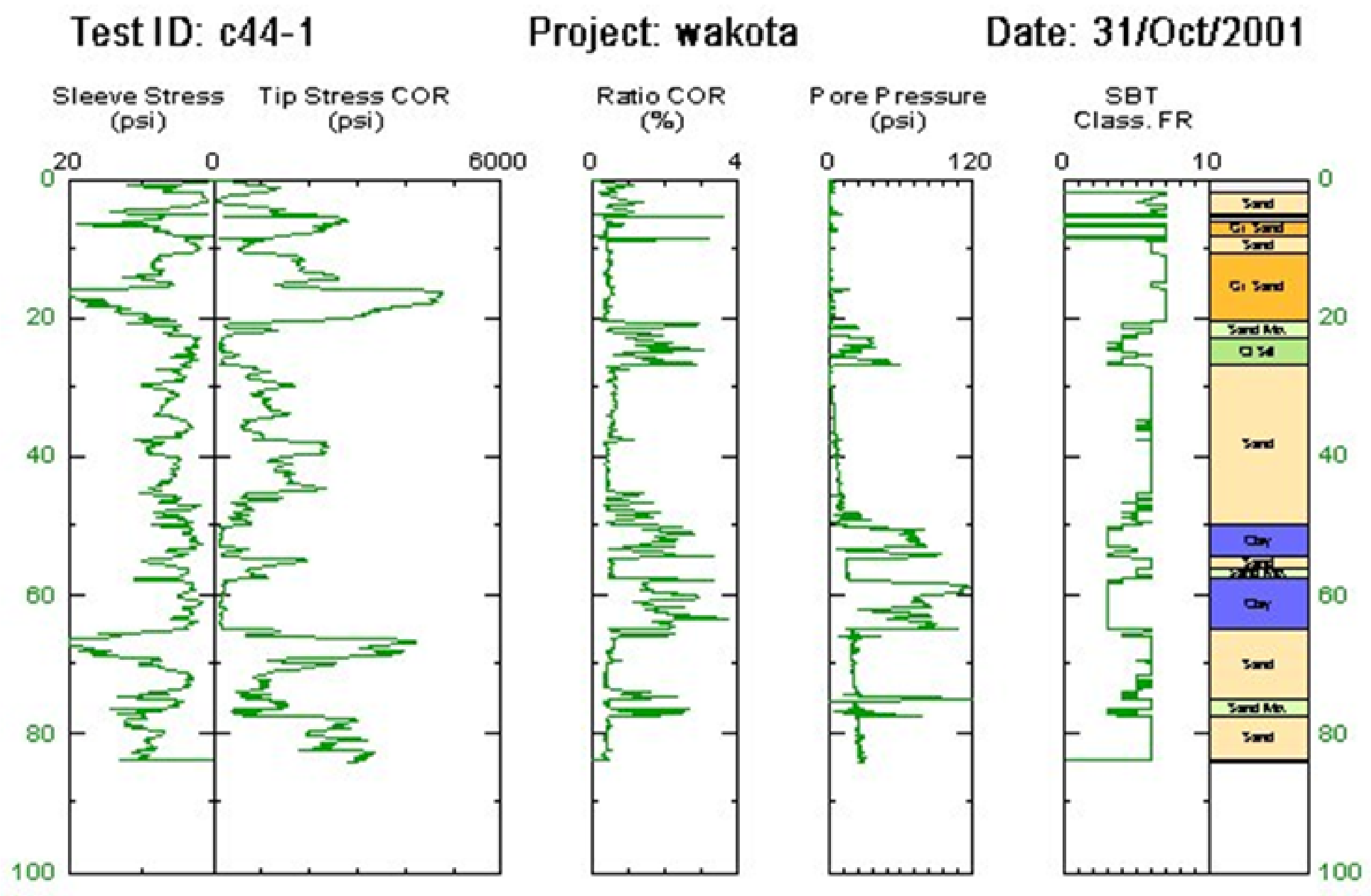

*Figure 28. A typical CPT output plot showing Sleeve Stress, Tip Stress, Friction Ratio, Pore Pressure and a calculated Soil Behavior Type (SBT). Figure courtesy of Minnesota Department of Transportation.*

The applicable US Standard is ASTM D5778-20, Standard Test Method for Electronic Friction Cone and Piezocone Penetration Testing of Soils

### Exposing CPT with the FROST Geotech Plugin

A typical CPT sounding includes the measurement of several soil properties at numerous closely spaced points as the CPT probe is advanced into the subsurface. To demonstrate how CPT measurements and and related metadata are stored and organized within the FROST server, we offer the following example of a simple CPT sounding where three propertis are measured:

#### GENERAL TEST INFORMATION

1. Test procedure: ASTM D3441-16, Standard Test Method for Mechanical Cone Penetration Testing of Soils
2. Sounding info: C-10, 15.03 ft sounding depth, top of sounding located at lan/lon 39.475026/-81.795909, elevation 252.61824 meters (WGS84).
3. Cone Serial Number: 128.074
4. Tip area: 15 cm2
5. Distance between tip and sleeve transducers: 15 cm
6. Probe Penetration Rate: 1 cm/s
7. Net Area Ratio Correction: 0.8

#### OBSERVATIONS MADE DURING THE TEST

Below is a sampling of the data generated. There are 263 sounding points total

| Depth (ft) | Tip Resistance (tsf) | Sleeve Friction (tsf) | u2 Pore Pressure (tsf) |
|---|---|---|---|
| 0.153 | 16.1 | 0 | 0.06 |
| 0.211 | 26.2 | 0 | 0.06 |
| 0.292 | 41.8 | 0.35 | 0.07 |
| .. | … | … | … |
| 3.357 | 32.2 | 1.2 | -0.1 |
| **3.41** | **32** | **1.19** | **-0.1** |
| 3.461 | 31.9 | 1.19 | -0.09 |
| 3.516 | 31.6 | 1.16 | -0.09 |
| … | … | … | … |
| 14.765 | 84.1 | 0.47 | 0.22 |
| 14.824 | 84.6 | 0 | 0.23 |
| 15.034 | 78.1 | 0 | 0.23 |

### Instance Diagram

Object instances and the associations required to properly expose the example test data with the FROST Geotech Plug-in are shown in the following object diagram. This shows data for only sounding point at depth of 3.41 ft:

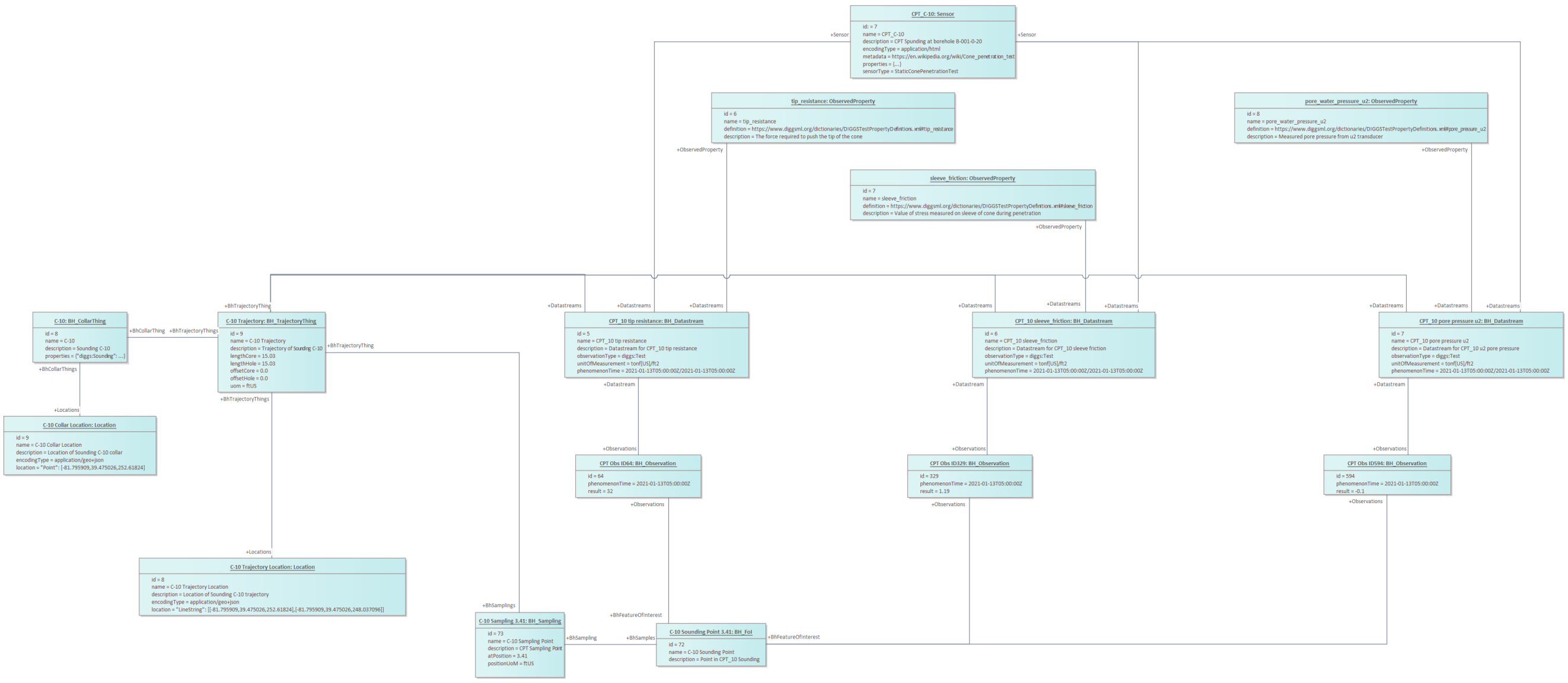

*Figure 29. CPT*

The following summarizes the various entities in the diagram:

#### Sensor

The Sensor object serves as the observing procedure in STA. One object instance is needed for this example (top center of diagram), and in this example holds the information about the test procedure, test parameters, equipment used and other metadata about the test, The general test data above (except for item 2) are all stored in the properties object of this Sensor instance. One Sensor instance is needed for all CPT tests conducted in one or more soundings provided that the same procedure, test parameters and test equipment is used for all tests.

#### ObservedProperty

The ObservedProperty object instances identify the properties that are observed by the CPT test. These are: - tip_resistance - sleeve_friction - pore_water_pressure_u2

As with Sensor, the ObservedProperty instances can be reused for multiple tests.

#### DataStream

All of the object instances in the diagram are linked to the Sensor and ObservedProperty instances via Datastream instances (below the ObservedProperty objects on the diagram), which serve to associate observation results obtained from a feature of interest to its observed property, observing procedure, and the sounding's trajectory.

Three Datastream instances are needed, one for each ObservedProperty instance.

#### BhCollarThing, BhTrajectoryThing and Location

The DataStreams all link to the souinding via its BhTrajectoryThing object instance. BhTrajectoryThing (left edge of diagram) represents the sounding's geometry and contains the sounding length and information for linear referencing. The trajectory's geometry is given in the associated Location instance. BhTrajectoryThing is associated with a BhCollarThing instance, which represents the sounding as a whole. All general metadata about the sounding is contained in the BhCollarThing object instance; it’s geometry is represented by an associated point Location object instance.

More detail about properties of BhCollarThing and BhTrajectoryThing can be found in the Borehole log discussion.

#### BhSampling and BhFeatureOfInterest

Sampling in the context of a CPT test or any in-situ test is the act of observing properties a a point or segment of the BhTrajectoryThing. The single BhSampling object instance (below and to the right of the BhTrajectoryThing in the diagram) holds the depth information of this sounding point (atPosition=3.41) and links to BhTrajectoryThing in order to affix the linear referenced sampling positions to the trajectory geometry.

BhSampling produces a BhFeatureOfInterest object, which represents the sampling location within the sounding (to the right of the BhSampling object instance in the diagram).

### Observation

The remaining entities on the diagram are Observation instances that provide the results for their associated observed properties. Each Observation instance links to the BhFeatureOfInterest (sampling location) and to the Datastream instance associated with the appropriate ObservedProperty. As the results of each observed property are independently measured by sensors on the CPT probe, they are independent and therefore there are no related observations.

To add data from more sounding points, this requires adding a new BH_Sampling and BH_FoI in addition to the three new Observations for each sounding point. In the diagram below, we have added an additional sounding point (the next one in the depth series). These additions are shown in grey to distinguish them from the objects already displayed in the simpler diagram above.

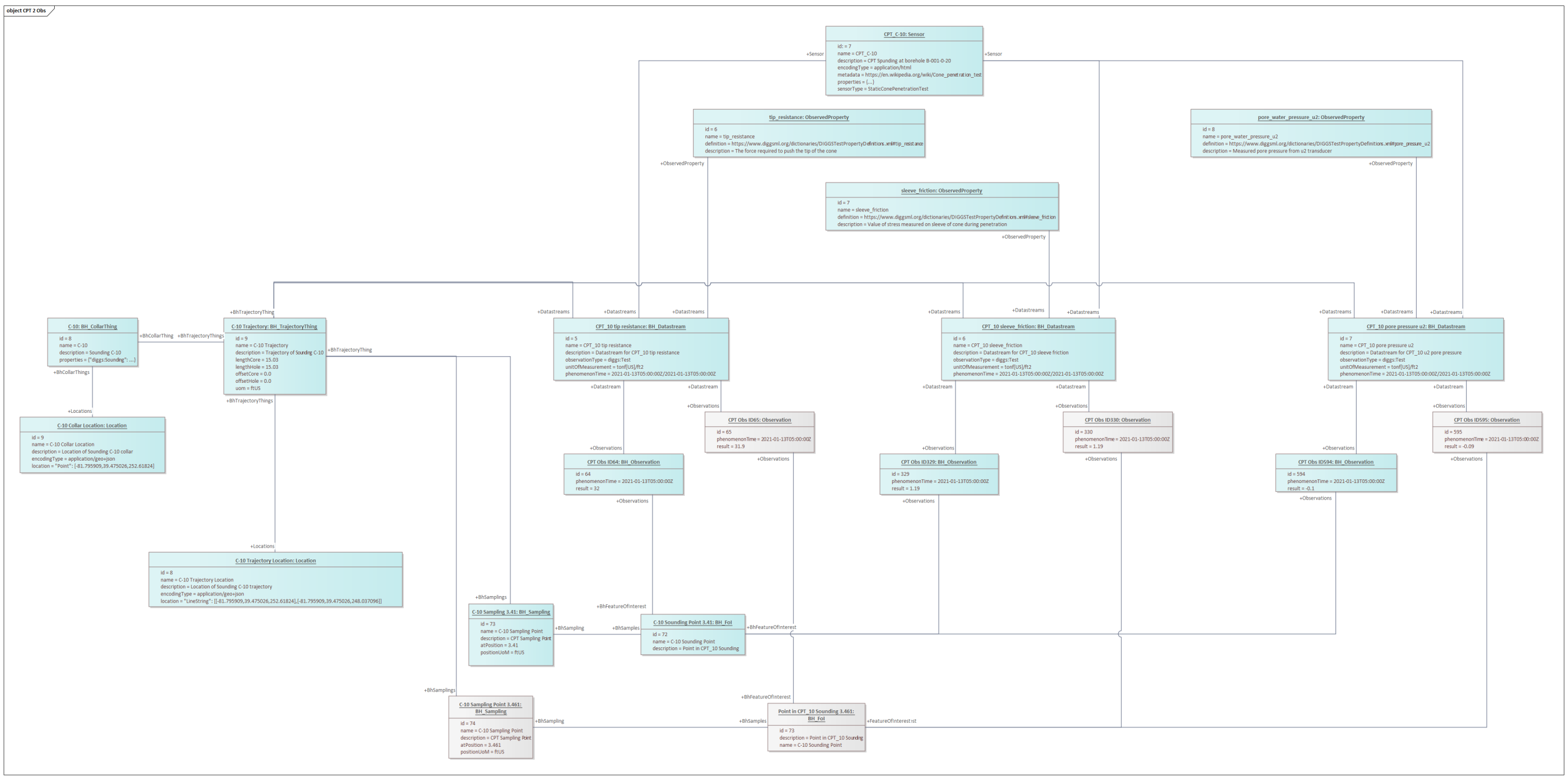

*Figure 30. CPT Observation*

### 5.5.3. Approach for SPT

The Standard Penetration Test (SPT) is a very widely used in-situ geotechnical test often applied in combination with other tests during the advance of a borehole by various rotary drilling techniques. The boreholes may also be advanced by sonic or direct push techniques, but these are far less common. The test is typically conducted at discrete intervals following separate drilling operations.

The test is valued for its ability to retrieve a sample and a simultaneous measurement of soil strength, as well as its relatively low cost, simple test protocol, wide-availability, and compact data set which can be easily manually recorded and transmitted. The sampler is relatively robust and can be used in nearly any geo-environment: soft to hard soils, cemented soils, regional geomaterials, urban fill, and soft rock.

The test is commonly used to evaluate the strength of soils through correlations to the angle of internal friction of granular materials and soil relative density. Over the decades of use of SPT, researchers have also amassed correlations to most every commonly used engineering parameter, with variable reliability, precision, and accuracy. Caution is advised when making interpretations about strength and other behaviors in environments where the test is not well suited, as described in governing standard test methods and other manuals of practice. While the SPT can collect samples in most materials it is not well suited to determining material strength properties of complex geotechnical materials such as peats, clays, and relatively large aggregates (as compared to the sampler size).

The SPT test is typically conducted across an interval of 18-inches [450 mm], creating a measurement associated with that sample length. Extremely soft or extremely hard materials may influence the test interval and measurement reporting. While back-to-back (continuous) SPT sampling is possible, the test is commonly conducted at intervals specified by the owner or agency. Often 2.5 ft., 5 ft., or 10 ft. intervals (or SI equivalent) during the advance of a borehole and is typically conducted at increasingly wider intervals as borings advance deeper owing to the need to alternate the drilling bit or tools with the sampler by pulling the complete string of drilling rods in and out of the hole.

The SPT test is considered a “disturbed sample” test as the sampler has a comparatively narrow aperture to collect the sample and thick outer wall. The basic mechanics of the test involve driving the soil sampler, at the base of the drill string, through the ground by a series of dynamic hammer drops applied to the top of the drill rods. The number of blows taken to drive the rods in measured depth intervals is recorded.

A variety of hammers and hammer systems can be used. Mechanical trip hammers, donut hammers, winch hammers, and safety hammers are in use worldwide. The test method prescribes a weight of 140 lbs [63.5 kg] and a drop height of 30 inches [750 mm]. Some hammer systems, such as rope and cathead hammers are operatory dependent and energy can vary widely. For improved performance and consistency, hammer systems should be calibrated to determine the energy transfer ratio (ETR), with the energy reported on boring logs for use in design [this is optional in the US standard].

For a typical test, a boring is advanced to the desired test depth (or elevation). The drilling tooling is pulled from the hole and the drill bit replaced with the SPT sampler. The sampler is lowered back into the borehole and the test begins. The top of the drill rod is marked off in three (3) 6-inch [750 mm] increments. The test consists of recording the number of blows to advance the sampler in each of three (3) consecutive 6-inch [150 mm] intervals. The sample tube is driven the first 6-inches and the number of blows needed to penetrate that interval is recorded. The procedure similarly repeats for the second 6-inch [150 mm] and third 6-inch [150 mm] intervals. After the sampler has advanced to a depth of 18-inches [450 mm] the test concludes. The three penetration “blow” numbers are recorded. A calculated value, the sum of the number of blows associated with the second 6-inch [150 mm] and third 6-inch [150 mm] intervals is reported as the “standard penetration resistance”, “N-value” or “SPT blow count”

Notably, there are a number of “special cases” where the N-value may be unusual. If the sampler is driven less than the full extent the number of blows for each full increment or partial increment is recorded. The number of blows in a given increment are counted until a test termination condition is reached:

- The sampler penetrates 6 inches (150 mm);
- The sampler is driven 50 blows in any of the 3 increments;
- The sampler is driven a total of 100 blows; or
- The sampler does not advance after 10 consecutive blows.

Additionally, if the sampler advances under the weight of the sampler and rods (or the sampler, rods, and hammer assembly) conditions of “weight of rods” or “weight of hammer” are noted rather than a “N-value.”

An additional component of the test is the “Sample Recovery” which is a visual observation and manual measurement of the sample recovered in the SPT sampler once the sampler is brought to the ground surface for recovery of the obtained sample within the SPT sampler.

A variety of metadata is required by the US and other standards. Applicable standards also specify the minimum precision associated with the reported values among other aspects of the testing procedures and equipment. Basic measurement output from a SPT test includes:

Basic measurement output from a SPT test includes:

- Depth of each test location along the extent of the borehole [measured by the operator based on the amount of drilling tooling in the ground and the ground elevation],
- The increment blow values of the 1st, 2nd, and 3rd increments, in addition to any special notations associated, such as the distance penetrated if less than 6-inches [150 mm]
- Notations if any of the termination criteria are reached other than penetration or if the sampler advances under the weight of rods or hammer
- The % recovery of each sample within the SPT sampler
- Other driller notations such as equipment and tooling, method of keeping the borehole open, depth of the water surface, size of casing, hammer systems used, etc.

In addition, the calculated N-value is the primary output from the test:

- The “standard penetration resistance”, “N-value” is calculated as the sum of the number of

blows associated with the second 6-inch [150 mm] and third 6-inch [150 mm] intervals.

- If the sampler is driven less than 18-inches [450 mm], the number of blows per each complete 6-inch [150 mm] increment and each partial increment is recorded. In this circumstance there is not an N-value associated with that depth.

The N-value is often reported along with a calculated value, N60, where the measured N-value is adjusted to a 60% drill rod energy transfer ratio. Depending on owner and agency test procedures, SPT measurements can further be corrected prior to their use in engineering calculations based on a variety of variables. Corrections may be applied based on:

- Rod type (A rod vs. N rod)
- Use of liners within the SPT sampler
- Short rod lengths
- Boring diameter
- Other factors

The SPT measurements (N-value or other partial penetration and blow measurements) are often recorded in the field on a boring log in a discrete column of data. This information is then supplemented with other field or lab data which may include the drilling operations, soil type by observation or identification or classification system evaluation techniques, and frequently moisture content evaluated from a portion of the sample recovered in the soil sampler. SPT measurements are commonly reported and associated on a boring log with other depth-based data at the same location:

- Soil moisture contents
- % sample recovery
- A text-based soil identification or classification value derived from the sample
- Symbology associated with a soil identification or classification
- A drilling operation (casing, plug drilling, etc.)

A challenge for this common test is that while the basic data is often relatively simple and concise, there are a large number of supporting informational elements either associated with the site, the borehole, or the test. Many of these details are required by the test standard in the report for a proper and complete understanding of the measurement data.

Driller and lab technician comments can be important to interpreting otherwise unusual values, when compared with other in-situ tests or laboratory tests on collected samples conducted in association with the SPT testing. As such, the numeric measurements associated with the SPT test are frequently associated with supplemental text-based notes and comments, which may not be consistent among organizations, drilling staff, or even an individual driller.

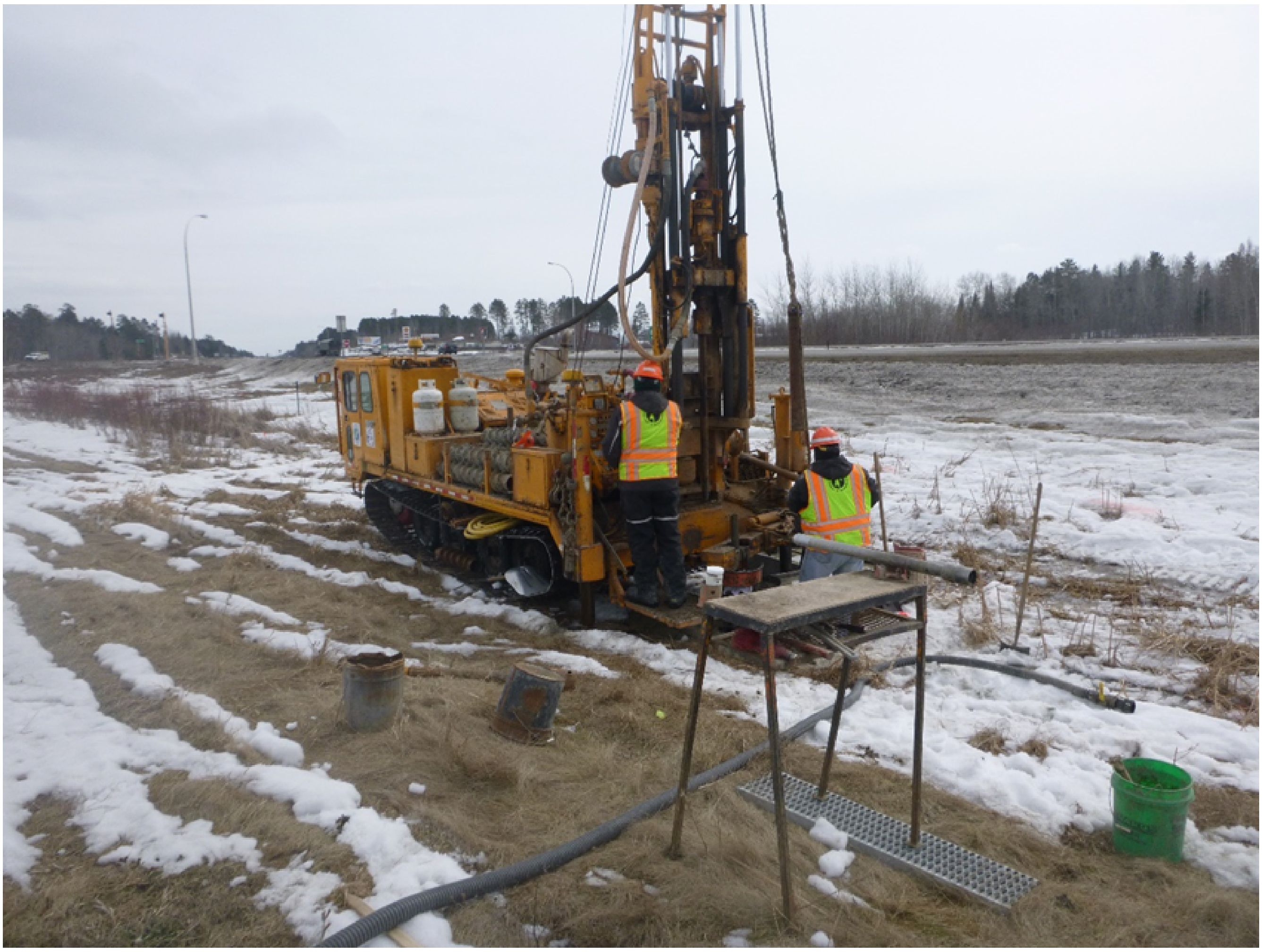

*Figure 31. A SPT drill rig on a project site; a work table with a SPT sampler is in the foreground. Photo courtesy of Minnesota Department of Transportation*

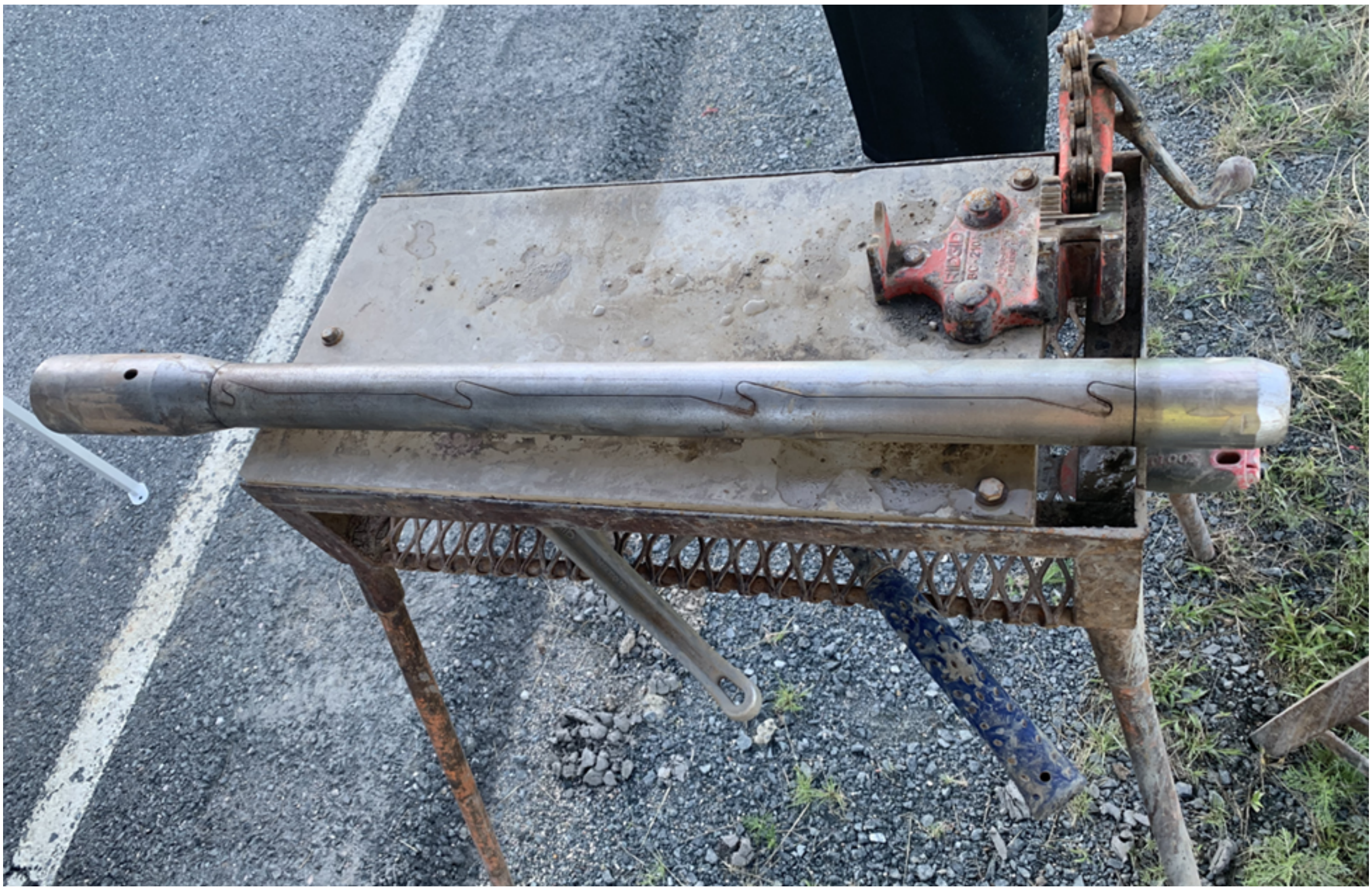

*Figure 32. A SPT sampler rests on a work table, ready to be attached to the drill rods when the drill tooling is removed at the desired sample depth. Photo courtesy of Minnesota Department of Transportation.*

The applicable US Standard is ASTM D1586/D1586M-18e1 Standard Test Method for Standard Penetration Test (SPT) and Split-Barrel Sampling of Soils.

**Exposing SPT with the FROST Geotech Plugin**

To demonstrate how the final results, interim test information and related metadata are stored and organized within the FROST server, we offer the following example SPT test for a single test interval. Relevant data are:

**GENERAL TEST INFORMATION**

1. Test procedure used: ASTM D1586/D1586M-18e1 Standard Test Method for Standard Penetration Test (SPT) and Split-Barrel Sampling of Soils
2. Location: Test run from 1.5 to 3 ft depth in borehole B-001-0-20
3. Borehole is is 41 ft deep and its collar is located at lat/lon 39.47466/-81.796858, elevation 249.50928 meters (WGS84)
4. Hammer used: CME Automatic
5. Hammer Efficiency: 84%

**OBSERVATIONS MADE DURING THE TEST**

| Drive Set Number | Blow Count | Penetration (ftUS) |
| --- | --- | --- |
| 1 | 9 | 0.5 |

| Drive Set Number | Blow Count | Penetration (ftUS) |
|---|---|---|
| 2 | 8 | 0.5 |
| 3 | 9 | 0.5 |

**FINAL REPORTED RESULTS * N-Value: 17 * N1-60: 24**

*Note: This example focuses only on the in-situ test procedure and its results; it does not provide any information regarding the recording of any material samples that may have been recovered as a result of the test. Use of the FROST plug-in for recording material sample information is discussed in* *Approach for Atterberg limits*

### Instance Diagram

Object instances and the associations required to properly expose the example test data with the FROST Geotech Plug-in are shown in the following object diagram:

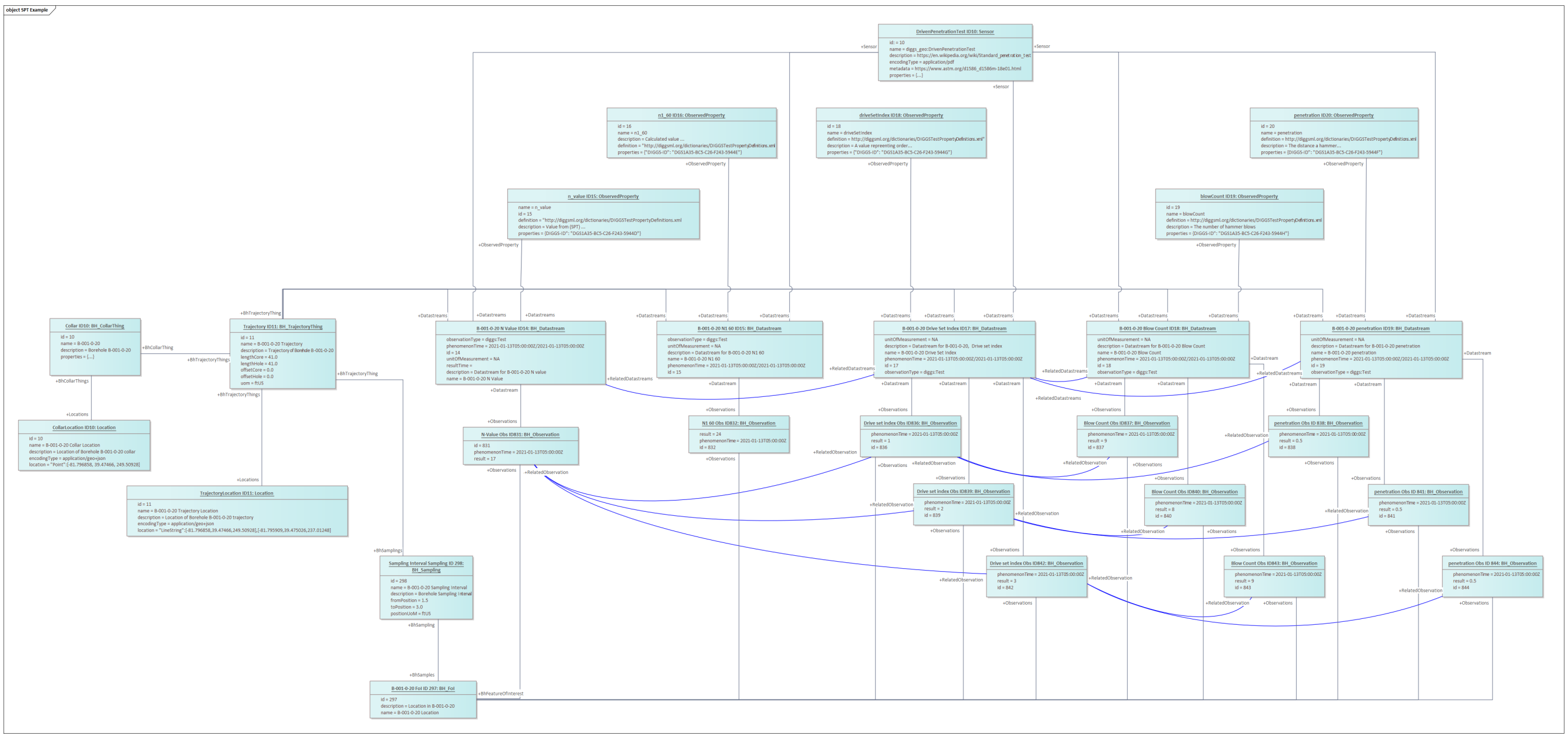

*Figure 33. SPT Example*

The following summarizes the various entities in the diagram:

#### Sensor

The Sensor object serves as the observing procedure in STA. One object instance is needed for this example (top center of diagram), and in this example holds the information about the test procedure, test parameters, equipment used and other metadata about the test. In this example, the hammer information and hammer efficiency data would be held in the properties object of the Sensor. One Sensor instance is needed for all SPT tests conducted in one or more boreholes provided that the same procedure, test parameters and test equipment is used for all tests.

#### ObservedProperty

The ObservedProperty object instances identify the properties that are observed by the SPT test. There are 5 ObservedProperty instances (top of diagram, below Sensor):

**two properties observed that constitute the final reported summary results of the test:**

1. n_value (standard penetration resistance)

2. n1_60 (standard penetration resistance corrected for hammer energy and overburden conditions

**three properties observed as the test is being run**

3. driveSetIndex (a counter for each set of hammer blows to drive the rod 150 cm)

4. blowCount (number of blows during the drive set (increment)

5. penetration (the distance the sampler travels during the drive set.

As with Sensor, the ObservedProperty instances can be reused for multiple tests.

#### DataStream

All of the object instances in the diagram are linked to the Sensor and ObservedProperty instances via Datastream instances (below the ObservedProperty objects on the diagram), which serve to associate observation results obtained from a feature of interest to its observed property, observing procedure, and the borehole.

Five Datastream instances are needed, one for each ObservedProperty instance.

#### BhCollarThing, BhTrajectoryThing and Location

The DataStreams all link to the borehole via its BhTrajectoryThing object instance. BhTrajectoryThing (left edge of diagram) represents the borehole's geometry and contains the borehole length and information for linear referencing. The trajectory's geometry is given in the associated Location instance. BhTrajectoryThing is associated with a BhCollarThing instance, which represents the borehole as a whole. All general metadata about the borehole is contained in the BhCollarThing object instance; it's geometry is represented by a point Location object instance.

More detail about properties of BhCollarThing and BhTrajectoryThing can be found in the Borehole log discussion.

#### BhSampling and BhFeatureOfInterest

Sampling in the context of an SPT test or any in-situ test is the act of observing properties in a segment of the hole. The single BhSampling object instance (below and to the right of the BhTrajectoryThing in the diagram) holds the depth information of the test run (fromPosition=1.5, toPosition=3) and links to BhTrajectoryThing in order to affix the linear referenced sampling positions to the trajectory geometry.

BhSampling produces a BhFeatureOfInterest object, which represents the sampling location within the borehole (below the BhSampling object instance in the diagram).

#### Observation

The remaining entities on the diagram are Observation instances that provide the results for their associated observed properties. Each Observation instance links to the BhFeatureOfInterest (sampling location) and to the Datastream instance associated with the appropriate ObservedProperty.

This SPT test consists of 11 individual observations. Two of them (results for n_value and n1_60) are reported summary results that derive from the driveSet (blowCount and penetration) results that are observed directly during the test. The links shown in the diagram that relate observations to each other provide the means for distinguishing between the driveSet and summary/derived results.

Each driveSet increment is a set of three observation results. These results are useless independently - for example, the blowCount in a driveSet has no meaning without the associated penetration result for that drivesetIndex, and vice versa. To model the driveSet, the driveSetIntex result is linked to its associated blowCount and penetration results, and the blowCount result is linked to its associated penetration result. These three observations, as a set, contribute to the determination of the final n_value result, and to model that association, the driveSetIndex observations are linked to the n-value observation. In this way, one can traverse from the n_value observation to access those blow count and penetration observations that contributed to the n_value result.

Finally, the n1_60 observation instance is linked to the n_value observation instance to demonstrate that the n1_60 result relies on n_value.

To provide the most flexibility for querying, Datastream object instances associated with the Observations are linked in the same manner as their associated Observations, as seen in the diagram.

The current STA model does not provide for one-way links where an association role can be assigned. Such capability would be useful in modeling evem more complex geotechnical test results.

### 5.5.4. Approach for Pressuremeter test

**Principle**

The principle of the test is to carry out an in-situ loading test, using an expandable cylindrical probe which is placed within the ground in a borehole. This probe, made up of three cells, is inflated with water and compressed air, thus exerting strictly uniform pressures on the wall of the borehole.

The strain of this wall are therefore accompanied by an increase in the volume of the probe which is then read, for each of the pressures, as a function of time. Thus, the pressuremeter allows to obtain a relationship between stress (applied pressure) and deformation (variation in the volume of water in the probe, directly linked to the strain of the soil).

Please note that unlike others in-situ test such as the CPT test, the pressuremeter test give «ponctual information», determined by the measurement step chosen by the experimenter. This measurement step is itself limited by the size of the probe, and it can be, generally, at least one meter or more.

The purpose and main interest of the pressurmeter test is to measure, trough the same test quantitative parameters of strength and deformation of the ground tested.

These parameters, calculated after the results of the test, are the following:

- _p_LM : Ménard pressuremeter limit pressure ;
- _p_f : p pressuremeter creep pressure ;
- _E_M : Ménard pressuremeter modulus.

These parameters could be used to do geotechnical calculations:

- Directly, through semi-empirical methods, example: foundation calculations; and
- Indirectly, trough the determination of others geotechnical parameters, for example, the Young Modulus of the soil used in numerical approach, which can be obtained from the Menard pressuremeter modulus

Indirectly, it is possible to link these parameters to a lithology, in particular the couple _p_LM _/ E_M, which could be used in the Pressiorama® (Baud, 2005), NF-P 94-262 annex B.

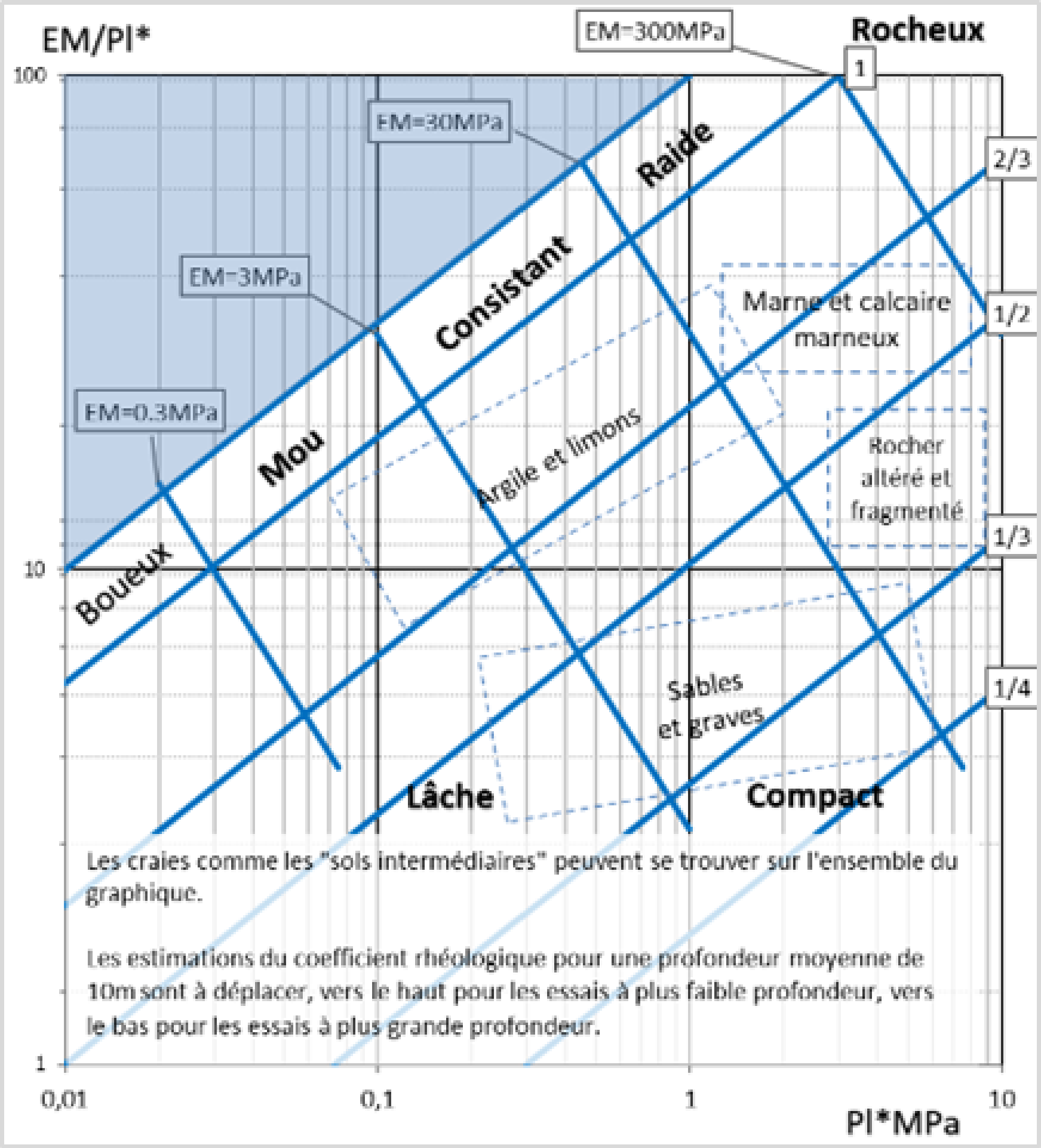

*Figure 34. Pressiorama® diagram (Baud, 2005, extracted from NF-P 94-262 annex B)*

Please note that the diagram is defined for a range of _p_LM up to 10 MPa, which corresponds to the actual application domain of the pressuremeter:

- _p_LM < 5 MPa : Range of values measurable by a « classic probe », corresponding to the values encountered in most soil type, to soft and weathered rocks.

- 5 MPa <_p_LM < 8 MPa : Range of values measurable by a « specific probe », corresponding to the values encountered in most soft and weathered rocks.
- _p_LM > 8 MPa : This range of values couldn't be measured with the actual equipment, and correspond to a rocky domain

In conclusion, keep in mind that the pressuremeter has a relatively wide range of application, from all types of soils to soft and weathered rocks. However, it could not be properly use for the characterisation of solid rocks.

**Equipment**

The following is mostly extracted from the standard NF EN ISO 22476-4, which is about the Ménard procedure of the pressurmeter test (the « classical » procedure). Please notice the existence of some variants of this procedure, depending principally of the way to instal the probe within the ground. These others specific procedures are exposed on the norms listed at the end of this page.

1. **General description**

The pressuremeter shown schematically in §2.2 shall include:

- the pressuremeter probe;
- the string of rods to handle the probe;
- the control unit (CU);
- The connecting lines between the control unit and the probe.

Some means of measuring the depth of the test with appropriate measurement error shall be provided.

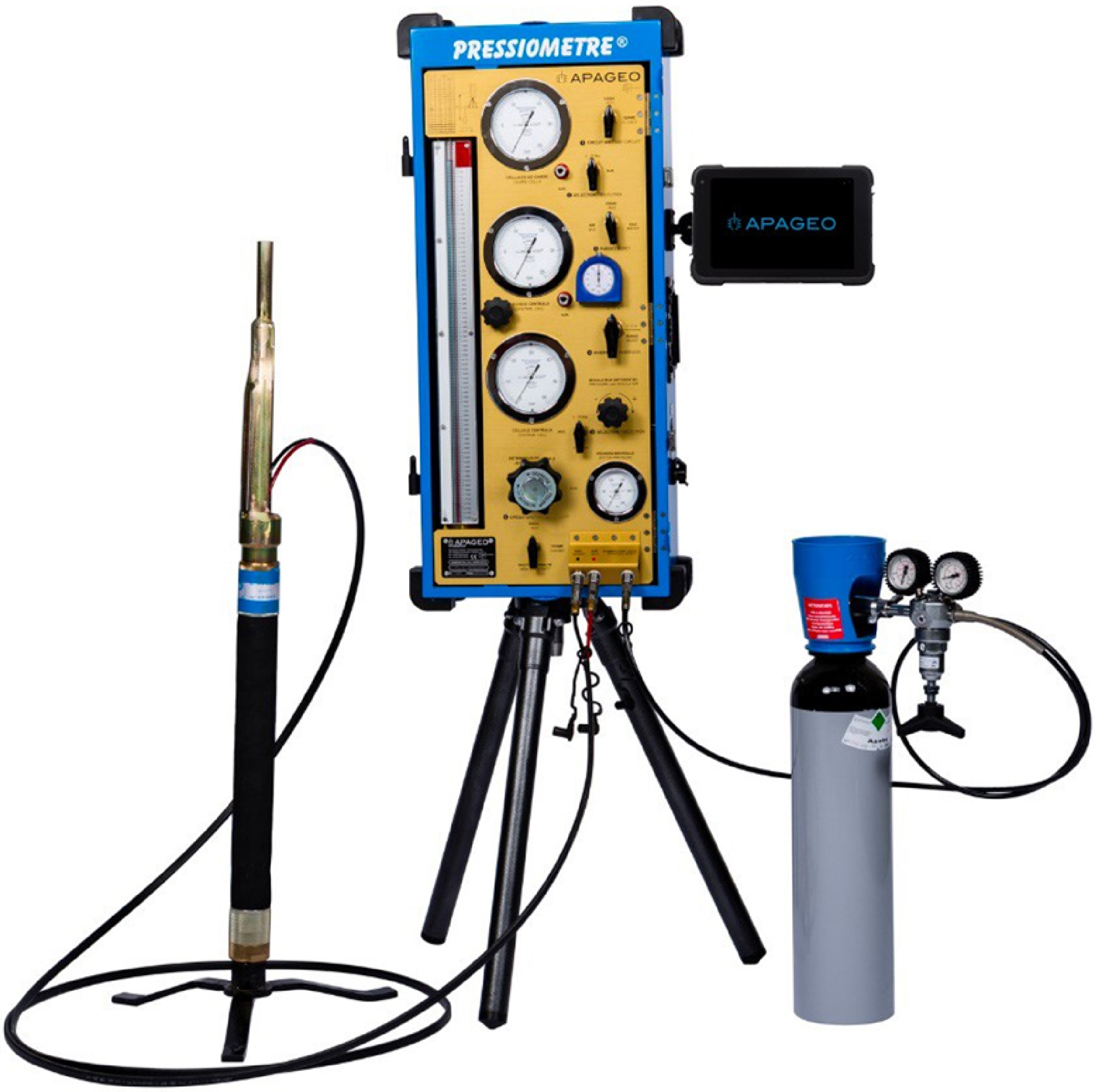

*Figure 35. Photograph of a pressuremeter set (extracted from https://www.apageo.com).*

2. **Probe description**

If expansion is followed by the volume of the measuring cell, the measuring cell shall be inflated by injecting a liquid of low compressibility. Alternatively air can be used to inflate the measuring cell and the expansion followed by displacement transducers.

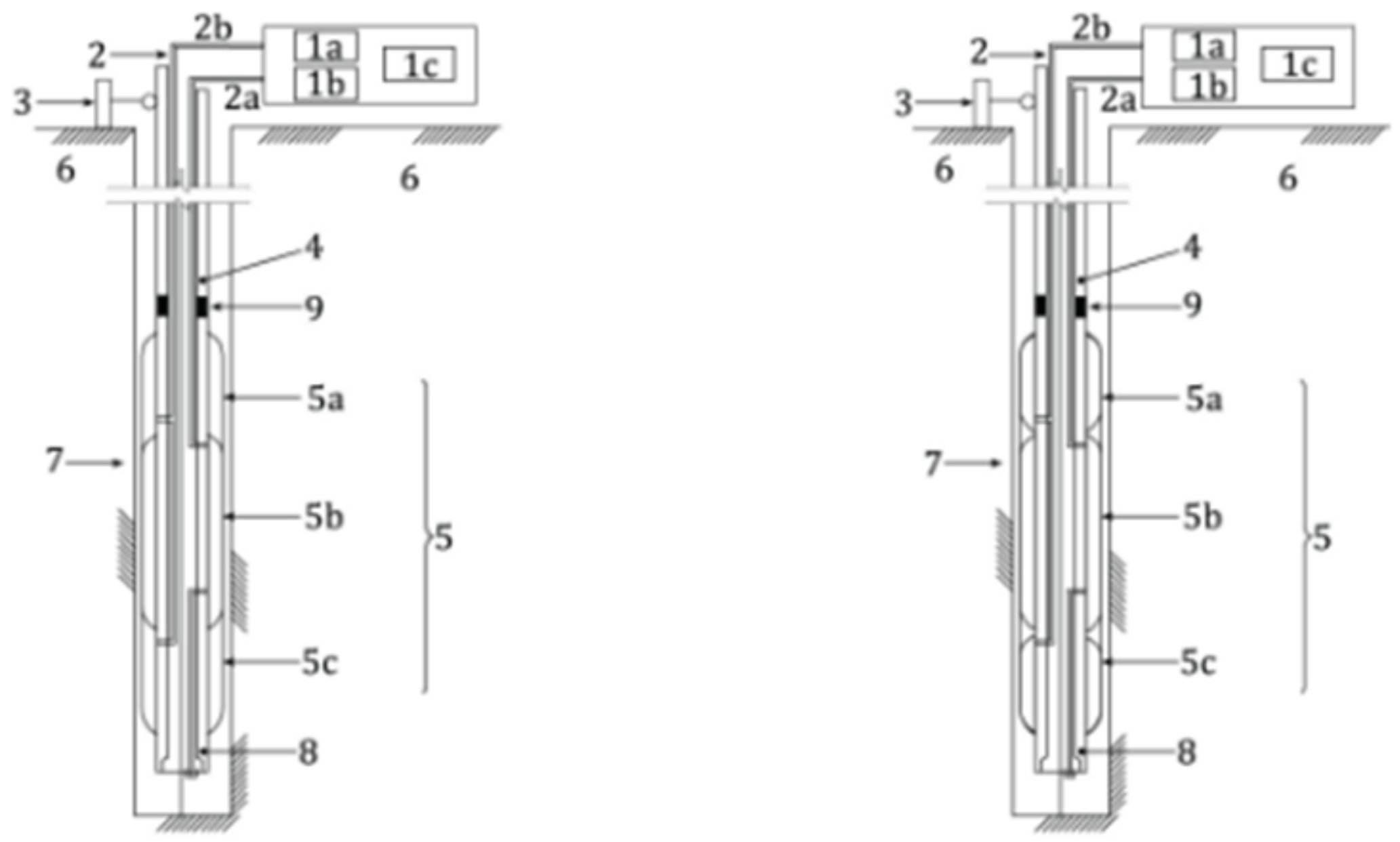

**a) Tri-cell probe, G type, with measurement of volume of the measuring cell** **b) Tri-cell probe, E type, with measurement of volume of the measuring cell**

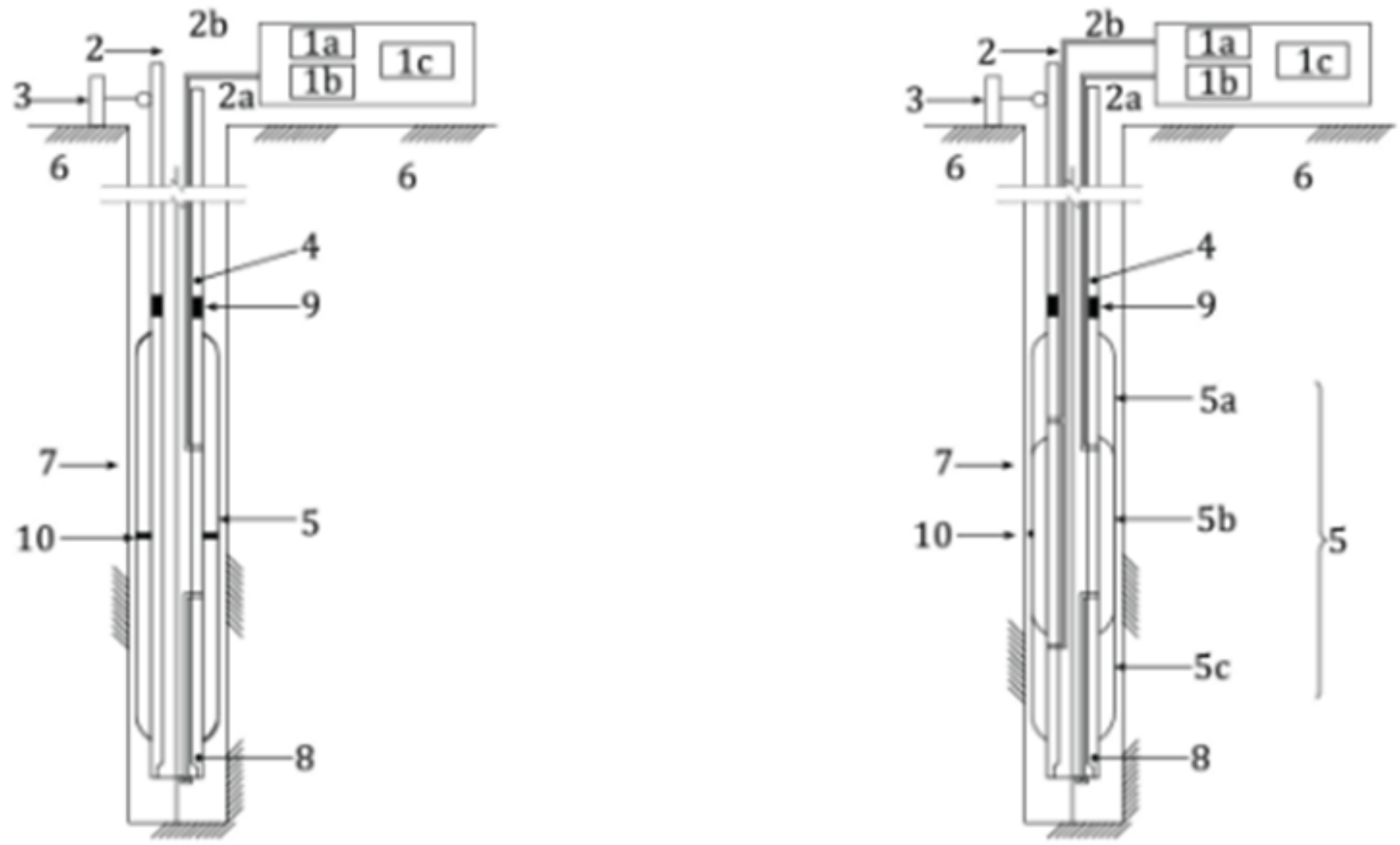

**c) Mono-cell probe, with measurement of radial displacement (optional)** **d) Tri-cell probe, with measurement of radial displacement**

*Figure 36. Diagram of a Ménard pressuremeter (extracted from NF EN ISO 22476-4).*

**Key**

- 1a pressurization, differential pressurization and injection devices
- 1b pressure and volume measuring devices
- 1c acquisition, storage and printing out of the data (required for CU type B and C)
- 2 connecting lines:
  - 2a line for liquid injection
  - 2b line for gas injection

- 3 depth measurement system 2b line for gas injection
- 4 rods
- 5 pressuremeter probe:
    - 5a upper guard cell
    - 5b central measuring cell
    - 5c lower guard cell
- 6 ground
- 7 pressuremeter test pocket
- 8 hollow probe body
- 9 probe rod coupling
- 10 transducers

3. **Pressure and volume control unit**

The control unit shall include:

- equipment to pressurize, and so to inflate the probe, and to maintain constant pressures as required during the test;
- equipment to maintain an appropriate pressure difference between the central measuring cell and the guard cells, if relevant;
- device which permits, according to the type defined in Table 1, the reading and recording of the parameters to be measured: time, pressure and volume.

| Type of control unit | Type of test regulation | Type of reading and recording |
|---|---|---|
| A | manual | manual |
| B | manual | automatic |
| C | automatic | automatic |

*Figure 37. Types of pressuremeter control units (Extracted from NF EN ISO 22476-4).*

The control unit shall control the probe cell expansion and permit the simultaneous reading of liquid and/or gas pressures and injected liquid volume or radius of the measuring cell as a function of time.

The pressurizing device shall allow:

- reaching the pressuremeter limit pressure or a pressure p r at least equal to the maximum pressure defined for the test;
- holding constant each loading pressure level in the measuring cell and in the guard cells during the set time;
- implementing a pressure increment of 0,5 MPa in less than 20 s as measured on the control unit;
- controlling the pressure difference between the measuring cell and the guard cells;
- injecting a volume of liquid in the measuring cell larger than at least its volume at rest V c,

i.e. 700 cm 3 for a 60 mm pressuremeter probe.

If volumetric measurement is used, a valve between the volumeter and the pressure measuring device shall allow stopping the injection.

4. **String of rod/connecting lines**

Not described here, to have information regarding this subject, please consult the norm.

**Test procedure**

The test procedure will not be detailed her, to consult the exact procedure, please consult the associate norm.

However, the principle is to establish a loading program in at least 8 to 10 steps, in which the volume of the probe increases, leading to a raising of the radial stress imposed to the ground. In each step, a constant stress is applied to the ground, during a certain time increment.

A template of loading program is exposed in the following figure, extracted from the norm.

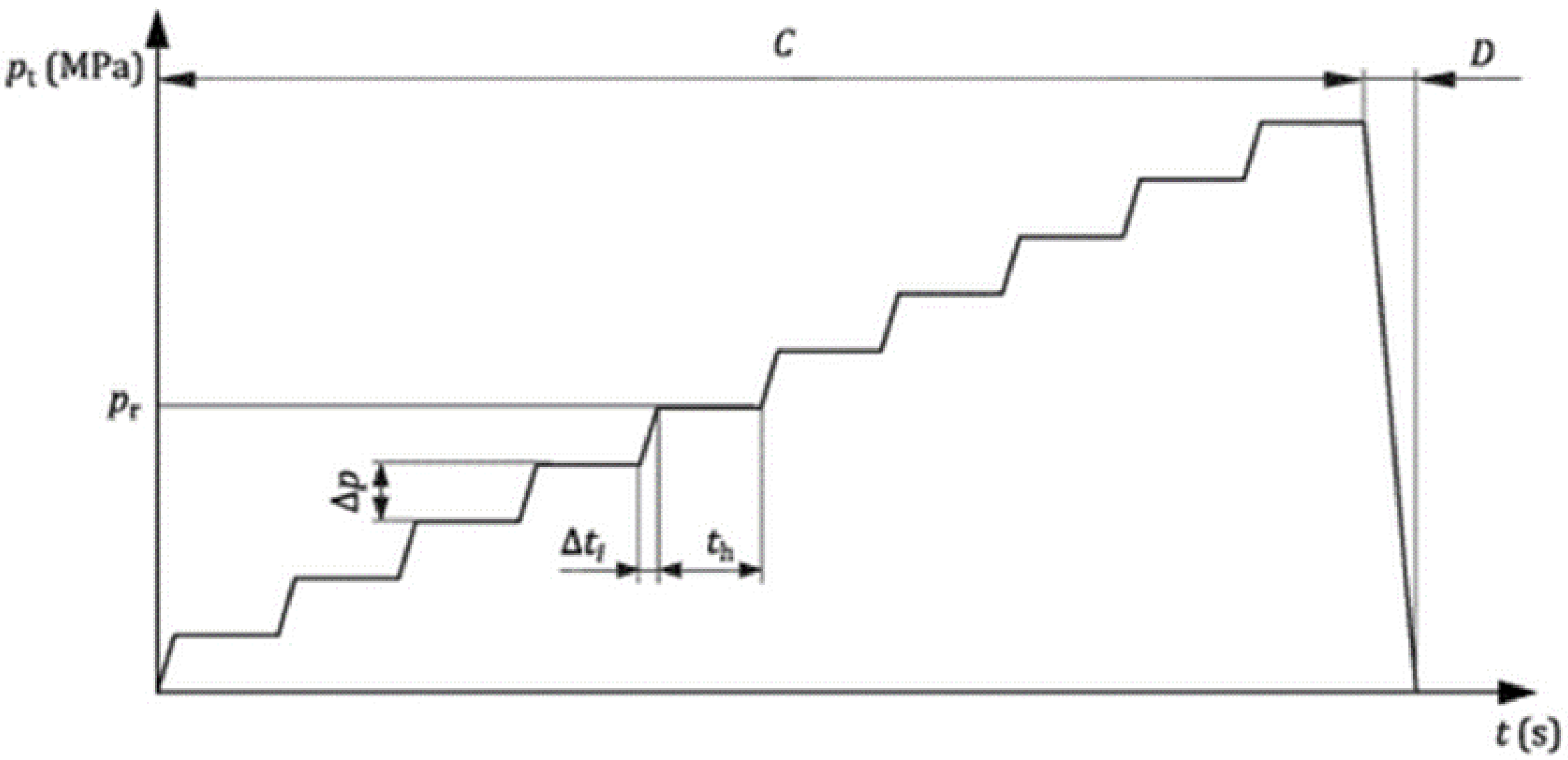

**Key**

***p*** $_t$ target pressures

**Δ*p*** pressure increment

***p*** $_r$ pressure step during $t_h$

**C** loading phase

$t$ time

$\Delta t_i$ pressure increment duration

$t_h$ duration of a pressure step

D unloading phase

*Figure 38. Pressuremeter test procedure*

**Results and interpretation**

1. **General**

The application of the loading programme results in a raw pressuremeter curve. Some correction should be applied to this raw curve in order to avoid experimental biases. The corrections procedure are well detailed in the norm.

After this step, the corrected pressuremeter curve is used to calculate the pressuremeter parameters described above. A typical pressuremeter curve is presented in the figure below.

At least three data points in the second group of readings and three data points in the third group shall be available to determine all three parameters p f, p LM and E M.

If in a test, one group of readings is incomplete or missing, the following effects on the determination of the three parameters shall be considered:

- when the pressuremeter curve includes only the second and third groups of readings and with fewer than two data points in the second group, values of E M and p f cannot be obtained;
- when the pressuremeter curve includes only the first and second groups of readings (i.e. only one or no points in the third group), p LM and p f cannot be obtained.

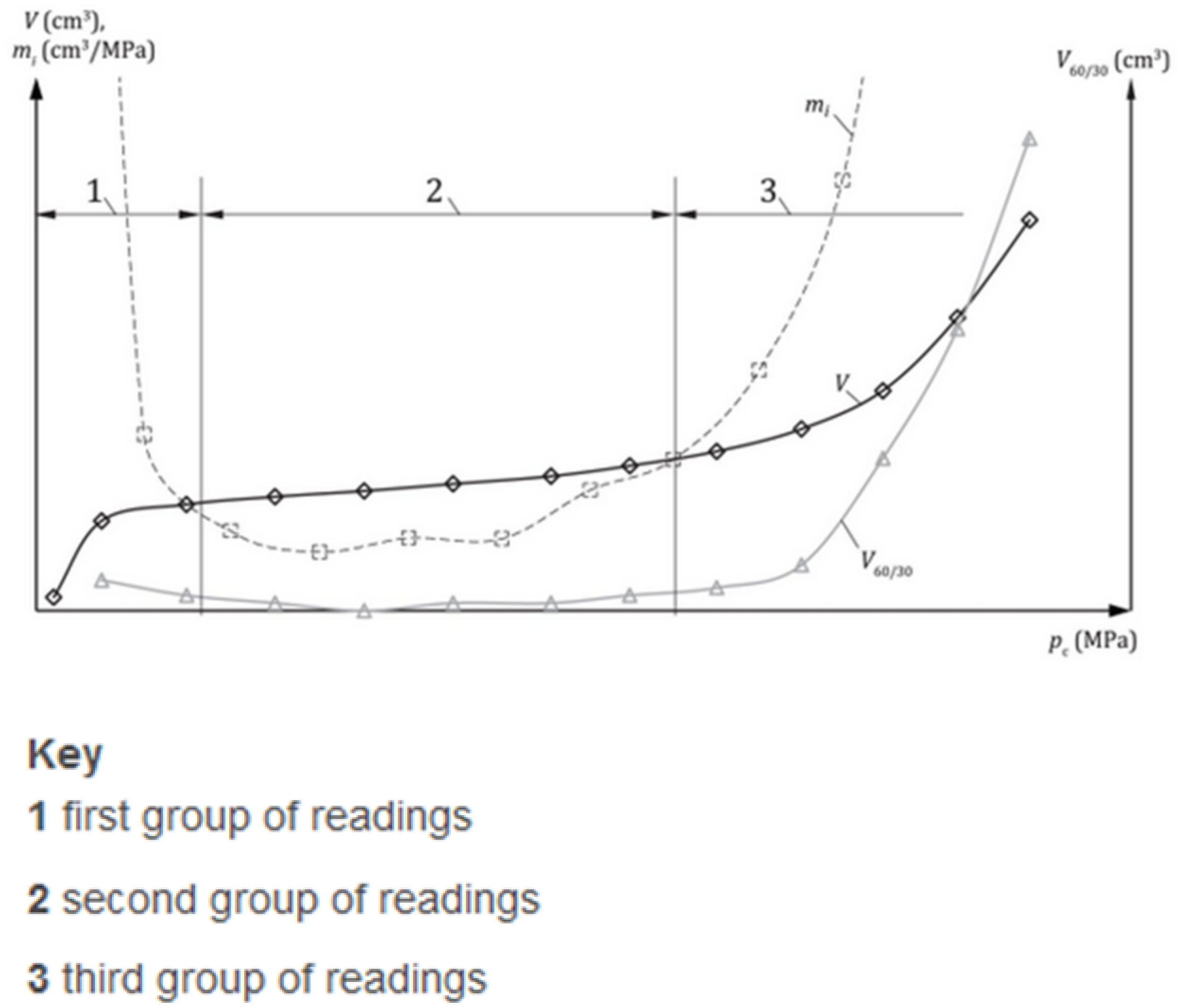

*Figure 39. Groups of pressuremeter readings*

In the following, the methods of determination of the parameters will be synthesized to give a brief overview of the process, and may not be exhaustive. Please consult the norm to have the complete procedure of determination

2. **Pressuremeter creep pressure p f**

If there are at least two sets of readings both in the second and in the third group, the creep pressure pf shall be estimated, using the following graphical analysis of the (p, V 60/30) diagram: 2 straight lines shall be drawn on the (p, V 60/30) graph, one involving the data points in the second group, the second one involving the data points in the third group, as illustrated on Figure D.4; the abscissa of the intersection of the 2 straight lines give p f.

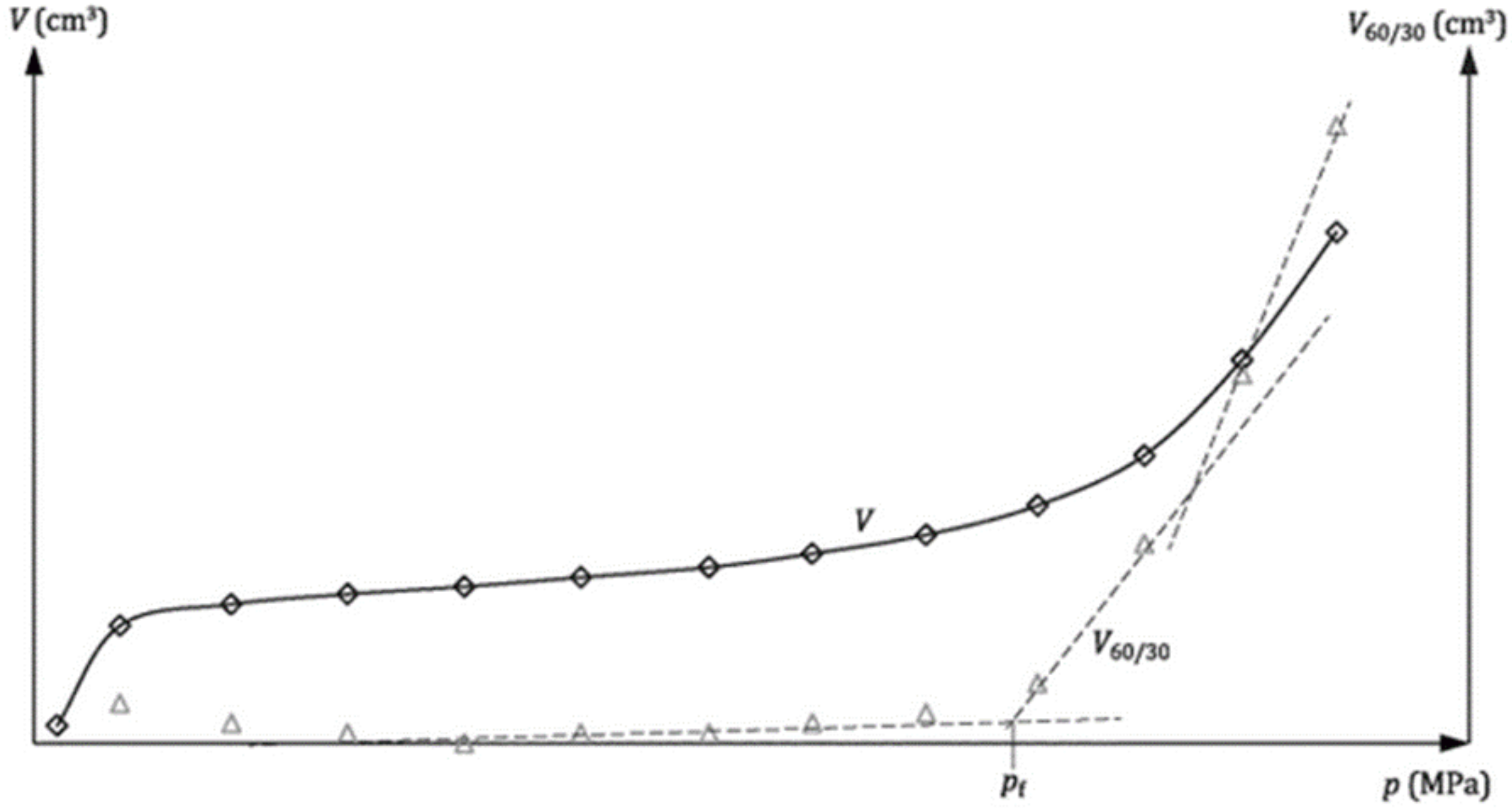

*Figure 40. Pressuremeter creep pressure determination (Extracted from NF EN ISO 22476-4).*

3. **Ménard pressuremeter limit pressure p LM**

The Ménard pressuremeter limit pressure is conventionally defined as the pressure leading to the doubling of the initial volume of the pocket.

It can be either obtained by direct measurement or determined using extrapolation methods.

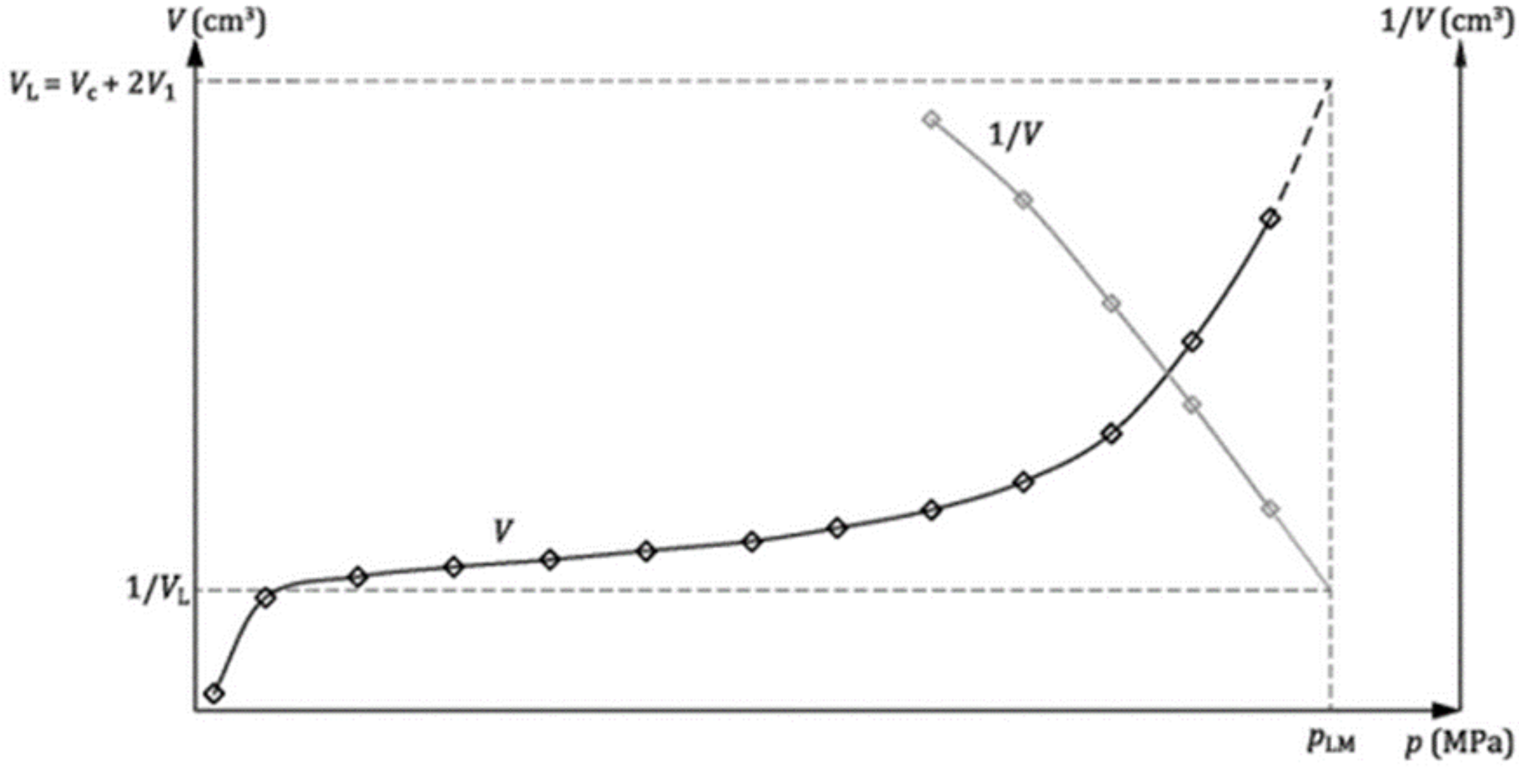

*Figure 41. Ménard pressuremeter limit pressure, with reciprocal fitting and extrapolation method (extracted from NF EN ISO 22476-4).*

4. **Ménard pressuremeter modulus E M**

The Ménard pressuremeter modulus is defined as the modulus of the pseudo-elastic part of the curve. The determination of this parameter should be done with great precaution and following the recommendation of the norm, that could not be detailed here.

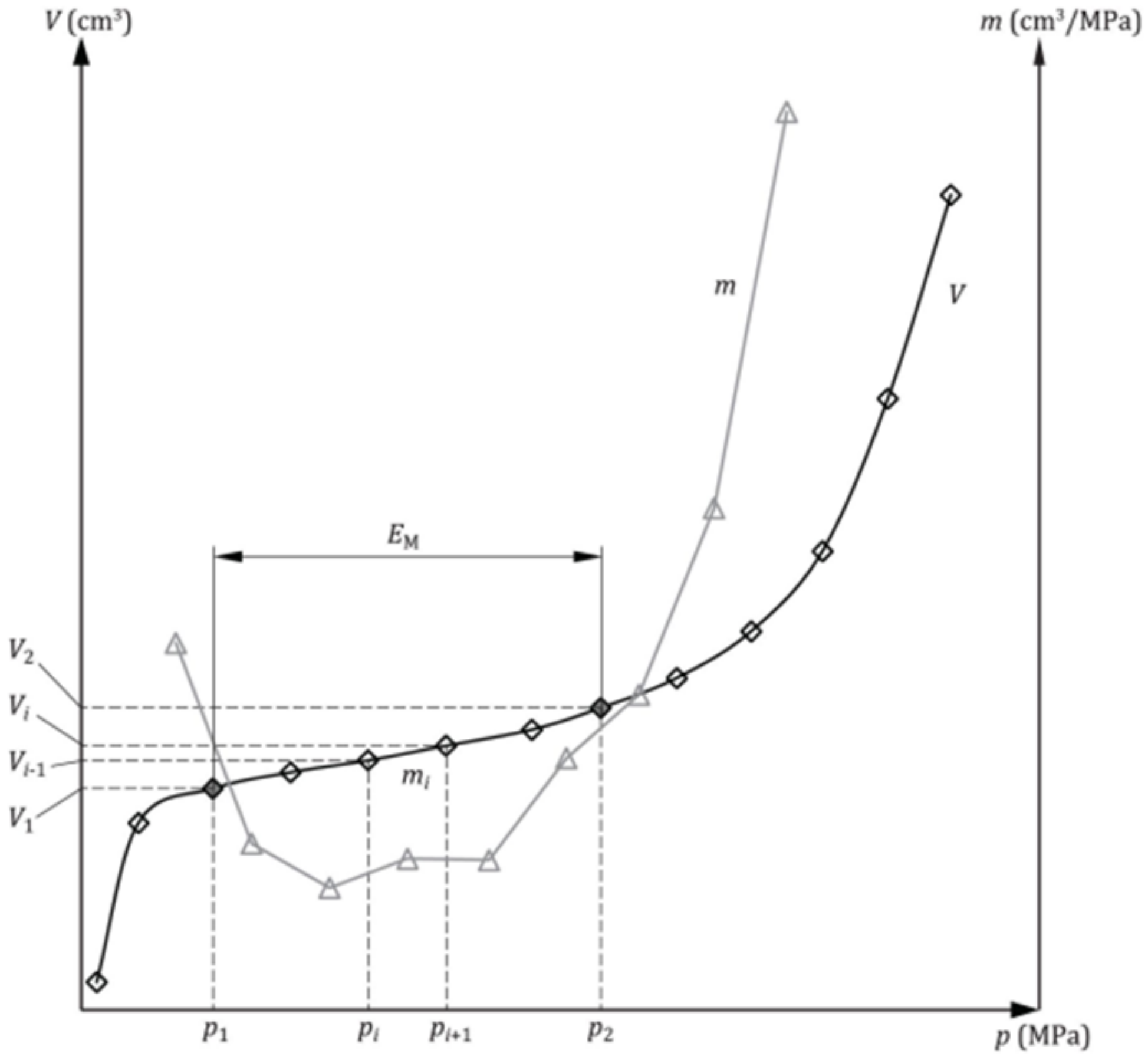

*Figure 42. Plot of a corrected pressuremeter curve, creep curve and slopes (Extracted from NF EN ISO 22476-4).*

Please note that the The Ménard pressuremeter modulus is not a Young Modulus of the tested soil. This Young Modulus could be calculated using the rheologic coefficient α, defined by Ménard, and depending on the soil type.

In addition, please note that the young modulus of soils depend on the strain level : it should be corrected to be properly use in geotechnical calculation.

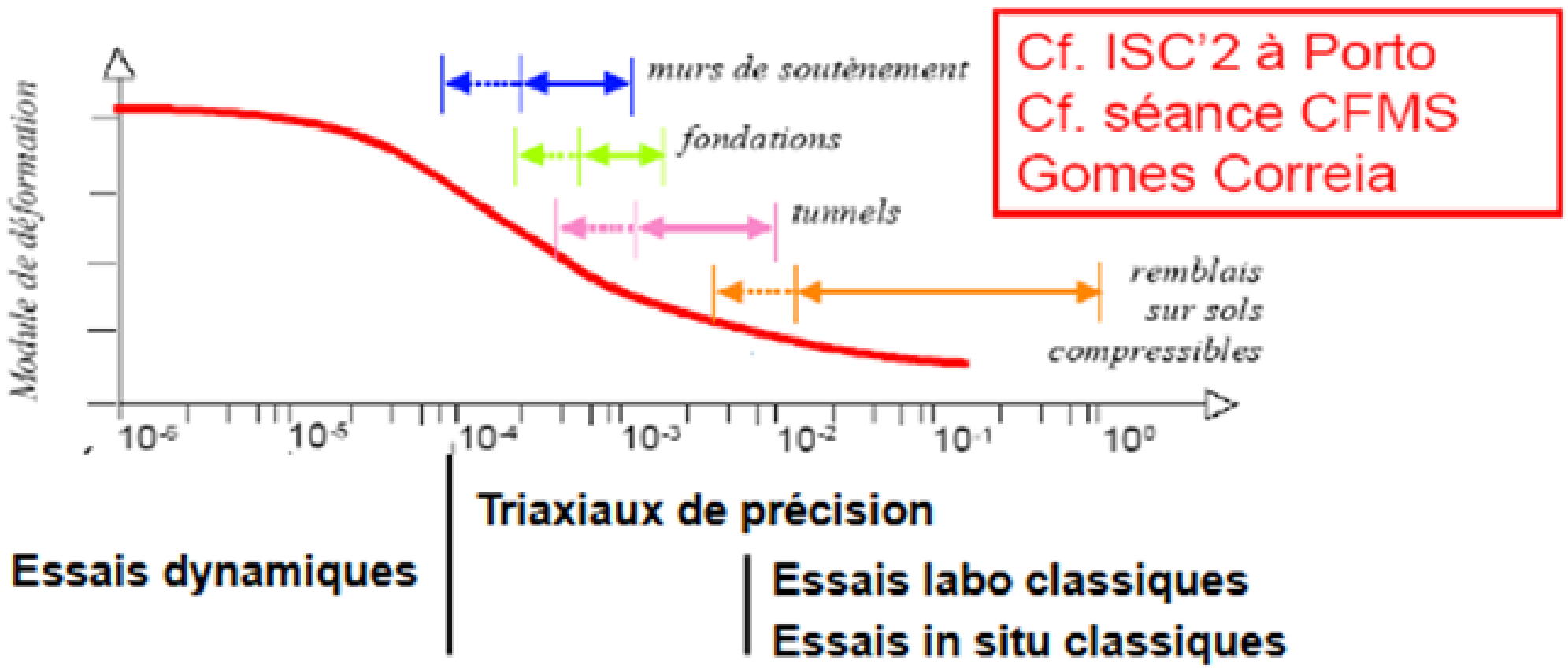

*Figure 43. Variation of the deformation modulus in function of the strain level (Reiffsteck 2002)*

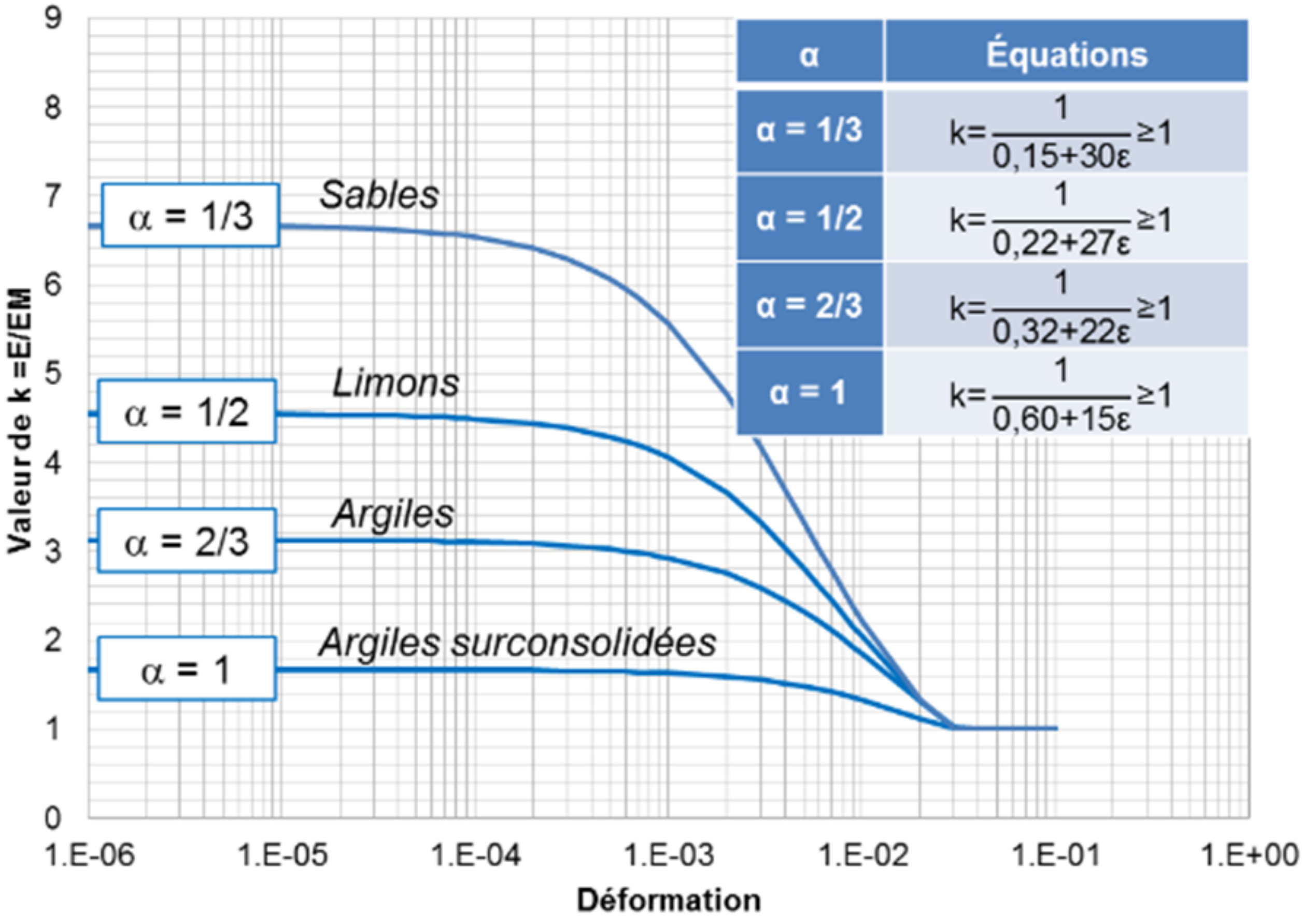

*Figure 44. Degradation laws E/EM = f(ε) (ARSCOP, 2017)*

**Normalization / Variants**

There are different variants of this kind of geotechnical test, depending mostly on the probe type and on the method to insert this probe within the ground. These variants are described in the following norms :

- NF EN ISO 22476-4, septembre 2021, Index : P 94-521-4, P 94-521-4 : Geotechnical investigation and testing - Field testing - Part 4 : prebored pressuremeter test by Ménard procedure
- NF EN ISO 22476-5, French Standard approved and published by AFNOR, 2023-04-12 : Geotechnical investigation and testing - Field testing - Part 5: prebored pressuremeter test
- NF EN ISO 22476-6, octobre 2019, Index P 94-521-6 , ICS : 93.020 : Geotechnical investigation and testing - Field testing - Part 6 : self boring pressuremeter test**
- NF EN ISO 22476-8, janvier 2018, Index : P 94-521-8, ICS : 93.020 : Geotechnical investigation and testing - Field testing - Part 8 : full displacement pressuremeter test

## Exposing Pressuremeter test with the FROST Geotech Plugin

### Sensor

The PressumeterTest instance has to be declared as a Sensor with a SensorType referring to its nature, eg. https://data.geoscience.fr/ncl/Proc/94

#### ObservedProperty and DataStream

For each PressuremeterTest performed, one DataStream shall be declared per ObservedProperty.

#### BhCollarThing and BhTrajectory

BhCollarThing and BhTrajectory shall be used to respectively describe the Borehole and its trajectory

#### BhSampling and BhFeatureOfInterest

BhSampling enable to declare each depth at which a test has been performed. For example, if a test has been performed from the Depth 1m to the Depth 10m every 1m, then there shall be BhSamplings for the Depths = 1m, 2m, …, 9m, 10m.

One BhFeatureOfInterest shall then be declared per Sampling.

As a PressuremeterTest is an in-situ test with ponctual measures (at depth), the BhFeatureTypes that shall be declared are : Hole & Point

### Observation and result

#### For intepreted or calculated values

In this case, there shall be one Observation and Result per combination of DataStream / FeatureOfInterest. For example, if there are 3 DataStreams (the different ObservedProperties) and 10 FeaturesOfInterests (the different depth), then there shall be 30 Observations declared.

#### For raw measurements

The Pressuremeter test imply several measurements of the same parameter at the same depth. Each measurement being associated to a step. In this case, there shall be for each FeatureOfInterest (the different depth) as much Observations as steps. For example, if the test imply 10 steps at one depth, then there shall be 10 observations associated to that depth. The number of the step shall be declared as a Parameter of the Observation.

## 5.5.5. Approach for Atterberg Limits

The collection of “Atterberg Limit” tests establish the moisture contents at which a fine-grained clay and silt soil sample transitions between behavior as a solid, semi-solid, plastic, and liquid material. Atterberg created a series of test procedures and definitions to provide reproducible results among soil samples. One of the principal uses of the system is to differentiate silty soils from clayey soils. The values obtained using the prescribed procedures can also be used to help characterize shrink-swell [expansive] soils, which are particularly problematic.

These values are important and common index tests in geotechnical engineering and are used to classify soils and compare the behaviors of different soils and relate soil plasticity behaviors to other performance qualities such as strength, stiffness, and permeability through empirical correlations. The tests are especially common for shallow foundations, roadway embankment construction, and pavement support. The tests are somewhat less common for deep foundation design and slope stability.

The tests are conducted on specimens taken from disturbed or undisturbed field samples and are usually conducted in a laboratory environment at a time following field sample acquisition, shipping, and subsequent lab extrusion or opening. The samples are remolded as part of the testing procedures; original intact samples are not required.

There are three (3) limits, each with a distinct corresponding laboratory test to establish the moisture content associated with a specific defined observed behavior of the sample. The tests are:

- **The Liquid Limit (LL)**
- **The Plastic Limit (PL)**
- **The Shrinkage Limit (SL)** *

The Shrinkage Limit test is less common and is not typically conducted or used in common soil identification and classification systems.

A calculated value, the Plasticity Index (PI), is included in the reporting of the Liquid Limit and Plastic Limit. The PI is the Liquid Limit minus the Plastic Limit, PI = (LL – PL). Values of all parameters are reported in whole numbers.

Importantly, the commonly termed “Atterberg Limits” (generally including the LL, PL, and PI values) are used in several formal soil classification systems to determine classifications associated with plastic and non-plastic clayey and silty soils.

- A plasticity chart is available for the Unified Soil Classification System (USCS) which uses a soil’s Plasticity Index (PI) and the Liquid Limit (LL) to determine its classification.
- The PI and LL are also used in the American AASHTO classification system used for highway construction. This system rates soils for their suitability for highway construction. Transportation and other agencies may have their own specifications with identification and classification systems.

The Liquid Limit (LL) is the water content at which the behavior of a clayey soil changes from the plastic state to the liquid state. Depending on the mineral composition, particle size, and nature of the pore fluid, the transition from plastic to liquid behavior occurs over a range of water contents. The definition of the liquid limit for geotechnical purposes is based on a standard test procedure developed by Arthur Casagrande. His standardized methods for determining the liquid, plastic, and shrinkage limits are used today.

Casagrande standardized Atterberg’s original liquid limit test by developing a hand cranked (or motor driven) specialty apparatus with a brass cup and resilient base. A rotating cam mechanism creates a reproducible drop height of 10 mm. A mass of soil is placed into the brass Casagrande cup portion of the device. A groove created using a standardized tool through the center of the specimen. The cup is repeatedly dropped at a standard rate of 2 revolutions per second. During the process the soil groove closes gradually at the base, as a result of the impacts. The number of blows necessary for the groove to close 1/2 -inch [13 mm] at 25 blows of the apparatus is desired.

The US Standard has both a single point and multipoint method to determine the Liquid Limit. Using the multipoint method, the test is progressively conducted at several moisture contents, and the moisture content which requires 25 blows to close the groove is interpolated from the test results. The single point test method also allows the use of a single test, where 20 to 30 blows are required to close the groove. A factor, k, provided in the standard, is applied to obtain the liquid limit from the moisture content. Two specimens are required for the single point test, with the Liquid Limit (LL) reported as the average value of the tests, reported to the nearest whole number.

The Plastic Limit (PL) is determined by hand rolling a small mass of a soil sample, creating a roughly cylindrical "thread" of soil on a flat ground glass plate. Plastic behavior is exhibited if the soil thread retains its shape when rolled progressively thinner to a diameter of 1/8-inch [3.2 mm]. At this diameter the rolling is discontinued, and the same sample is remolded into a larger thread and the test procedure repeated, rolling the newly remolded thread to progressively smaller diameters. As the moisture content lessens due to evaporation and loss of moisture along the technician's hands and the glass plate, the thread will begin to break apart at larger diameters. At the point the thread can no longer be rolled into 1/8-inch [3.2 mm] thread, after having previously been rolled to that diameter, the test is considered concluded The sample is collected and placed in a container with other tested samples until two specimen containers of at least 6 grams of material are obtained and the moisture content can be evaluated. The water content of the soil is measured using approved test methods. A soil sample is reported as non-plastic (NP) if a thread cannot be rolled out down to 1/8-inch [3.2 mm] at any moisture condition.

The Shrinkage Limit (SL) is the water content where additional loss of specimen moisture will not result in additional soil volume reduction. Determination of the shrinkage limit is a more involved, and far less common process than the other two limits. Information on test procedures is in the associated standards.

Final reported [calculated] properties are typically:

- Liquid Limit (LL),
- Plastic Limit (PL),
- Plasticity Index (PI), and in much less frequent cases,
- Shrinkage Limit (SL)

Values are in whole numbers, unless the test result is that the soils are non-plastic, in which case rather than the numeric values a classification of "NP" is assigned.

Reporting of the final test values*, following application of the defined test procedures, is generally straightforward. Calculated result values are expressed as whole numbers. Unlike the SPT test, values for Plastic Limit (PL), Liquid Limit (LL) and the calculated Plasticity Index (PI) are generally well-behaved numeric data without non-numeric possibilities, other than the potential outcome of a non-plastic (NP) soil. If either the liquid limit (LL) or plastic limit (LL) could not be determined, or if the plastic limit is equal to or greater than the liquid limit, the soil is designated non-plastic (NP).

It is common to have these test results reported in conjunction with other test results from a single extracted soil sample (such as moisture content, direct shear, and unconfined compression) and a soil description- often obtained from grain size analysis and Atterberg Limits testing. Different specimens (subsets of the sample) are prepared for each type of testing.

*Typically, only final calculated values are reported and transmitted in log formats. Original measured values, calculations, and interim testing information is usually preserved in the original test records but not included as part of geotechnical design memoranda (although it could appear

in comprehensive report appendices). The information may be included in geotechnical data reports which may contain complete field and laboratory testing data.

The applicable US Standard is ASTM D4318-17e1 Standard Test Methods for Liquid Limit, Plastic Limit, and Plasticity Index of Soils The applicable US Standard is ASTM D4943-18 Standard Test Method for Shrinkage Factors of Cohesive Soils by the Water Submersion Method

### Exposing Atterberg limits with the FROST Geotech Plugin

To demonstrate how the final results, interim test information and related metadata are stored and organized within the FROST server, we offer the following example Atterberg limits test for a single sample. Relevant data are:

#### GENERAL TEST INFORMATION

1. Test procedure used: ASTM D4318-17, Method A (multipoint method)
2. Name of sample tested: 1
3. Sample location: Collected from a depth of 1.5 to 3 ft from borehole B-001-0-20
4. Borehole is is 41 ft deep and its collar is located at lat/lon 39.47466/-81.796858, elevation 249.50928 meters (WGS84)
5. Device used to collect sample: Geoprobe Interlocking Split-spoon sampler
6. Length of sample recovered from the hole: 1.08 ft (72% recovery).
7. Specimen preparation method : Wet Preparation Method described in Sec. 10.1, ASTM Standard D4318-17
8. Weight of specimen prior to testing: 65 grams, wet weight.

#### CASAGRANDE TEST RESULTS FOR DETERMINING LIQUID LIMIT

| Increment Number | Blow Count | Water content (%) |
|---|---|---|
| 1 | 16 | 35.2 |
| 2 | 22 | 28.6 |
| 3 | 27 | 23.1 |
| 4 | 32 | 17.4 |

#### PLASTIC LIMIT WATER CONTENT MEASUREMENTS

| Increment/Container Number | Water content (%) |
|---|---|
| 1 | 11.9 |
| 2 | 11.7 |
| 3 | 11.4 |

*Note: A third trial was run because of concerns that the second container was not properly sealed before measurement*

### FINAL REPORTED RESULTS

- **Liquid limit: 25**
- **Plastic limit: 12**
- **Plasticity Index: 13**

### Instance Diagram

Object instances and the associations required to properly expose the example test data with the FROST Geotech Plug-in are shown in the following instance diagram:

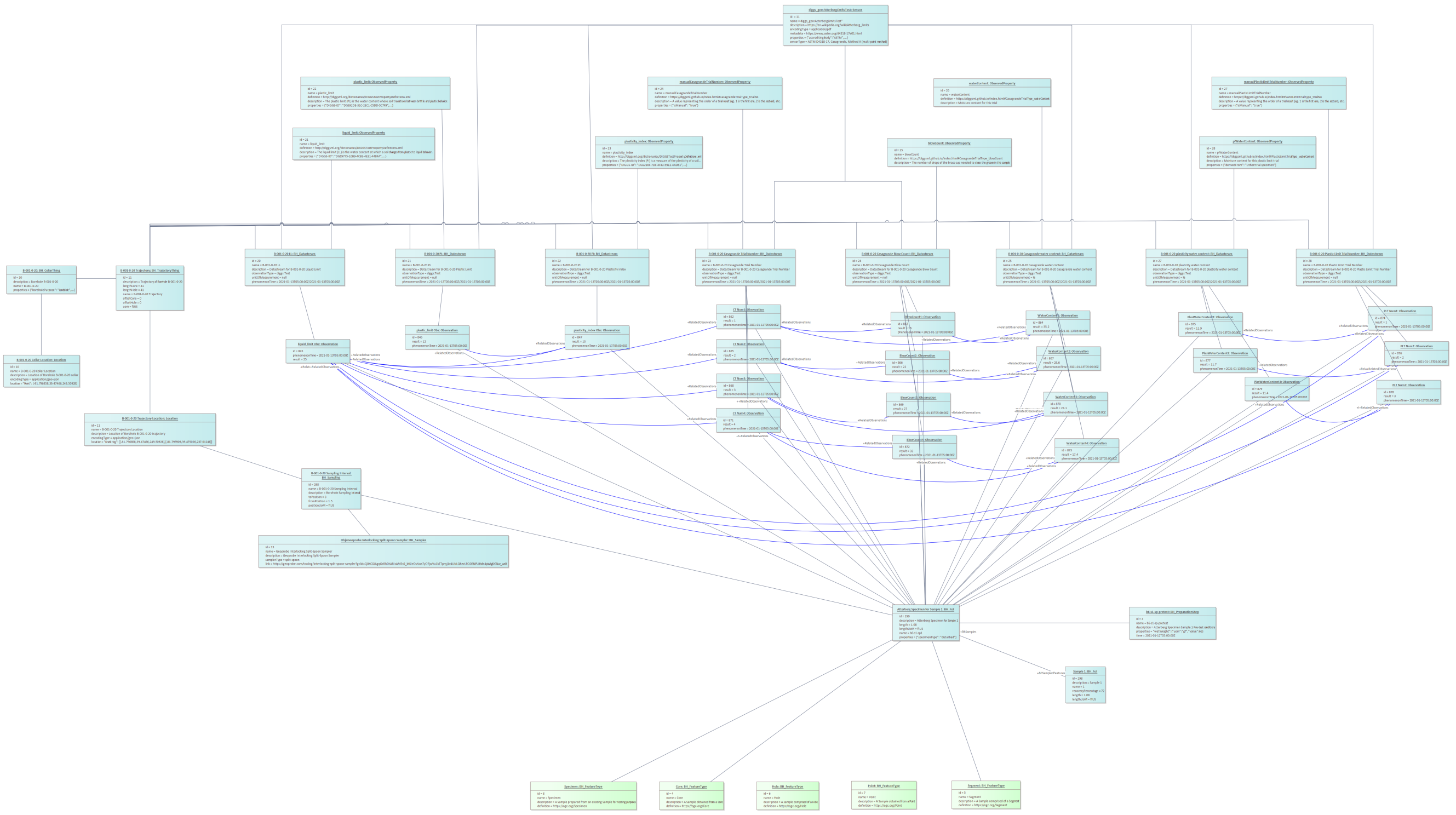

*Figure 45. Atterberg Object Diagram*

The following summarizes the various entities in the diagram:

### Sensor

The Sensor object serves as the observing procedure in STA. One object instance is needed for this example (top center of diagram), and in this example holds the information about the test procedure used. One Sensor instance is needed for Atterberg limits tests and the test procedure is specified in its sensorType property and fully referenced in its metadata property. As constructed, this Sensor instance can be reused for multiple Atterberg tests.

### ObservedProperty

The ObservedProperty object instances identify the properties that are observed by the Atterberg limits test. There are eight ObservedProperty instances (top of diagram, below Sensor):

**3 properties observed that constitute the final reported results of the test:**

1. liquid_limit

2. plastic_limit
3. plasticity_index

**3 properties observed for the Casagrande (liquid limit) measurements:**

4. manualCasagrandeTrialNumber (the Increment Number column in the above table)
5. blowCount
6. waterContent

**2 properties observed for the plastic limit measurements:**

7. manualPlasticLimitTrialNumber (the Increment/Container Number column in the above table)
8. plWaterContent

*Note: The plWaterContent observed property is not necessary (and probably should not be used) as it is identical to waterContent, but is included in this example to better distinguish the water contents determined for liquid limit from those for plastic limit*

As with Sensor, the ObservedProperty instances can be reused for multiple tests.

#### DataStream

All of the object instances in the diagram are linked to the Sensor and ObservedProperty instances via Datastream instances (below the ObservedProperty objects on the diagram), which serve to associate observation results obtained from a feature of interest to its observed property, observing procedure, and the borehole.

Eight Datastream instances are needed, one for each ObservedProperty instance.

*Note: if the plWaterContent ObservedProperty were not used, there would still need to be 8 Datastream instances. Two DataStreams would link to the waterContent ObservedProperty instance. One of these DataStreams would link to the liquid limit water content Observation instances, the other to the plastic limit water content Observation instances.*

#### BhCollarThing, BhTrajectoryThing and Location

The DataStreams all link to the borehole via its BhTrajectoryThing object instance. BhTrajectoryThing (left edge of diagram) represents the borehole's geometry and contains the borehole length and information for linear referencing. The trajectory's geometry is given in the associated Location instance. BhTrajectoryThing is associated with a BhCollarThing instance , which represents the borehole as a whole. All general metadata about the borehole is contained in the BhCollarThing object instance; it's geometry is represented by a point Location object instance.

More detail about properties of BhCollarThing and BhTrajectoryThing can be found in the Borehole log discussion.

**BhSampling and BhFeatureOfInterest**

The act of collecting a sample from the borehole for testing is represented by the single BhSampling object instance (below and to the right of the BhTrajectoryThing in the diagram). BhSampling holds the sample depths (fromPosition=1.5, toPosition=3) and links to BhTrajectoryThing in order to affix the linear referenced sample positions to the trajectory geometry. In addition, BhSampling also links to a BhSampler object instance which identifies the type of sampler used in the sampling act (eg. the Geoprobe split-spoon sampler).

BhSampling produces a BhFeatureOfInterest object, which represents Sample 1 that is collected from the borehole (below and to the right of the BhSampling instance object). This object holds the sample length and recovery percentage of the sample. That this BhFeatureOfInterest is a physical material sample from the borehole is given by the associated BhFeatureType Core and Segment instances.

As part of the Atterberg limits test procedure, a 65 gram specimen of Sample 1 is prepared using the “Wet method”. This is represented by an additional BhFeatureOfInterest object instance linked to Specimen, Core and Segment feature types. The linked BhPreparationStep object instance holds the specimen weight, and the further linked BhPreparationProcedure object carries the procedure method used to obtain the specimen.

**Observation**

The remaining entities on the diagram are Observation instances that provide the results for their associated observed properties. Each Observation instance links to the prepared specimen that was tested and to the Datastream instance associated with the appropriate ObservedProperty.

This Atterberg limits test consists of 18 individual observations but only three of them (results for liquid limit, plastic limit and plasticity index) are the primary reportable results. The other observations are interim observations, made as part of the test procedure, that result in the determination of the primary results. The links shown in the diagram relating observations to each other provide the means for distinguishing among the various types of observations.

Starting with the Observation instances associated with the Casagrande liquid limit measurements, each test increment is a set of three observation results. These results are useless independently - for example, the blow count in a Casagrande trial has no meaning without the associated water content observation, and vice versa. To model the observation set, the increment result is linked to its associated blow count and water content results, and the blow count result is linked to its associated water content result. These three observations, as a set contribute to the determination of the final liquid limit result, and to model that association, the increment observations are linked to the liquid limit observation. In this way, one can traverse from the liquid limit observation to access only those blow count and water content observations that contributed to the liquid limit result.

A similar link structure is also made for the interim water content observations that contribute to the plastic limit result.

Finally, the plasticity index observation instance is linked to both the liquid limit and plastic limit observation instances to demonstrate that plasticity index relies on those values for its results.

Note that the liquid limit and plastic limit observation instances are correctly not linked to show that those results are derived independently from each other.

To provide the most flexibility for querying, Datastream object instances associated with the Observations are linked in the same manner as their associated Observations as seen in the diagram.

The current STA model does not provide for one-way links where an association role can be assigned. Such capability would be useful in modeling even more complex geotechnical test results.

### 5.5.6. AGS & DIGGS mapping to STA

This chapter provides a road map for how to map AGS and DIGGS data to the OGC SensorThingsAPI data model.

#### Overview of the AGS format / data organization

The AGS format is a data transfer format for ground data that is routinely used in the United Kingdom. It has also been adopted and adapted for use in some other countries including Hong Kong and Australia.

AGS was created and is maintained by the Association of Geotechnical and Geoenvironmental Specialists, a UK based industry organization. Documentation can be found here.

The scope of the AGS data transfer format is limited to factual ground investigation data (typically 'Book A' data, adopting the terminology used in this Wiki). However, interpretative/design data and ground models are covered by AGSi, a separate relatively new transfer format created by AGS.

The AGS data transfer format adopts a CSV-like format for output. The data structure does not adopt an object model approach as such, but is broadly comparable comprising Groups (–> objects / database tables) and Headings ( –> attributes / table fields) and utilizing parent-child relationships between groups. The AGS 'data dictionary' is very comprehensive, covering almost all field and laboratory test data that may be published in a factual ground investigation report. There are about 150 Groups and over 2000 Headings.

AGSi (for ground models and interpretative/design data) is object model based and the output is JSON. AGSi adopts a concise but very flexible data model. It does not include schema for complex geometry, instead allowing users to adopt existing standards for model geometry.

#### Overview of the DIGGS format / data organization

DIGGS is an XML-based schema and associated dictionaries for the storage and transfer of the properties of real-world objects and activities, amd their associations, that occur within the geotechnical and geoenvironmental domains. DIGGS is designed as a GML application schema insofar as all objects within the DIGGS namespace derive from GML abstract types and encoding follows GML's object-property convention. Additionally, it incorporates GML 3.3 objects to support linear referencing for linear sampling features like boreholes, and vector linear referencing for 2D and 3D objects. DIGGS also relies on another XML vocabulary, WITSML, developed for the oil industry, for measure types and units symbology which allows for schema validation of unit symbols for measure type properties.

DIGGS has been adopted as the "Provisional Standard Practice for Digital Interchange of Geotechnical Data" by AASHTO. It builds off of the AGS data dictionary for its test procedure objects, but is capable of handling more complex sampling feature geometries, testing and sampling protocols, and construction activities.

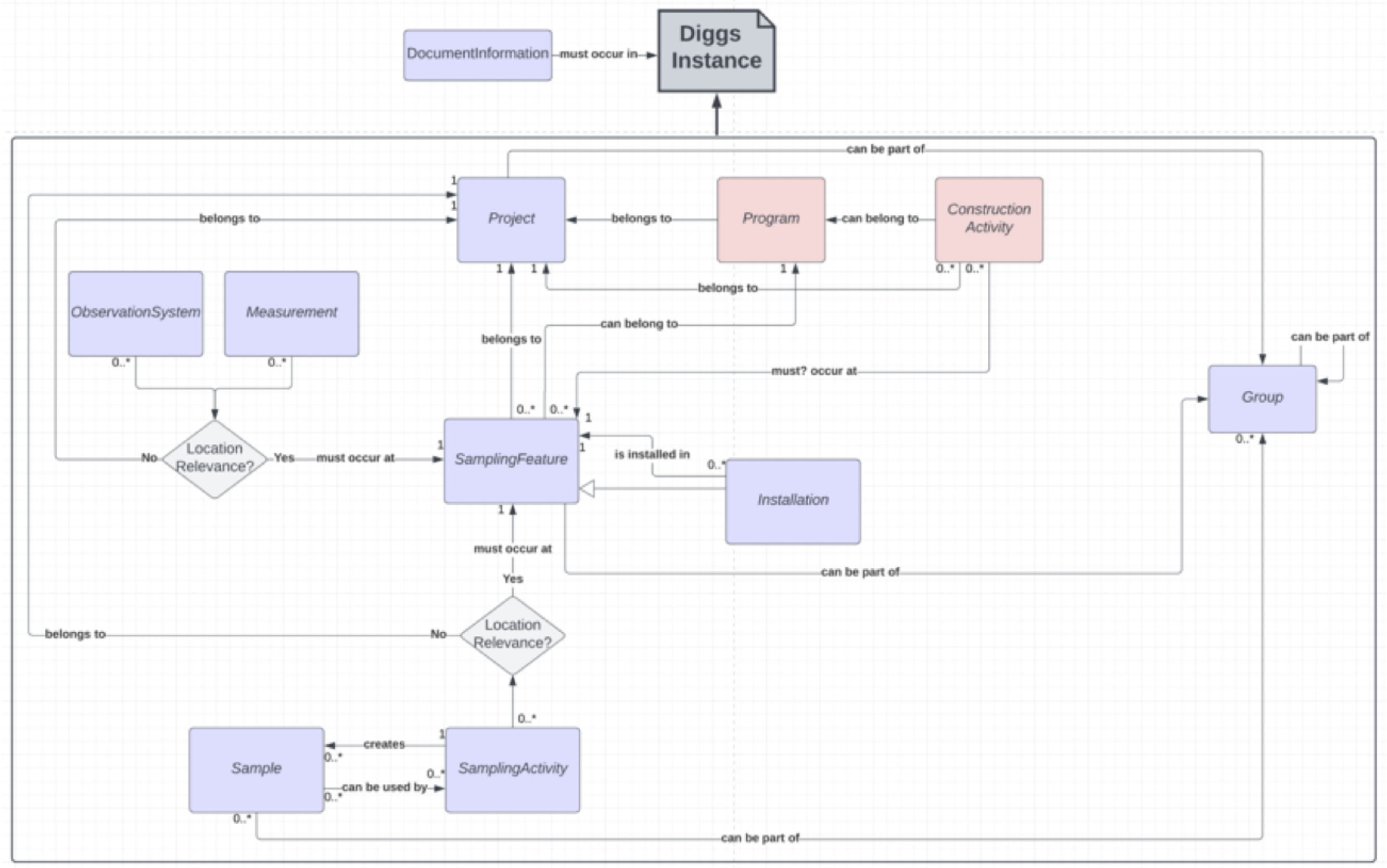

*Figure 46. Generalized DIGGS data model showing the major object classes. These classes are of abstract type and they serve as the head of substitution groups that allow for specialized concrete features.*

The DIGGS data model consists of a set of object classes that share some similarity, conceptually, with OMS objects. Most relevant to interfacing with OGC's Geotech efforts are:

**SamplingFeature** - A physical object or location through which we observe or measure properties of an investigation target, or perform some type of activity. It serves to define the dimensionality, extent, and local spatial reference system for zones where observations or activities take place. DIGGS has developed a number of concrete sampling feature types, including Borehole, Sounding, TrenchWall.

**SamplingActivity** - The action taken to obtain or produce a material sample, even if the activity does not produce a sample (eg. a core run that produces no recovery). This activity typically occurs at a location on a sampling feature, or could occur elsewhere (eg. a laboratory) in the case of test or blank samples, or can produce aggregate samples where the location of the samples produced have no meaning. All material samples must refer to a SamplingActivity.

**Sample** - A material sample, solid, fluid, or gas that is obtained as a result of a sampling activity, for the purpose of observation and/or testing.

**Measurement** - An observation where the result derives principally from the use of testing equipment and/or sensors. Results are typically objective with high precision results as long as procedures are followed and equipment is operating appropriately. Results are usually numeric values, although category assignments can be calculated from numeric results.

**Observation** - An observation where the result derives from direct human interaction, judgment or analysis. These results are typically subjective; and often category based. Typical observation results are descriptive, interpretive, or involve generalization or modeling.

DIGGS' distinction between observations and measurements has resulted in these two object classes to be structurally quite different. DIGGS measurements are structured such that measurement components (results, observed properties, features of interest) map in a fairly straightforward way to the OGC SensorThings API data model extension for Geotechnics. Mapping DIGGS observation results to the OGC SensorThings API data model poses more of a challenge although quite feasible.

### Mapping information

Mapping from the various AGS and DIGGS formats to a proposed harmonized GeoTech Model can be illustrated by the following mapping to STA.

#### Mapping from AGS data

AGS data can be mapped to the STA data model by adopting the following general rules:

For field observations/in situ tests:

- Nature of observation/test, e.g., SPT = Sensor: name = AGS Group, e.g., ISPT
- LOCA_ID = BHCollarThing: name
- Depth (e.g., to test):
  - BHSampling: from/to, and
  - BHFeatureOf Interest
- fields providing metadata about the test or observation = Sensor: properties comprising a JSON object that uses the AGS Headings as keys
- other data, i.e., results = Observation:
  - result from the DataStream that has as ObservedProperty: name = AGS Heading, e.g., ISPT_NVAL

Geology information (GEOL, DETL) and borehole field observations (e.g., PTIM, CHIS) also adopt this general approach.

Note that BHSampling is used for both insitu and lab test, despite in situ tests not using sampling as such.

For a lab test, the mapping is similar to the above but with the following differences: - BHSampling along with BHSampler and BHSamplingProcedure are be used for the sample (SAMP data) - BHFeatureofInterest used for the specimen (defined in lab test results in AGS) but note that it is not currently possible to map the specific from/to depths of the specimen reported by AGS data.

Where there is a set of 'child' readings within a 'parent' result, e.g., compaction test points (CMPT) within a compaction test (CMPG) then the results for each point are provided as sets of RelatedObservations where the first Observation would logically be the test/point number.

The approach described here is in keeping with the intent of SensorThings. However, there is a difficulty in practice as a judgement has to be made about whether a field represents test metadata or a test result. Sometimes this is obvious, but quite often it is not and this introduces the risk of an inconsistent in approach between users. This could be addressed by the publication of an official mapping, but it would be an onerous tasks to cover the entire AGS data dictionary as no such thing exists at present. Applications would also need to be able to deal with this programmatically.

An alternative approach would be to map all metadata as results, i.e., everything is treated as an Observation. This would make implementation easier, for both loading and querying data. However, it is not true to the intent of the SensorThings model.

**Mapping from DIGGS data**

Sample XML instances with proposed mapping of elements to the SensorThings API objects are provided here for DIGGS Test objects (a type of DIGGS measurement) for CPT, SPT and Atterberg limits tests. Although some effort has gone in to mapping a limited portion of DIGGS' ObservationSystems for category-based data to STA (specifically for Lithology observations), considerably more effort is needed. These observations ,as well as mapping to additional test types such as compaction tests, and triaxial tests will continue to be developed.

As can be seen from the mapping examples, the DIGGS Test object, which handles all in-situ and laboratory based measurements that are not time-domain measurements (those are handled in a similarly structured Monitor object) are comprised of three main components 1. Measurement metadata with time of test and roles 2. Outcome, which further is subdivided into: Location and ResultSet objects. The Location object carries the feature of interest's location information while the ResultSet holds the observed properties and observation results. 3. A procedure property that can hold any number of Procedure objects that describe how the test was performed but also contain metadata and interim observations that are procedure-specific.

In general, for the Test object:

- Test metadata properties will map to Sensor, unless the same information is provided in the procedure object;
- Location information will map to BhSampling and BhFeatureOfInterest;
- ResultSet properties map to Observed Property and Datastream (for units of measure);
- ResultSet datavalues property maps to Observation; and
- Procedure object maps to Sensor, except for included interim observations, where those elements will map to ObservedProperty and Observations.

As for other DIGGS objects:

- DiGGS linear sampling features such as boreholes and soundings map to properties in BhCollarThing, BhTrajectoryThing and the associated Location object;
- DIGGS SamplingActivity maps to BhSampling and BhSamplingProcedure;
- DIGGS Sample maps to BhFeatureOfInterest; and

- DIGGS Specimen objects map to BhFeatureOfInterest.

### 5.5.7. STA endpoints

#### Fraunhofer Endpoint

https://ogc-demo.k8s.ilt-dmz.iosb.fraunhofer.de/FROST-GeoTech/v1.1/

#### BRGM Endpoint

https://frost4geotech.brgm-rec.fr/FROST-Server/v1.1/

### 5.5.8. Play with the FROST Geotech plugin

The SensorThings API has very powerful query features that keep easy things easy, but still allow very complex requests. A good starting explanation on how to query the SensorThings API can be found at the Getting Data section on the FROST-Documentation website.

This page is about relevant queries you may do with the FROST Geotech plugin

#### Simple queries

##### Lists

The simplest queries are those that fetch a list of all entities of a type. For instance, getting a list of all the geotechTests (Mapped to Sensor in STA):

```
https://[FrostGeotechServerAddress]/v1.1/Sensors
```

Or to get all borehole collars:

```
https://[FrostGeotechServerAddress]/v1.1/BhCollarThings
```

##### Single Entities

To get a single entity from a list, add `(<id>)` to the URL of the list. To get the geoTechTest (Sensor) with id `1`:

```
https://[FrostGeotechServerAddress]/v1.1/Sensors(1)
```

Or to get the borehole collar with id `42`:

```
https://[FrostGeotechServerAddress]/v1.1/BhCollarThings(42)
```

#### Related Lists

To get a list of entities related to a certain entity, one can navigate along relations. for instance, to get all Datastreams for Sensor (geoTechTest) with id 1:

```
https://[FrostGeotechServerAddress]/v1.1/Sensors(1)/Datastreams
```

Or to get the Trajectories related to borehole collar 1:

```
https://[FrostGeotechServerAddress]/v1.1/BhCollarThings(1)/BhTrajectoryThings
```

Or to get the Datastreams related to a Trajectory 1:

```
https://[FrostGeotechServerAddress]/v1.1/BhTrajectoryThings(1)/Datastreams
```

And from there, to get the Observations related to a Datastream 1:

```
https://[FrostGeotechServerAddress]/v1.1/Datastreams(1)/Observations
```

#### Filtering Lists

All lists can be filtered using the filter parameter. To get a list of all the geotechTests that are of type CPT (sensorType `https://data.geoscience.fr/ncl/Proc/86`):

```
https://[FrostGeotechServerAddress]/v1.1/Sensors?]filter=sensortype eq
'`https://data.geoscience.fr/ncl/Proc/86`'
```

### More complex queries

Get the results of the available DataStreams associated to my Geotechtest organised this way : (Depth, Result for property 1, Result for property 2, ...)

```
https://ogc-demo.k8s.ilt-dmz.iosb.fraunhofer.de/FROST-
GeoTech/v1.1/BhTrajectoryThings(9)?
$select=name& $expand=BhSamplings( $orderBy=atPosition;
$select=atPosition,time; $expand=BhSamples( $top=1; $select=name;
$expand=Observations( $top=3; $select=result,phenomenonTime;
$expand=Datastream( $select=unitOfMeasurement; $expand=ObservedProperty(
$select=name ) ) ) ) )
```

Get this same result in csv

```
_Currently not possible unfortunately, but plans to add this functionality are in
```

```
place._
```

Get the CPT tests in this perimeter that match this lithology (Rock)

Given that BhTrajectoryThings(9) is the trajectory with the CPT test:

```
https://ogc-demo.k8s.ilt-dmz.iosb.fraunhofer.de/FROST-
GeoTech/v1.1/BhTrajectoryThings(9)/BhSamplings?
$filter=BhTrajectoryThing/BhCollarThing/RelatedBhCollarThings/BhTrajectoryThings/BhSam
plings/BhSamples/Observations/result
eq '`Rock`' and atPosition gt
BhTrajectoryThing/BhCollarThing/RelatedBhCollarThings/BhTrajectoryThings/BhSamplings/f
romPosition
and atPosition lt
BhTrajectoryThing/BhCollarThing/RelatedBhCollarThings/BhTrajectoryThings/BhSamplings/t
oPosition
```

Get the results of those same CPT tests

```
https://ogc-demo.k8s.ilt-dmz.iosb.fraunhofer.de/FROST-
GeoTech/v1.1/BhTrajectoryThings(9)/BhSamplings?
$filter=BhTrajectoryThing/BhCollarThing/RelatedBhCollarThings/BhTrajectoryThings/BhSam
plings/BhSamples/Observations/result
eq '`Rock`' and atPosition gt
BhTrajectoryThing/BhCollarThing/RelatedBhCollarThings/BhTrajectoryThings/BhSamplings/f
romPosition
and atPosition lt
BhTrajectoryThing/BhCollarThing/RelatedBhCollarThings/BhTrajectoryThings/BhSamplings/t
oPosition&
latexmath:[expand=BhSamples(]expand=Observations) ```
```

# 5.6. Semantics for geotech

Controlled vocabularies are a key component of interoperability to ensure each terms are non ambiguous.

In geotechnics, multiple controlled vocabularies can be set. This includes: - Procedure: this is to indicate a particular method or technic that have been applied to get a result, - Observable Property: this is to indicate which parameter was measured or observed, - Units of measurements: this is to characterize the (generally numerical) value that come with the result.

## 5.6.1. Registries

Several registries already exist and may help you to describe the test you made.

#### BRGM GeoScience Linked Data Registry

https://data.geoscience.fr/ncl/?lang=en

Codelist for Methods, as described in ObservingProcedure: https://data.geoscience.fr/ncl/Proc?lang=en

Codelist for Observable Properties, as described in ObservableProperties: https://data.geoscience.fr/ncl/ObsProp?lang=en

#### GeoScience EU Registry

https://data.geoscience.earth/ncl/

### 5.6.2. DIGGS Dictionaries/Schemas

***Observation Procedures:*** For DIGGS measurements (observations with numerical results), procedures are defined in the XML schema as concrete objects (XML elements of complex type) that derive, ultimately from gml:AbstractGMLType. These objects are defined in the following schema file: https://www.diggsml.org/schemas/2.5.a/Geotechnical.xsd.

For DIGGS observations (category-based results), the observing procedure can be described using a Specification object for the entire ObservationSystem (a collection of observations of the same class), although this does not require a controlled term, or by using the howDetermined attribute of an individual Observation (see: https://diggsml.github.io/index.html#DescriptorMethodEnumType).

***Observed Properties:*** For DIGGS measurements (observations with numerical results), observed properties are defined in an XML dictionary file: https://www.diggsml.org/dictionaries/DIGGSTestPropertyDefinitions.xml

For DIGGS observations (category-based results), observed properties are hard-coded XML elements, dependent on the class of observation. These can be found in the following schema file: https://www.diggsml.org/schemas/2.5.a/Kernel.xsd.

***Units of Measurement:*** All DIGGS elements of measure type (consisting of a numerical value plus a unit of measure) use units symbology defined in the Energistics Units of Measurement dictionary described here: https://www.energistics.org/energistics-unit-of-measure-standard.

# Chapter 6. Conclusions and future work

## 6.1. Plans for Future Work

“Paths are made by walking” - Franz Kafka

### 6.1.1. GeoSciML Update

The O&M based GeoSciML was last updated to V4.1 in 2017. This standard should be revisited in light of the developments in both O&M (updated to OMS) as well as the use of SensorThings API.

### 6.1.2. GroundWaterML2 Update

While GroundWaterML (OGC WaterML 2: Part 4 - GroundWaterML 2 (GWML2)) was last updated in 2019, this was before the introduction of OMS. Revisiting this standard in order to integrate the new concepts from OMS would be valuable.

### 6.1.3. JSON-LD Context for properties

Adding JSON-LD features to the SensorThings API has been a feature request for a long time, but making the entire API compatible with JSON-LD is not likely to happen very soon. A possible intermediate approach may be to add a JSON-LD context to the properties of each entity type. This won’t make the entire response JSON-LD compatible, but may help with interpreting the free-form properties objects available in the SensorThings API.

### 6.1.4. Thing vs. FeatureOfInterest

In this edition of the GeoTech IE (assuming there will be more) borehole collars and trajectories are mapped to Things, while all samples are mapped to FeatureOfInterest. Although this split works, it is not ideal, and thus in the WaterQuality IE an experiment is being done with mapping all domain features into the FeatureOfInterest entity type, together with a new FeatureType entity type. This moves slightly further away from the STA core model, but results in a cleaner model and is currently considered as a basis for the SensorThings API version 2.0.

The Sampling, Sampler, SamplingProcedure, PreparationStep and PreparationProcedure entity types from this GeoTechIE have been taken up and improved by the WaterQualityIE, and are very likely to also be adopted as standard extensions to the SensorThings API v2.0.

The many extra self-relations that were added to many entities (RelatedBhCollarThings, RelatedDatastreams, RelatedObservations and RelatedFeatures) have also been improved upon in the WaterQualityIE and will most likely also be added as a standard extension to version 2 of the SensorThings API.

### 6.1.5. Geophysics, Profiles and 2D sections, 3D subsurface models

This work was mainly semantic focused with the principal intention to harmonize geotech concepts. A next focus should be made on the provision of geometry, including notably the capacity to share and query cross sections and 3D models. This topic, as well as Geophysics are in discussion as possible new activities of the OGC GeoScienceDWG.

### 6.1.6. Semantic transposition

OMS allows for provision of explicit information on observational or measurement data, providing detailed information on how these observations were created. Such additional measurement metadata is essential in order to determine if a specific dataset is fit for purpose within a specific use case. However, the complexity of the underlying data models can make interaction with OMS based systems difficult for inexperienced users, who can easily get lost in the many options. Based on these reflections, we came to the insight that within individual use cases, there are often two very different user groups interacting with OMS data: - Thematic experts analyzing the data for usefulness within the UC. These experts require access to the full complexity of the underlying OMS data model to understand if the data being provided is truly fit for purpose. - Data analysts or developers utilizing an already vetted dataset no longer need the details provide by the OMS model, for their purposes all they need is a geometry and some numbers, as shown in the example below.

```
{
    "type": "Feature",
    "geometry": {
        "type": "Point",
        "coordinates": [
            125.6,
            10.1
        ]
    },
    "properties": {
        "O3": {
            "value": 24.3,
            "uom": "mg/L"
        },
        "NO2": {
            "value": 16.8,
            "uom": "ppb"
        }
    }
}
```

The challenge in such a transformation pertains to how to transfer the content of the ObservableProperty (name="O3" or name="NO2") to the data structure of the simplified view, as this is transferring information from data content to structure. For this purpose, the term "Semantic Transposition" has been coined, defined in View points on data points: A shared vocabulary for cross-domain conversations on data and metadata

Within the current work on updating SensorThings API to V2.0, integration of Semantic Transposition will be investigated.

### 6.1.7. Simple import options for complex models

In order to enable simple import to the complex models defined utilizing OMS and STA, once the structure for standard tests has been defined, simplified tools enabling data providers to easily provide their data must be investigated. Simple wizards could be of great help in enabling data provision.

### 6.1.8. Tabular options for complex models

Related to the points above on Semantic Transposition and simplified import, a further approach to simplified interaction with STA endpoints pertains to tabular views. This approach can be useful both in provided data as well as accessing data.

- On data provision, work is ongoing at the European Environment Agency (EEA) on how to enable provision of OMS data via tabular views, especially in cases where most of the observational metadata has default values, as common under reporting obligations (e.g., the UoM and measurement methodology are predefined.
- On data access, once the observational metadata has been checked by the relevant experts, a simple tabular download format would be valuable for further analysis as well as portrayal.

### 6.1.9. Visualization options for results

Within this IE, the focus was on utilizing the OMS model as well as the STA implementation for sharing of borehole data. A next step involved implementing various visualization options for better access to the retrieved data.